\documentclass[iop, numberedappendix]{emulateapj}

\usepackage{ifthen}% setup ifthen

\usepackage{natbib, latexsym}
\usepackage{graphics}
\usepackage{txfonts}
\usepackage{color}
\usepackage{graphicx}% Include figure files
\usepackage{enumitem}

\def\mum{${\rm \mu m}$}
\def\arcsec{\hbox{$^{\prime\prime}$}}

\shorttitle{PAH emission in NGC~2023}
\shortauthors{Peeters et al.}
\begin{document}
\title {The PAH emission characteristics of the reflection nebula NGC~2023}
\author{Els Peeters\altaffilmark{1,2}, Charles W. Bauschlicher, Jr.\altaffilmark{3}, Louis J. Allamandola\altaffilmark{4}, Alexander G.G.M. Tielens\altaffilmark{5}, Alessandra Ricca\altaffilmark{2},  Mark G. Wolfire\altaffilmark{6}}

\altaffiltext{1}{Department of Physics and Astronomy, University of Western Ontario, London, ON N6A 3K7, Canada;
epeeters@uwo.ca}
\altaffiltext{2}{Carl Sagan Center, SETI Institute, 189 N. Bernardo Avenue, Suite 100, Mountain View, CA 94043, USA}
\altaffiltext{3}{Entry Systems and Technology Division, Mail Stop 230-3, NASA Ames Research Center, Moffett Field, CA 94035, USA}
\altaffiltext{4}{NASA-Ames Research Center, Space Science Division, Mail Stop 245-6, Moffett Field, CA 94035, USA}
\altaffiltext{5}{Leiden Observatory, PO Box 9513, 2300 RA Leiden, The Netherlands}
\altaffiltext{6}{Astronomy Department, University of Maryland, College Park, MD 20742, USA}


\begin{abstract}
We present 5-20 \mum\, spectral maps of the reflection nebula NGC~2023 obtained with the Infrared Spectrograph SL and SH modes on board the {\it Spitzer} Space Telescope which reveal emission from polycyclic aromatic hydrocarbons (PAHs), C$_{60}$, and H$_{2}$ superposed on a dust continuum. 
We show that several PAH emission bands correlate with each other and exhibit distinct spatial distributions revealing a spatial sequence with distance from the illuminating star. 
We explore the distinct morphology of the 6.2, 7.7 and 8.6 \mum\, PAH bands and find that at least two spatially distinct components contribute to the 7--9 \mum\, PAH emission in NGC~2023. We report that the PAH features behave independently of the underlying plateaus. 
We present spectra of compact oval PAHs ranging in size from C$_{66}$ to C$_{210}$, determined computationally using density functional theory, and investigate trends in the band positions and relative intensities as a function of PAH size, charge and geometry. 
Based on the NASA Ames PAH database, we discuss the 7--9 \mum\, components in terms of band assignments and relative intensities. 
We assign the plateau emission to very small grains with possible contributions from PAH clusters and identify components in the 7--9 \mum\, emission that likely originates in these structures. 
Based on the assignments and the observed spatial sequence, we discuss the photochemical evolution of the interstellar PAH family as they are more and more exposed to the radiation field of the central star in the evaporative flows associated with the PDRs in NGC~2023.

\end{abstract}

\section{Introduction}
\label{intro}

The mid-infrared (IR) spectra of many astronomical objects are dominated by strong emission bands at 3.3, 6.2, 7.7, 8.6, 11.3 and, 12.7 \mum, generally attributed to polycyclic aromatic hydrocarbon molecules (PAHs) and related species. Despite the significant progress made in our understanding of these emission bands (further referred to as PAH bands), the specifics of their carrier remain unclear. Indeed, no single PAH molecule or related species has been firmly identified to date. This lack of detailed knowledge hampers our understanding of the PAH emission bands and their use as a diagnostic tool for probing the local physical conditions. 

The PAH emission bands show clear variations in peak positions, shapes and (relative) intensities, not only between sources, but also spatially within extended sources \citep[e.g.][]{Hony:oops:01, Peeters:prof6:02, SmithJD:07, Galliano:08}.  
In particular, it is well established that the 3.3 and 11.2 \mum\, PAH bands correlate with each other and that the 6.2, 7.7 and 8.6 \mum\, PAH bands correlate tightly with each other. 
Laboratory and theoretical PAH studies have long indicated that the main driver behind these correlations is the PAH charge state: emission from neutral PAHs dominates at 3.3 and 11.2 \mum\, while ionized PAHs emit strongest in the 5-10 \mum\, and are responsible for the 11.0 \mum\, band \citep[e.g.][and references therein]{Hudgins:04}. 
However, \citet{Whelan:13} and \citet{Stock:14} recently reported that the strong correlation between the 6.2 and 7.7 \mum\, PAH bands breaks down on small spatial scales towards the giant star-forming region N66 in the Large Magellanic Cloud and towards the massive Galactic star-forming region W49A. This suggests that in addition to PAH charge, other parameters such as molecular structure influence the astronomical 6.2 and 7.7 \mum\, PAH emission. A dependence on PAH parameters such as charge, molecular structure, size can be probed by investigating theoretical calculations of PAH spectra \citep{Langhoff:neutionanion:96, Bauschlicher:97, Ellinger:98, Pauzat:02}. Advances in computing power and computational methods now allow spectra for larger PAHs and for a much wider range of PAH species to be calculated and thus for a better and larger scan of the parameter space \citep{Mulas:06, Malloci:database, Bauschlicher:vlpahs1, Bauschlicher:vlpahs2, Ricca:10, Ricca:12, Candian:14, Candian:15}.
Hence, a systematic study of fully resolved PAH emission at high spatial scale over the full mid-IR bandwidth of a large sample of objects in combination with a systematic investigation of theoretical PAH spectra promises to reveal a more detailed and nuanced view of the characteristics of the emitting PAH population.

The high sensitivity and broad wavelength coverage of the \textit{Spitzer} Space Telescope has provided a unique opportunity to study the variations of the PAH features within extended sources that have been spectrally mapped.  In particular, reflection nebulae (RNe) are very interesting in this respect as they are known to be excellent laboratories to determine the characteristics of interstellar dust and gas due to their high surface brightness and relatively simple geometry. Moreover, in contrast to H~{\sc ii} regions, observations of RNe are not confused by diffuse emission due to recombination and cooling of ionized gas, simplifying their analysis greatly. 
Consequently, the well-known reflection nebulae NGC~7023 and NGC~2023 have been prime targets for PAH studies \citep[e.g.][]{Abergel:02, Werner:04, Sellgren:07, Compiegne:08, Fleming:10, Peeters:12, Pilleri:12, Boersma:13, Boersma:14, Shannon:15, Stock:16}.

In this paper, we present \textit{Spitzer}-IRS spectral maps of NGC~2023 in the 5-20 \mum\, region and theoretical data from oval, compact PAHs in order to investigate and interpret the spatial behaviour of the PAH emission components, in particular in the 6 to 9 \mum\, region. 
Section \ref{data} discusses the observations, the data reduction, typical spectra and the continuum and flux determination. The data are analyzed in terms of features' emission maps and correlations in Section \ref{analysis} and theoretical data from oval, compact PAHs are presented in Section \ref{lab}. These are combined and discussed in Section \ref{discussion}. Finally, we end with a short summary in Section \ref{conclusion}.

\section{The data}
\label{data}

\subsection{Observations}
\label{obs}

The observations were taken with the Infrared Spectrograph \citep[IRS,][]{houck04} on board the \textit{Spitzer} Space Telescope \citep{werner04} and were part of the Open Time Observations PIDs 20097, 30295 and 50511.

\begin{figure}[tb]
    \centering
\resizebox{\hsize}{!}{%
  \includegraphics{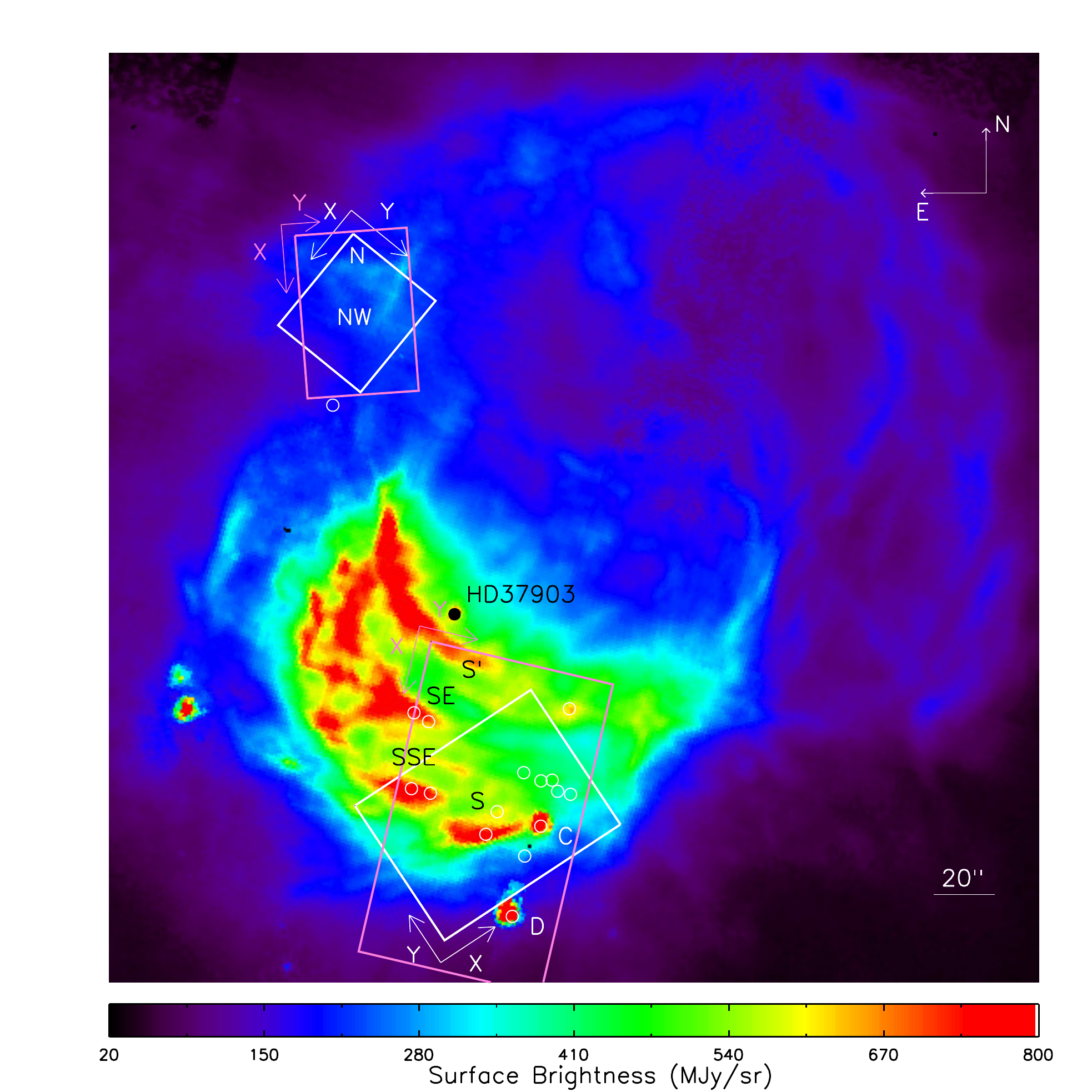}}
\caption{The IRAC [8.0] image of NGC~2023 with the SL and SH FOV shown (pink and white lines respectively) for both the north and south positions studied here. The star HD~37903 is indicated by a black circle, sources C and D are from \citet{Sellgren:83}, and the white circles indicate 2MASS point sources located inside the apertures. S refers to the southern ridge located in the middle of the FOV of the south map, S' to the southern ridge at the top of the FOV of the south SL map, SE to the easternmost ridge in the SL FOV and SSE to the south-southeastern ridge. Moving to the northern FOV, N refers to the north ridge and NW to the northwestern ridge. Maps shown in this paper use the pixel orientation denoted by the vectors (X, Y). Figure adapted from paper I.}
\label{fov}
\end{figure}

 We obtained spectral maps for two positions in the reflection nebula NGC~2023 (see Fig.~\ref{fov}): towards the dense shell south south-west (-11\arcsec, -78\arcsec) of the exciting star HD 37903 corresponding to the H$_2$ emission peak, and towards a region to the north (+33\arcsec, +105\arcsec) of the exciting star \citep{Burton:98}. The north position is characterized by a much lower density ($\sim$10$^{4}$ cm$^{-3}$) than that at the south position, where densities exceed 10$^{5}$ cm$^{-3}$ \citep{Burton:98, Sheffer:11}. Likewise, the radiation field between 6 and 13.6 eV in the northern region ($\sim$500\nolinebreak~G$_0$\footnote{G$_0$ is the integrated 6 to 13.6 eV radiation flux in
units of the Habing field = 1.6x10$^{-3}$ ergs/cm$^2$/s.}) is lower than the UV field  where the H$_2$ peaks in the southern region \citep[$\sim$10$^4$ G$_0$,][]{Burton:98, Sheffer:11}. 

The spectral maps are made with the short-low and short-high modes (respectively SL and SH). The SL mode covers a wavelength range of 5-15 \mum\, at a resolution ranging from 60 to 128 in two orders (SL1 and SL2) and has a pixel size of 1.8\arcsec , a slit width of 3.6\arcsec and a slit length of 57\arcsec (resulting in 2 pixels per slit width). The SH mode covers a wavelength range of 10-20 \mum\, at a resolution of $\sim$ 600 and has a pixel size of 2.3\arcsec , a slit width of 4.7\arcsec and a slit length of 11.3\arcsec (resulting in 2 pixels per slit width). We obtained background observations in SH for the north position and in SL for the south position. Table \ref{log} gives a detailed overview of the observations. The 15-20 \mum\, SH spectra have been presented in \citet{Peeters:12}, hereafter paper I.

%%%%%%%%%%%%%%%%%%%%%%%%%%%%%%%%%%%%%%%%%%%%%%%%%%
%%%%%%%%%%%%%%%%%%%%%%%%%%%%%%%%%%%%%%%%%%%%%%%%%%

\begin{table}[tb]
\small
\caption{\label{log} Observation log}
\begin{center}
\begin{tabular}{c|cc|cc}
 & \multicolumn{2}{c|}{north position}  & \multicolumn{2}{c}{south position} \\
 \hline
 \hline
 & & & & \\[-5pt]
map coordinates$^a$ & \multicolumn{2}{c|}{5:41:40.65, -2:13:47.5} &  \multicolumn{2}{c}{5:41:37.63, -2:16:42.5}\\

mode & SL & SH & SL & SH \\
PID & \multicolumn{2}{c|}{50511}   &  30295 & 20097\\
AORs &  \multicolumn{2}{c|}{26337024}& 17977856 &   14033920 \\
cycle x ramp time& 2x14s & 2x30s & 2x14s & 1x30s \\
pointings $\parallel$ & 1 & 7 & 3 & 12\\
step size $\parallel$ & 26\arcsec & 5\arcsec &  26\arcsec & 5.65\arcsec\\
pointings $\bot$ & 20 & 15 & 18 & 12\\
step size $\bot$ & 1.85\arcsec & 2.3\arcsec & 3.6\arcsec & 4.7\arcsec\\
Background$^{a,b}$ & \multicolumn{2}{c|}{5:42:1.00, -2:6:54.5} & \multicolumn{2}{c}{5:40:26.21, -2:54:40.4} \\
\hline
 \multicolumn{4}{c}{} \\[-5pt]
\end{tabular}
$^a$ $\alpha, \delta$ (J2000) of the center of the map; units of $\alpha$ are hours, minutes, and seconds, and units of $\delta$ are degrees, arc minutes, and arc seconds. The illuminating star, HD~37903, has coordinates of 05:41:38.39, -02:15:32.48.\\ 
$^b$ in SH for the north position and in SL for the south position.
\end{center}
\end{table}

\subsection{Reduction}
\label{reduction}
The SL raw data were processed with the S18.7 pipeline version by the \textit{Spitzer} Science Center. The resulting bcd-products are further processed using {\it cubism} \citep{cubism} available from the SSC website\footnote{http://ssc.spitzer.caltech.edu}. 

As discussed in detail in paper I, the background adds only a small contribution to the on-source PAH flux except for source D from \citet{Sellgren:83}. Consequently, we did not apply a background subtraction and exclude source D from the PAH analysis for the remainder of this paper. 

We applied {\it cubism}'s automatic bad pixel generation with $\sigma_{TRIM} = 7$ and Minbad-fraction = 0.50 and 0.75 for the global bad pixels and record bad pixels. We excluded the spurious data at the extremities of the SL slit by applying a {\it wavsamp} of 0.05 to 0.95 for the north map and of 0.06 to 0.94 for the south map. Remaining bad pixels were subsequently removed manually. 

Spectra were extracted from the spectral maps by moving, in one pixel steps, a spectral aperture of 2x2 pixels in both directions of the maps. This results in overlapping extraction apertures. Slight mismatches in flux level were seen between the different orders of the SL module (SL1 and SL2). These were corrected by scaling the SL2 data to the SL1 data. The applied scaling factor averaged to 5 and 7\% for the south and north SL map respectively. Around 8\% of the spectra in the south SL map have scaling factors of more than 20\%, which are all located at high x-values (i.e. at the south side of the map, below the S ridge). Hence, the applied scaling does not influence our results. 

We've noticed an apparent small (of the order of a few $\times$ 0.01 \mum) wavelength shift in some 2x2 SL spectra, in particular for the south map (see Appendix \ref{wavelengthshifts}).  While its effects can be noticed in the top row of the north map and in the bottom two rows of the south map in  most feature intensity maps presented in this paper (Figs.~\ref{fig_slmaps_s}, \ref{fig_slmaps_n}, \ref{fig_maps_decomp}, \ref{fig_movie}, \ref{fig_pahfit_s}, \ref{fig_pahfit_n}, and \ref{fig_PAHTAT_maps}), it does not change the conclusions of this paper.\\

For the reduction of the SH data, we refer to paper I for a detailed discussion. Spectra were extracted from the spectral maps by moving, in one pixel steps, a spectral aperture of 2x2 pixels in both directions of the maps; the same procedure as for the SL data.

%%%%%%%%%%%%%%%%%%%%%%%%%%%%%%%%%%%%%%%%%%%%%%%%%%
\begin{figure}[tb]
    \centering
\resizebox{\hsize}{!}{%
  \includegraphics{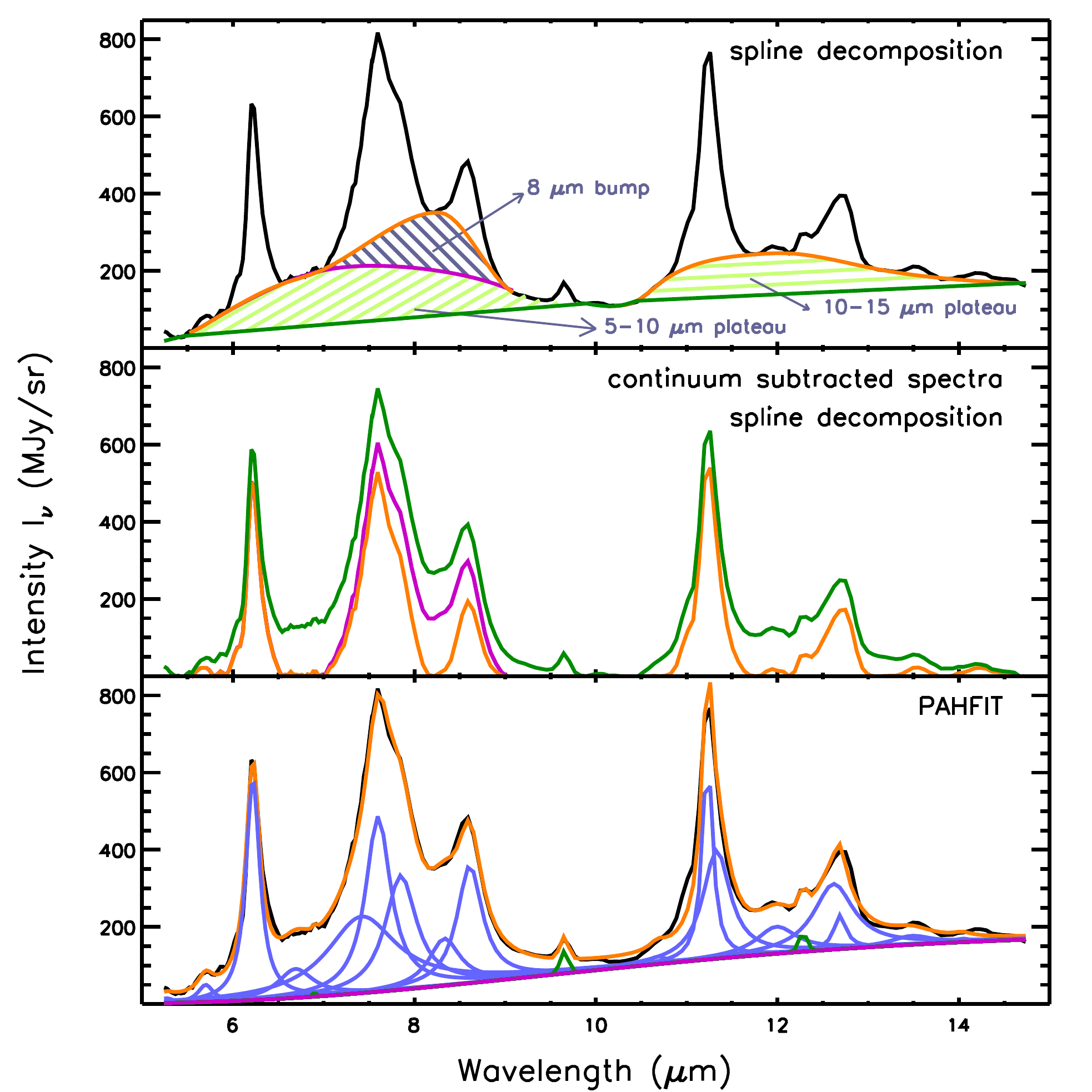}}
\caption{A typical SL spectrum towards NGC~2023 shown with the different SL continuum extraction methods discussed here. The top panel exemplifies the spline decomposition methods applied in this paper: the plateau continuum (dark green), the global continuum (GS, magenta), and the local continuum (LS, red). The 5-10 and 10-15 \mum\, plateaus are shown by the light-green-lined regions and the 8 \mum\, bump by the blue-lined region. The middle panel shows the resulting continuum subtracted spectra when using the plateau, global and local continuum (respectively, in dark green, magenta and red). For comparison, the decomposition obtained with PAHFIT is shown in the bottom panel: fit (red), PAH features (blue), H$_2$ lines (green) and continuum
(magenta).}
\label{fig_sp_sl}
\end{figure}
%%%%%%%%%%%%%%%%%%%%%%%%%%%%%%%%%%%%%%%%%%%%%%%%%%

%%%%%%%%%%%%%%%%%%%%%%%%%%%%%%%%%%%%%%%%%%%%%%%%%%
\begin{figure}[tb]
    \centering
\resizebox{\hsize}{!}{%
  \includegraphics{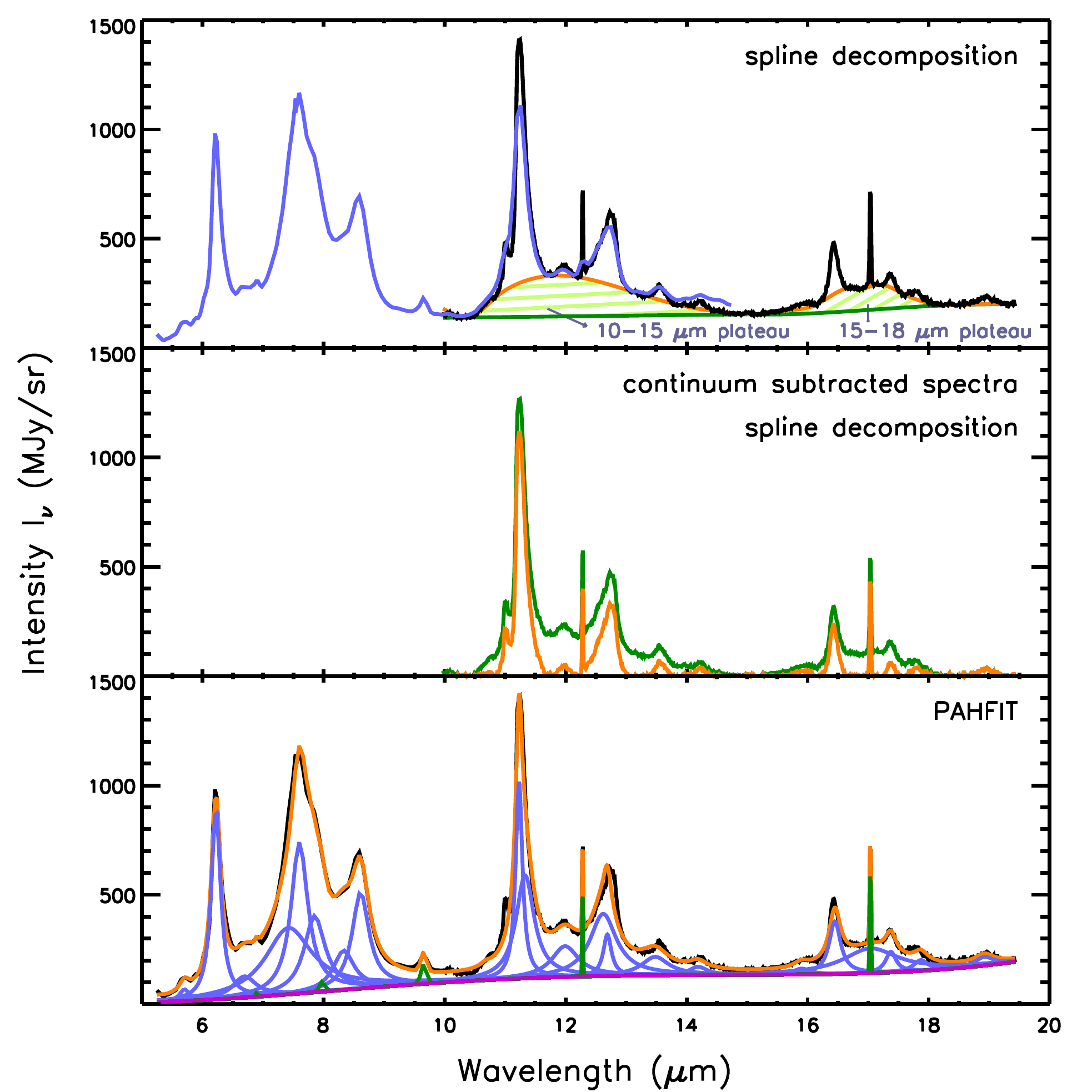}}
\caption{A typical SL+SH spectrum towards NGC~2023 (SL in blue and SH in black;  both taken with the same aperture) shown with the applied SH continuum (top panel): the local spline continuum (LS, red) and the plateau continuum (green). Note that SL is scaled to match the SH flux between 10 and 13 $\mu$m (excluding the H$_2$ line) and no correction for the different spatial PSF of SL and SH is made.  The middle panel shows the resulting continuum subtracted spectra when using the plateau, and local continuum (respectively, in dark green and red). For comparison, the decomposition obtained with PAHFIT is shown in the bottom panel: fit (red), PAH features (blue), H$_2$ lines (green) and continuum
(magenta).}
\label{fig_sp_all}
\end{figure}
%%%%%%%%%%%%%%%%%%%%%%%%%%%%%%%%%%%%%%%%%%%%%%%%%%

\subsection{The spectra}
Figs. \ref{fig_sp_sl} and \ref{fig_sp_all} show typical spectra towards NGC~2023. The complete 5-20 \mum\, spectra reveal a weakly rising dust continuum, H$_2$ emission lines, C$_{60}$ emission, and a plethora of PAH emission bands. In addition to the main PAH bands at 6.2, 7.7, 8.6 and 11.2 \mum, weaker bands are detected at 5.7, 6.0, 11.0, 12.0, 12.7, 13.5, 14.2, 15.8, 16.4, 17.4 and, 17.8 \mum. These PAH bands are perched on top of broad emission plateaus from roughly 5--10, 10--15 and,  15--18 \mum. C$_{60}$ exhibit bands at 7.0, 8.6, 17.4 and 19 \mum\, \citep{Cami:10, Sellgren:10}. The latter two are clearly present in these spectra (paper I).  

NGC~2023 contains a cluster of young stars \citep{Sellgren:83}.
YSO source D is located in the southern quadrant of the south SL map and just outside the south SH map (see Fig.~\ref{fov}). The spectra of regions close to source D show PAH emission features as well as characteristics typical of
YSOs, i.e., a strong dust continuum and a (strong) CO$_2$ ice feature near
15 \mum. Furthermore, the background contribution to the 11.2
PAH flux is significant for a large fraction of the spectra containing ice-features. As in paper I, we therefore excluded these spectra in the analysis. The 15-20 \mum\, spectrum of YSO source C is found in both the SL and SH south map  (see Fig.~\ref{fov}). This YSO has the same spectral characteristics as the spectra across NGC~2023 but with enhanced surface brightness. However, its SL2 spectrum suffers from instrumental effects (i.e. a ÒsawtoothÓ pattern likely due to the undersampled IRS PSF). Hence, we excluded this source in the analysis of the SL data but included it for the analysis of the SH data.

\subsection{Continuum subtraction and band fluxes}
\label{cont}

The extinction towards NGC~2023 is estimated to be small: $A_K=0.2-0.65$ mag \citep[for a detailed overview, see][]{Sheffer:11}.  Moreover, little variation in the extinction is found across the south map except around source C and D \citep{Pilleri:12, Stock:16}. An extinction $A_K=0.4$ mag results in extinction corrections of 11 \% for the 8.6 and 11.2 \mum\, PAH bands (because of its overlap with the silicate band) and between 4-7 \%  for the other features considered here. In this paper, we determined the PAH fluxes assuming zero extinction. \\ 

\begin{figure}[tb]
    \centering
\resizebox{\hsize}{!}{%
  \includegraphics{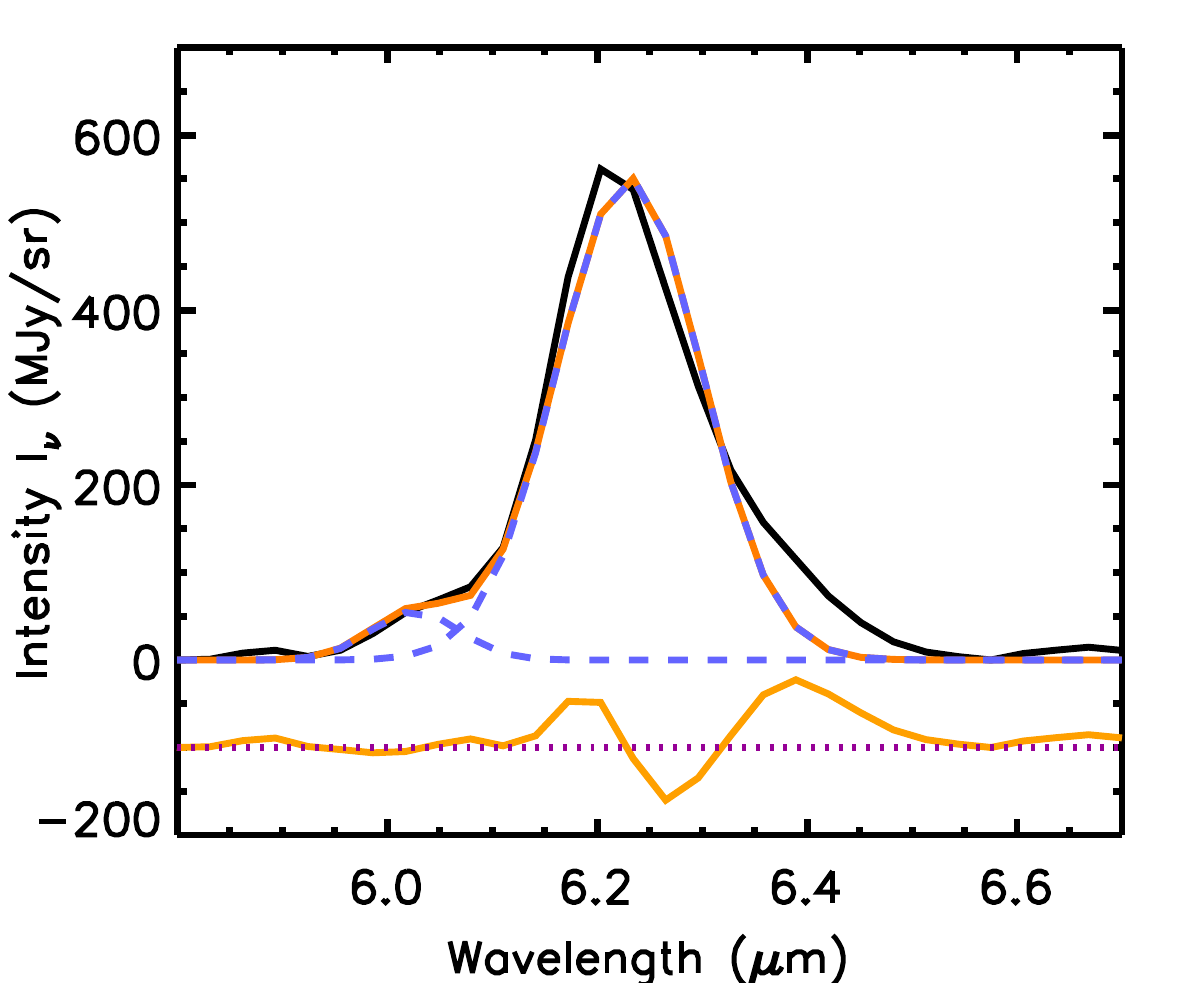}
  \includegraphics{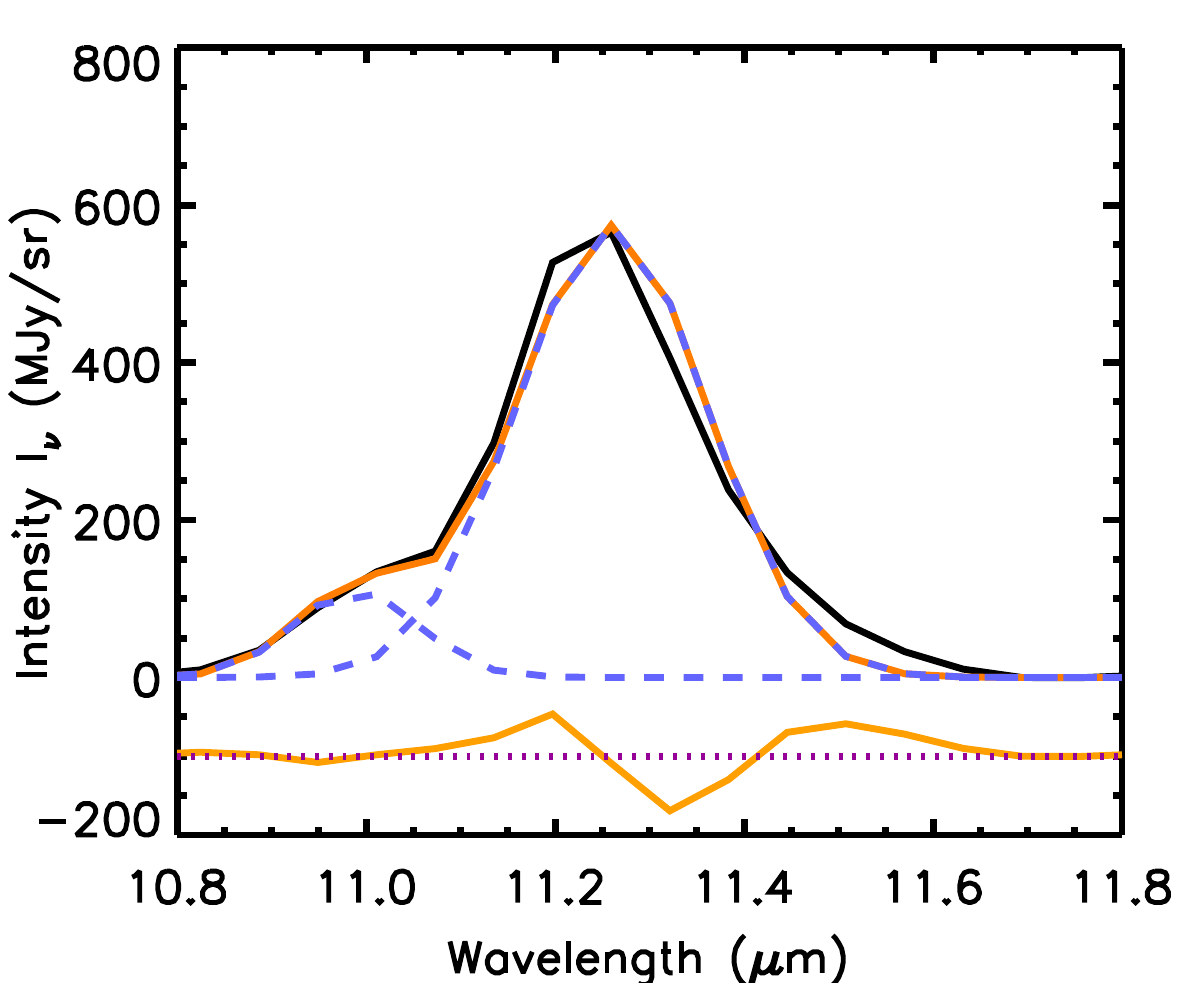}
  \includegraphics{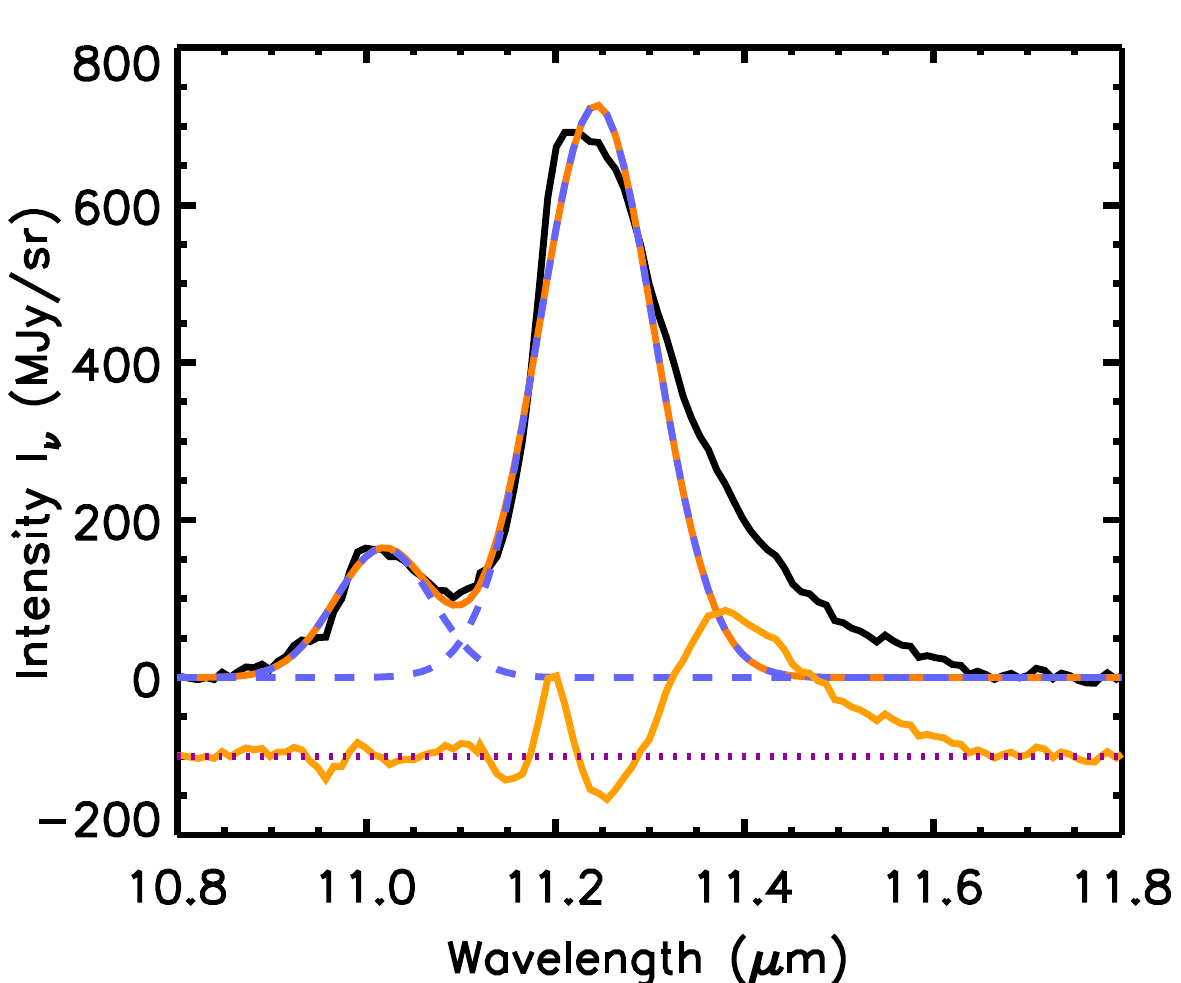}}
\caption{A typical Gaussian decomposition to extract the 6.0 \mum\, PAH band (left) and the 11.0 \mum\, PAH band (SL: middle, SH: right). The data are shown by a solid black line, the fit by a solid red line, the two individual Gaussians by striped cyan lines, and the residuals by a solid gold line (offset by -100 MJy/sr). See Sect. \ref{cont} for details on the composition. }
\label{fig_decomp6}
\end{figure}

We applied three different decomposition methods to the SL data. 
For the first method, we subtract a local spline (LS) continuum from the spectra, consistent with the method of \citet{Hony:oops:01} and \citet{Peeters:prof6:02} as shown in Fig.~\ref{fig_sp_sl}. This continuum is determined by using anchor points at roughly 5.4, 5.5, 5.8, 6.6, 7.0, 8.2, 9.0, 9.3, 9.9, 10.2, 10.5, 10.7, 11.7, 12.1, 13.1, 13.9, 14.7, and 15.0 \mum.  The fluxes of the main PAH bands are then estimated by integrating the continuum subtracted spectra while those of the weaker bands and the H$_2$ lines are measured by fitting a Gaussian profile to the band/line.  The central wavelength of the Gaussian used to measure the H$_2$ line fluxes was allowed to vary, to correct for the wavelength shifts.  The 6.0, 11.0, and 12.7 \mum\, features need special treatment because of blending. Since the 6.0 and 6.2 \mum\, PAH features are blended and given the low resolution of the SL data, we extracted the 6.0 band intensity by fitting two Gaussians with  $\lambda$ (FWHM) of 6.026 (0.099) and 6.229 (0.1612) \mum\, respectively to the data (excluding the red wing of the 6.2 PAH band). These values were obtained by taking the average over all spectra when fitted by two Gaussians having peak positions and FWHM that were not fixed. As can be seen in Fig.~\ref{fig_decomp6}, this decomposition works reasonably well except, of course, for the red wing of the 6.2 \mum\, band.  The 6.0 \mum\, PAH flux is then subtracted from the integrated flux of the 6.0+6.2 \mum\, PAH bands to obtain the 6.2 \mum\, PAH flux. A similar decomposition method can be applied for the 11.0 and 11.2 \mum\, bands by fitting two Gaussians with $\lambda$ (FWHM) of 10.99 (0.154) and 11.26 (0.236) \mum\, respectively (Fig.~\ref{fig_decomp6}, middle panel). Analogously to the 6 \mum\, region, this decomposition works remarkably well except for the red wing of the 11.2 \mum\, band. In order to obtain the fluxes of the 12.7 \mum\, PAH band and the 12.3 \mum\, H$_2$ line, we fit a second order polynomial to the local `continuum' (i.e. the blue wing of the 12.7 \mum\, PAH band) and a Gaussian profile to the H$_2$ line. The flux of the 12.7 \mum\, PAH band was then obtained by subtracting the H$_2$ flux from that obtained by the (combined) integrated flux of the 12.7 \mum\, PAH band and the H$_2$ line. 
We estimated the signal-to-noise ratio of the features as follows: 
$S/N = F~/~(rms \times \sqrt{N} \times \Delta \lambda$) with $F$ the feature's flux [W/m$^2$/sr], $rms$ the rms noise, $N$ the number of flux measurements within the feature and $\Delta \lambda$ the wavelength bin size as determined by the spectral resolution. 
Although the rms noise is a measure of how accurate the continuum can be determined, this method does not take into account the error in the continuum measurement.

%%%%%%%%%%%%%%%%%%%%%%%%%%%%%%%%%%%%%%%%%%%%%%%%%%
%%%%%%%%%%%%%%%%%%%%%%%%%%%%%%%%%%%%%%%%%%%%%%%%%%

%%%%%%%%%%%%%%%%%%%%%%%%%%%%%%%%%%%%%%%%%%%%%%%%%%
\begin{figure*}[tb]
    \centering
\resizebox{\hsize}{!}{%
  \includegraphics[angle=266.4]{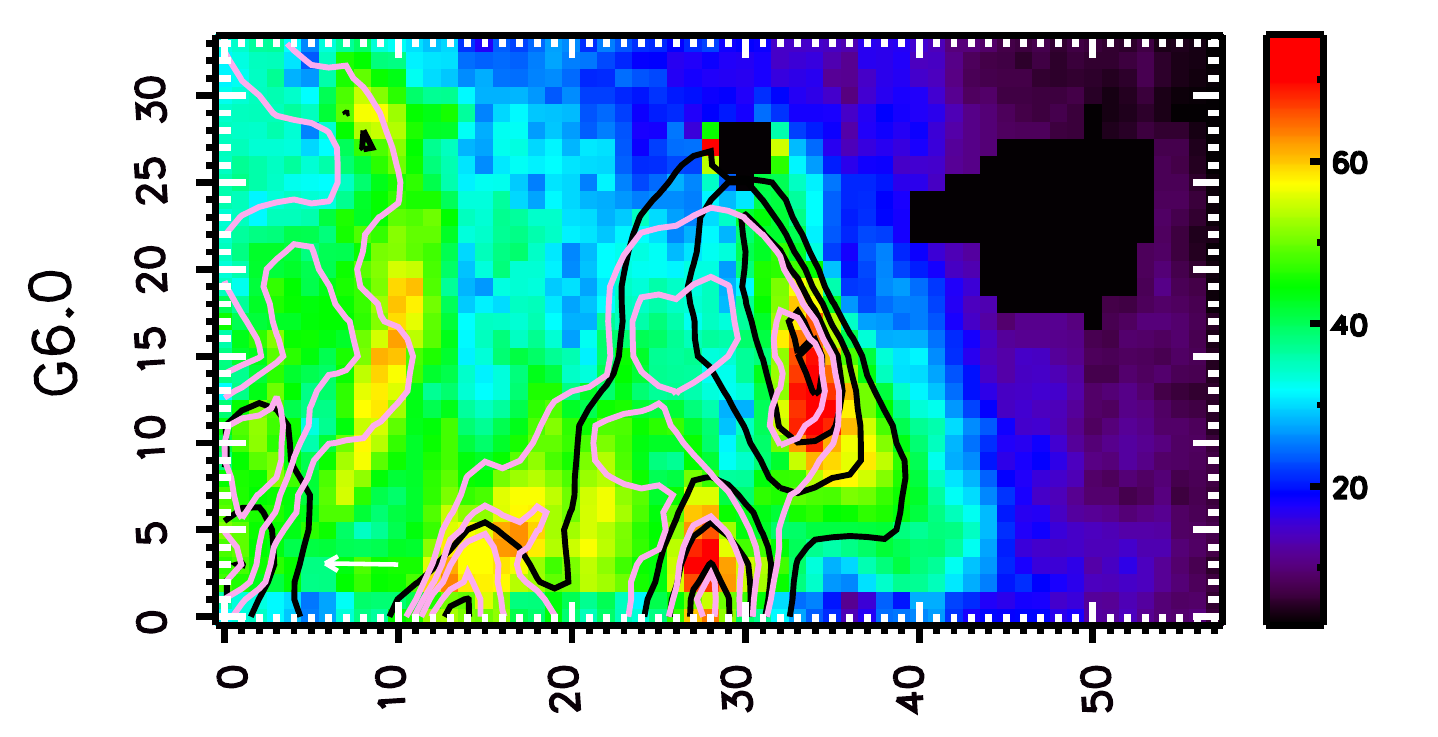}
  \includegraphics[angle=266.4]{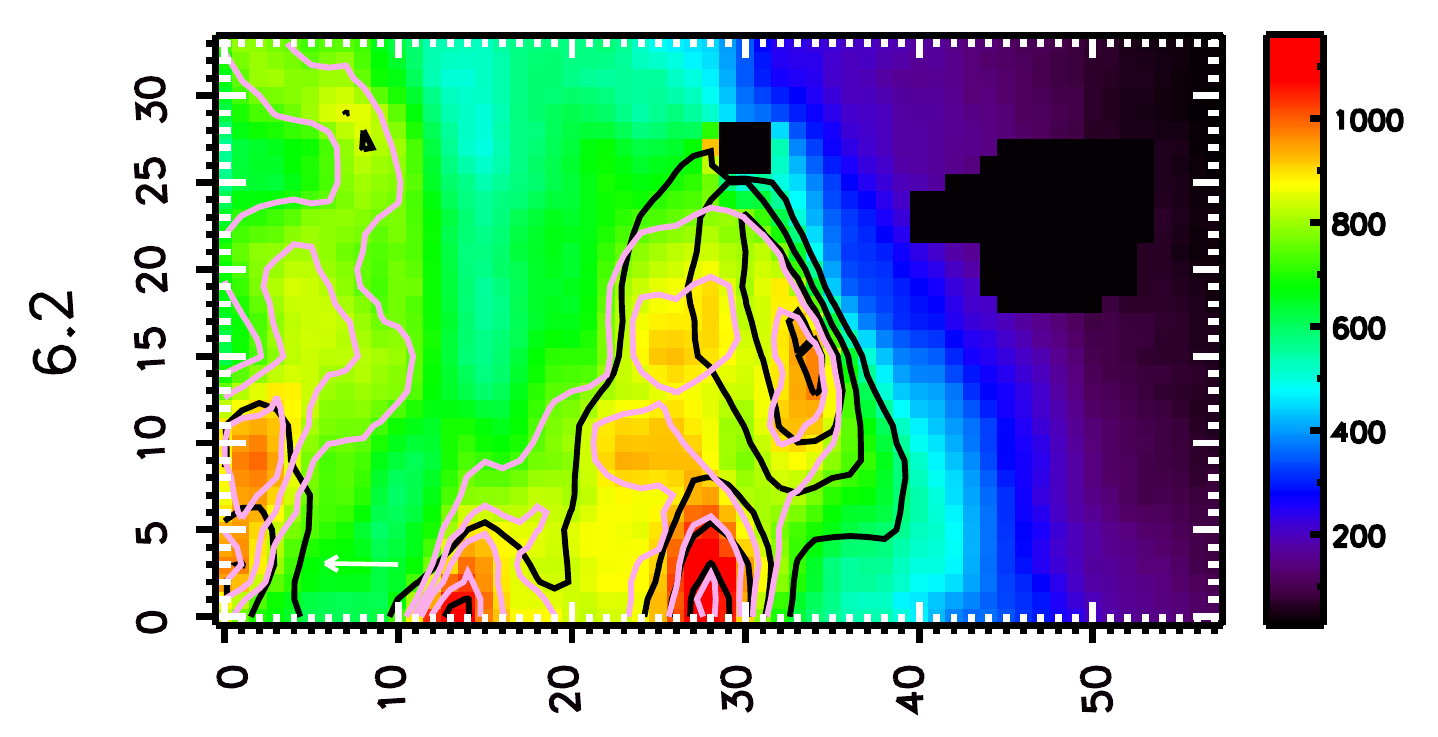}
  \includegraphics[angle=266.4]{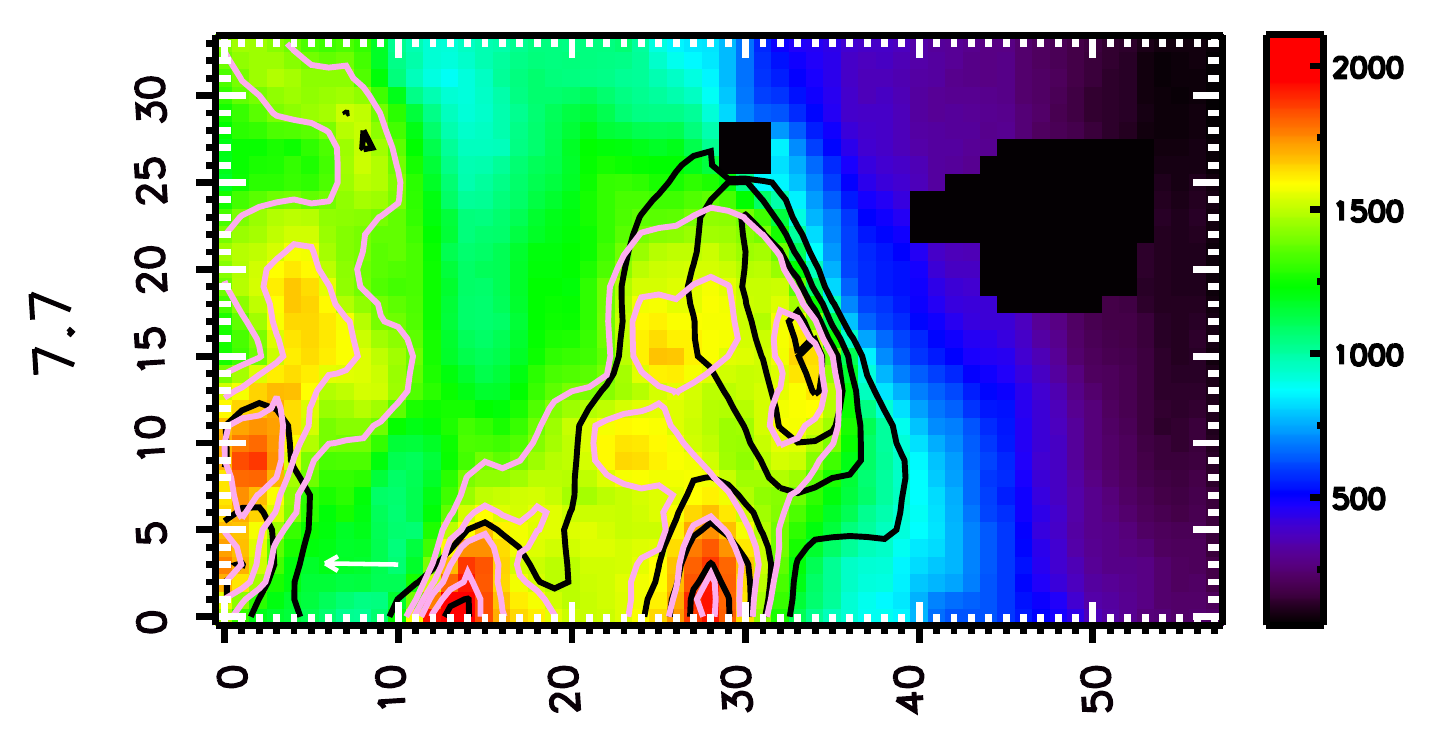}
  \includegraphics[angle=266.4]{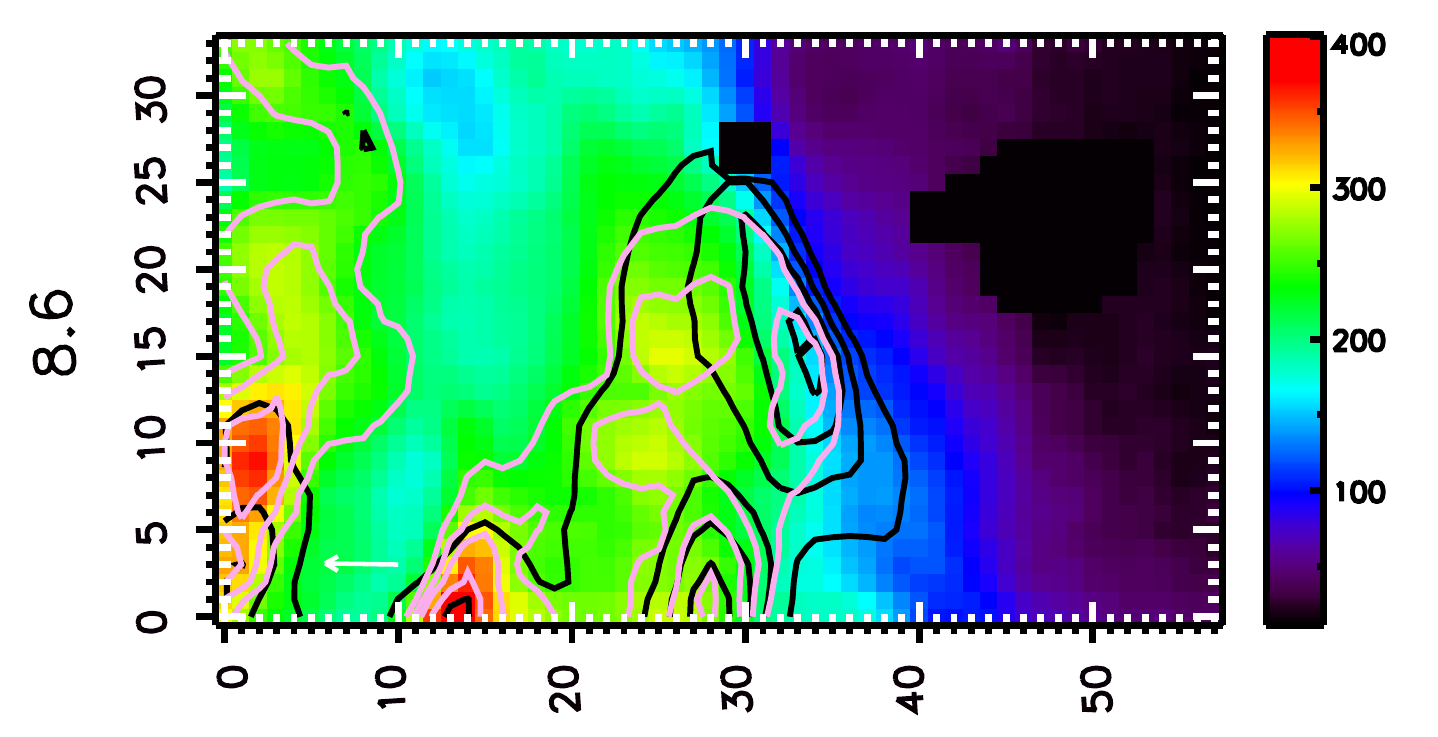}
  \includegraphics[angle=266.4]{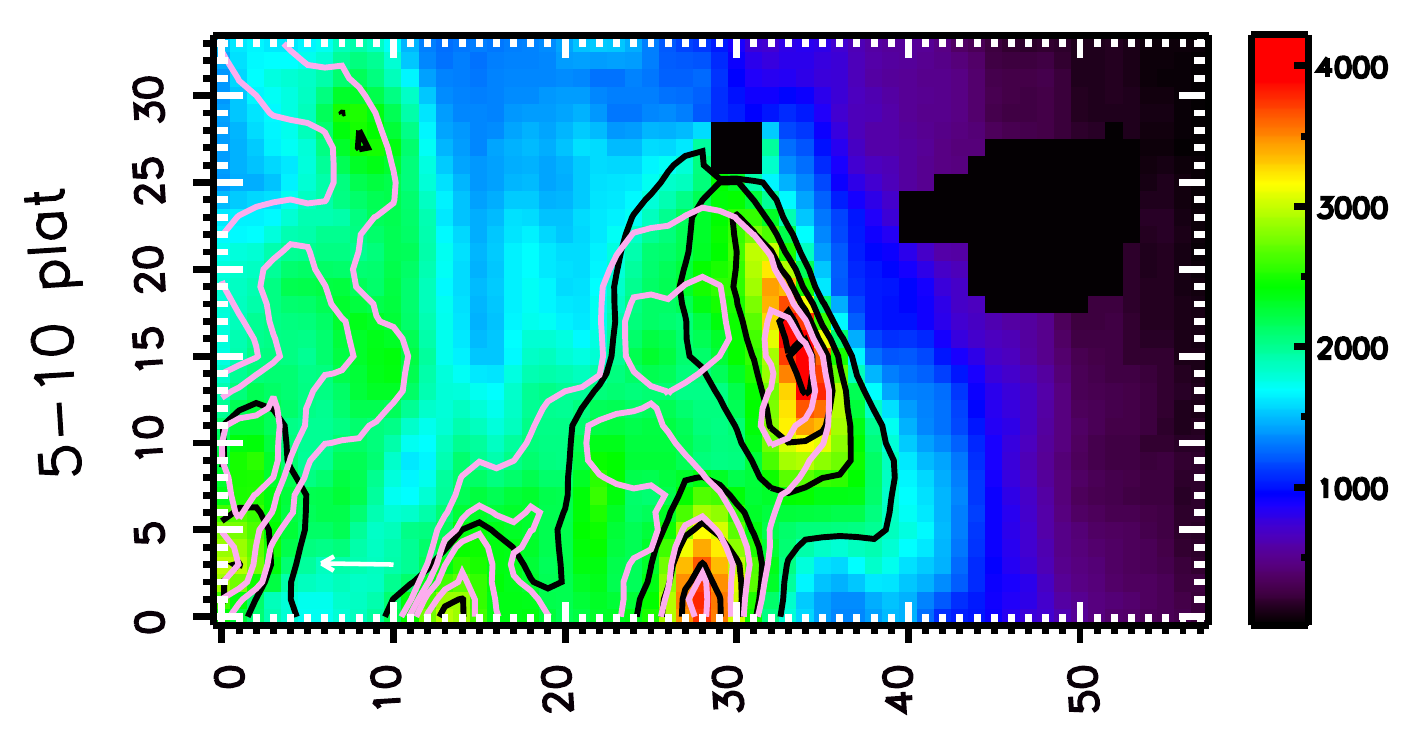}}
\resizebox{\hsize}{!}{%
     \includegraphics[angle=266.4]{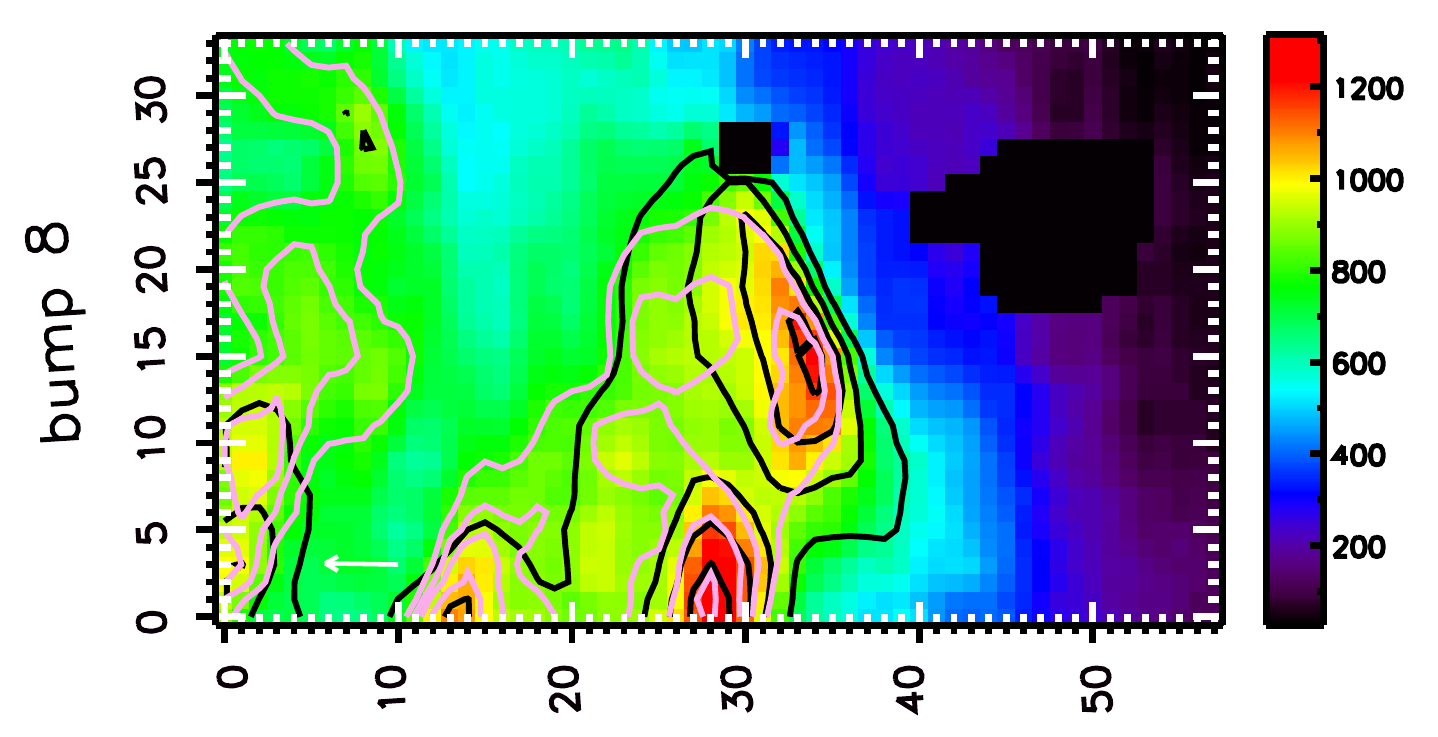}
    \includegraphics[angle=266.4]{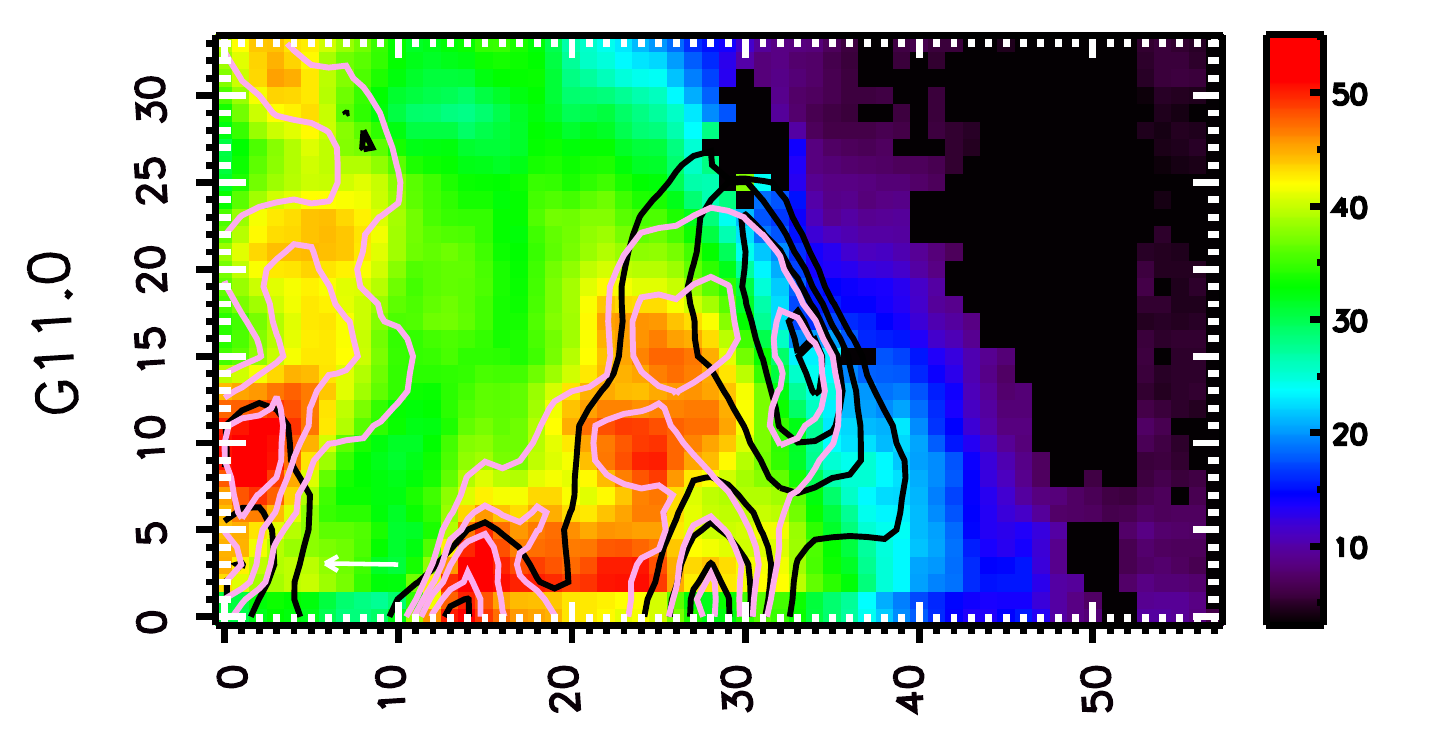}
   \includegraphics[angle=266.4]{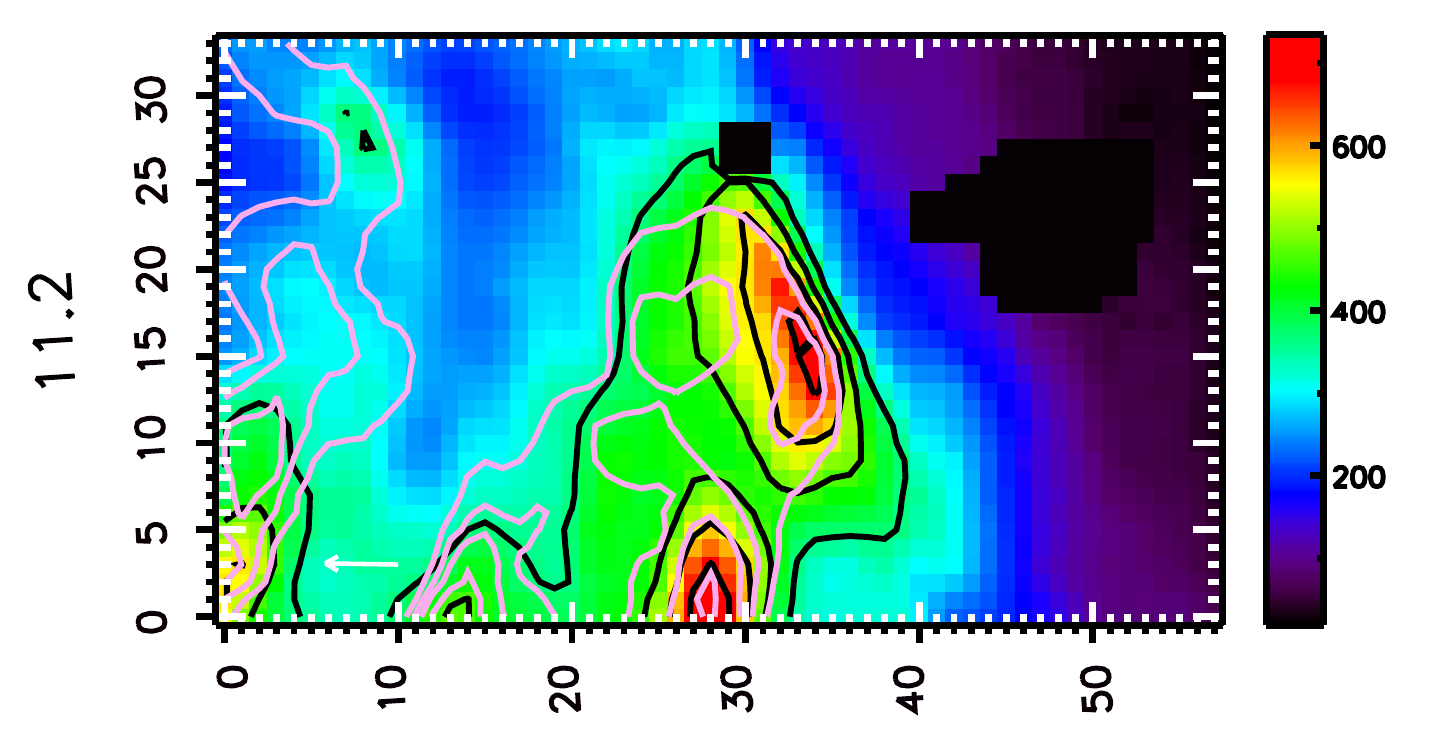} 
   \includegraphics[angle=266.4]{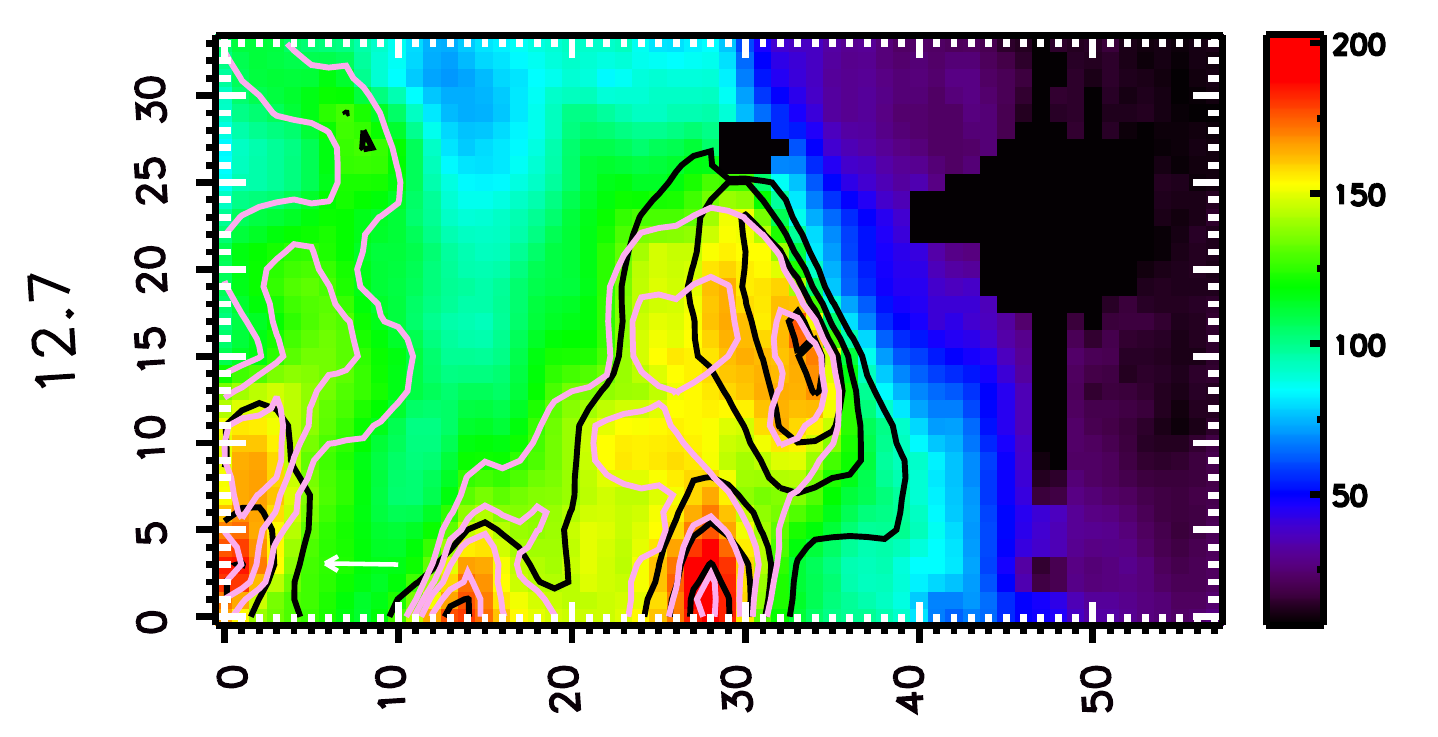}
   \includegraphics[angle=266.4]{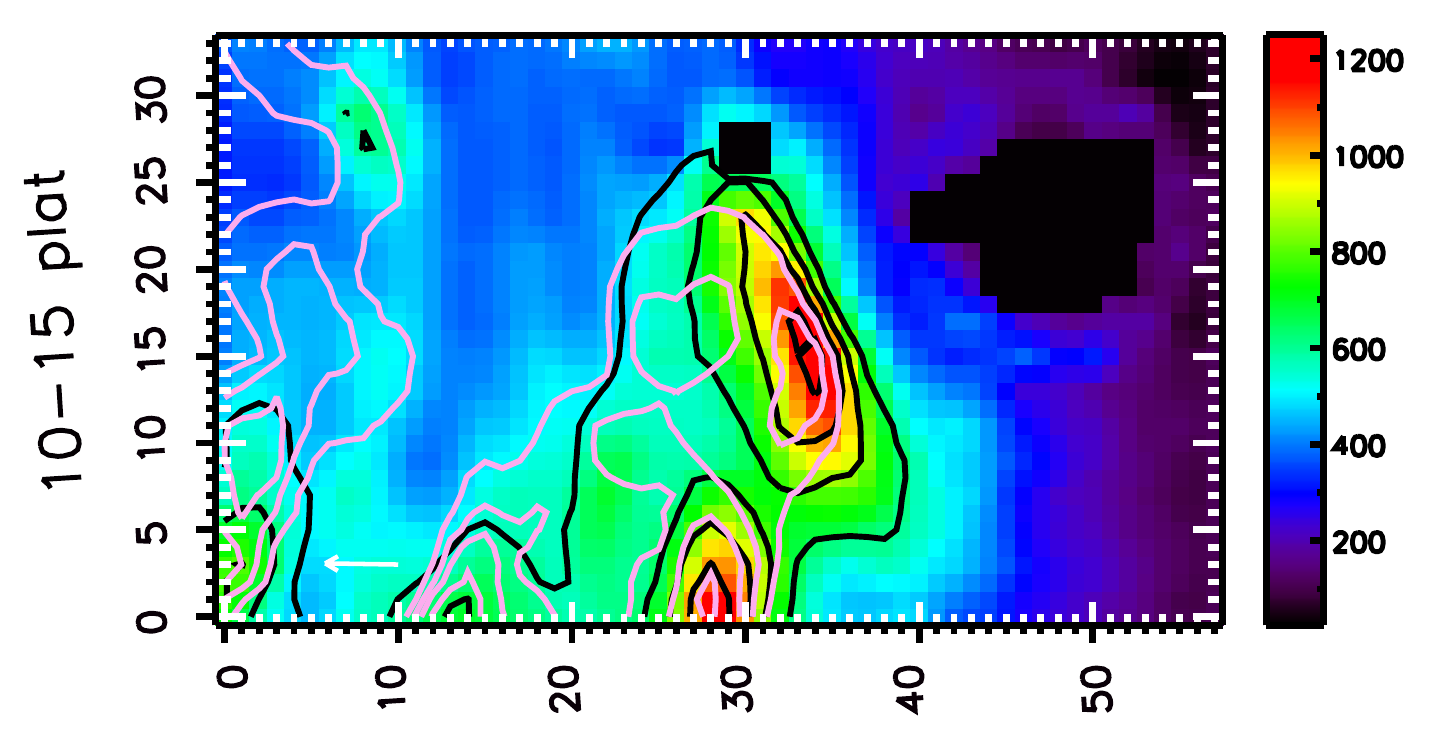}}
\resizebox{\hsize}{!}{%
   \includegraphics[angle=266.4]{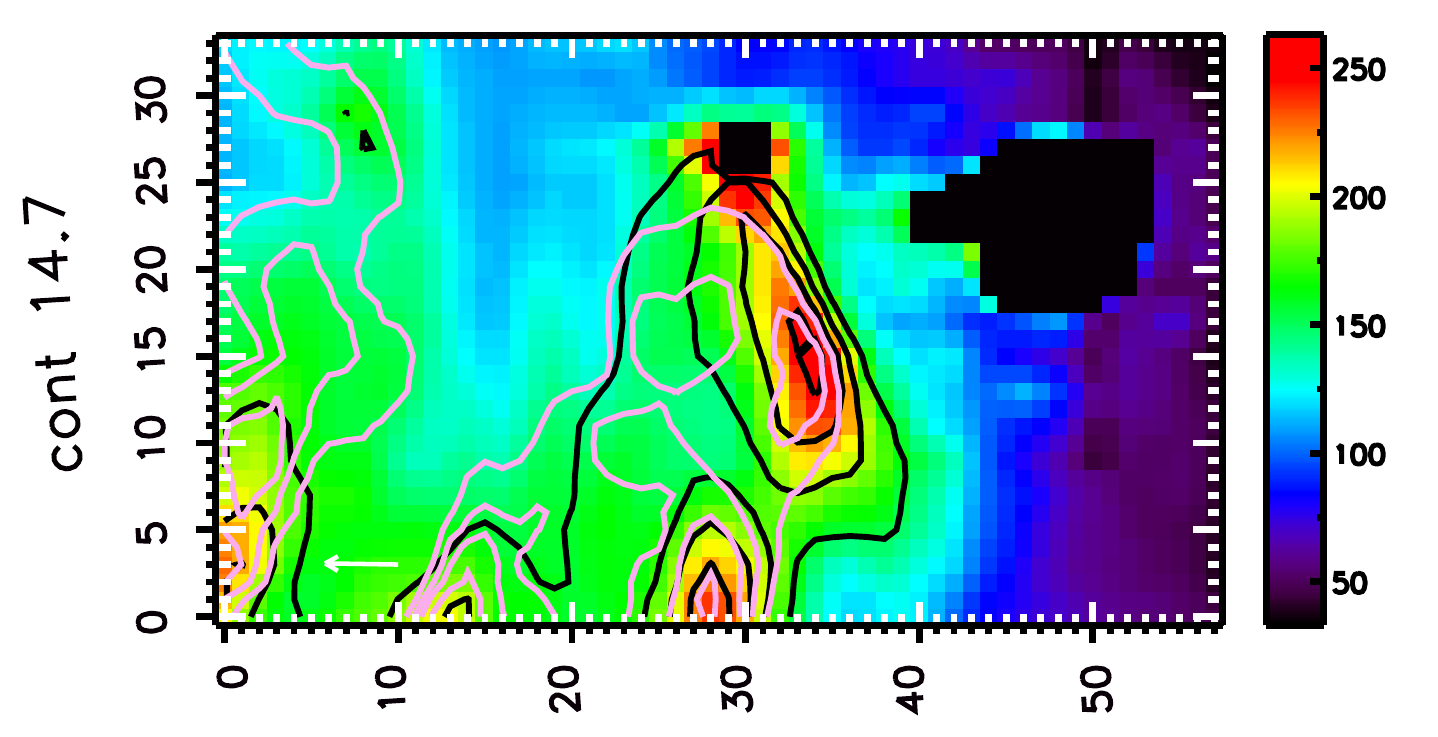}
   \includegraphics[angle=266.4]{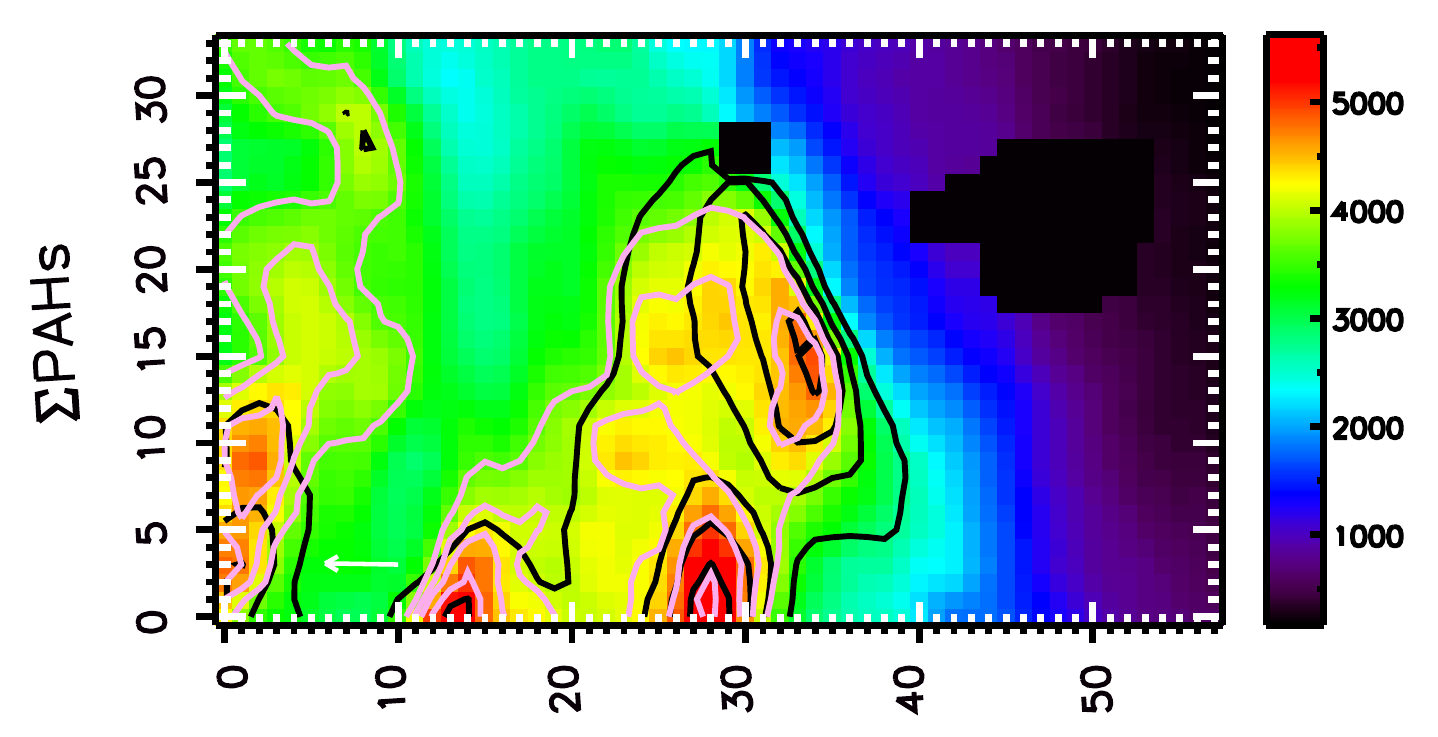}
   \includegraphics[angle=266.4]{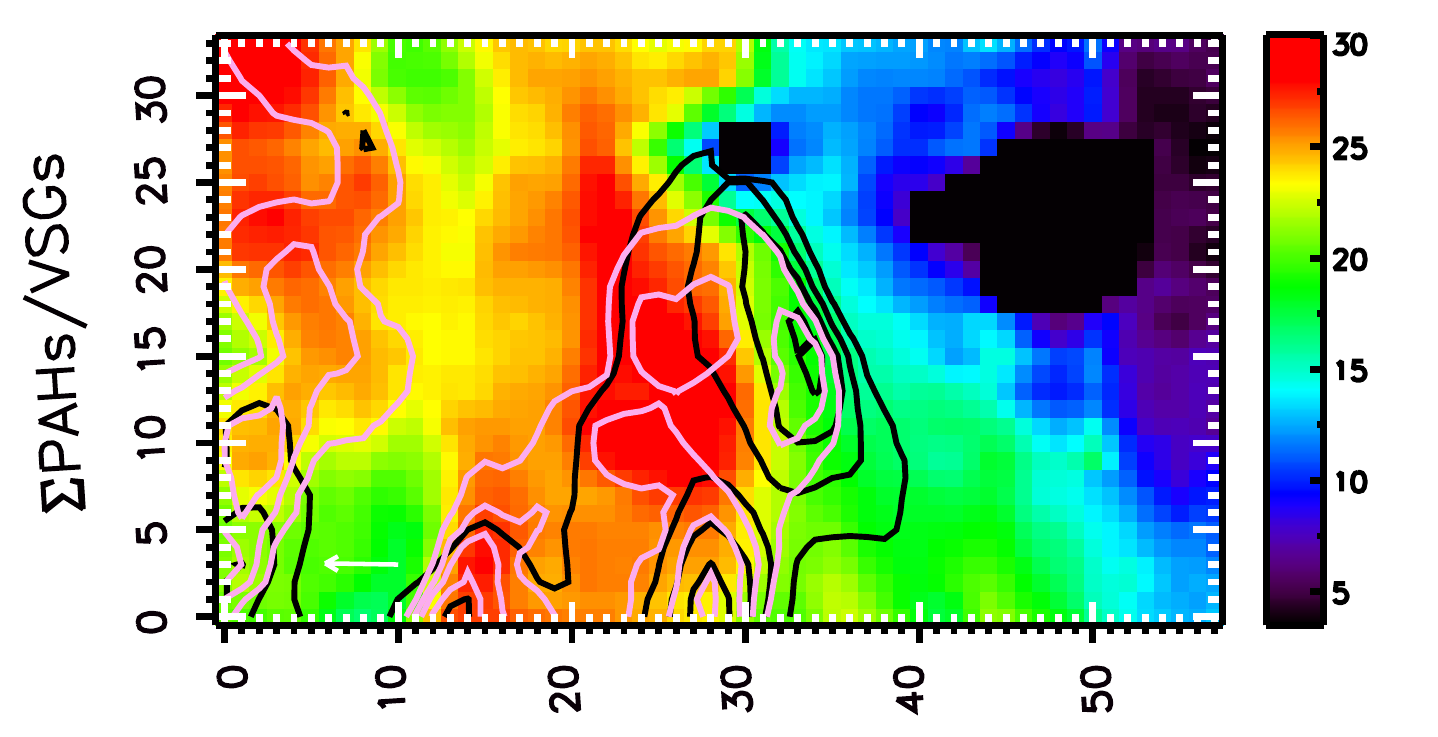}
   \includegraphics[angle=266.4]{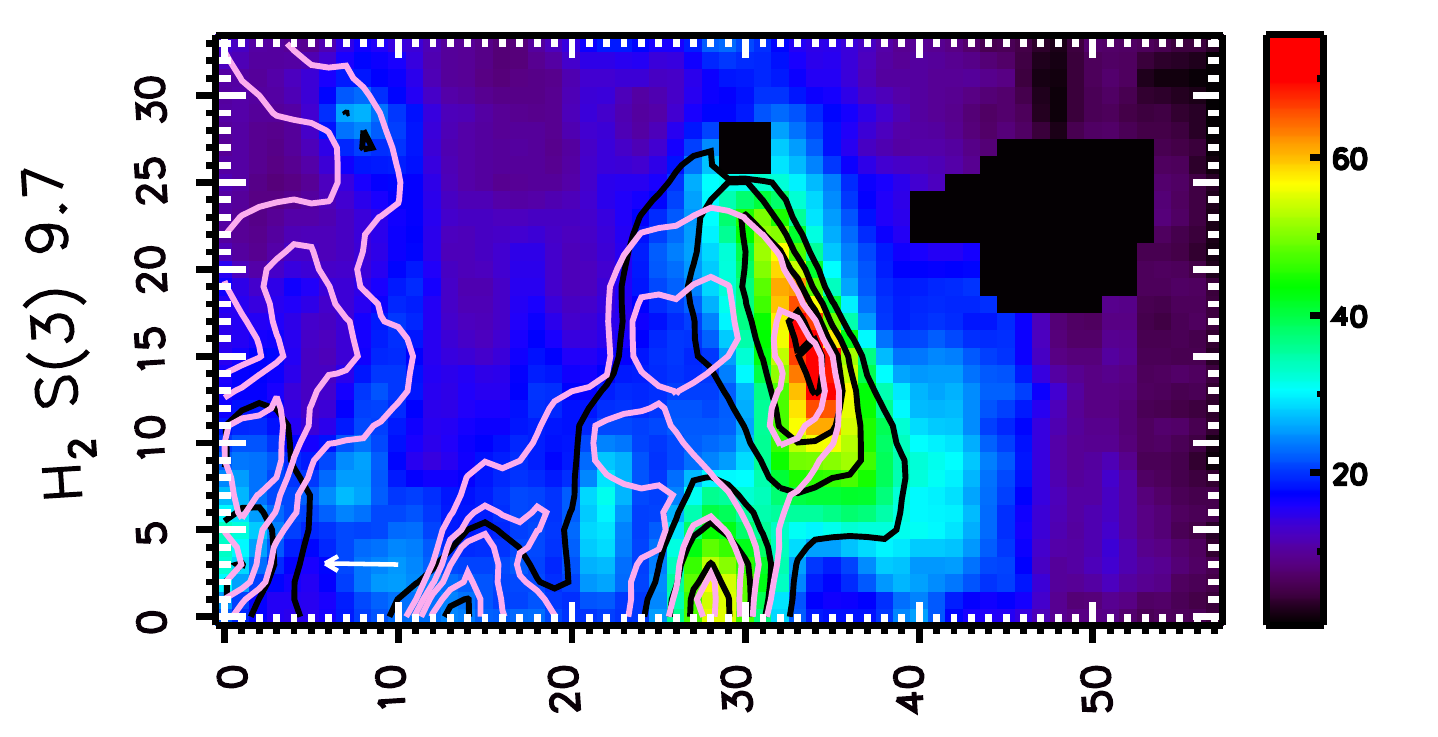}
   \includegraphics[angle=266.4]{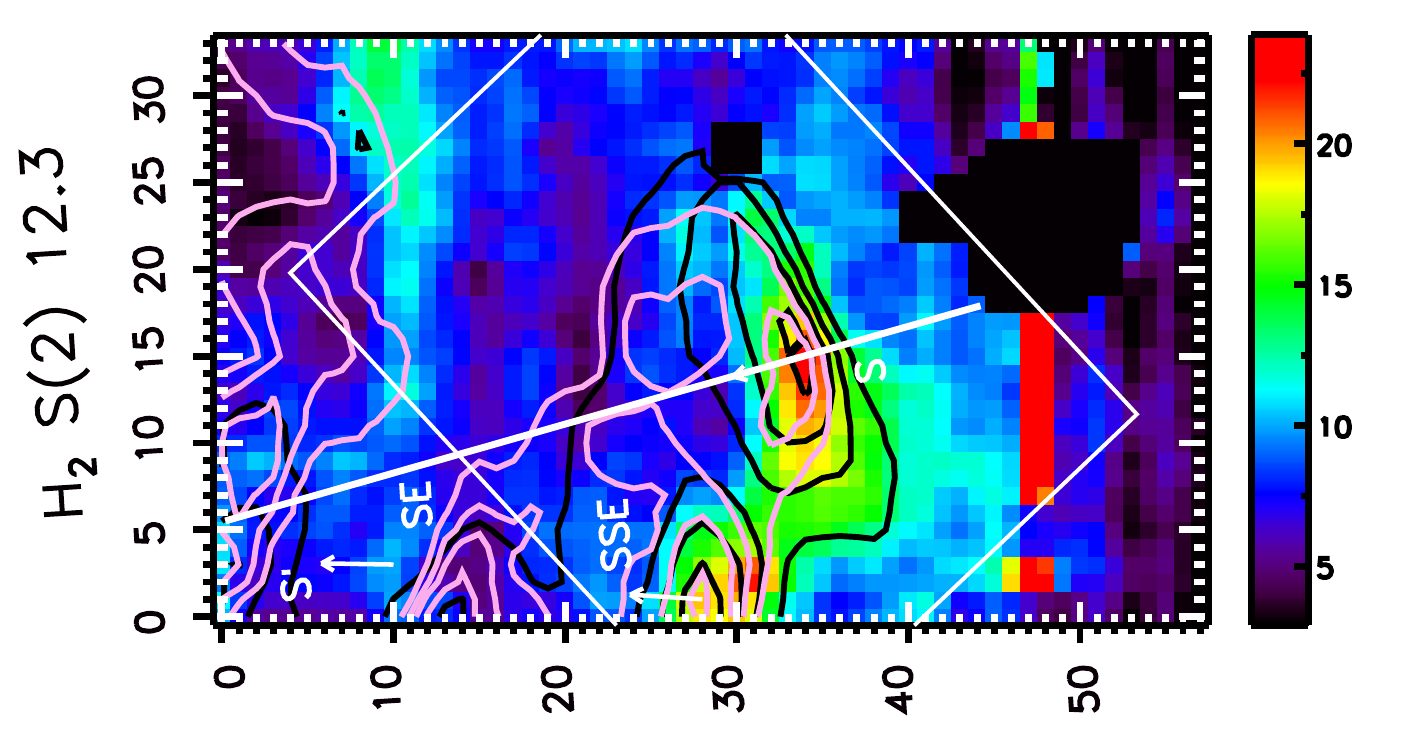}}
\caption{South map: Spatial distribution of the emission features in the 5-15 \mum\, SL data towards NGC~2023 (using a cutoff value of 2 sigma and applying a local spline continuum except for the 5--10 \mum\, plateau). $\Sigma$PAHs refers to the combined flux of all PAH features and the 8 \mum\, bump (i.e. excluding the 5-10 and 10-15 \mum\, plateaus) and VSG refers to the continuum flux at 14.7 \mum. Band intensities are measured in units of 10$^{-8}$ Wm$^{-2}$sr$^{-1}$ and continuum intensities in units of MJysr$^{-1}$. As a reference, the intensity profiles of the 11.2 and 7.7 \mum\, emission features are shown as contours in respectively black (at 3.66, 4.64, 5.64 and, 6.78 10$^{-6}$ Wm$^{-2}$sr$^{-1}$) and pink (at 1.40, 1.56, 1.70 and, 1.90 10$^{-5}$ Wm$^{-2}$sr$^{-1}$). The maps are orientated so N is up and E is left. The white arrow in the top left corners indicates the direction towards the central star. Arrows are shown separately for the S, SE, and SSE ridges in the bottom right map. The white line across the FOV represents the line cut used in Figs~\ref{linecuts},~\ref{linecuts_PAHFIT}, and~\ref{linecuts_PAHTAT}. The axis labels refer to pixel numbers. Regions near source C and D excluded from the analysis are set to zero. The nomenclature and the FOV of the SH map are given in the bottom right panel (see also Fig.~\ref{fov}). }
\label{fig_slmaps_s}
\end{figure*}
%%%%%%%%%%%%%%%%%%%%%%%%%%%%%%%%%%%%%%%%%%%%%%%%%%

The second method is a modification of the first method. A global spline (GS) continuum is determined by using the same anchor points as in the first method except for the continuum point at roughly 8.2 \mum\, (Fig.~\ref{fig_sp_sl}). This affects the band profiles and intensities of the 7.7 and 8.6 \mum\, PAH bands and the underlying plateau. This also creates a new plateau underneath only the 7.7 and 8.6 \mum\, PAH bands that is then defined by the difference of the local spline continuum and the global continuum and is further referred to as the 8 \mum\, bump. The plateau continuum (Fig.~\ref{fig_sp_sl}) is obtained by using continuum points at roughly 5.5, 9.9, 10.2, 10.4, and 14.7 \mum. The underlying plateaus between 5-10 and 10-15 \mum\, are then defined by the difference of this continuum with the GS and LS continua respectively. The fluxes are obtained in the same way as discussed above for the LS continuum.

Finally, the third approach employs PAHFIT to analyze the data \citep[Fig.~\ref{fig_sp_sl},][]{SmithJD:07}. The PAHFIT decomposition results in components representing the dust continuum emission (which is a combination of modified blackbodies), the H$_2$ emission and the PAH emission. In particular, the PAH emission is fit by a combination of Drude profiles. A detailed discussion of this decomposition method and its results can be find in Appendix \ref{pahfit}. \\

%%%%%%%%%%%%%%%%%%%%%%%%%%%%%%%%%%%%%%%%%%%%%%%%%%
\begin{figure*}[tb]
    \centering
\resizebox{\hsize}{!}{%
  \includegraphics[angle=274.1]{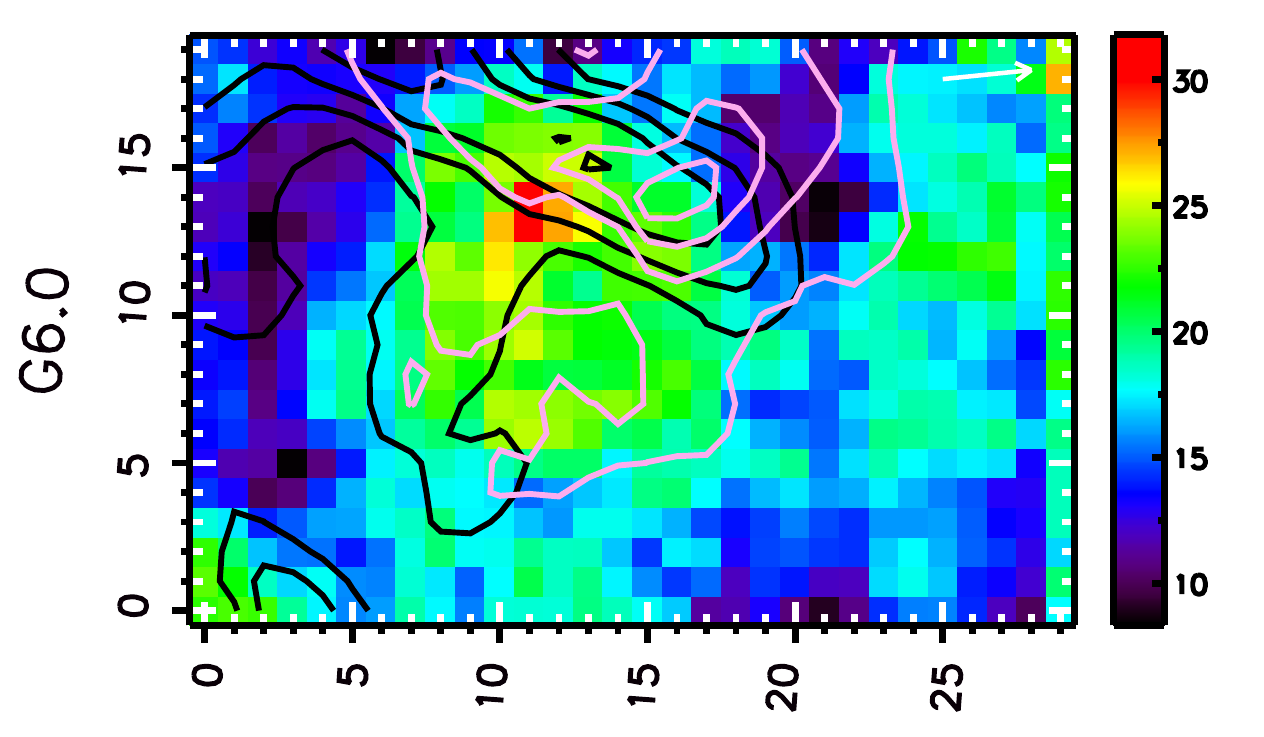}
  \includegraphics[angle=274.1]{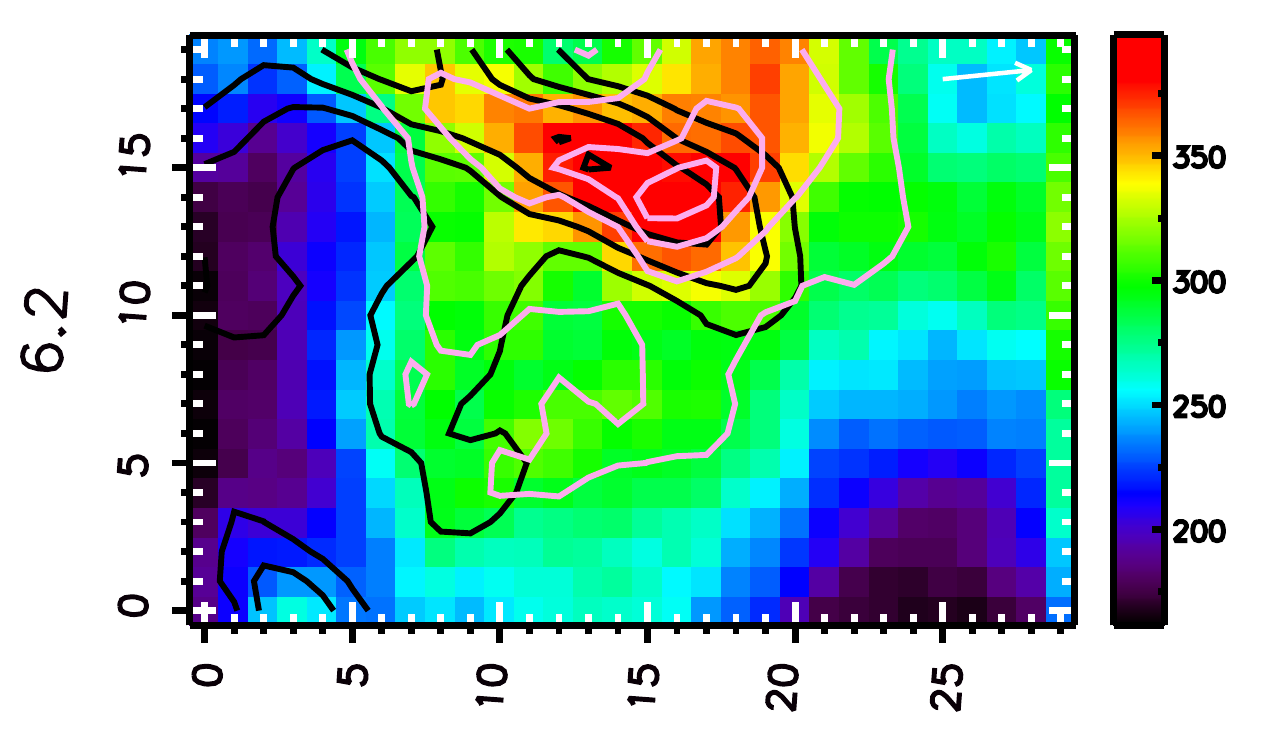}
  \includegraphics[angle=274.1]{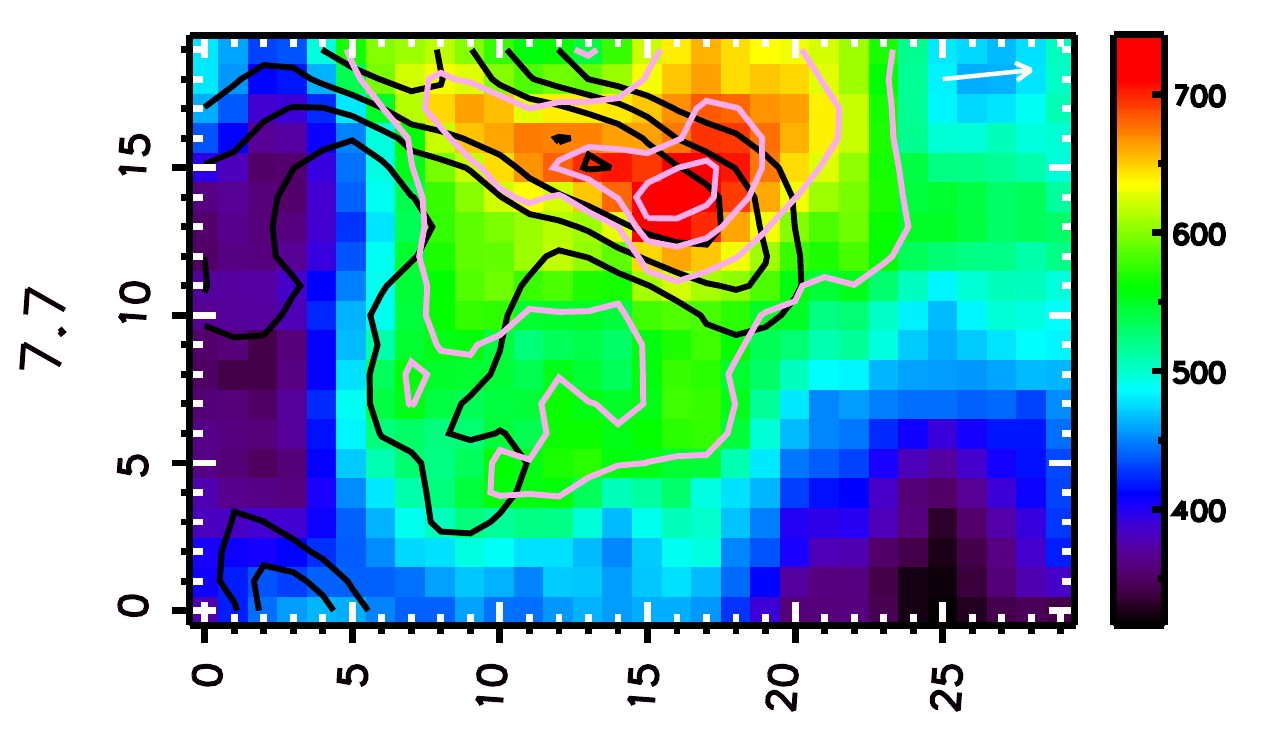}
  \includegraphics[angle=274.1]{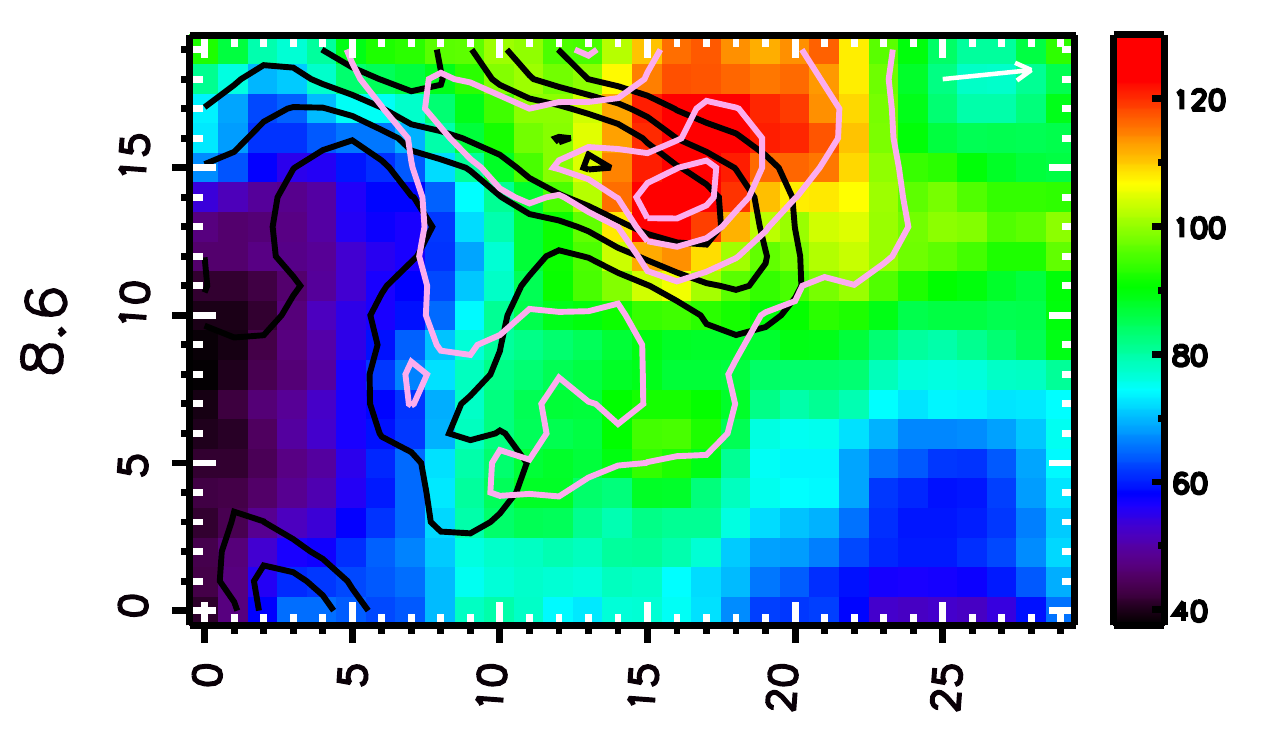}
  \includegraphics[angle=274.1]{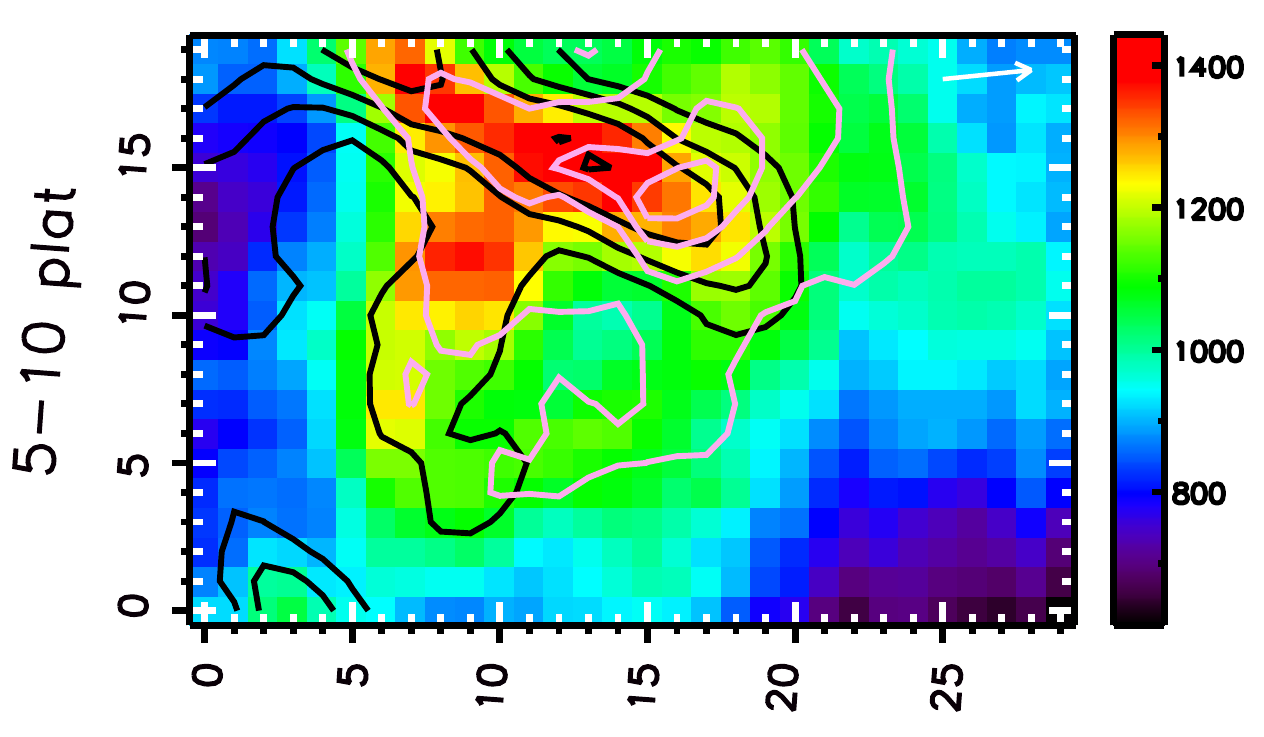}}
\resizebox{\hsize}{!}{%
   \includegraphics[angle=274.1]{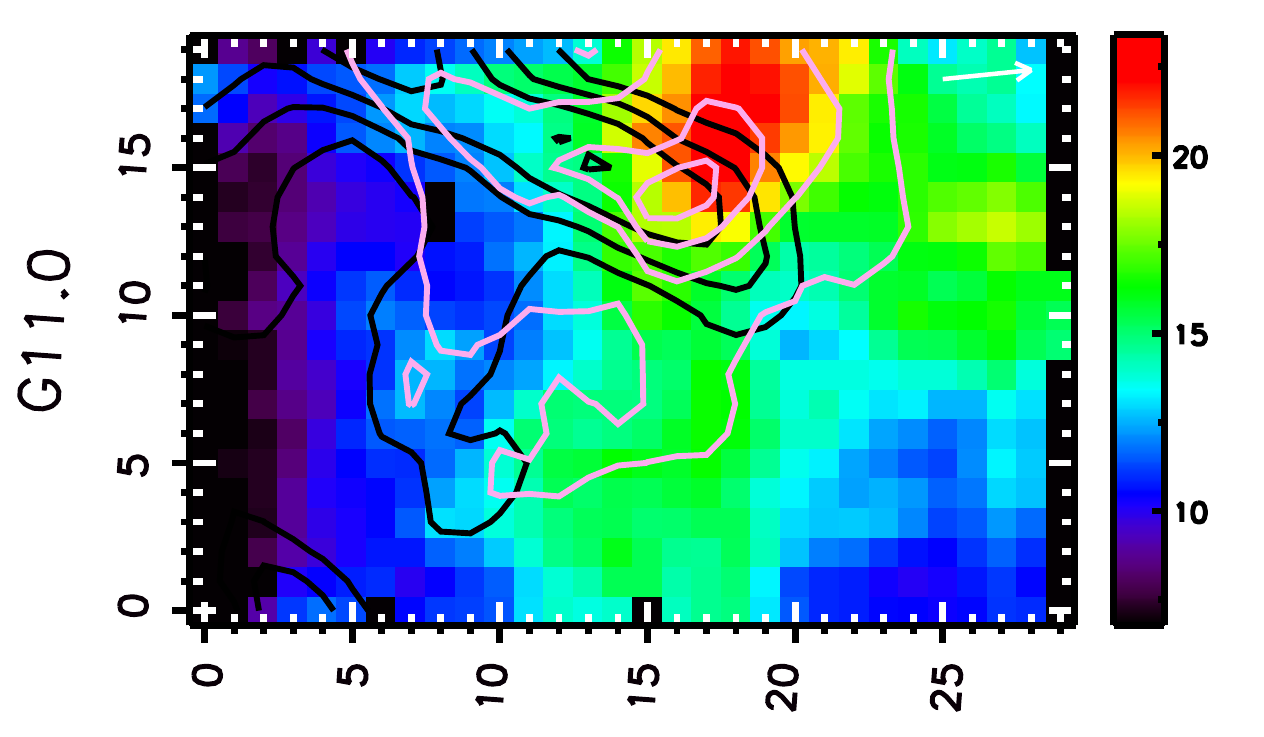}
   \includegraphics[angle=274.1]{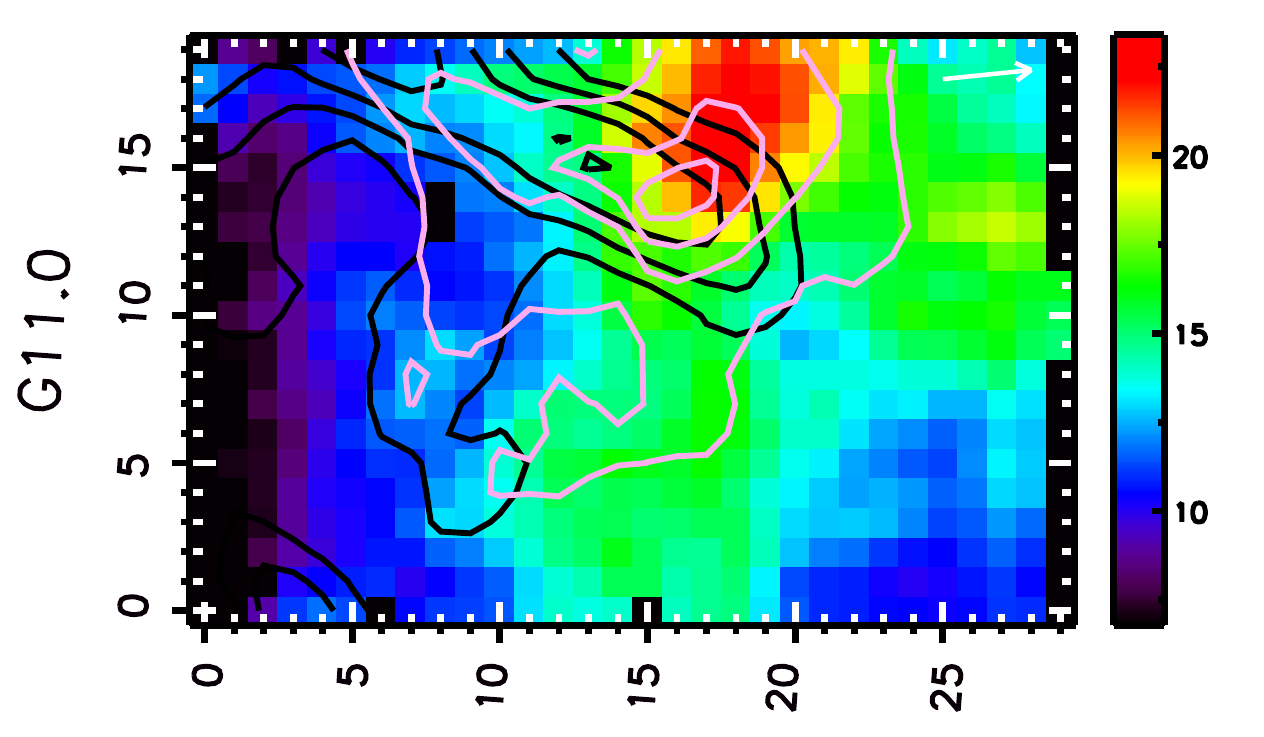}
   \includegraphics[angle=274.1]{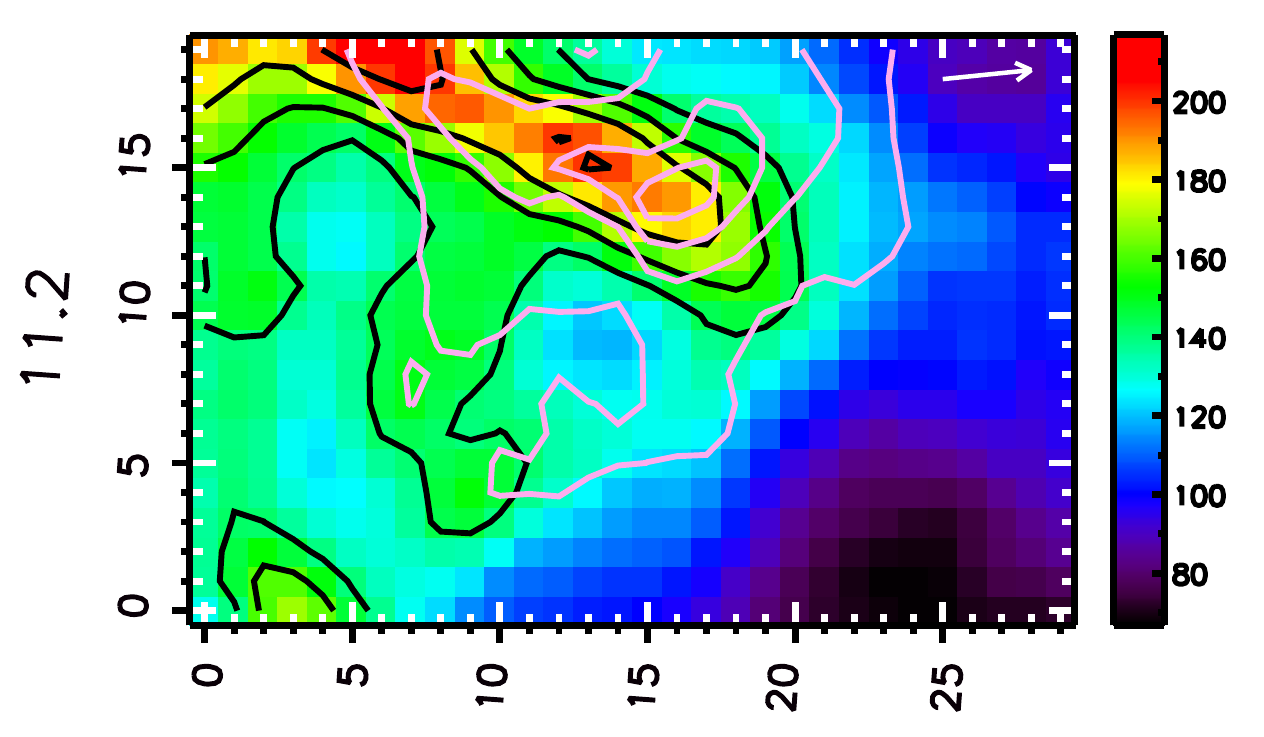}
   \includegraphics[angle=274.1]{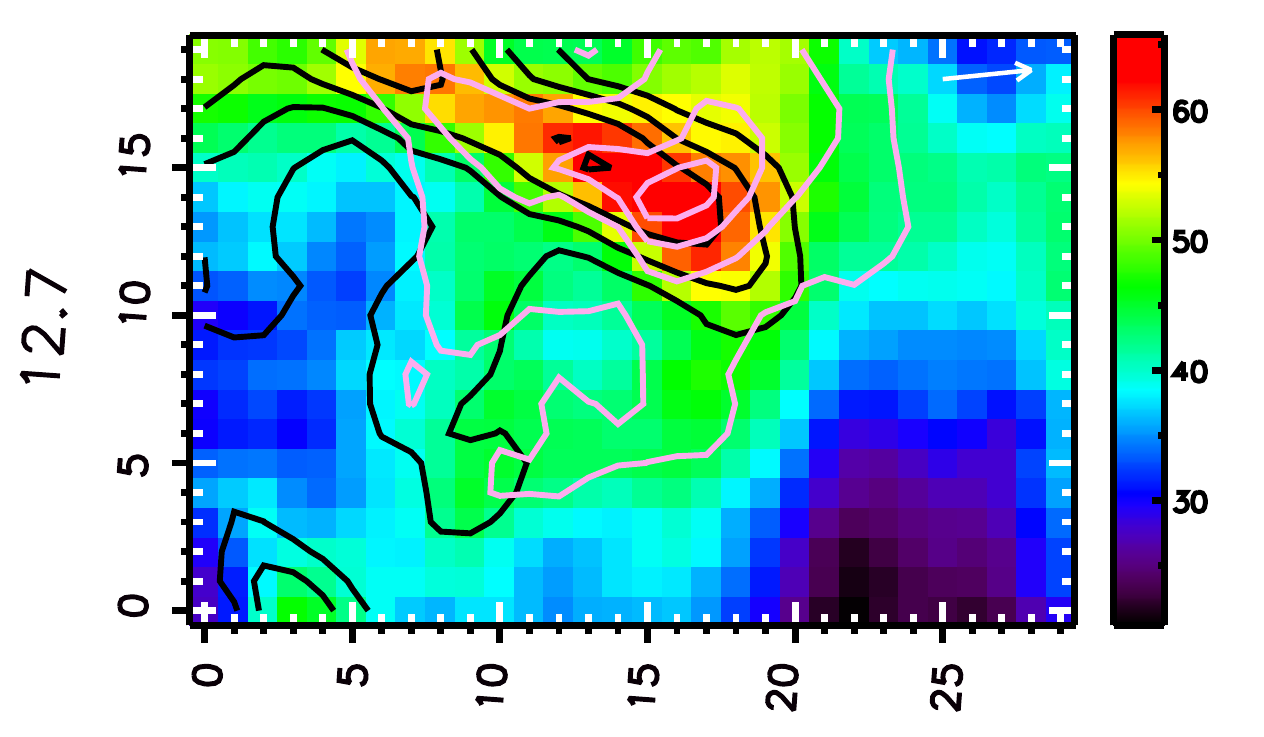}
   \includegraphics[angle=274.1]{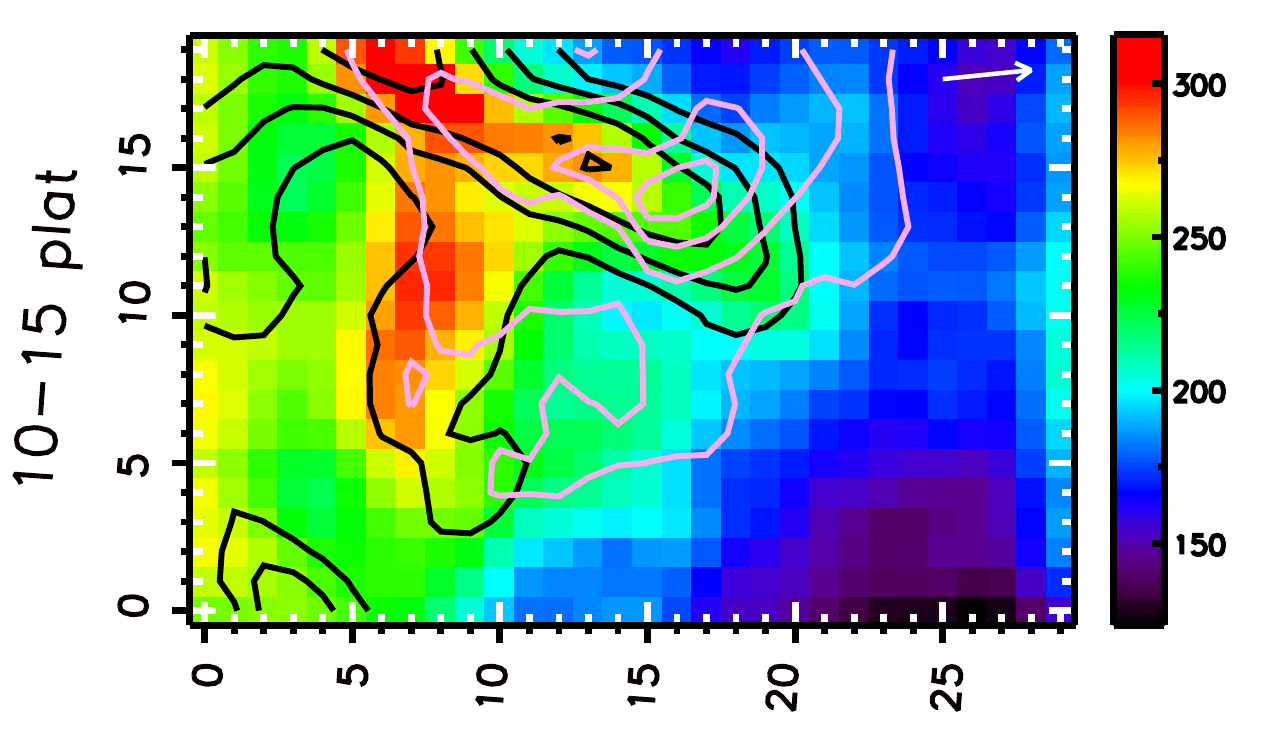}}
\resizebox{\hsize}{!}{%
   \includegraphics[angle=274.1]{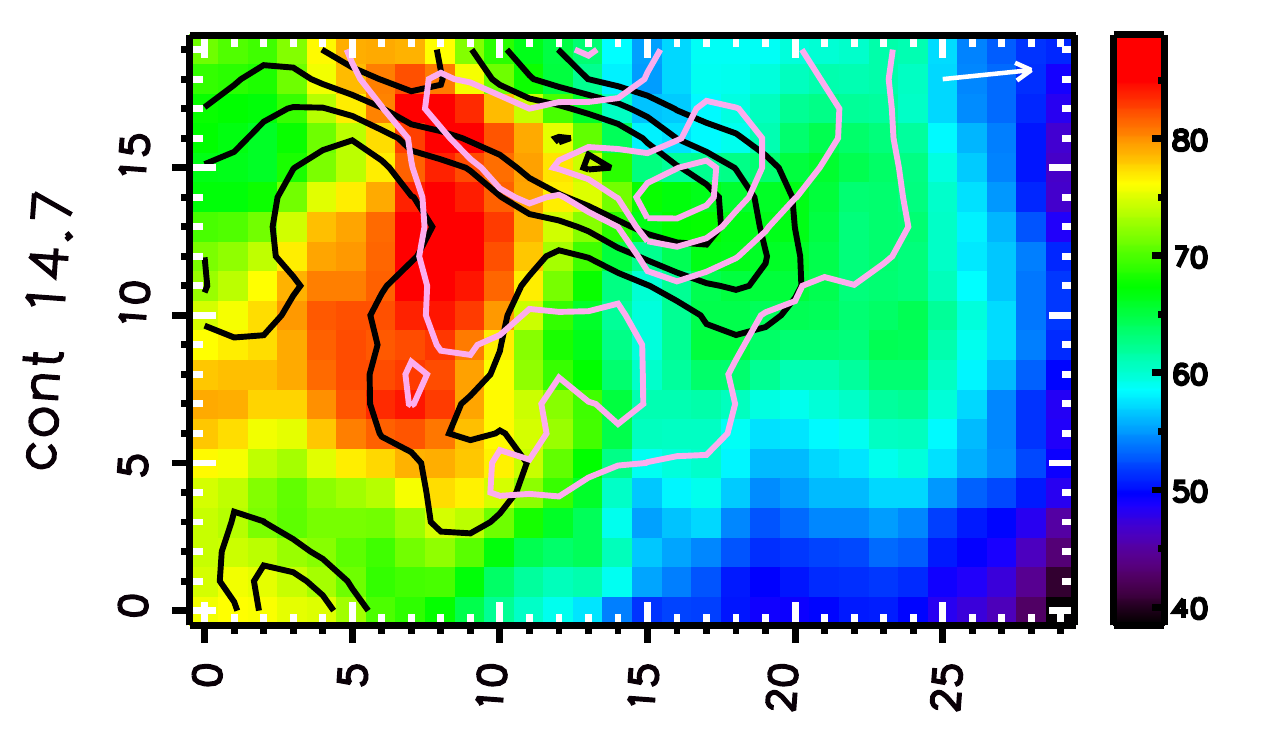}
   \includegraphics[angle=274.1]{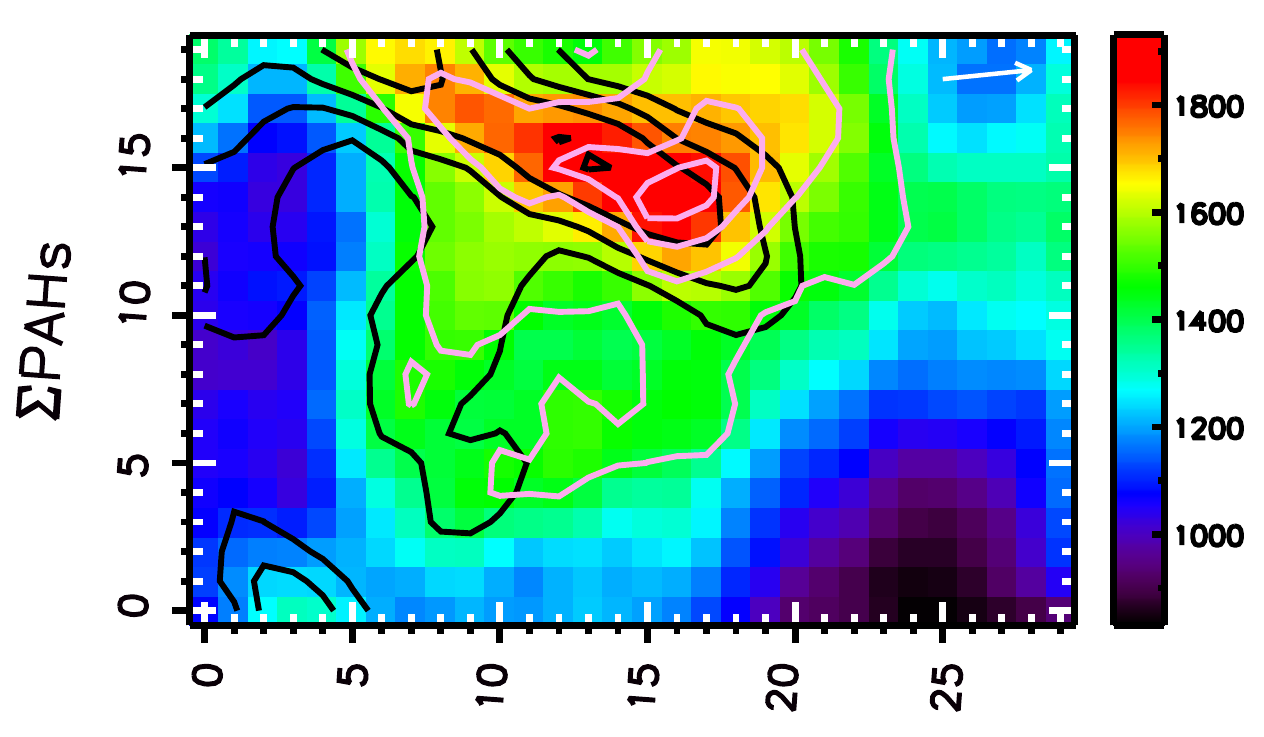}
   \includegraphics[angle=274.1]{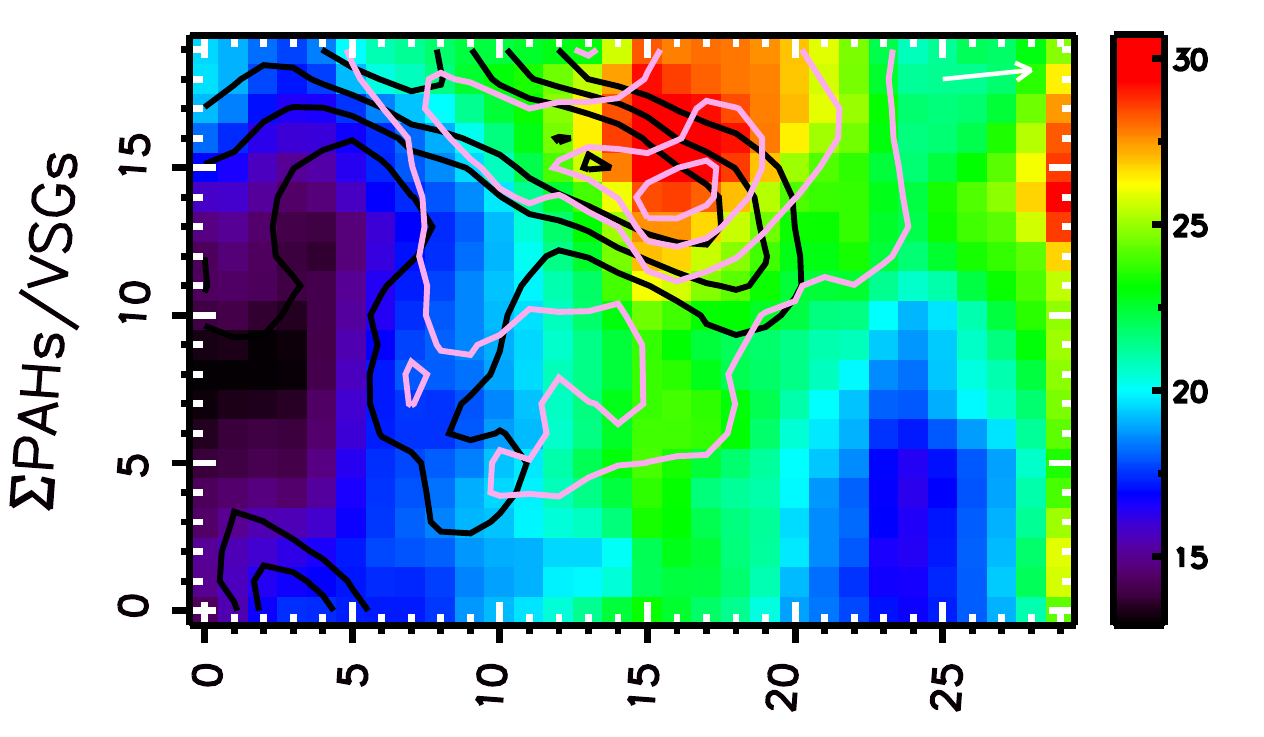}
   \includegraphics[angle=274.1]{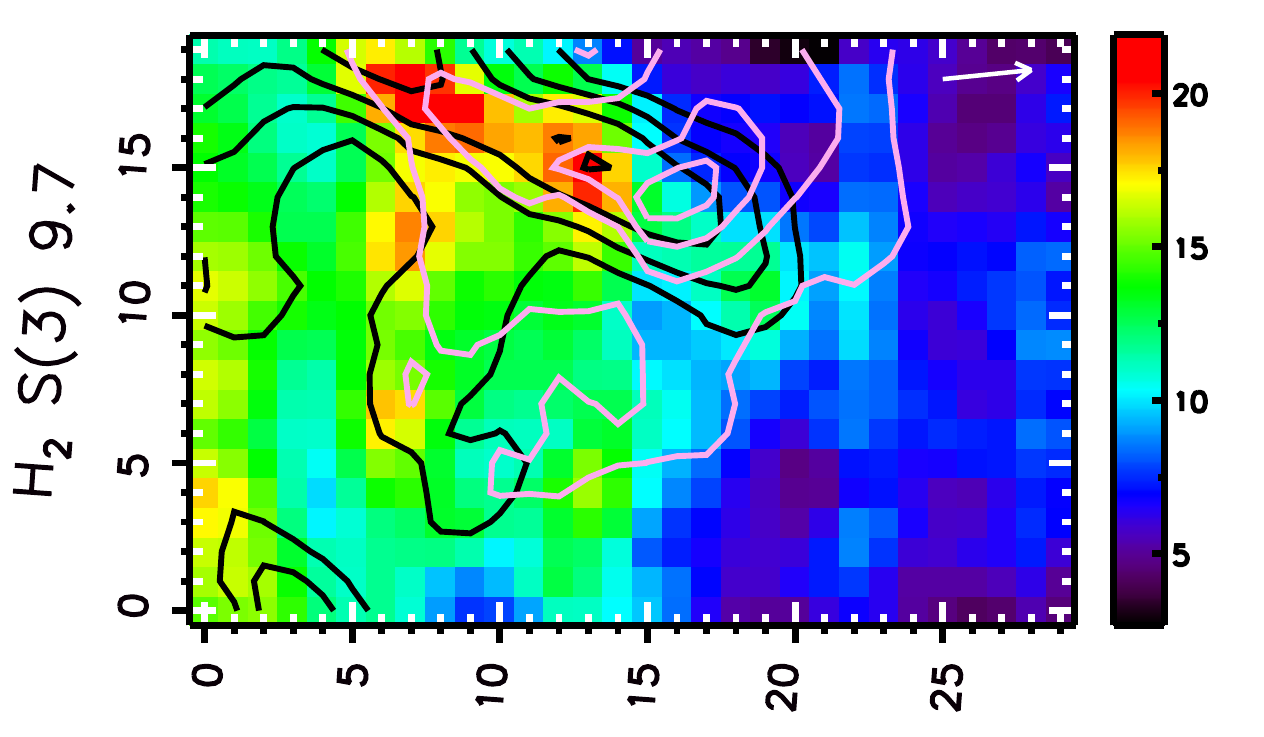}
   \includegraphics[angle=274.1]{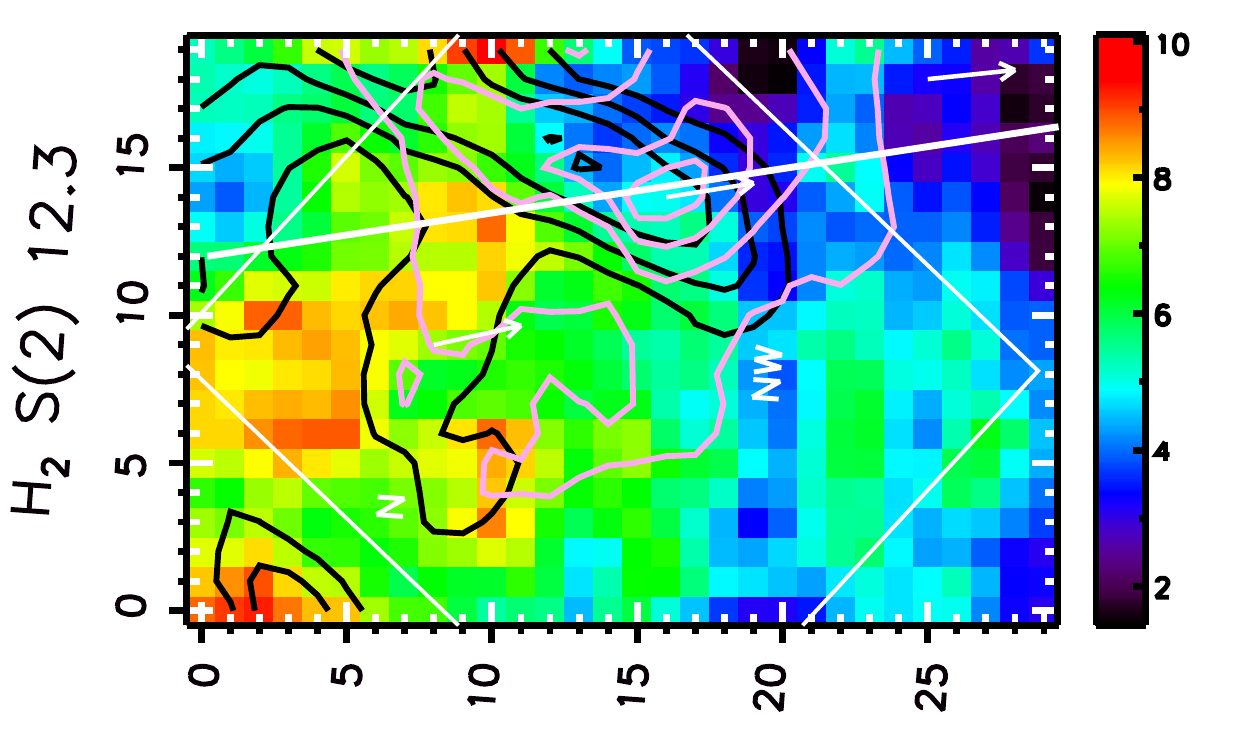}}
\caption{North map: Spatial distribution of the emission features in the 5-15 \mum\, SL data towards NGC~2023 (using a cutoff value of 2 sigma and applying a local spline continuum except for the 5--10 \mum\, plateau). $\Sigma$PAHs refers to the combined flux of all PAH features and the 8 \mum\, bump (i.e. excluding the 5-10 and 10-15 \mum\, plateaus) and VSG refers to the continuum flux at 14.7 \mum. Band intensities are measured in units of 10$^{-8}$ Wm$^{-2}$sr$^{-1}$ and continuum intensities in units of MJysr$^{-1}$. As a reference, the intensity profiles of the 11.2 and 7.7 \mum\, emission features are shown as contours in respectively black (at 1.42, 1.57, 1.75 and, 1.99 10$^{-6}$ Wm$^{-2}$sr$^{-1}$) and pink (at 5.54, 6.30, 6.80 and, 7.20 10$^{-6}$ Wm$^{-2}$sr$^{-1}$). The maps are orientated so N is up and E is left. The white arrow in the bottom right corners indicates the direction towards the central star.  Arrows are shown separately for the N and NW ridges in the bottom right map. The white line across the FOV represents the line cut used in Figs~\ref{linecuts},~\ref{linecuts_PAHFIT}, and~\ref{linecuts_PAHTAT}. The axis labels refer to pixel numbers. The nomenclature and the FOV of the SH map are given in the  bottom right panel 
(see also Fig.\,\ref{fov}). }
\label{fig_slmaps_n}
\end{figure*}
%%%%%%%%%%%%%%%%%%%%%%%%%%%%%%%%%%%%%%%%%%%%%%%%%%

For the SH data, we applied the local spline continuum by using anchor points at roughly  10.2, 10.4, 10.8, 11.8, 12.2, 13.0, 13.9, 14.9, 15.2, 15.5, 16.1, 16.7, 16.9, 17.16/17.19 (for the south and north maps,
respectively), 17.6, 18.14/18.17 (for the north and south maps,
respectively), 18.5, 19.3, and 19.4 \mum\, and the plateau continuum by using anchor points at 10.4, 15.2, 18.5 and 19.4 \mum. The fluxes of the 11.0, 11.2, 12.7 \mum\, PAH bands and the 12.3 \mum\, H$_2$ line are determined in the same way as for the SL data. The parameters for the 11.0 and 11.2 Gaussians are $\lambda$ (FWHM) of 11.018 (0.1205) and 11.243 (0.144) \mum\, respectively (Fig.~\ref{fig_decomp6}, right panel). The fluxes of the weaker 12.0 and 13.5 \mum\, bands are determined by fitting a Gaussian. The 14.2 \mum\, PAH displays a weaker blue shoulder. These two components were fitted by a Gaussian ($\lambda$ (FWHM) of 13.99 (0.1178) and 14.230 (0.1830) \mum\, respectively).  \\

The SL and SH fluxes for features in the 10-14 \mum\, region tend to differ. This difference arises from the following reasons. Firstly, we do not regrid the SL data to the coarser SH grid and do not correct for the different spatial PSFs. However, when the SL data are regridded to the SH grid and the SL data are scaled to the SH data but no correction for the different spatial PSFs are made, it is clear that the features' strength differ to varying degrees in the SL and SH data (see Fig.~\ref{fig_sp_all}). Secondly, the continuum is less accurately determined in the SL data due to blending of the emission features (11.2 PAH, 12.0 PAH, H$_2$ and 12.7 PAH) which strongly influences the fluxes of the weaker bands. Hence, we will refrain from comparing the SL and SH fluxes directly with each other and we will not use the weaker 12.0, 13.5 and 14.5 \mum\, bands in the SL data.

The applied continuum is clearly not unique. Hence neither is the decomposition of the PAH emission, nor the calculated band strengths. Thus, for comparison, we did a full analysis of the SL data using these three methods. In the remainder of the paper, the default applied continuum is the local spline continuum for the intensities of the PAH bands (excluding the 5--10 \mum\, plateau\footnote{Defined as the difference between the plateau continuum and the global spline continuum.}) unless stated otherwise (e.g. Section \ref{decomp7}). We will discuss the influence of the continuum and decomposition on the results where necessary. 

\section{Data analysis}
\label{analysis}

Here we investigate the relationship between individual PAH emission bands, the underlying plateaus, H$_2$ emission, C$_{60}$ emission, and the dust continuum emission present in the 5-20 \mum\, region. The north map is characterized by lower flux levels and therefore larger scatter is present in the maps/plots tracing weaker PAH features.  Hence, when discussing the spatial distribution of the 14.2, 15.8 and 17.8 \mum\, bands, we restrict ourselves to the south map. 

\subsection{SL data}
\label{sldata}

The spatial distribution of the various emission components in the 5-15 \mum\, SL data are shown in Figs. \ref{fig_slmaps_s} and \ref{fig_slmaps_n} (the range in colours is set by the minimum and maximum intensities present in the map, and a local spline continuum is applied except for the 5--10 \mum\, plateau) and feature correlations in Fig.~\ref{fig_corr} (with a local spline continua applied except for the 5--10 \mum\, plateau and the Gaussian components as discussed in Section \ref{decomp7}).  To exclude the influence of PAH abundance and column density in the correlation plots, we normalized the band fluxes to that of another PAH band.  

\subsubsection{Overall appearance} 
\label{generalimpression}

The following trends are derived from the south spectral maps shown in  Fig.~\ref{fig_slmaps_s}. The 11.2 \mum\, feature, the 5-10 and 10-15 \mum\, plateaus and the continuum emission show very similar spatial morphology with distinct peaks at the S and SSE positions. The 8 \mum\, bump exhibits a similar morphology but shows more enhanced emission at the SE ridge and west of the S' ridge. The PDR front is well traced by the H$_2$ emission, which also clearly peaks at the S and SSE positions and is heavily concentrated along these two ridges only. In contrast, the distribution of the 6.2, 7.7, 12.7 \mum\, features are displaced towards the illuminating star and away from the S ridge, with the loss of emission at the S position accompanied by a rise at the S' and SE ridges. Put another way, the 6.2, 7.7 and 12.7 \mum\, features show very similar spatial morphology with distinct peaks at the SSE and SE ridges and with weaker peaks at the S and S' positions. In addition, they show broad, diffuse emission NW of the line connecting the S and SSE ridges. The 8.6 \mum\, band is further displaced towards the illuminating star: it peaks at the SE and S' position but does {\it not} peak at the S and SSE ridges as do the 6.2, 7.7 and 12.7 \mum\, emission. In fact, it lacks emission in the S ridge.  Similar to the 6.2, 7.7 and 12.7 \mum\, emission, it does show a broad, diffuse plateau NW of the line connecting the S and SSE ridges.  As does the 8.6 \mum\, PAH emission, the 11.0 \mum\, PAH emission also peaks at the SE and S' ridges and lacks emission in the S ridge. But it is also very strong at the broad, diffuse plateau N and NW of the line connecting the S and SSE ridges. The 6.0 \mum\, PAH emission peaks at the S and SSE ridges but its peaks seem to be displaced towards the east compared to the 11.2 \mum\, PAH emission. It has weaker emission maxima in the form of an arc south of the S' ridge and west of the SE ridge. Hence, 6.0 \mum\, PAH emission seems to be somewhat unique.

%%%%%%%%%%%%%%%%%%%%%%%%%%%%%%%%%%%%%%%%%%%%%%%%%%
\begin{figure*}[tb]
    \centering
\resizebox{\hsize}{!}{%
  \includegraphics{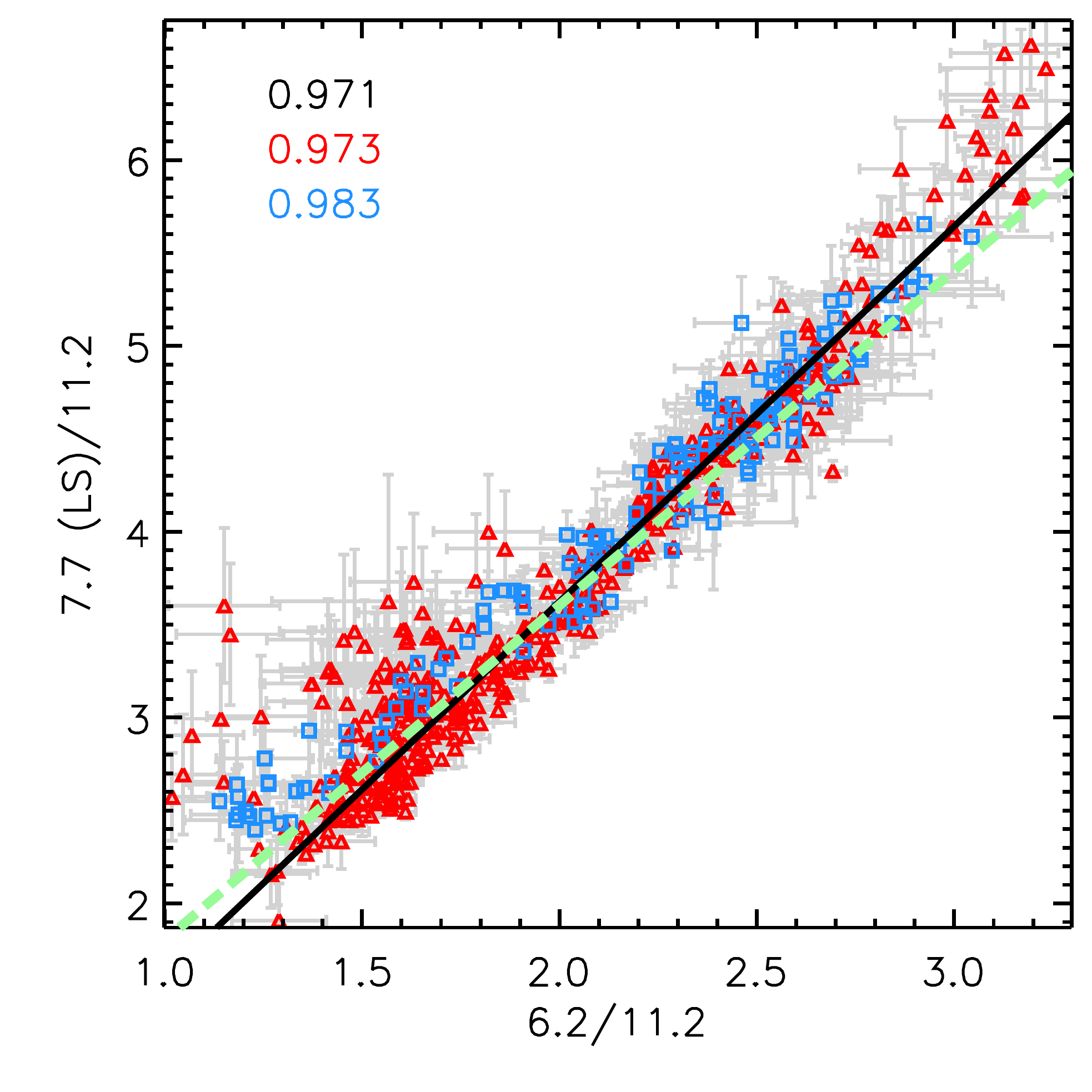}
  \includegraphics{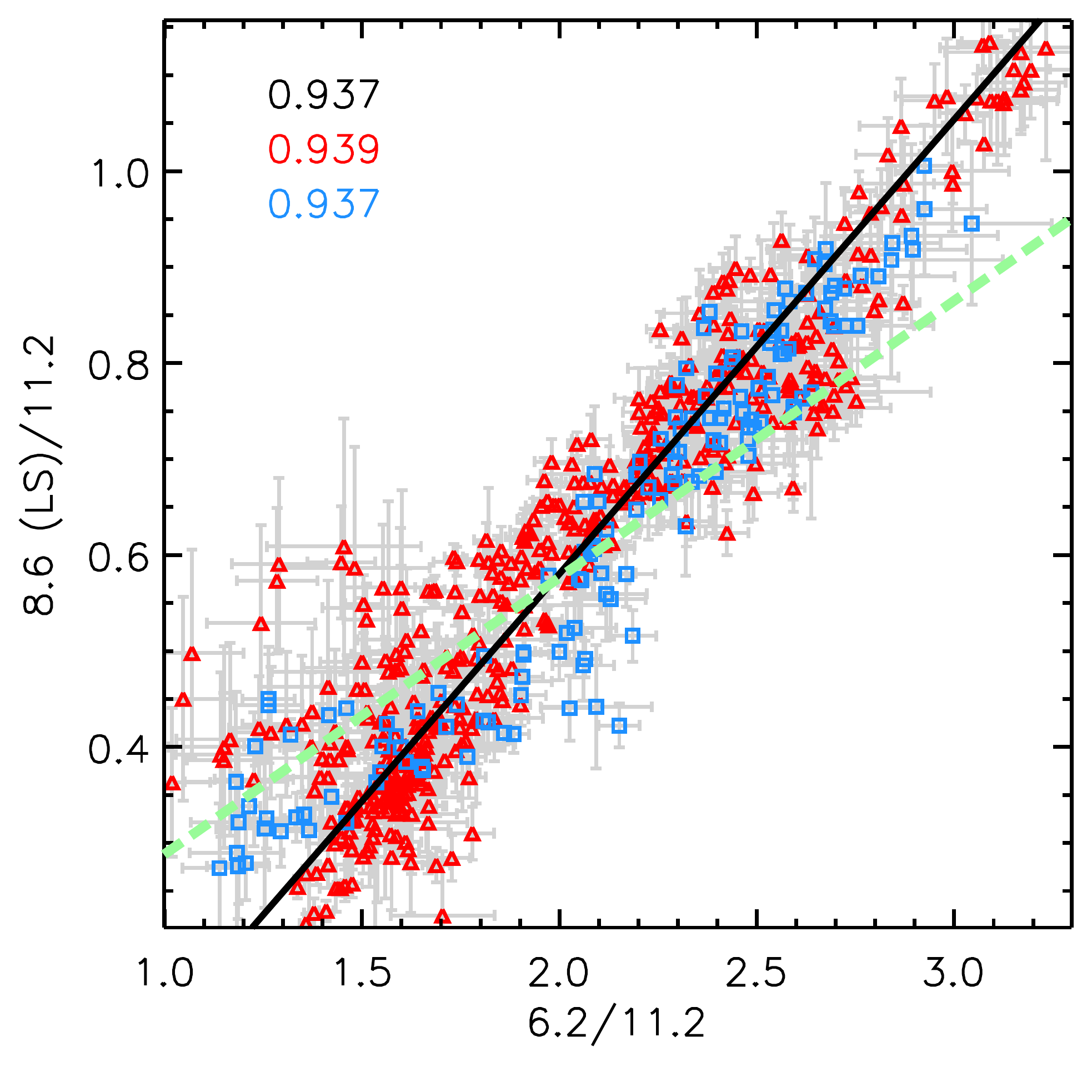}
  \includegraphics{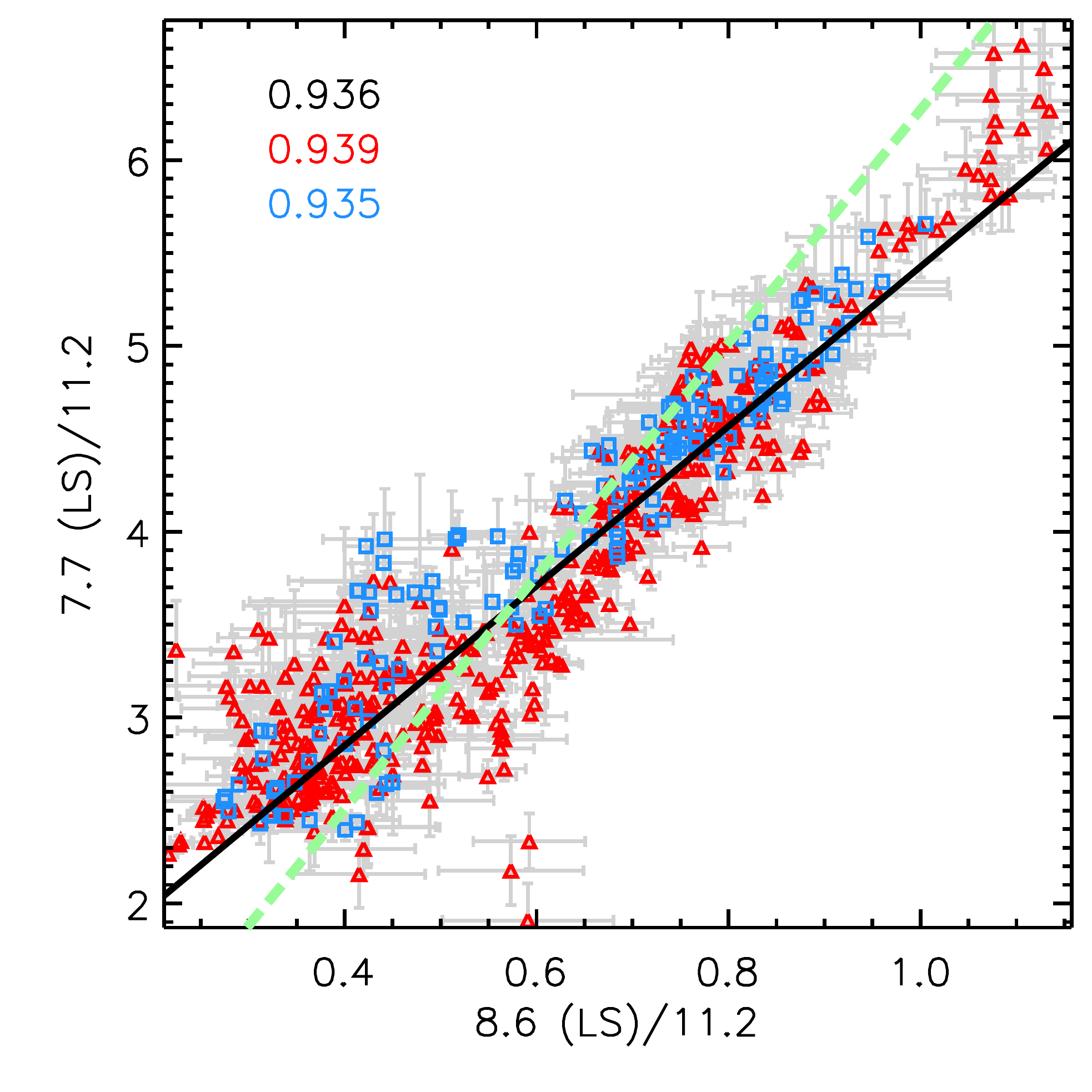}
  \includegraphics{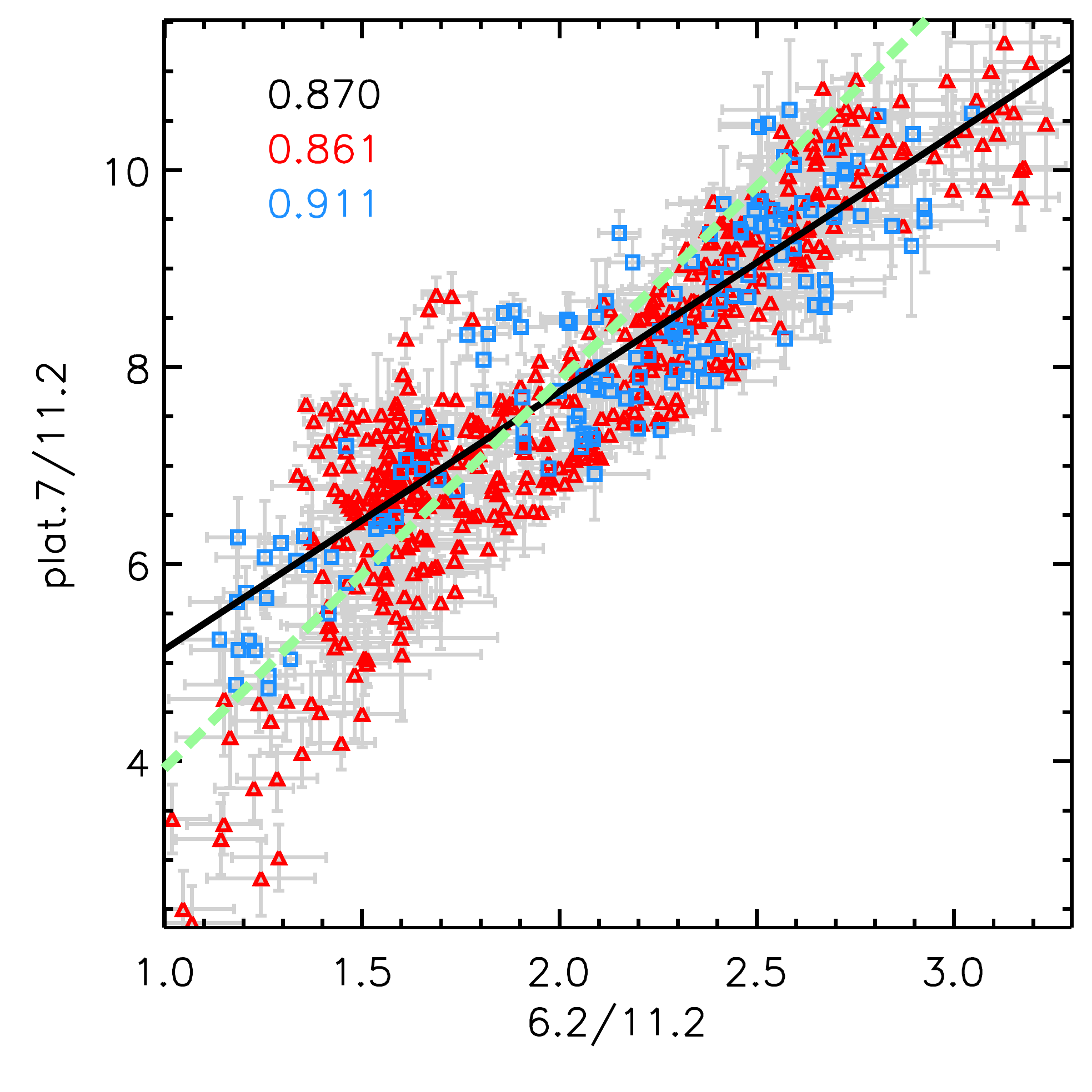}}
\resizebox{\hsize}{!}{%
  \includegraphics{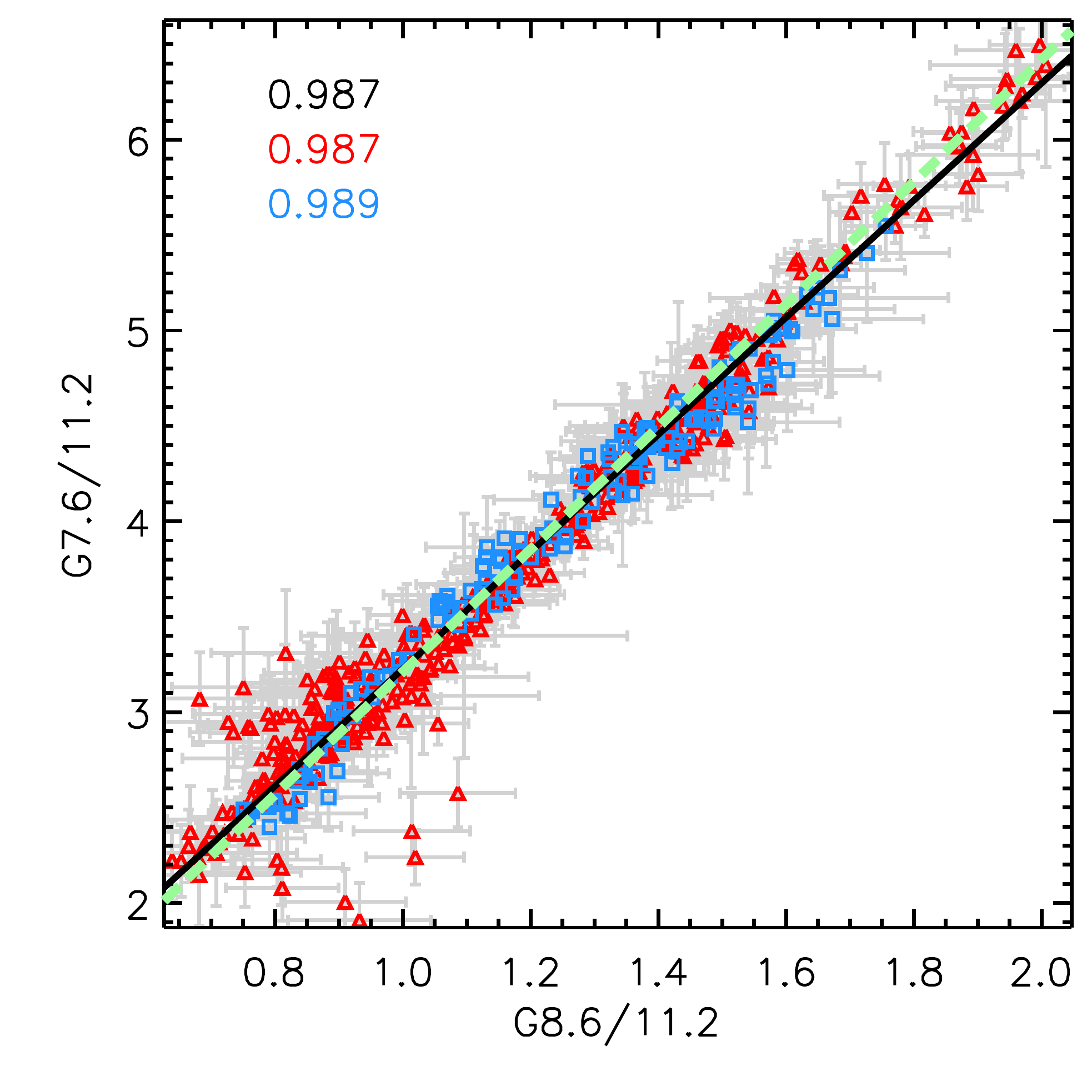}
  \includegraphics{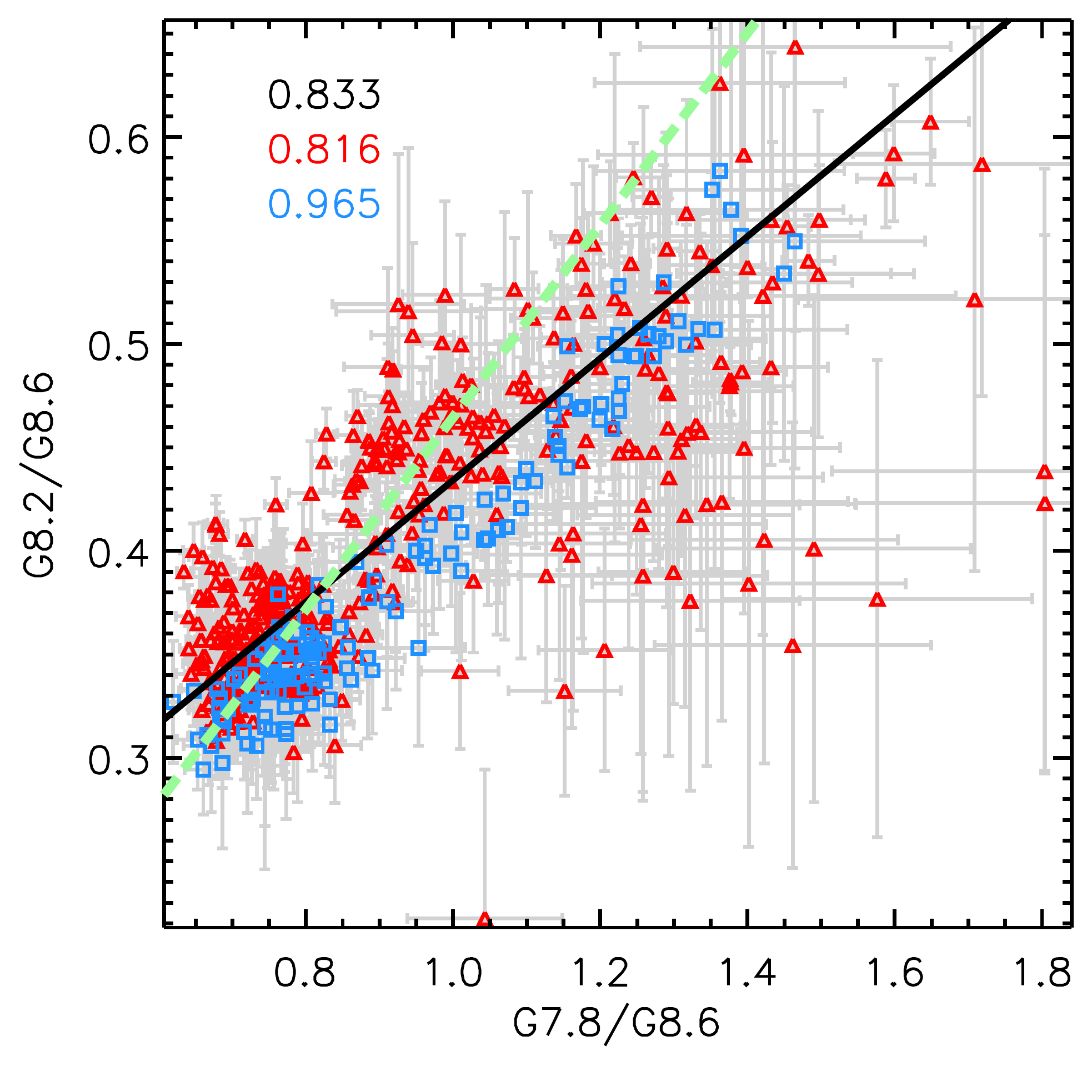}
  \includegraphics{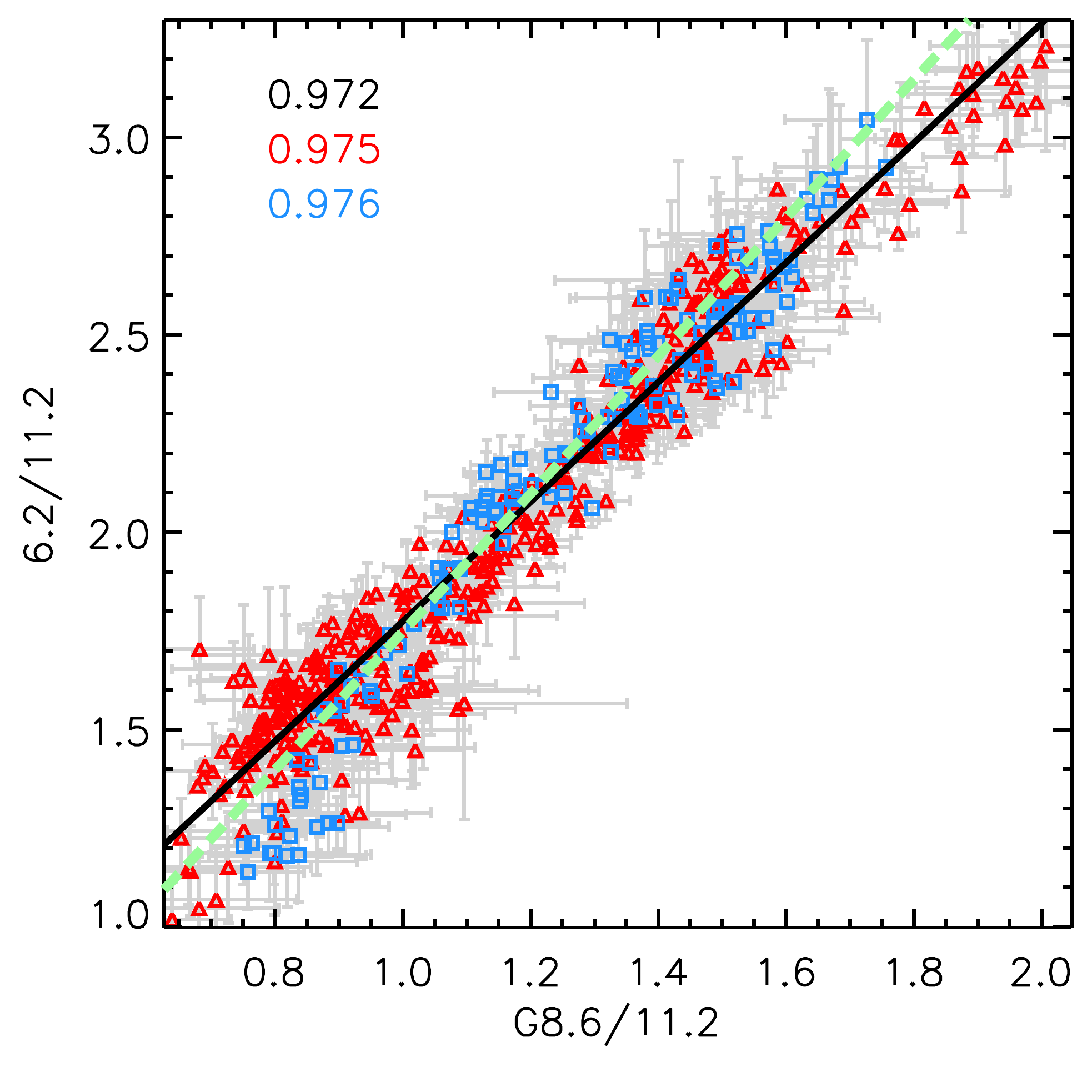}
  \includegraphics{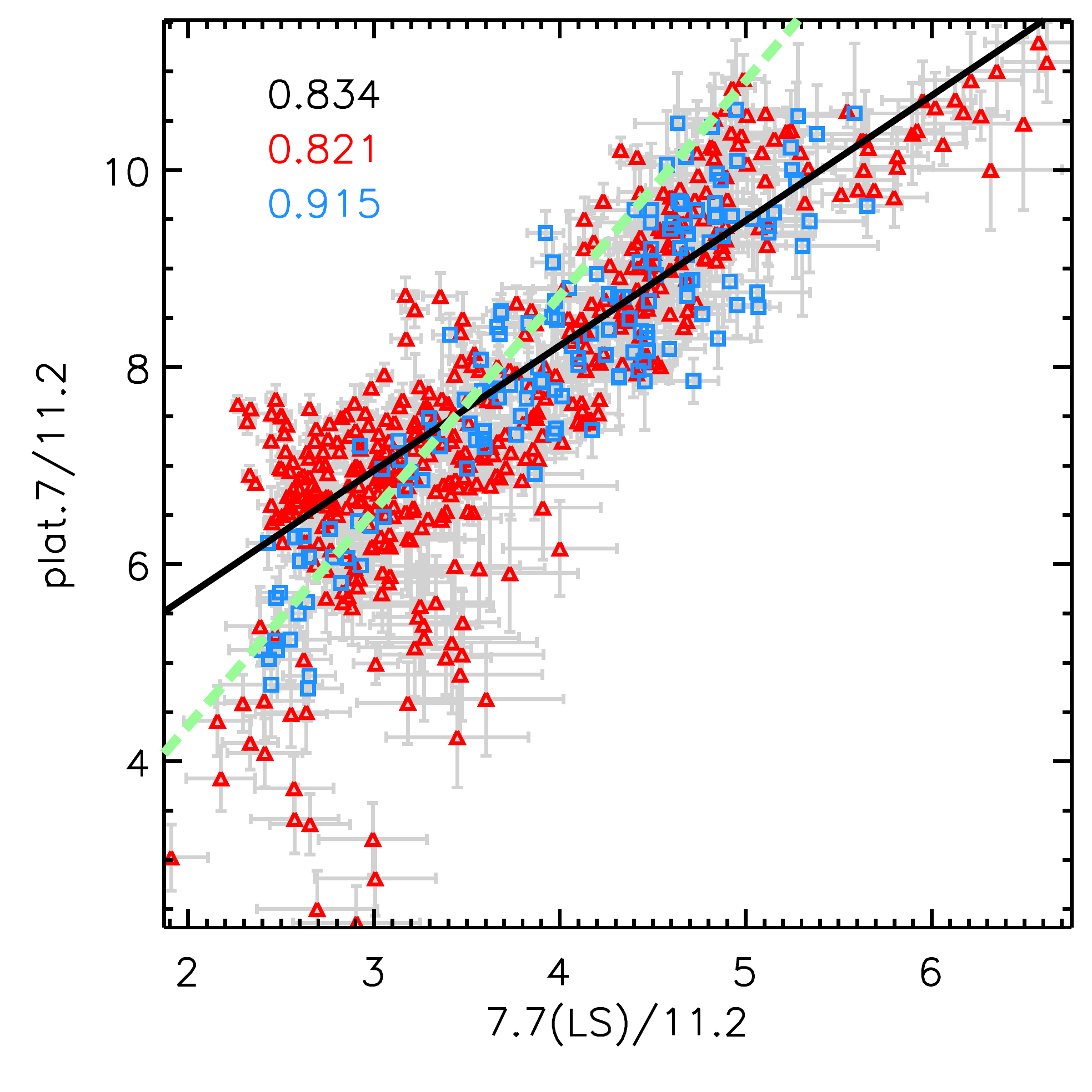}}
\resizebox{\hsize}{!}{%
  \includegraphics{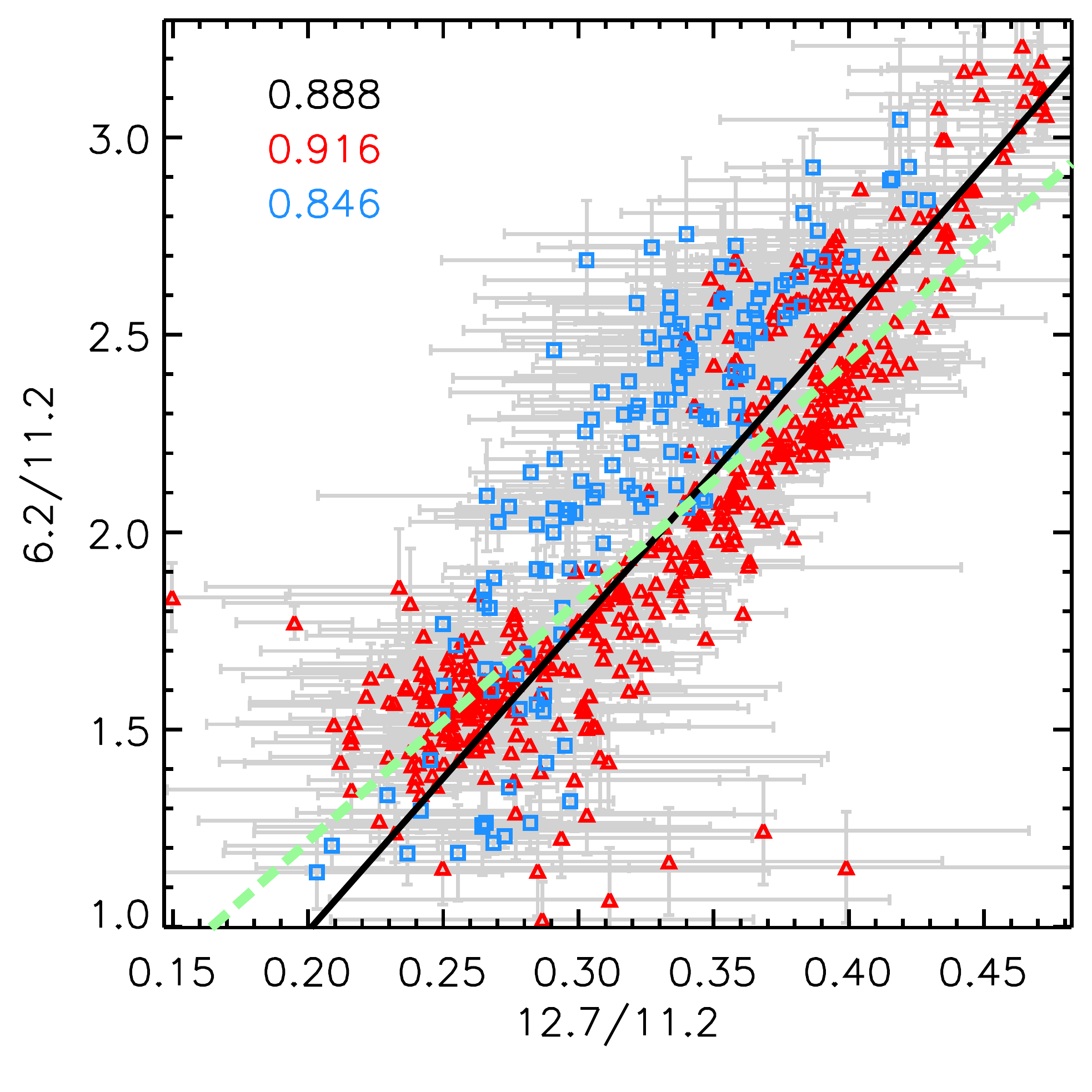}
  \includegraphics{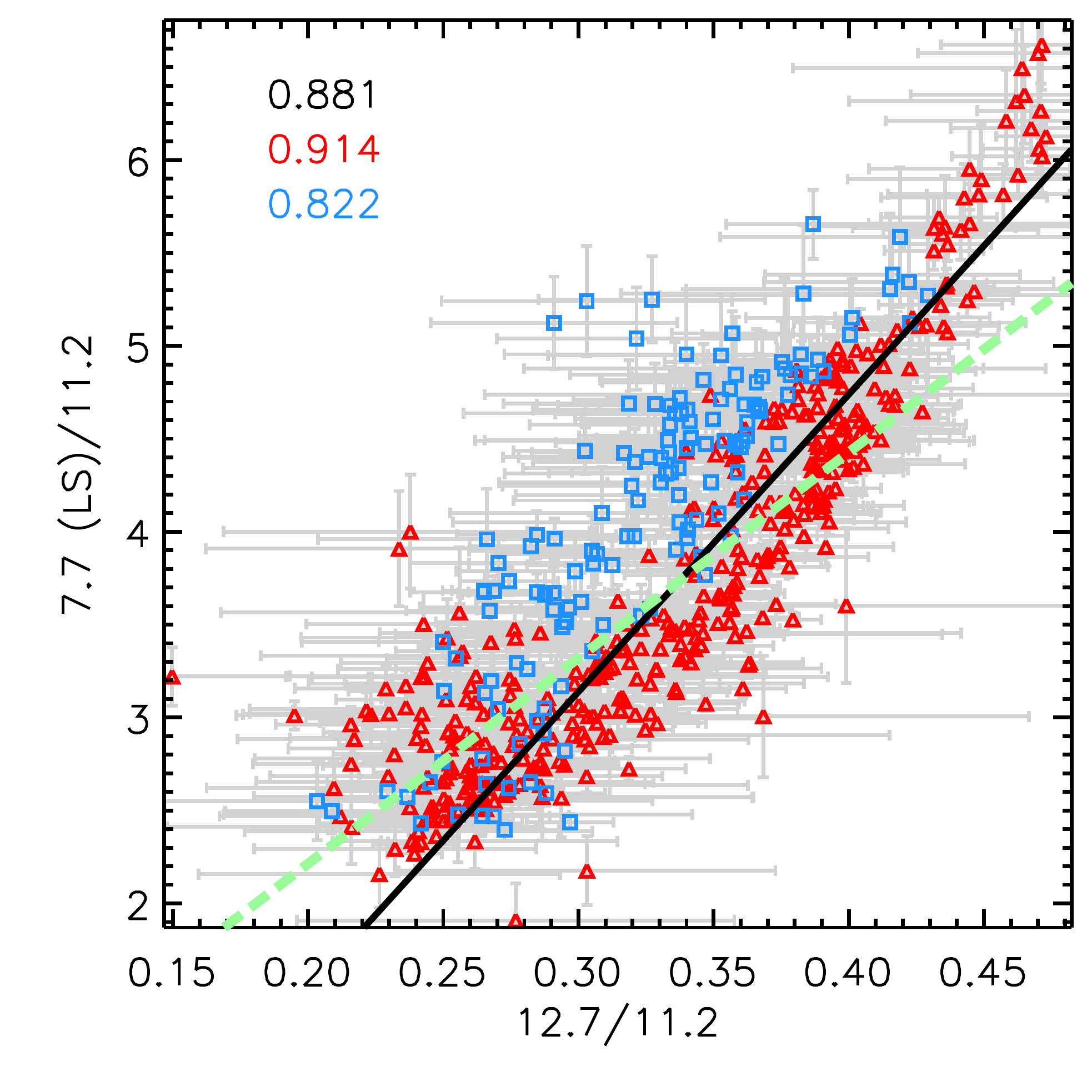}
  \includegraphics{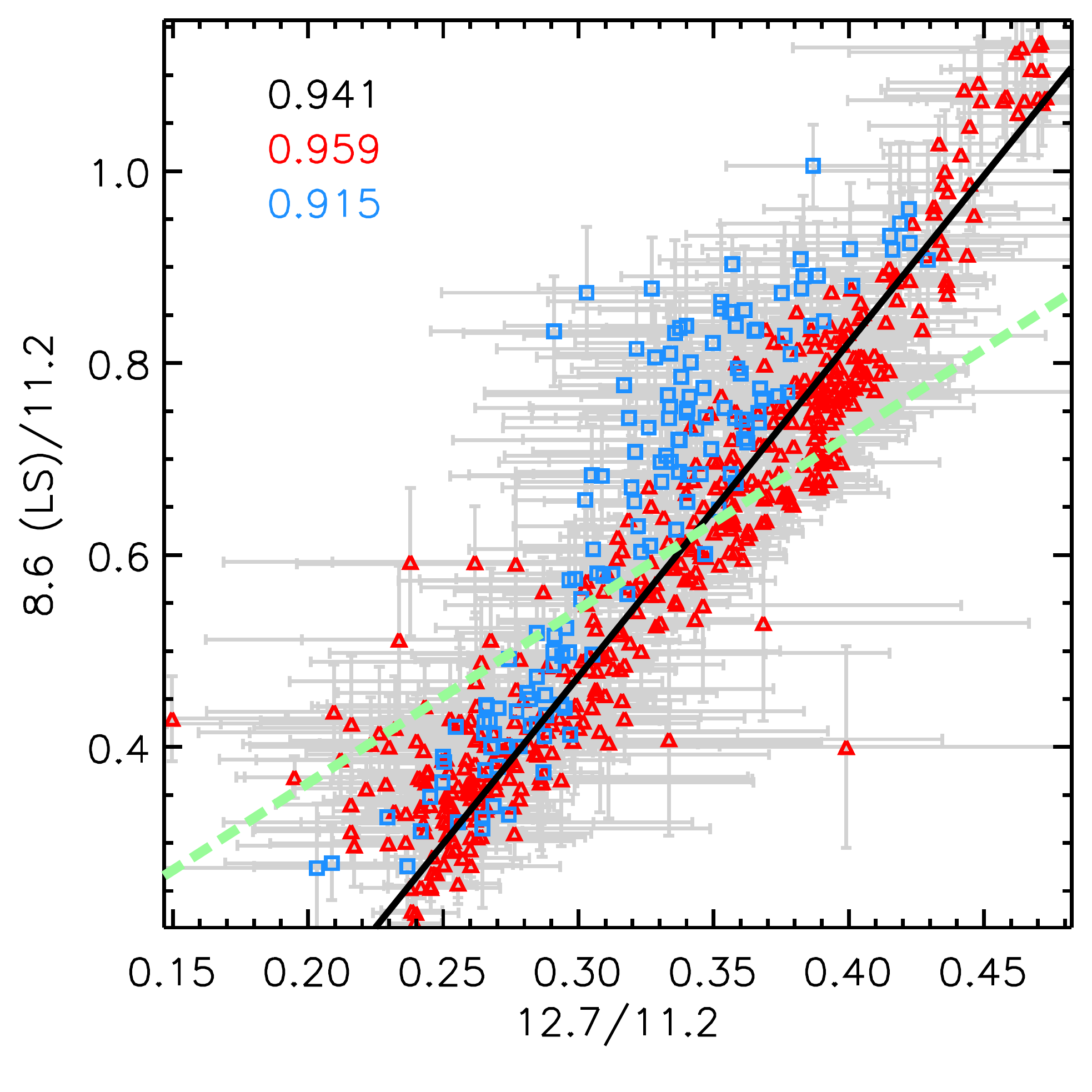}
  \includegraphics{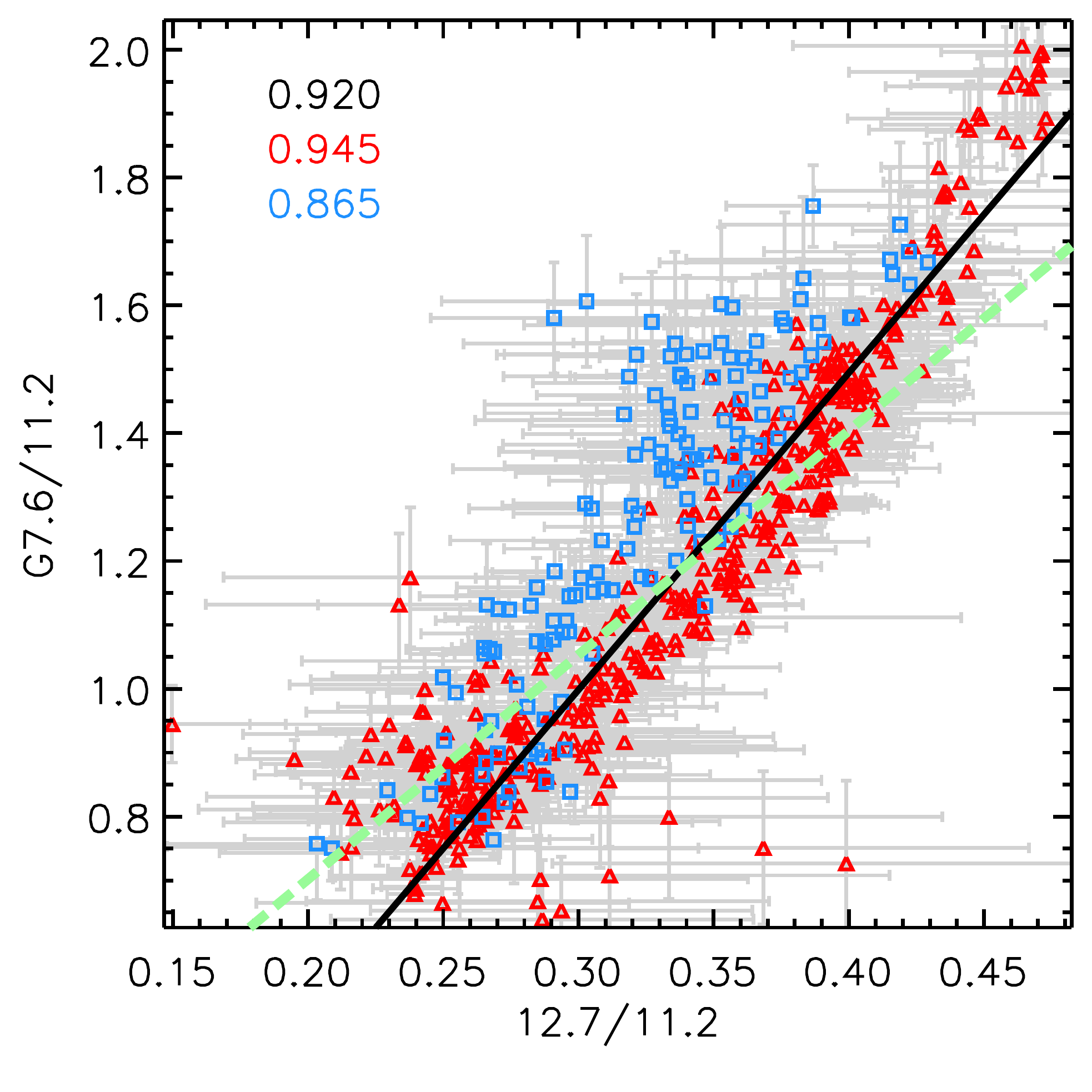}}
\resizebox{\hsize}{!}{%
 \includegraphics{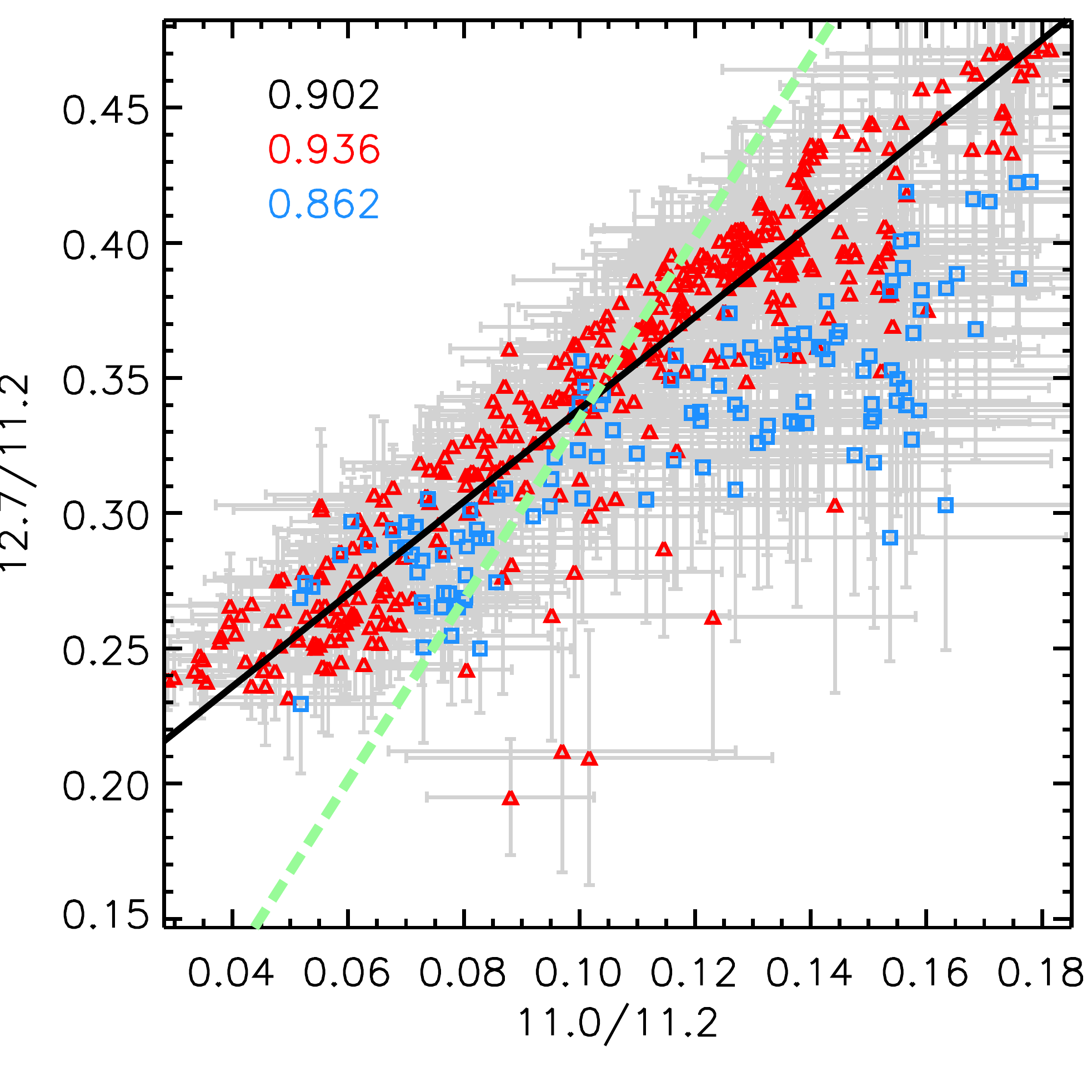}
  \includegraphics{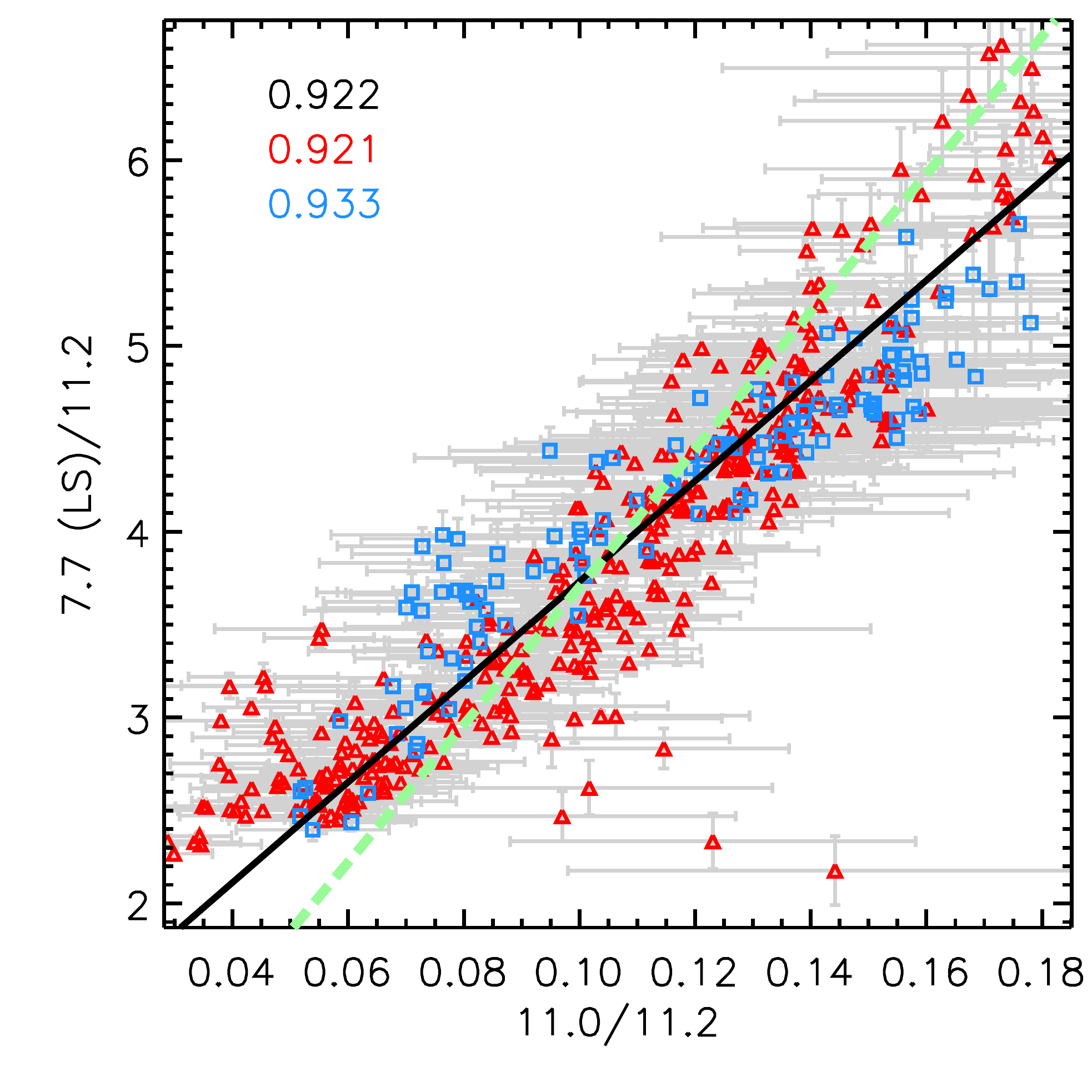}
  \includegraphics{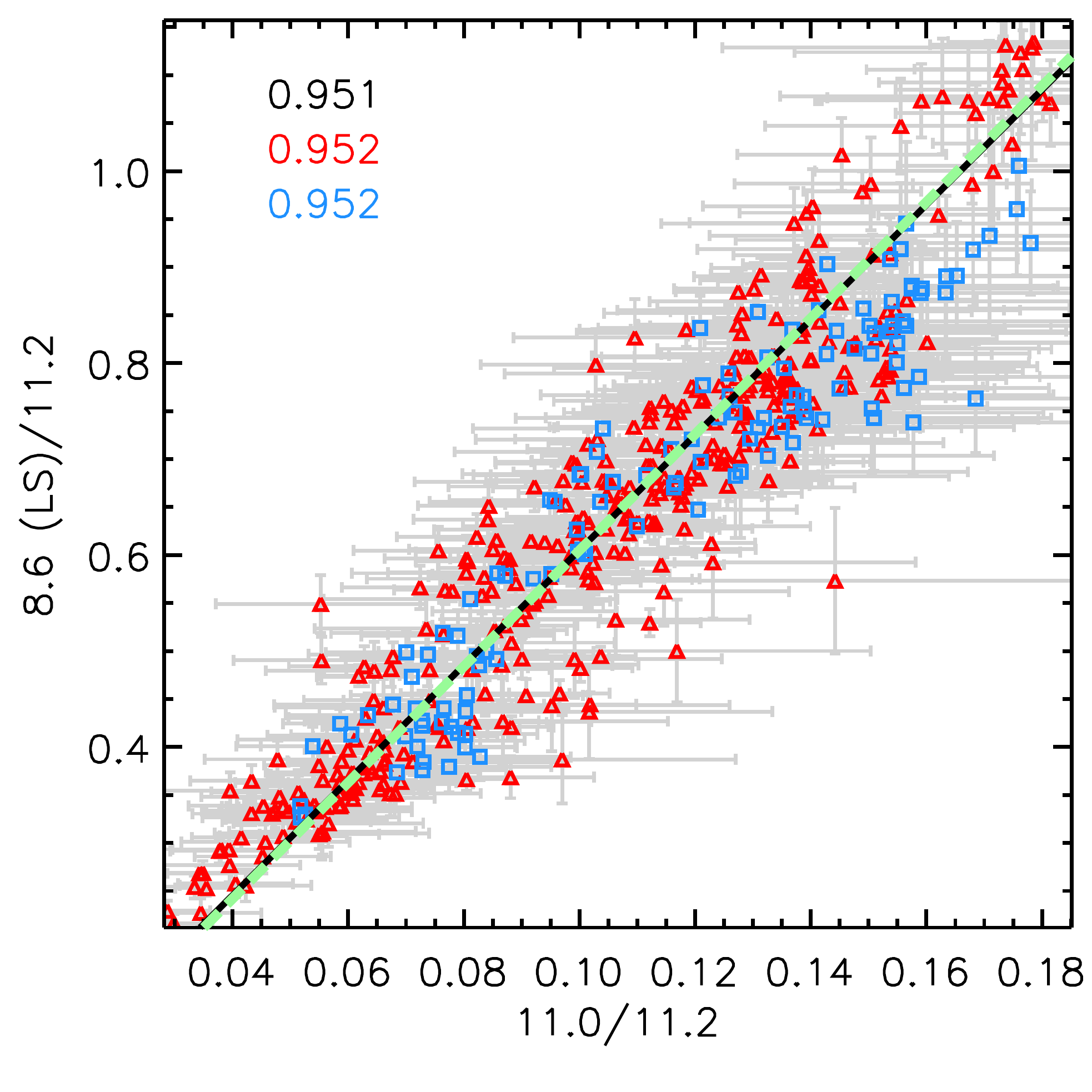}
  \includegraphics{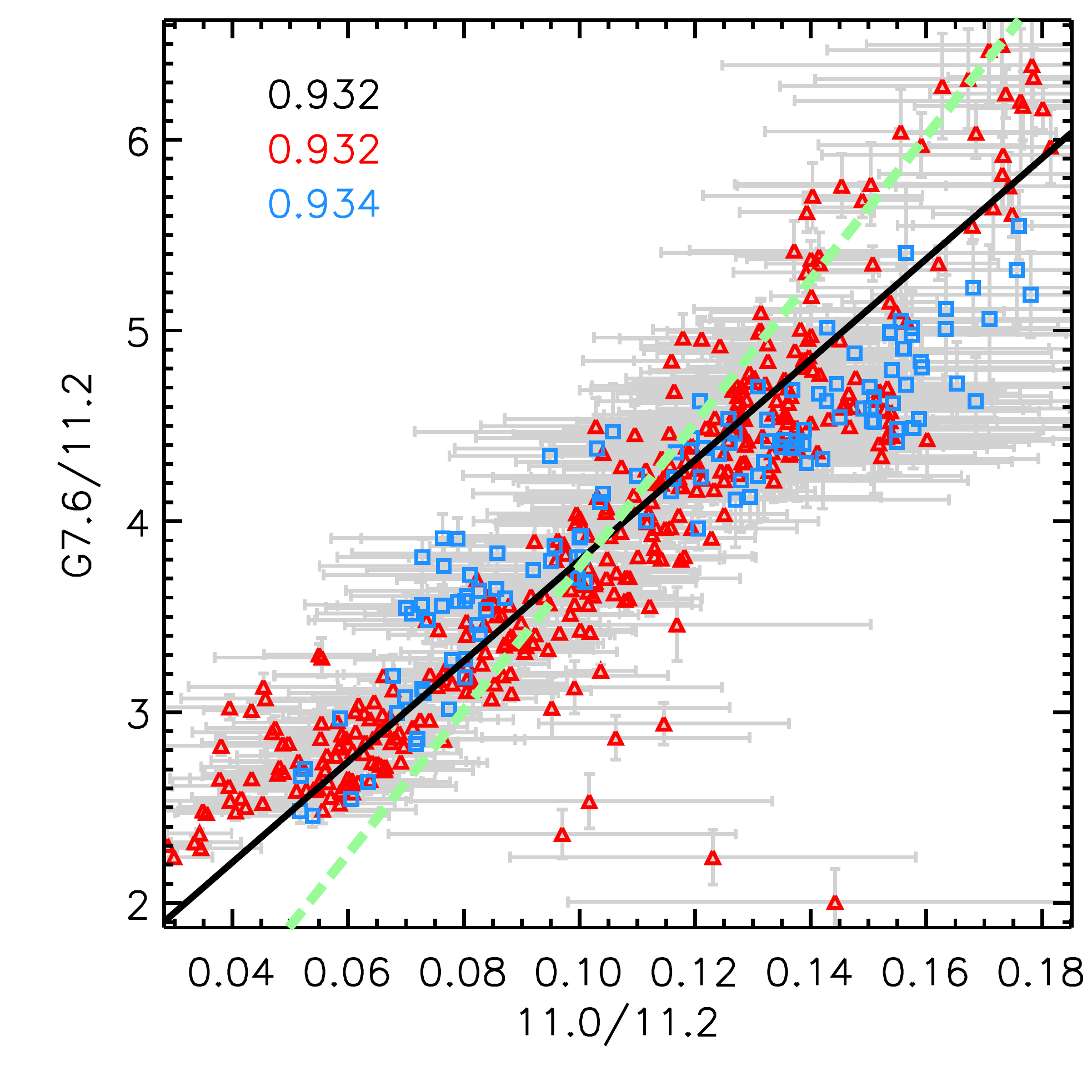}}
\caption{Correlations in PAH SL intensity ratios across NGC~2023. The red squares are for the south map and the blue triangles for the north map. We applied a cutoff of 3 sigma on the individual features. The weighted correlation coefficient is given in the top left corner for both the north and south map (black), for the south map (red) and for the north map (blue). The Levenberg-Marquardt least-squares minimization fit to the N + S data is shown as a solid line. The dashed line represents a fit to the data forced through (0,0). The fit parameters can be found in Table \ref{fit_parameters}. Note that we changed the normalization for the G7.8 and G8.2 components from the 11.2 \mum\, band to the G8.6 component since the former two components exhibit a spatial morphology similar to that of the 11.2 \mum\, band in the south map. 7.7 (LS) and 8.6 (LS) refer to the 7.7 and 8.6 \mum\, PAH bands when applying a local spline (LS) continuum and plat. 7 refers to the 5--10 \mum\, plateau. }
\label{fig_corr}
\end{figure*}
%%%%%%%%%%%%%%%%%%%%%%%%%%%%%%%%%%%%%%%%%%%%%%%%%%
	
The variety in the spatial distribution of the different emission components is even more pronounced in the north map (Fig.~\ref{fig_slmaps_n}). 
The 11.2 \mum\, PAH emission peaks in the NW ridge. In contrast with the south map, the continuum flux and the 10-15 \mum\, plateau are displaced from the 11.2 \mum\, PAH emission and peaks in the N ridge while the 5-10 \mum\, plateau and the 8 \mum\, bump trace both the N and (part of) the NW ridge.  The 6.2 and 7.7 \mum\, PAH emission are again very similar and differ from the 11.2 \mum\, PAH emission. While they also peak in the NW ridge as does the 11.2 \mum\, PAH feature, they show decreased emission in the northern part of this ridge and extend towards the west in the southern part of this ridge (note the difference in the black and pink contours which trace the 11.2 and 7.7 \mum\, PAH bands respectively). The morphology of the 12.7 \mum\, PAH emission is a bit of both that of the 11.2 and 7.7 \mum\, PAH emission. The 8.6 \mum\, emission is distinct from the 6.2 and 7.7 \mum\, PAH emission and peaks slightly west of the southern part of the NW ridge. This is also seen in the spatial distribution of the 11.0 \mum\, PAH which peaks even further west of the southern part of the NW ridge compared to the 8.6 \mum\, PAH emission. The 6.0 \mum\, PAH emission in the north map, as for the south map, has a unique morphology which is highly concentrated and peaks at the intersection of the N and NW ridge. The spatial distribution of the H$_2$ line intensities varies as well: the S(3) [9.7 \mum] intensity peaks between the maxima in the 11.2 and 7.7 \mum\, PAH intensities along NW ridge and the S(2) at 12.3 \mum\ intensity peaks in N ridge. These S(2) and S(3) distributions are confirmed by the maps obtained with PAHFIT for both the north and south FOVs (see Appendix \ref{pahfit}).   

Fig.~\ref{fig_corr} shows observed intensity correlations; their fit parameters and correlation coefficients can be found in table~\ref{fittingparameters} and their line cuts in Fig.~\ref{linecuts_ratios}. The well known, very tight correlation between the 6.2 and 7.7 \mum\, PAH bands is also observed within our sample, consistent with their similar spatial morphology. Note that the observed correlation is close to a 1:1 correlation (i.e. through (0,0)).  Surprisingly, the 8.6 \mum\, band correlates with the 6.2 and 7.7 \mum\, bands despite their differing spatial distributions. However, this correlation is not as tight as that between the 6.2 and 7.7 \mum\, bands, but overall, in keeping with their close but differing spatial distributions.  The enhanced scatter in the correlation of the 8.6 \mum\, band with the 6.2 and 7.7 \mum\, bands has regularly been attributed to the influence of extinction and/or the larger uncertainty in determining the 8.6 \mum\, band intensity.  However, we found that larger deviations from the line of best fit are found in locations where the spatial distribution of the bands is different, i.e., the observed scatter originates in the distinct spatial distributions. The 11.0 \mum\, PAH band correlates best with the 8.6 \mum\, PAH band although not as tight as the 6.2 with the 7.7 \mum\, PAH bands. The 11.0 vs. 8.6 correlation exhibits exactly a 1:1 dependence (i.e. the best fit goes through (0,0); see Fig.~\ref{fig_corr}) and has a high correlation coefficient. This may not be immediately clear from their spatial distribution for the south map as shown in Fig.~\ref{fig_slmaps_s} which is attributed to the bottom two rows (in y-direction; see Appendix ~\ref{scaled_maps} for details). The 11.0 \mum\, PAH band also correlates with the 6.2 and 7.7 \mum\, PAH bands but with slightly more scatter. Here as well, the 11.0 \mum\, PAH emission has a different spatial distribution than the 8.6, 6.2 and 7.7 \mum\, PAH emission resulting in enhanced scatter in these correlation plots and thus a lower correlation coefficient (ranging from 0.949 to 0.958 compared to 0.978 for the 6.2 vs. 7.7 correlation). The 12.7 \mum\, PAH band also shows a dependence on the 6.2, 7.7, 8.6 and 11.0 \mum\, PAH bands but this connection is clearly weaker than those amongst the 6.2, 7.7, 8.6 and 11.0 \mum\, bands (with correlation coefficients ranging between 0.930 to 0.945).   
 
\setcounter{figure}{8}
 %%%%%%%%%%%%%%%%%%%%%%%%%%%%%%%%%%%%%%%%%%%%%%%%%%
\begin{figure*}[t]
    \centering
\resizebox{\hsize}{!}{%
  \includegraphics[angle=266.4]{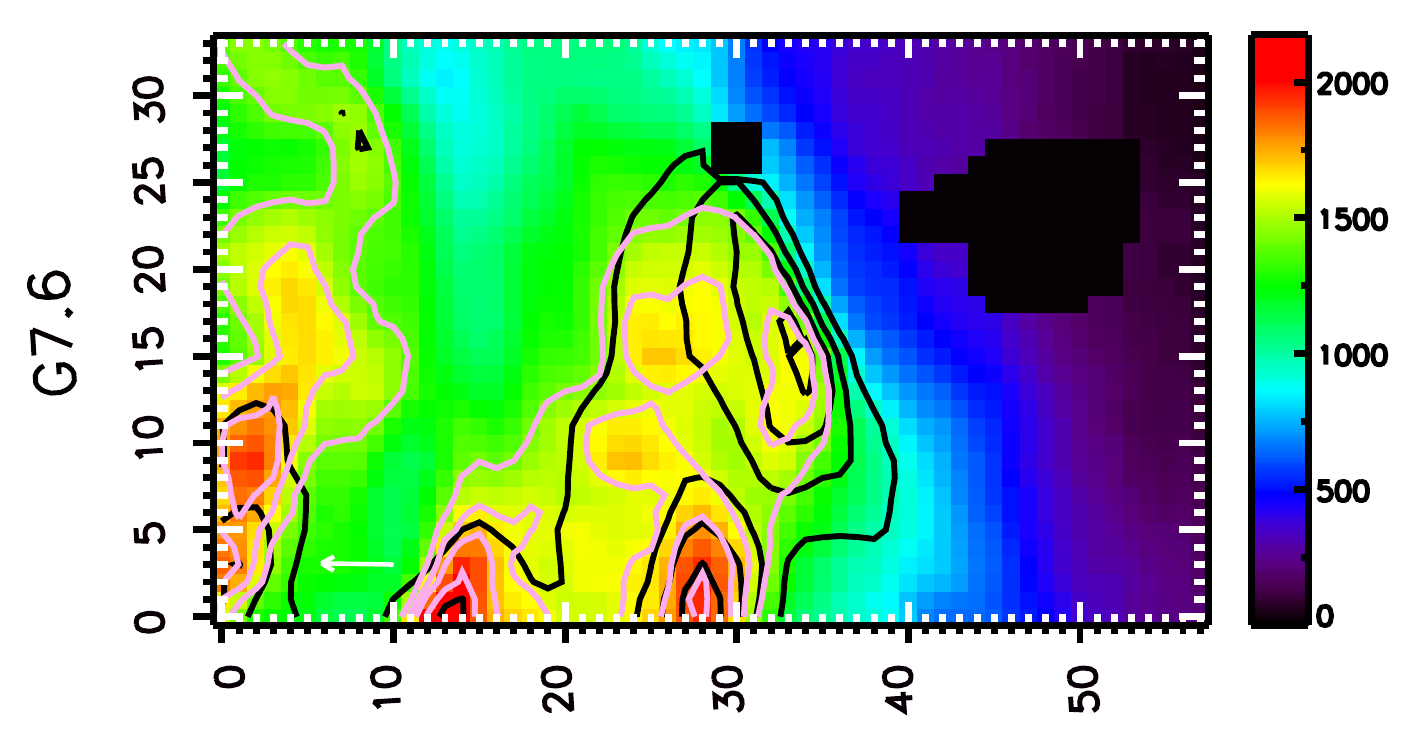}
  \includegraphics[angle=266.4]{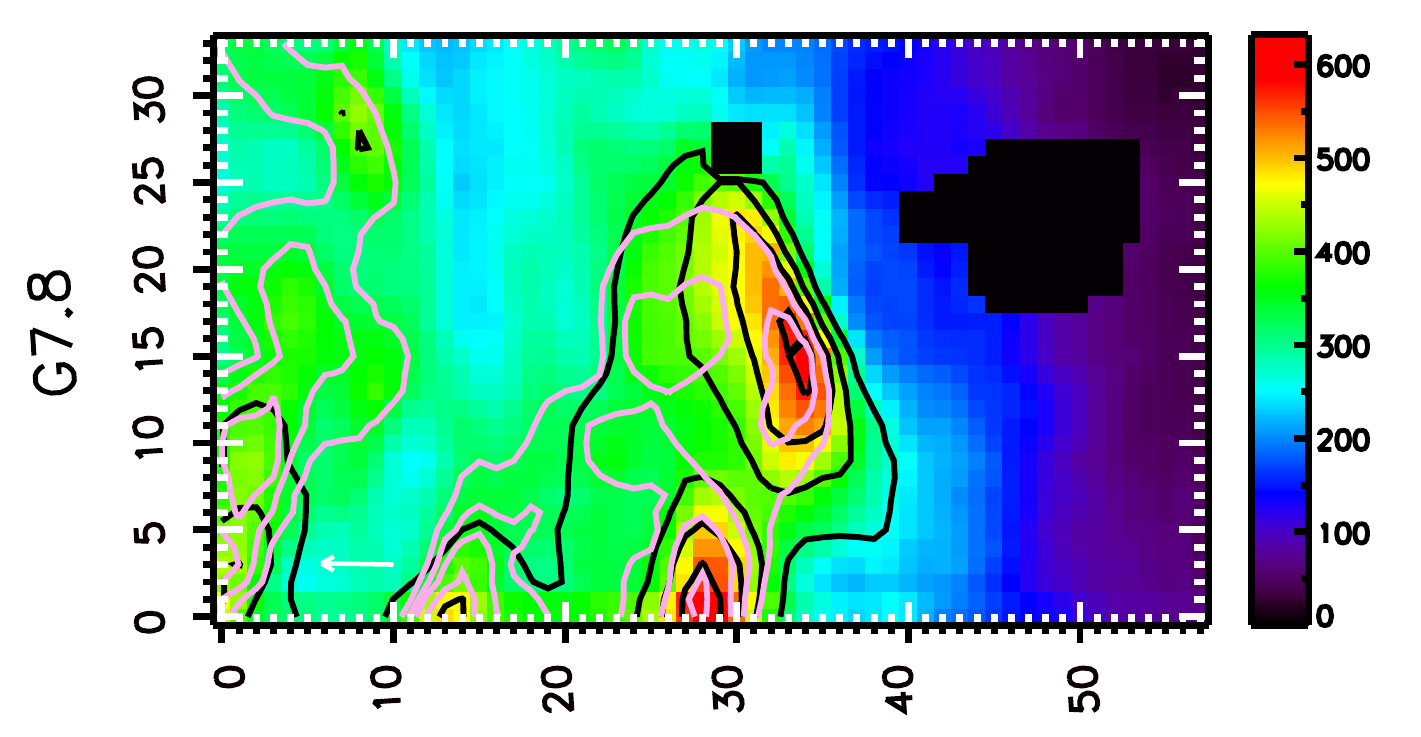}
  \includegraphics[angle=266.4]{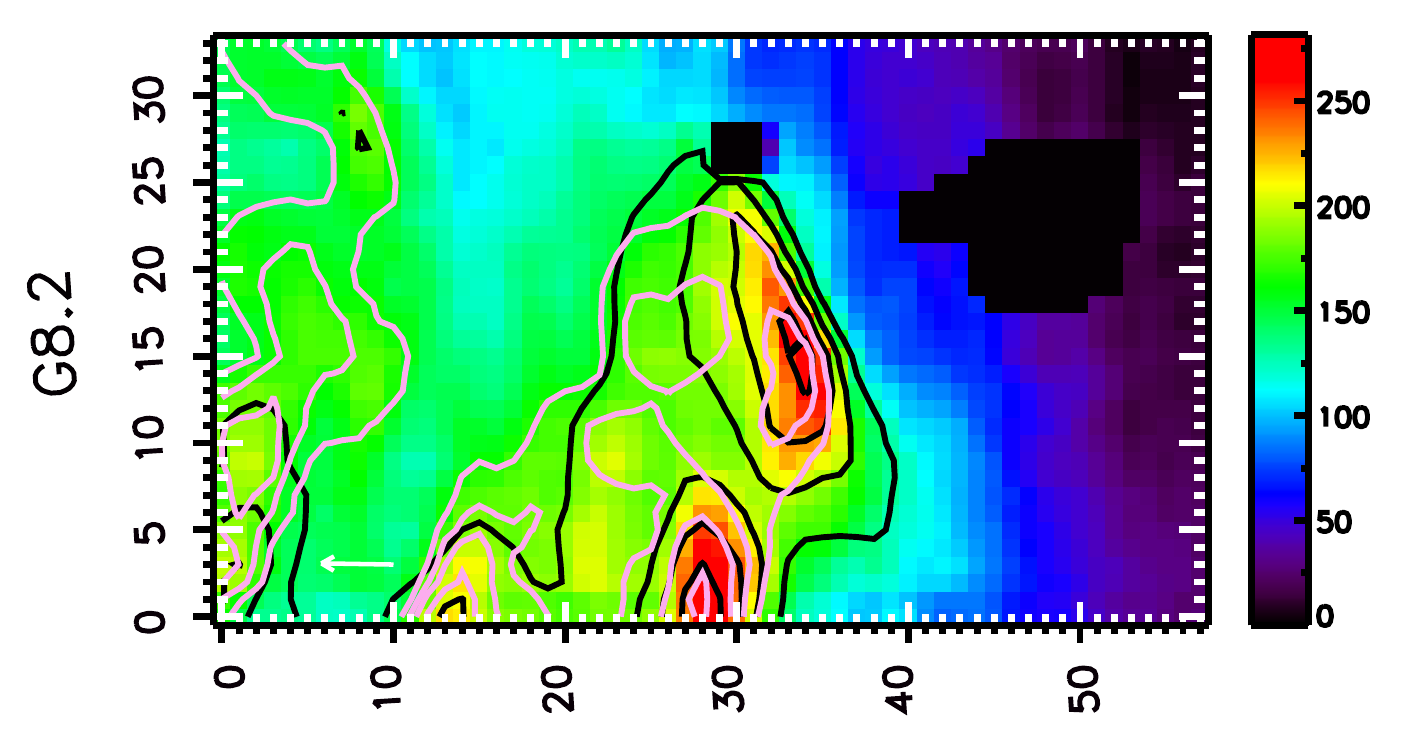}
    \includegraphics[angle=266.4]{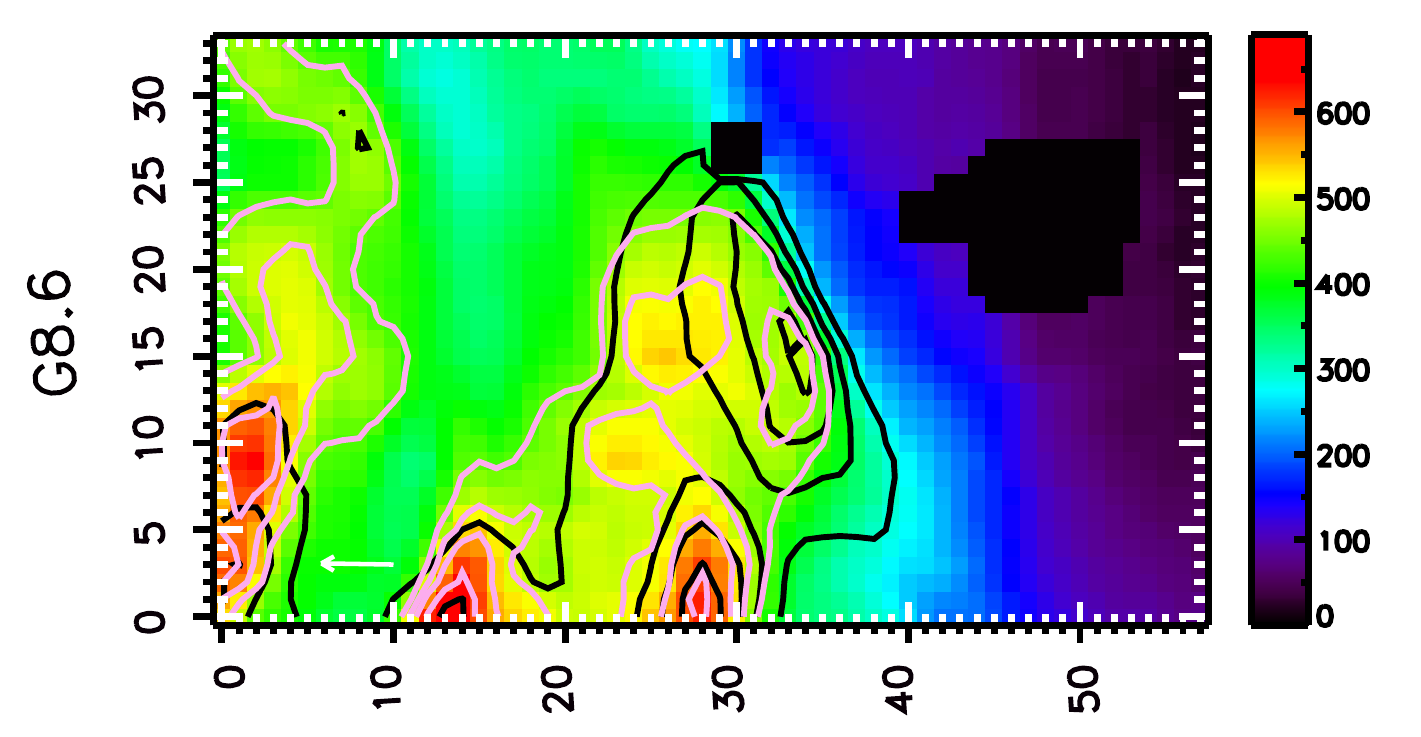}
    \includegraphics[angle=266.4]{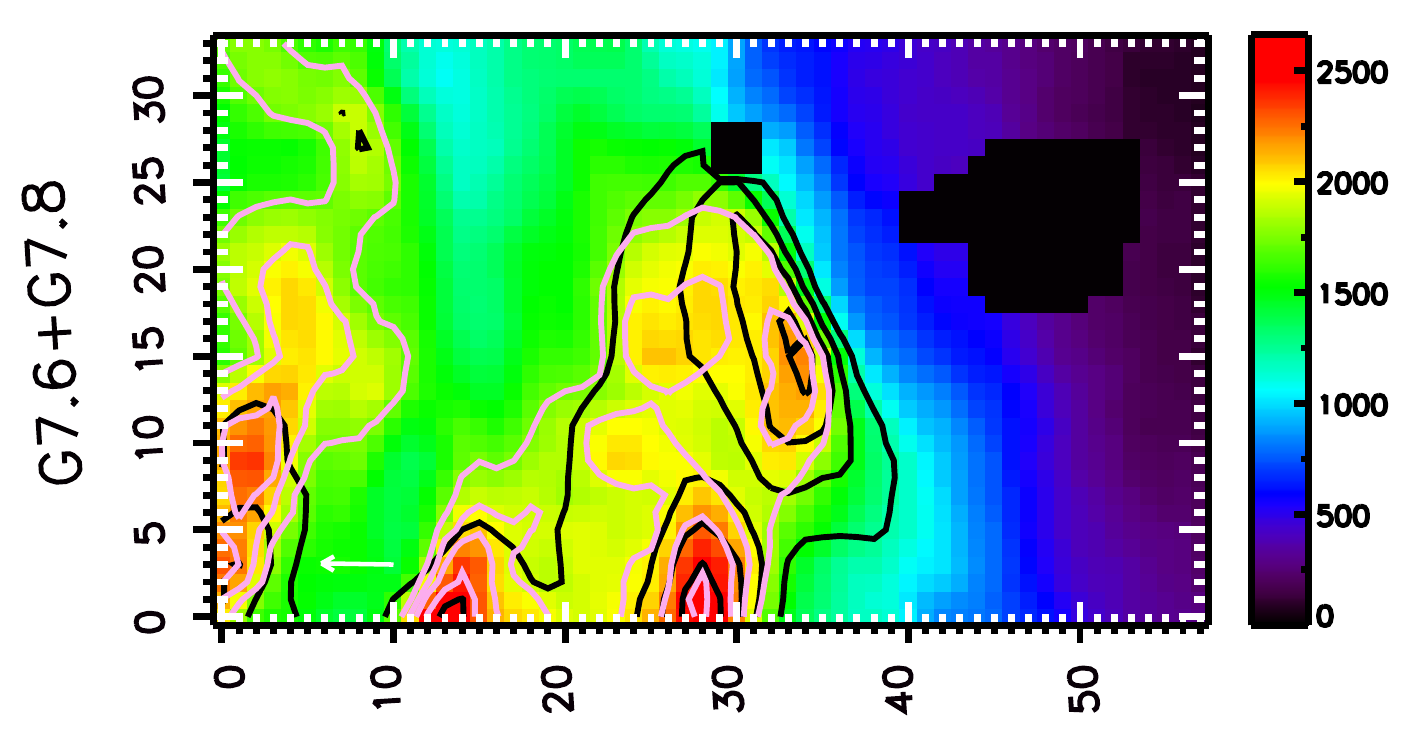}}
\resizebox{\hsize}{!}{%
  \includegraphics[angle=274.1]{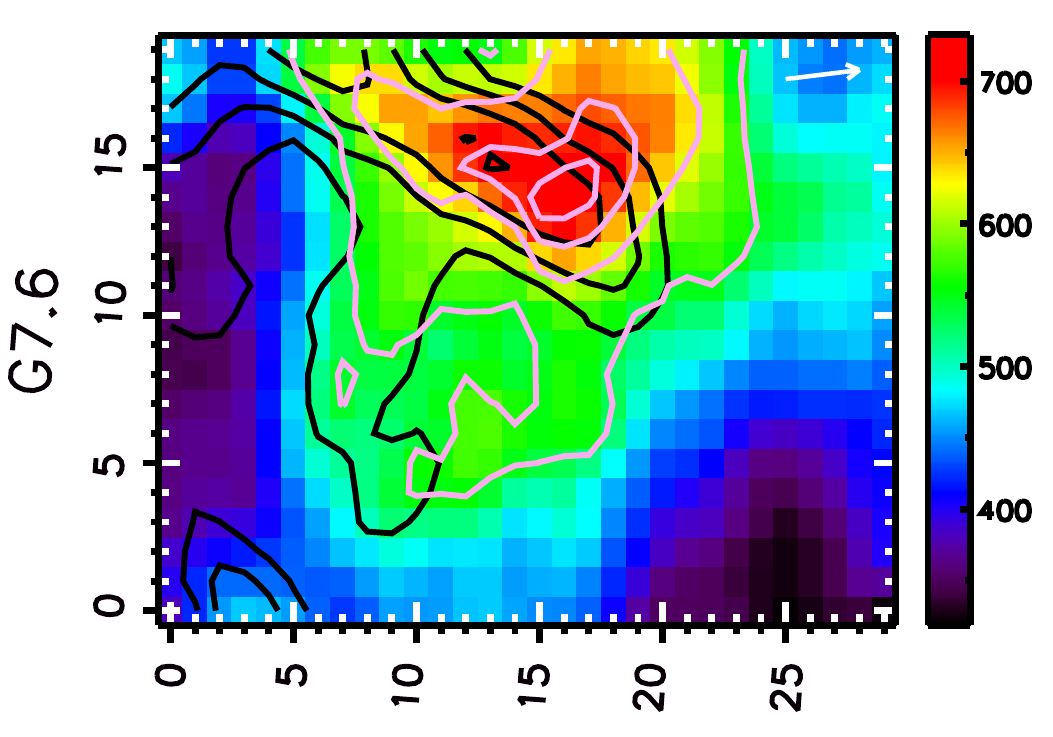}
  \includegraphics[angle=274.1]{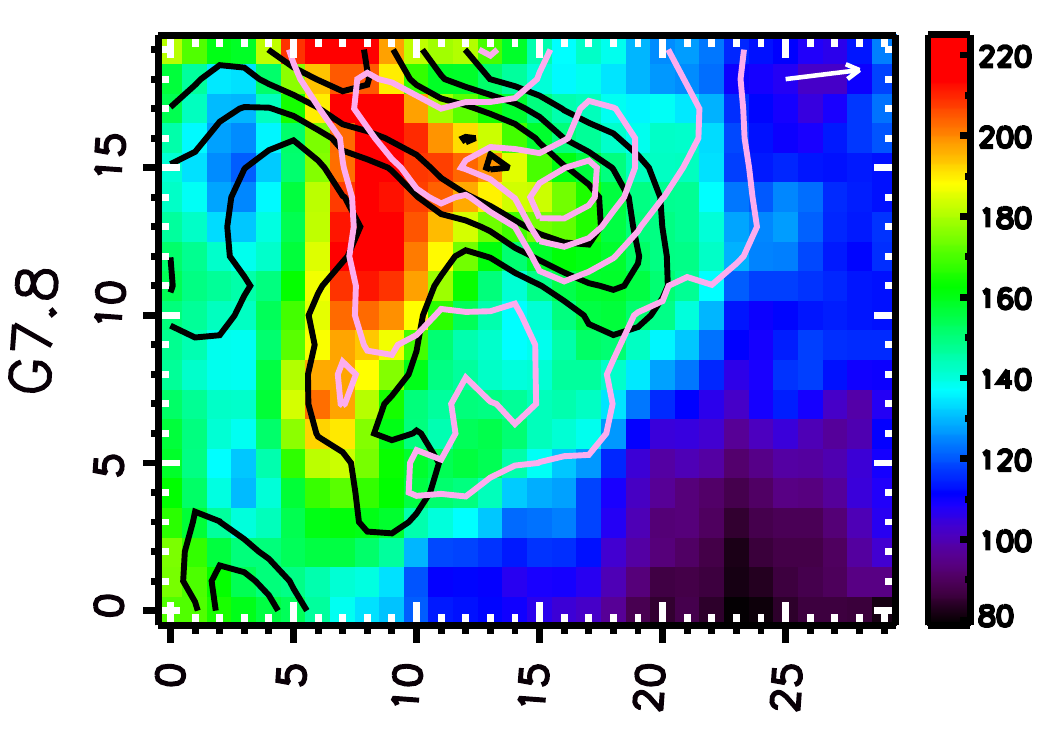}
  \includegraphics[angle=274.1]{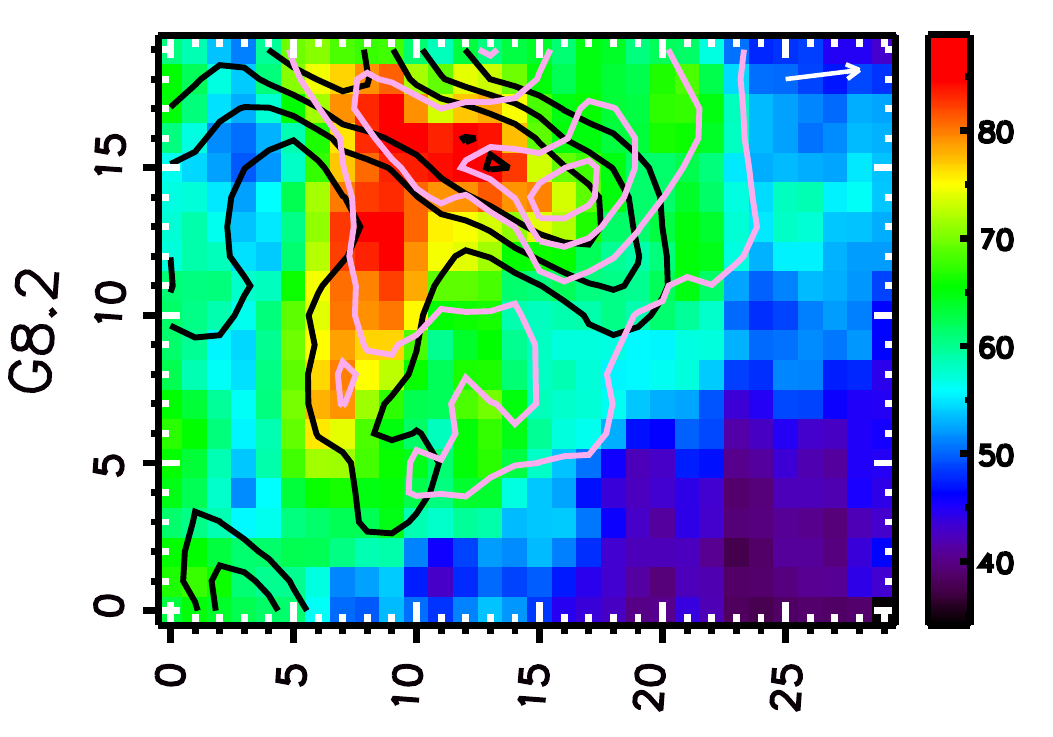}
     \includegraphics[angle=274.1]{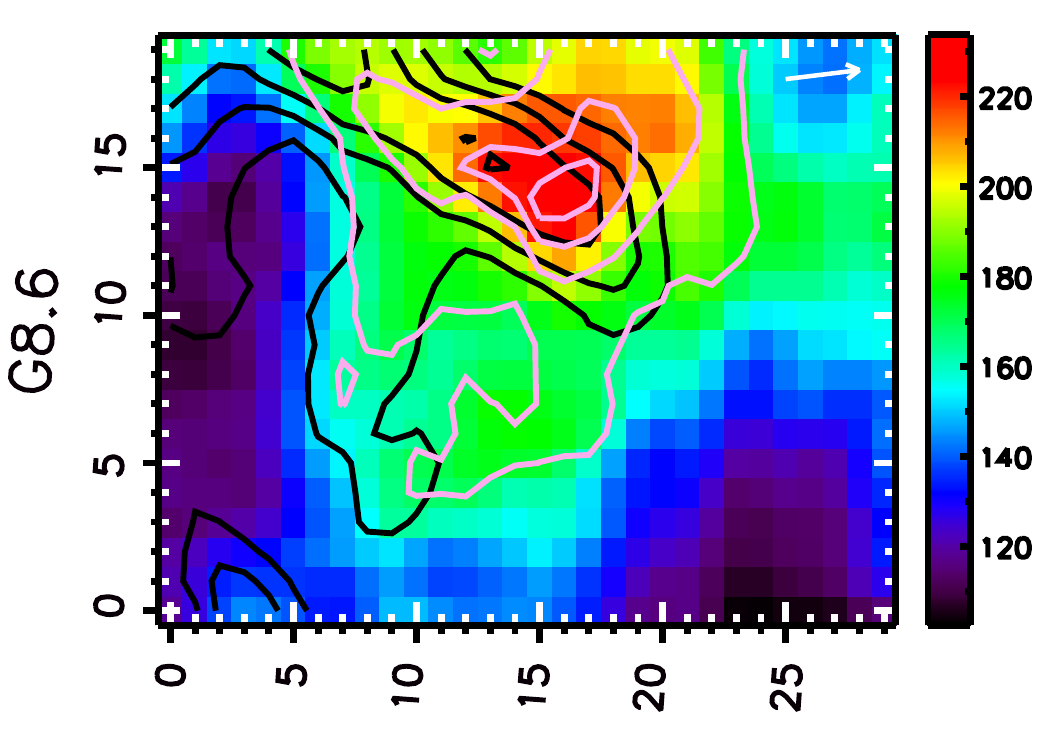}
      \includegraphics[angle=274.1]{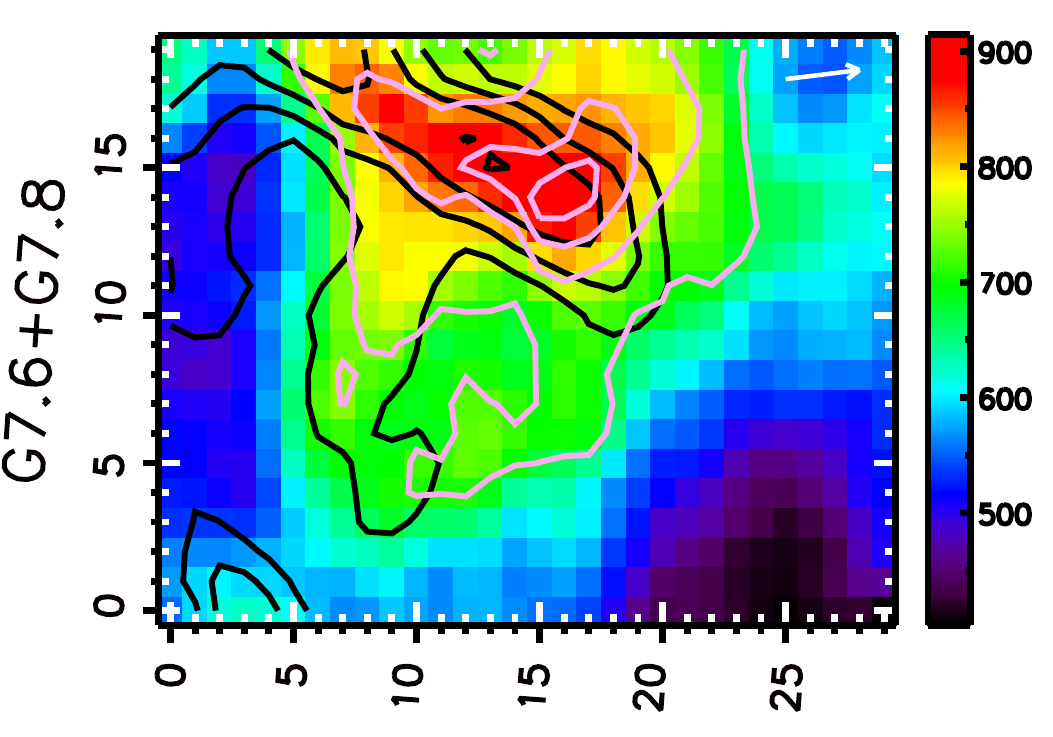}}
\caption{Spatial distribution of the four Gaussian components used in the decomposition of the 7 to 9 \mum\, region  when applying a global spline continuum for the south map (top panels) and north map (bottom panels). As a reference, the intensity profiles of the 11.2 and 7.7 \mum\, emission features are shown as contours in black and pink, respectively. Units, map orientation and symbols are the same as in Figs. \ref{fig_slmaps_s} and \ref{fig_slmaps_n}. }
\label{fig_maps_decomp}
\end{figure*}
%%%%%%%%%%%%%%%%%%%%%%%%%%%%%%%%%%%%%%%%%%%%%%%%%%

\setcounter{figure}{7}
%%%%%%%%%%%%%%%%%%%%%%%%%%%%%%%%%%%%%%%%%%%%%%%%%%
\begin{figure}[t]
    \centering
\resizebox{5cm}{!}{%
   \includegraphics{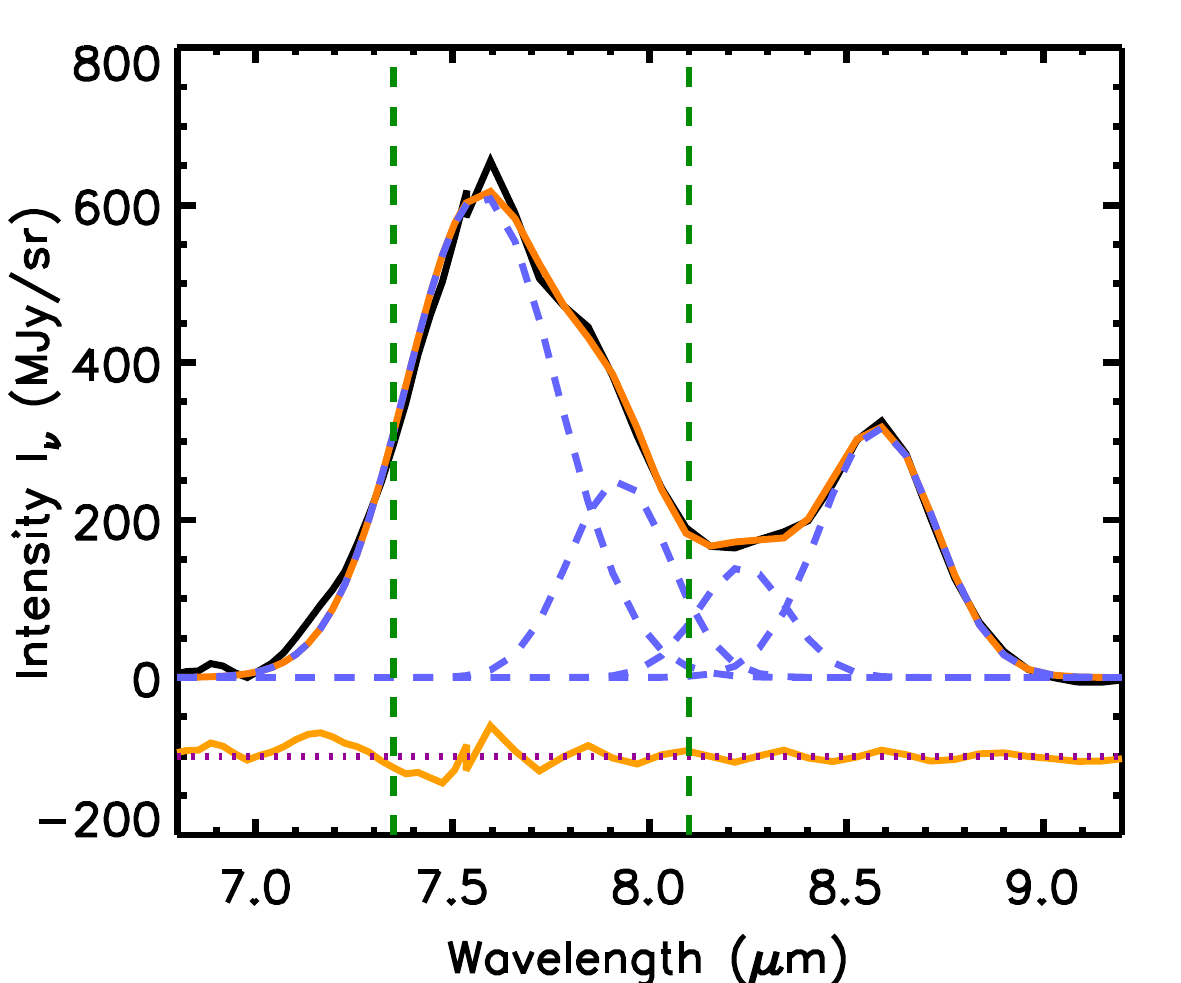}}
\caption{A typical Gaussian decomposition of the continuum subtracted spectrum (using the global spline continuum) in the 7 -- 9 \mum\, region into the G7.6, G7.8, G8.2 and G8.6 components. The data are shown by a solid black line, the fit by a solid red line, the four individual Gaussians by striped cyan lines, and the residuals by a solid gold line. The latter are offset by -100\,MJy/sr. Green striped lines indicate 7.35 and 8.1 \mum. See Sect. \ref{decomp7} for details on the composition. The four Gaussians have a peak position (FWHM) of 7.59 (0.45), 7.93 (0.3), 8.25 (0.27), and 8.58 (0.344) \mum\, respectively. }
\label{fig_decomp7}
\end{figure}
%%%%%%%%%%%%%%%%%%%%%%%%%%%%%%%%%%%%%%%%%%%%%%%%%%
\setcounter{figure}{9}

%%%%%%%%%%%%%%%%%%%%%%%%%%%%%%%%%%%%%%%%%%%%%%%%%%
\begin{figure}[t]
    \centering
\resizebox{6.7cm}{!}{%
  \includegraphics[angle=266.4]{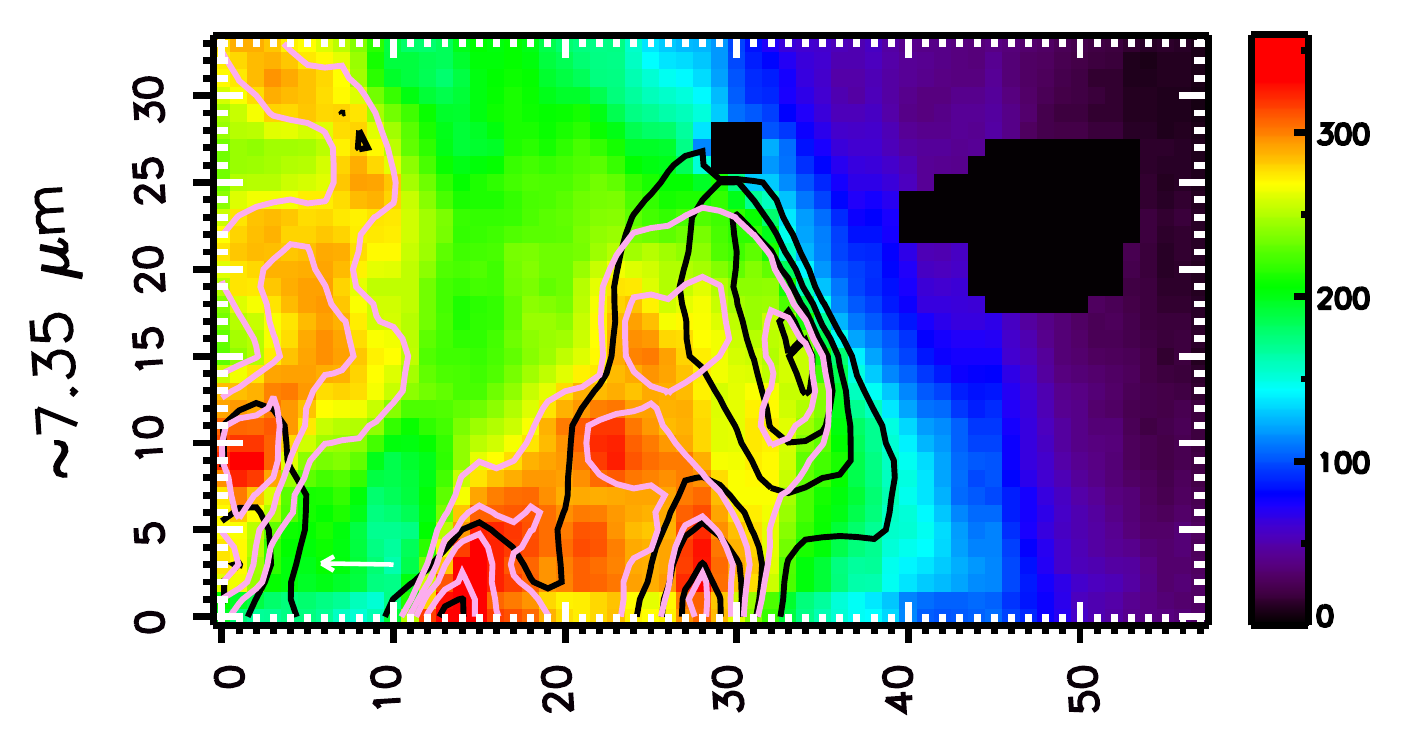}
  \includegraphics[angle=266.4]{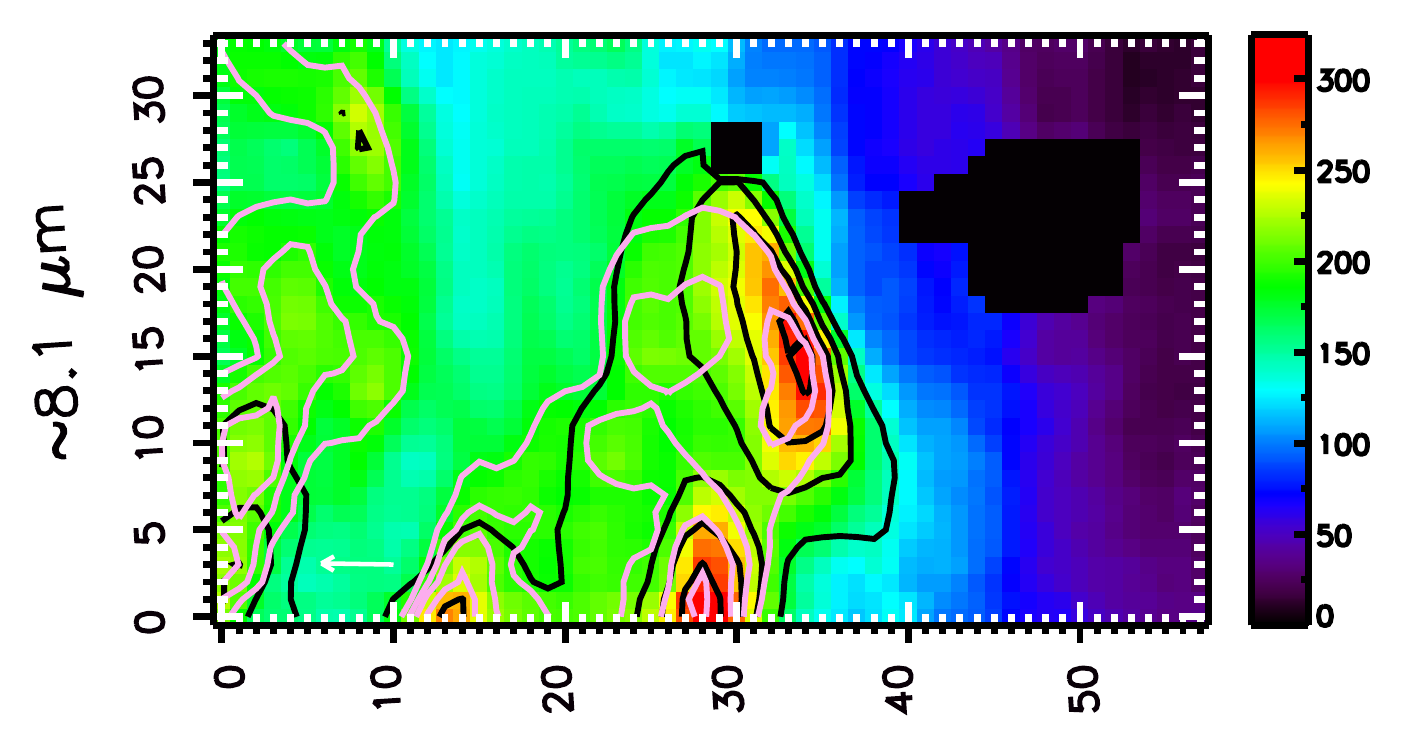}}
 \resizebox{6.7cm}{!}{%
 \includegraphics[angle=274.1]{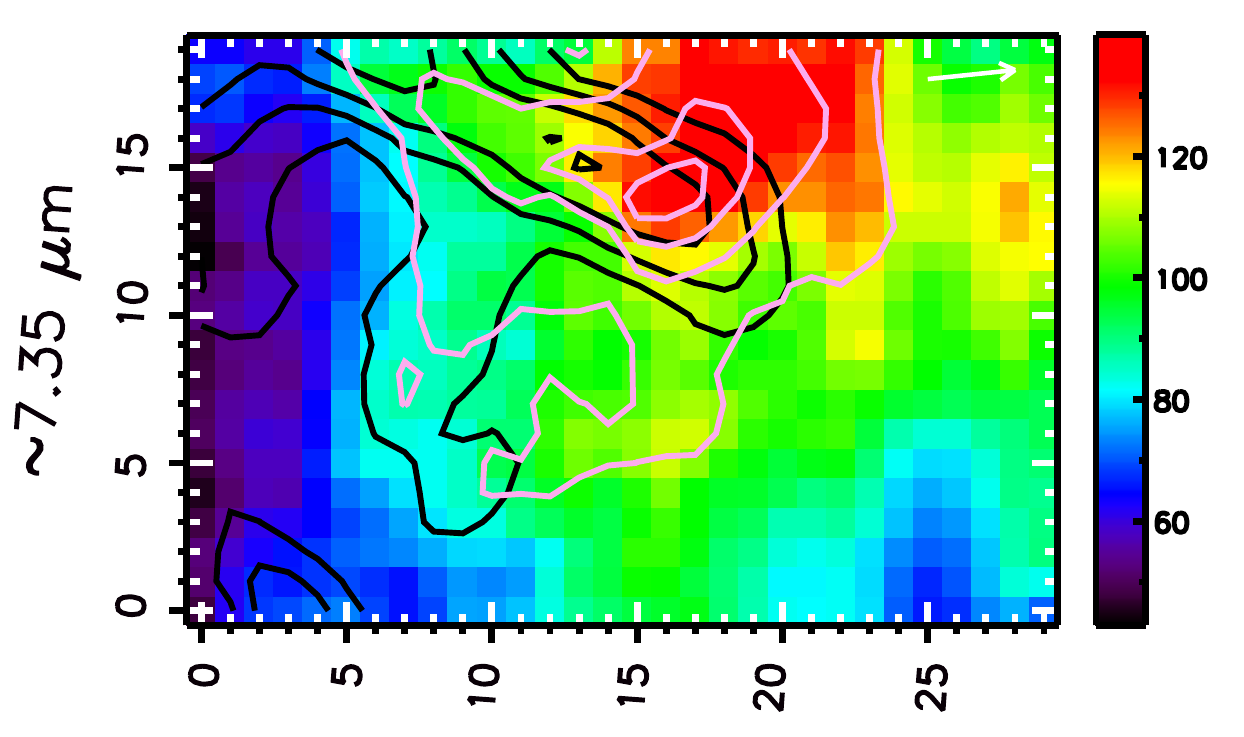}
  \includegraphics[angle=274.1]{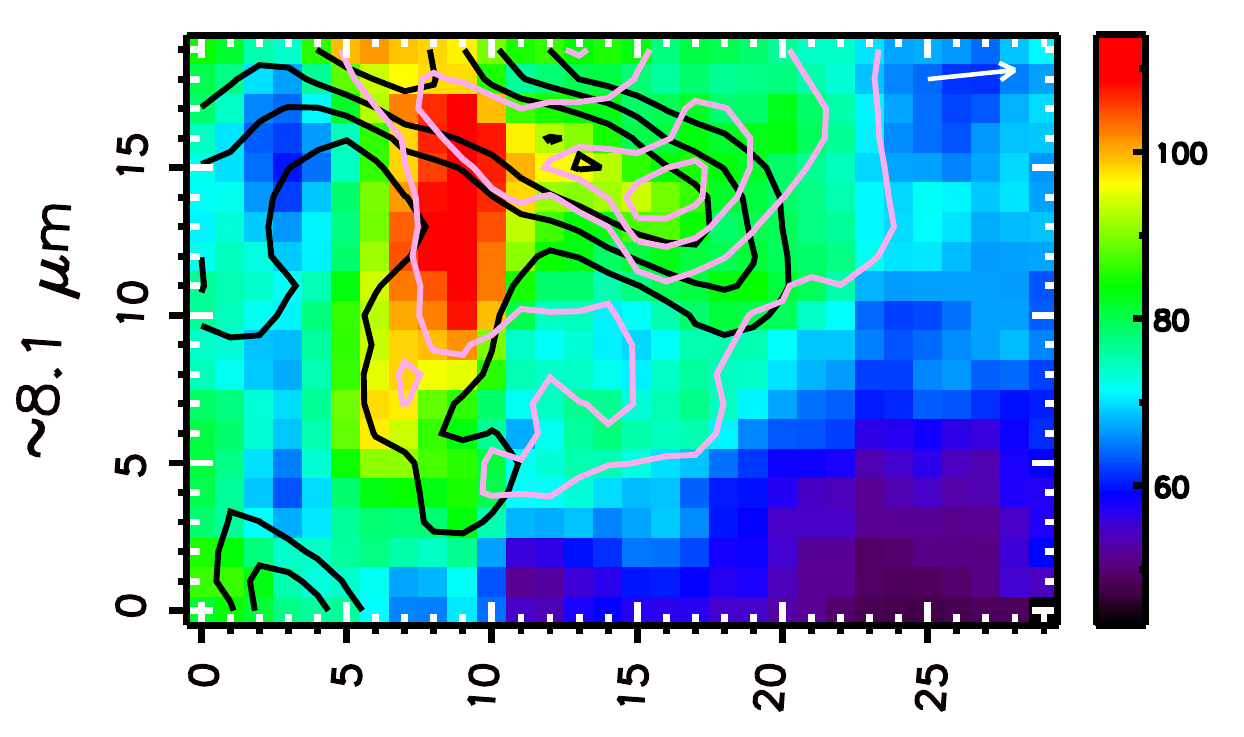}}
\caption{A set of still frames from an animation showing spatial maps of the
    intensity $I_\nu$ of the continuum-subtracted spectra (applying the
    global spline continuum) as a function of wavelength for the south map
    (top panels) and north map (bottom panels). The two images show the two
    extreme spatial distributions found in the 7 to 9 \mum\, region. As a
    reference, the 11.2 and 7.7 \mum\, band intensities are shown as
    contours in black and pink, respectively. Units, map orientation and
    symbols are the same as in Figs. \ref{fig_slmaps_s} and
    \ref{fig_slmaps_n}. The full animated south and north maps are available 
    in the online version of the Journal. In the animated maps the individual 
    6-7, 7-9 and 10.8-13 \mum\, maps begin at the t=0, t=14 and t=50sec 
    timestamps, respectively.}
\label{fig_movie}
\end{figure}
%%%%%%%%%%%%%%%%%%%%%%%%%%%%%%%%%%%%%%%%%%%%%%%%%%

\subsubsection{Decomposition of the 7 to 9 \mum\, region}
\label{decomp7}
The distinct spatial distribution of the 7.7 and 8.6 \mum\, PAH bands prompts further investigation. As discussed in Sect. \ref{cont}, the chosen local spline continuum clearly influences band intensities. However, if an emission feature is due to a single carrier or distinct subset of the PAH population, the spatial distributions of its sub-components should all be identical, independent of how these sub-components have been defined. {\it Hence, the distinct spatial distribution of the 7.7 and 8.6 \mum\, PAH emission indicates that they originate in multiple carriers or loosely related PAH subpopulations.}  
In an attempt to resolve the sub-components of the 7 to 9 \mum\, PAH emission, each related to a single carrier or subset of the PAH population, we subtracted the global spline  continuum (GS, see Fig.~\ref{fig_sp_sl}, magenta line) from the spectra and decomposed the remaining PAH emission into four Gaussians with $\lambda$  (FWHM) of 7.59 (0.450), 7.93 (0.300), 8.25 (0.270), and 8.58 (0.344) \mum\, respectively (see Fig.~\ref{fig_decomp7})\footnote{Note that the G8.2 component is not present when using the local continuum (LS, see Fig.~\ref{fig_sp_sl}, orange line) as for this continuum, an anchor point at 8.2 \mum\, is taken and this component thus becomes part of the 8 \mum\, bump. When using the plateau continuum (see Fig.~\ref{fig_sp_sl}, green line), the G8.2 component sits on top of the plateau emission.}.  These values were obtained  
by taking the average over all spectra when fitted by four Gaussians having peak positions and FWHM that were not fixed but were constrained to fit these four bands. These components are further referred to as G7.6, G7.8, G8.2 and G8.6 bands. We chose four components to represent the 8.6 \mum\, PAH band and the 7.6 and 7.8 \mum\, subcomponents of the 7.7 \mum\, complex, whose ratio, 7.6/7.8, exhibit spatial variation with distance from the star \citep{Bregman:05, Rapacioli:05, Boersma:14}. A fourth component is needed to obtain a good fit in the 7 -- 9 \mum\, region.  We did not include a Gaussian at 7.4 \mum\, because none of the spectra in the map show a 'feature' near 7.4 \mum\, like that observed in the ISO-SWS data of NGC 7023 \citep{Moutou:c60:99}. The  `nominal' 7.7 \mum\, PAH complex is dominantly comprised of the G7.6 component and only originates by a relatively small fraction in the G7.8 component (see Appendix~\ref{ratiomaps}). The G7.6 and G8.6 \mum\, components exhibit almost identical spatial distributions indicating there is no significant contribution from an additional, spatially distinct component in either of the G7.6 and G8.6 \mum\, components (Fig. ~\ref{fig_maps_decomp}). Both components peak at the SE, SSE and S' ridges in the south map. In the northern FOV, they peak at the southern and centre portion of the NW ridge, with an extension towards the west in the southern part of this ridge. 
The G7.8 and G8.2 \mum\, components also have a similar spatial distribution, though discrepancies are present (Fig. ~\ref{fig_maps_decomp}). They peak at the S and SSE ridges in the south map, similar to the 11.2 \mum\, emission and the 5-10 and 10-15 \mum\, plateau emission. In the north map, they trace both the N, centre and northern portion of the NW ridge, which is similar to the 10-15 \mum\, plateau and the 10.2 \mum\, continuum emission. As reported for the ratio of the 7.6 and 7.8 \mum\, subcomponents \citep{Bregman:05, Rapacioli:05, Boersma:14}, the ratio of G7.6/G7.8 traces the different environments very well: it is clearly smallest where H$_2$ emission peaks and inside the molecular cloud (see Appendix~\ref{ratiomaps}).

These results are also found in the correlation plots (Fig.~\ref{fig_corr}). The G7.6 and G8.6 \mum\, components exhibit a very tight correlation (best correlation coefficient of 0.987) which closely resembles a 1:1 relation (i.e. it goes through (0,0)).
In contrast, the correlation of the G7.8 and G8.2 \mum\, Gaussian components shows more scatter resulting in a correlation coefficient of 0.833. Remarkably, this enhanced scatter largely originates in the south map. Indeed, the correlation coefficient for the south map is 0.817 while that for the north map is a whopping 0.965\footnote{The same holds when normalized to the 6.2 \mum\, band with correlation coefficients of 0.823, 0.787 and 0.961 for respectively both the north and south maps}. This scatter originates in the slight mismatch of their spatial distributions and either indicates the shortcomings of our decomposition or is due to the fact that they arise in different PAH sub-populations.

Despite the arbitrariness of this decomposition (by assuming a decomposition into four Gaussians), we can conclude that at least 2 spatially distinct components contribute to the PAH emission in the 7 to 9 \mum\, region. 
In this respect, it is very enlightening to watch an animation showing the change in spatial distribution, as a function of wavelength, of the continuum-subtracted emission (applying the global spline continuum,  Fig.~\ref{fig_movie}; animation is available online). The spatial distribution of the PAH emission in the 7 to 9 \mum\, region continuously varies between two extremes which are found at around $\sim$7.35 and $\sim$8.1 \mum\, in both the north and south map\footnote{Note that the morphology of the two extremes depends on the chosen continuum (here, the global spline continuum). Using instead the local spline continuum or the plateau continuum will remove the contribution captured by the 8 \mum\, bump in the former case and add the contribution of the 5-10 \mum\, plateau in the latter case. These components have a different morphology as that of the 7.35 \mum\, extreme and are more similar to that of the 8.1 \mum\, extreme. }. For the south map, the extreme $\sim$7.35 \mum\, emission is spatially very similar to that of the 11.0 \mum\, PAH emission (Fig.~\ref{fig_slmaps_s}): it traces the S', SE and SSE ridges, the horizontal filaments in the northern region of the map and the broad, diffuse plateau NW of the line connecting the S and SSE ridges. The extreme $\sim$8.1 \mum\, emission is similar to the G8.2 component (Fig.~\ref{fig_maps_decomp}) and peaks at the S and SSE ridges. So does the 11.2 \mum\, PAH emission (Fig.~\ref{fig_slmaps_s}) but the extreme $\sim$8.1 \mum\, emission has enhanced emission in the horizontal filaments in the northern region of the map and the broad, diffuse plateau compared to the 11.2 \mum\, PAH emission.
For the north map, the extreme distribution at $\sim$7.35 \mum\, peaks at the southern part of the NW ridge and the extension towards the west of the NW ridge like the 8.6 and 11.0 \mum\, PAH emission (Fig.~\ref{fig_slmaps_n}) and the G8.6 component (Fig.~\ref{fig_maps_decomp}). In contrast, the extreme $\sim$8.1 \mum\, emission peaks in the N ridge as does the 10.2 \mum\, continuum (Fig.~\ref{fig_slmaps_n}) and the G7.8 component (Fig.~\ref{fig_maps_decomp}).

\subsubsection{Implications for other PAH bands}

The morphology of the PAH emission changes continuously with wavelength for all major PAH bands (i.e. the 6.2, 11.2 and, 12.7 \mum\, bands), as for the south map, albeit to a considerable lesser extent (Fig.~\ref{fig_movie}): none of the wavelengths exhibit a morphology as extreme as that of the $\sim$7.35 \mum\, or the $\sim$8.1 \mum\, extremes. 

At $\sim$6.0 \mum, the morphology is similar to that of 6.0 PAH shown in Figs. \ref{fig_slmaps_s} and \ref{fig_slmaps_n} for both maps. For the south map, with increasing wavelength from 6.0 to 6.2 \mum, emission increases in the SE ridge, followed by increased diffuse emission NW of the line connecting the SSE and S ridge, and increased emission in the S' ridge. This is accompanied by decreased emission in the S ridge resulting in a spatial distribution very similar to that of the integrated 6.2 PAH emission map. Subsequently, for longer wavelengths ($>\, \sim$6.25 \mum), the diffuse emission and the emission in the S' ridge decreases and the S ridge becomes a tiny bit stronger while the peak emission is found in the SE and SSE ridge. Considering the north map, at $\sim$6.1 \mum, the emission peaks in the NW ridge. With increasing wavelength up to 6.2 \mum, emission in the southern part and west of the southern part of the NW ridge increases while the northern part of the NW ridges fades a little, resulting in a morphology somewhat between the 6.2 PAH and the G8.6 emission. Subsequently, the NW ridge becomes a bit stronger and eventually, the emission west of the southern part of the NW ridge fades away.

The most dramatic change in the 11 \mum\, region is found from 11.07 to 11.14 \mum\, with the switch from the 11.0 to 11.2 PAH emission. The morphologies near 11.0 and 11.2 \mum, as in the 6 \mum\, region, are well represented by that of the 11.0 PAH emission and 11.2 PAH emission respectively. Once beyond the peak of the 11.2 PAH band, the morphology continues to vary, albeit to a less dramatic extent, and becomes more sharply peaked. In the south map, the emission peaks sharply at the S and SSE ridges (and weaker at S' ridge), very similarly to the $H_2$ emission. In contrast, for the north map, the emission peaks in the northern part of the NW ridge while the remaining of the NW ridge has similar intensity as the N ridge. 

 \begin{figure}[tb]
    \centering
\resizebox{\hsize}{!}{%
  \includegraphics{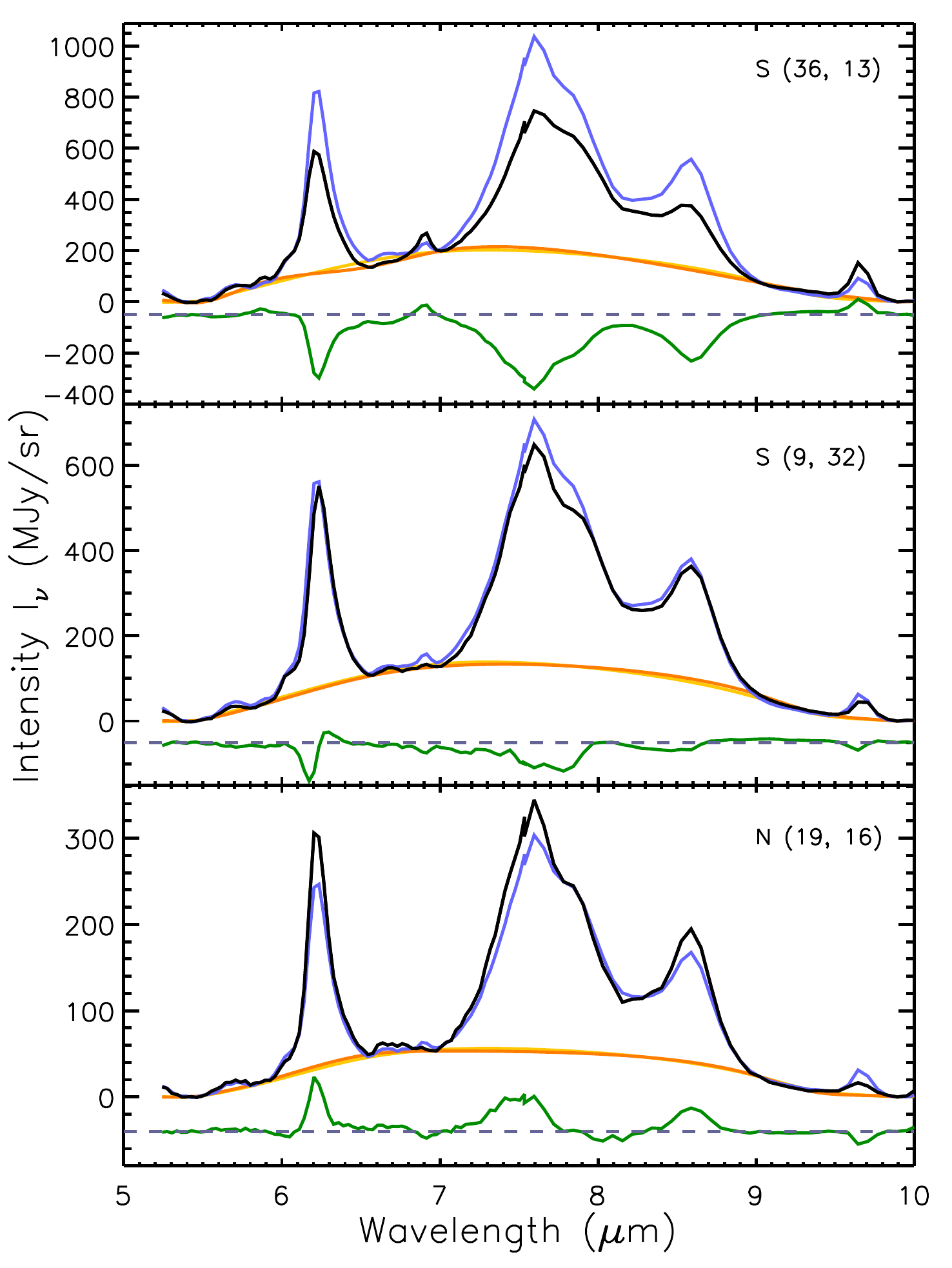}}
\caption{\label{5to10plat} 
5--10 \mum\, plateau. Shown are spectra from different pixels (black),
the {\it average} PAH emission spectrum scaled such that its 5-10 \mum\,  plateau flux matches the 5--10 \mum\,
plateau flux of the black spectra (blue), the global spline continuum for the spectra (orange), and the global spline continuum for the scaled average PAH emission (gold), and the residuals (green). Note that the continuum for the average PAH emission is mostly identical to that of the spectra with the most noticeable difference around 6.7 \mum. For all but the residuals, the dust continuum emission as fitted by the plateau continuum (see Sect. \ref{cont}) is subtracted. Note that the strength of the features varies independently of the 5--10 \mum\,
plateau and of each other.} 
\end{figure}

 \begin{figure}[tb]
    \centering
\resizebox{\hsize}{!}{%
  \includegraphics{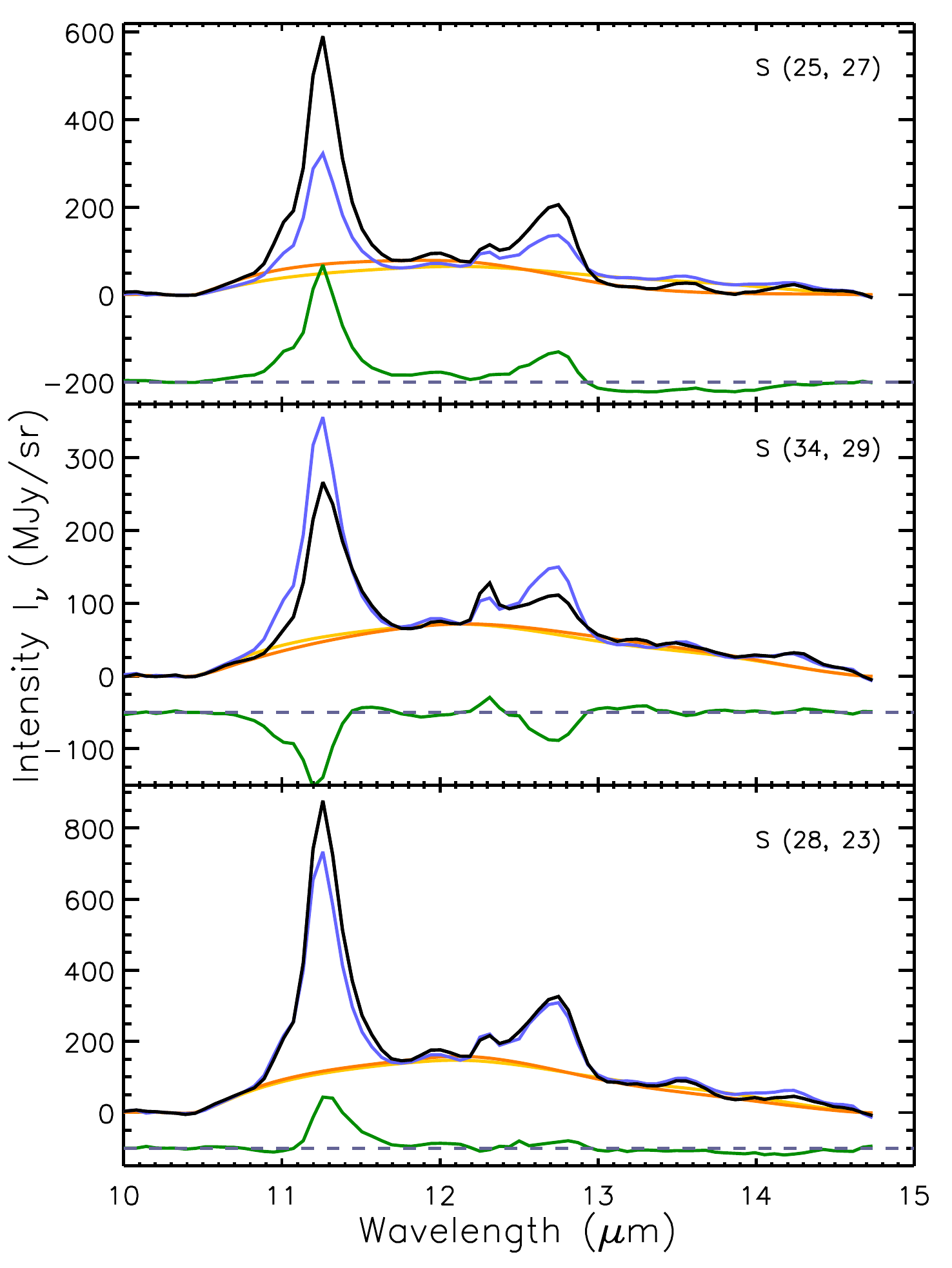}}
\caption{\label{10to15plat} 
10--15 \mum\, plateau. Shown are spectra from different pixels (black),
the {\it average} PAH emission spectrum scaled such that its 10-15 \mum\,  plateau flux matches the 10--15 \mum\,
plateau flux of the black spectra (blue), the local spline continuum for the spectra (orange), and the local spline continuum for the scaled average PAH emission (gold), and the residuals (green). Note that the continuum for the average PAH emission is very similar to that of the spectra. For all but the residuals, the dust continuum emission as fitted by the plateau continuum (see Sect. \ref{cont}) is subtracted. Note that the strength of the features varies independently of the 10--15 \mum\,
plateau and of each other.} 
\end{figure}

The morphology of the integrated 12.7 PAH band closely resembles that at 12.7 \mum. Going to shorter wavelengths, emission is enhanced at locations where the ions peak (as traced by the 8.6 and 11.0 \mum\, emission). This is most noticeable in the south map: emission enhances in the SE ridge, the S' ridge (in particular the western part) and NW of the line connection the SSE and S ridge (earlier referred to as the diffuse emission). In addition, the local emission peak in the S ridge becomes displaced towards the north to eventually being merged with the diffuse emission NW of the line connection the SSE and S ridge.  In the north map, this is seen by a slight enhancement of the emission in the extension to the W of the southern part of the NW ridge (but the peak of the emission remains in the NW ridge).

\subsubsection{PAH features versus plateaus}

In the north map, the PAH emission and dust continuum emission are displaced from each other (i.e. they peak at different ridges), giving a unique opportunity to further explore the relationship between the PAH emission features and the underlying plateaus. Indeed, in the north map the 10--15 \mum\, plateau resembles the dust continuum emission (at 10 and 14.7 \mum) but not the PAH emission (of any of the PAH bands), the 5--10 \mum\, plateau emission exhibit a spatial distribution between that of the continuum and that of the PAH emission and the 15-18 \mum\, plateau emission follows the PAH emission (Sections ~\ref{generalimpression} and ~\ref{shdata}; paper I). 
In paper I (Fig. 8), we reported that the 15-18 \mum\, PAH bands vary independently of the 15-18 \mum\, plateau. Figures \ref{5to10plat} and \ref{10to15plat} show a similar exercise for the 5--10 and 10--15 \mum\, features and their underlying plateaus respectively: we compared the PAH emission in each pixel with the average PAH emission when normalizing the strength of the plateaus. While small variations are present in the shape of the plateaus amongst the different pixels\footnote{Specifically, the largest variation in the 5--10 \mum\, plateau is found between the 6.2 and 7.7 \mum\, features (Fig.~\ref{5to10plat}, top panel). In case of the 10--15 \mum\, plateau, the most noticeable change is the ratio of the plateau strength underneath the 11.2 and 12.0 \mum\, bands versus the plateau strength underneath the 13.5 and 14.2 \mum\, bands (Fig.~\ref{10to15plat}, top panel).}, it is clear in both cases that the features behave independently of their underlying plateau and of each other. After investigating the possible influence of the applied decomposition on this result (see Appendix~\ref{influence_decomp_plat}), we conclude and confirm earlier reports \citep[][paper I]{Bregman:orion:89, Roche:orion:89} that {\it the plateaus are distinct from the features.}

%%%%%%%%%%%%%%%%%%%%%%%%%%%%%%%%%%%%%%%%%%%%%%%%%%
\begin{figure*}[tbp]
    \resizebox{\hsize}{!}{%
  \includegraphics[angle=33.4]{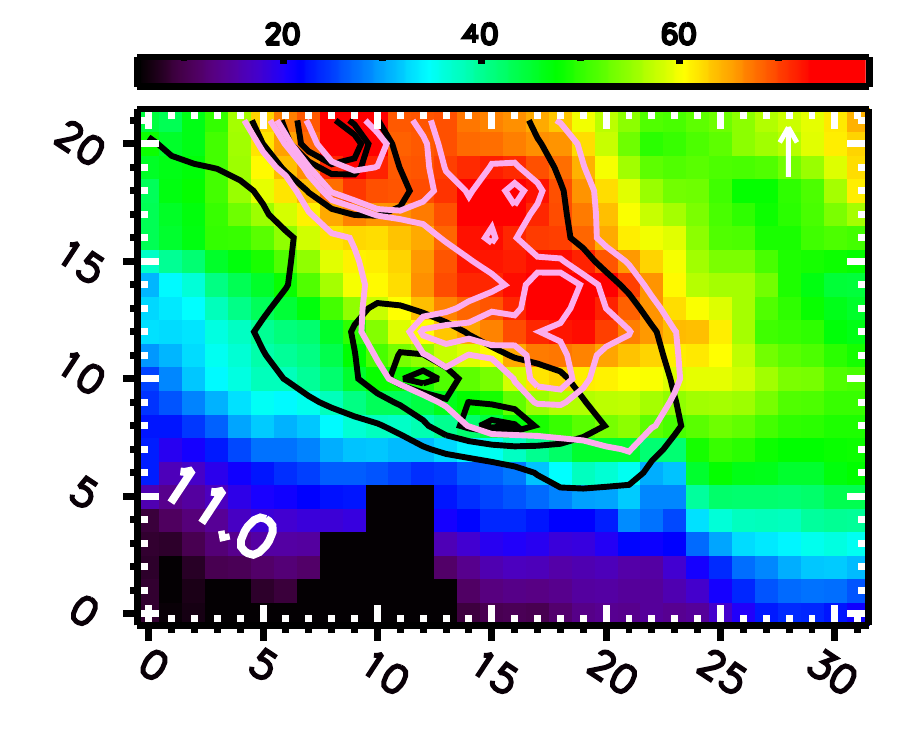}
  \includegraphics[angle=33.4]{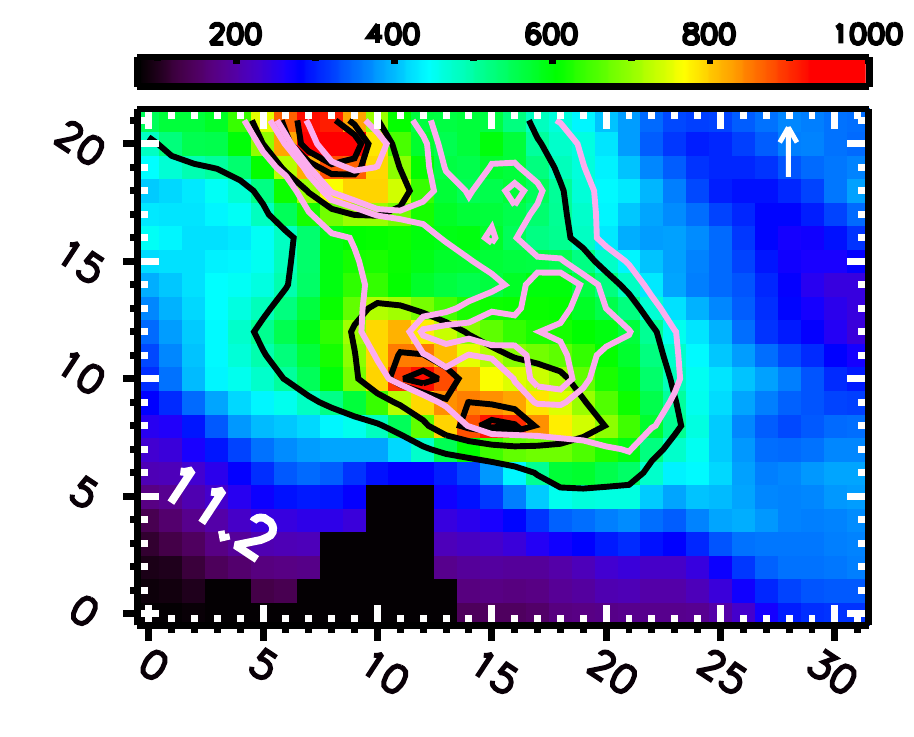}  
  \includegraphics[angle=33.4]{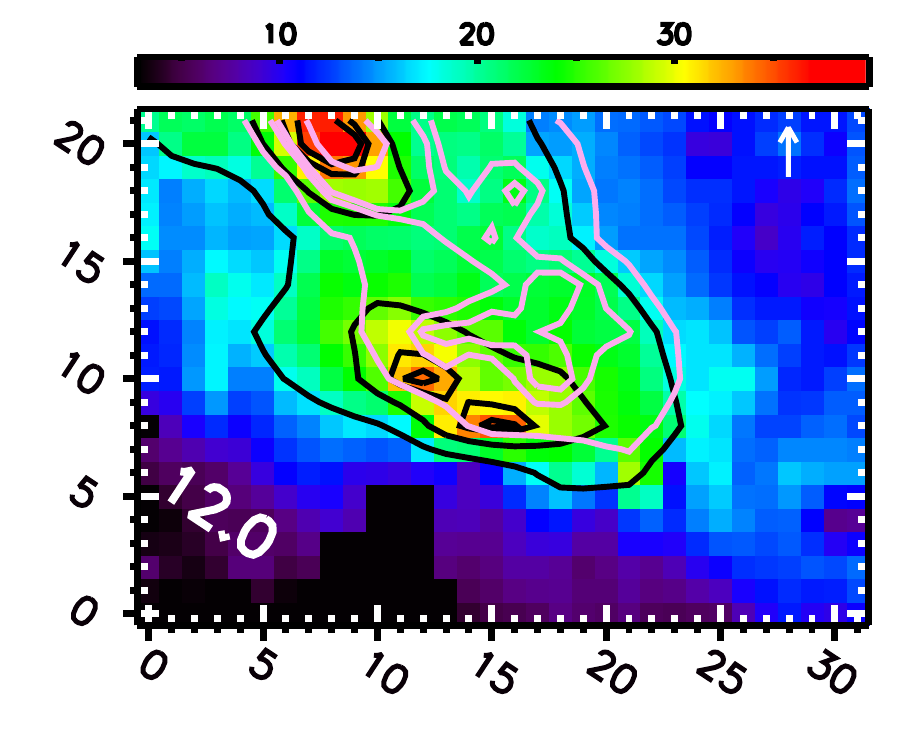}
  \includegraphics[angle=33.4]{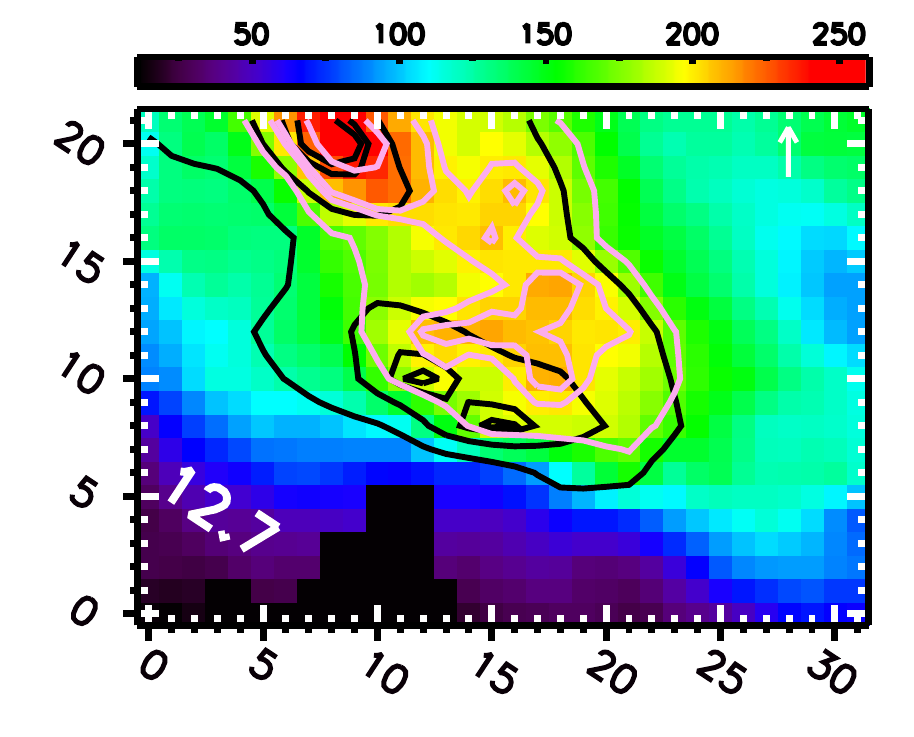}}
  \resizebox{\hsize}{!}{%
  \includegraphics[angle=33.4]{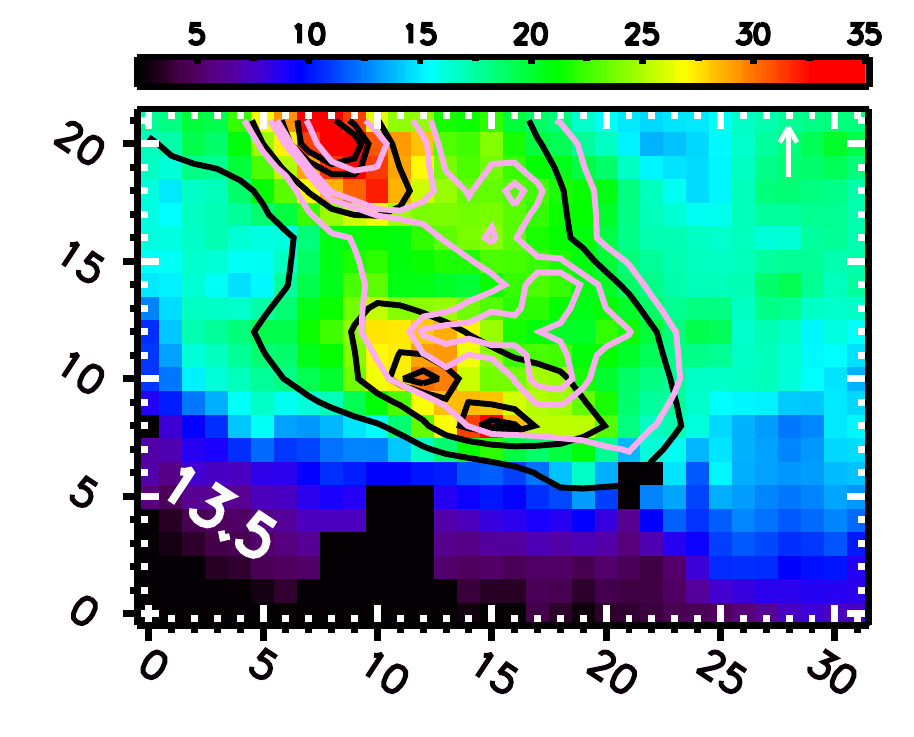}
  \includegraphics[angle=33.4]{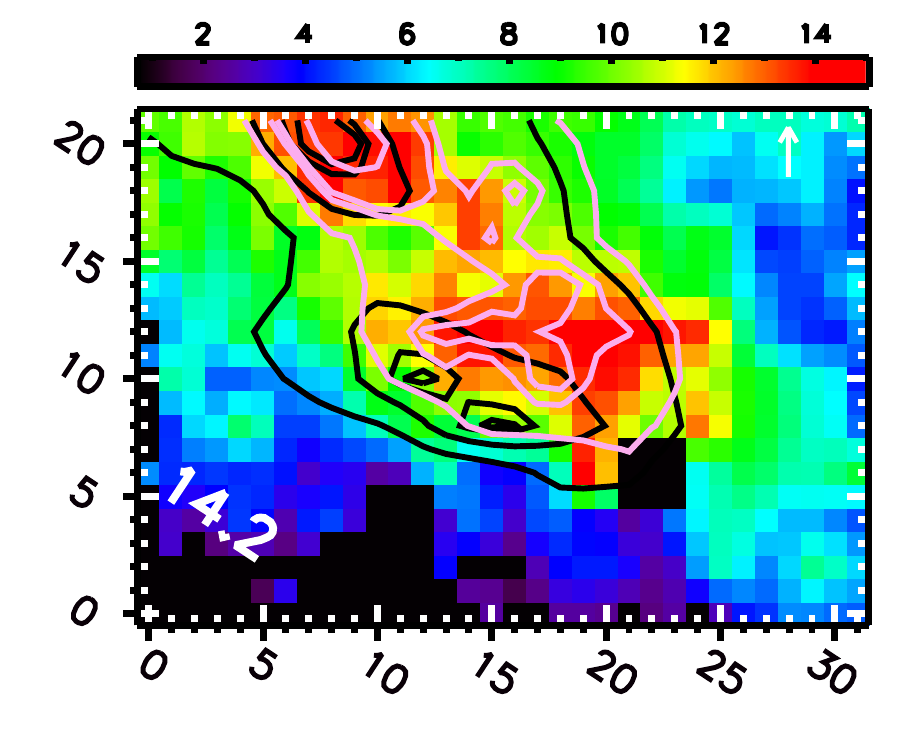} 
  \includegraphics[angle=33.4]{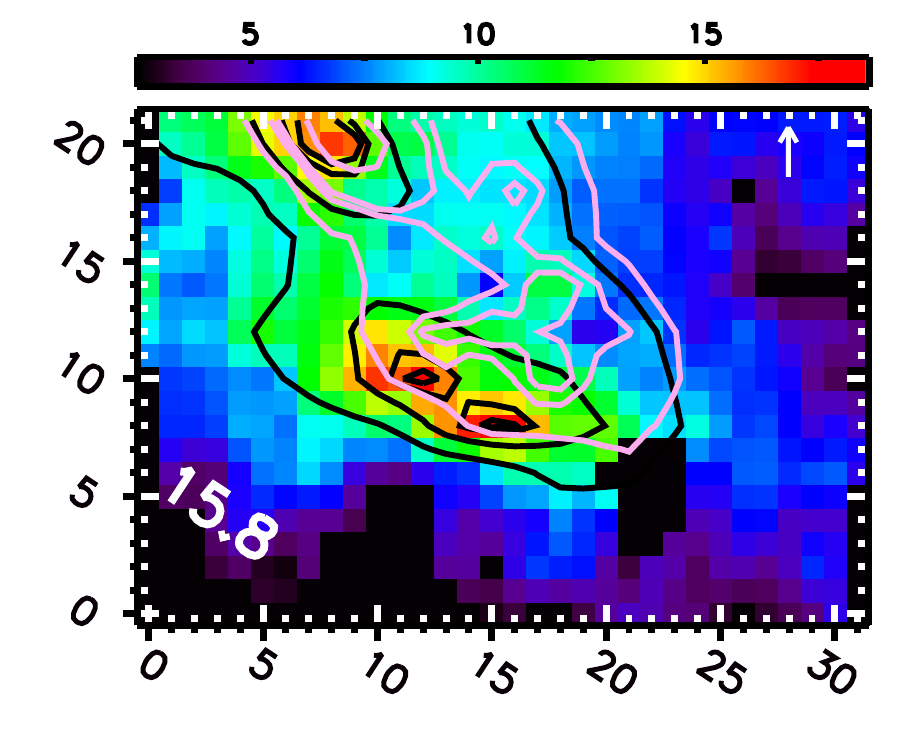}
  \includegraphics[angle=33.4]{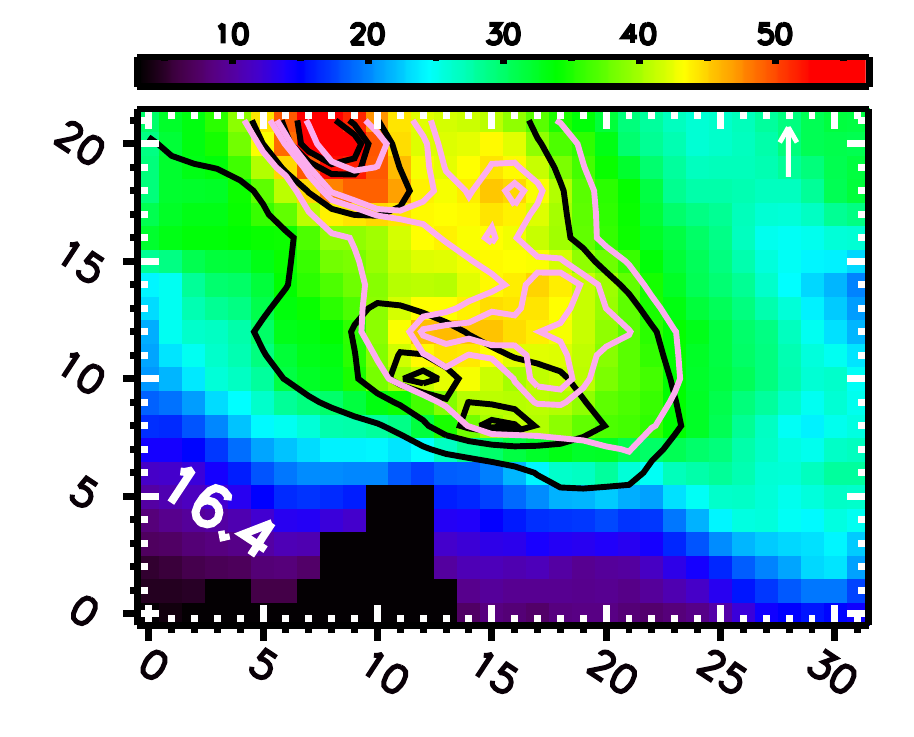}  }    
  \resizebox{\hsize}{!}{%
 \includegraphics[angle=33.4]{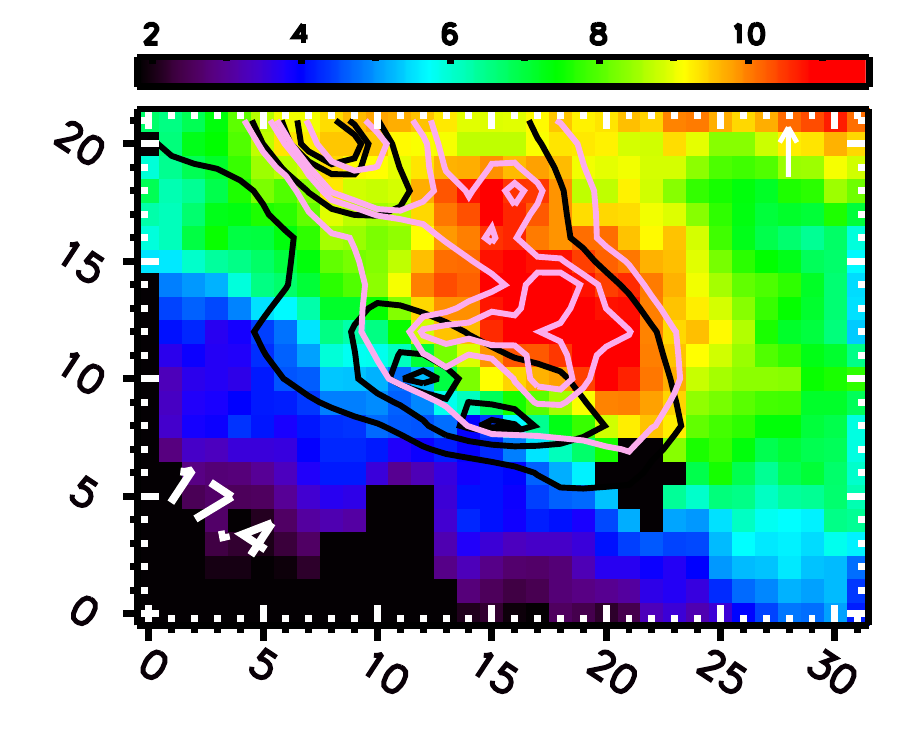}
\includegraphics[angle=33.4]{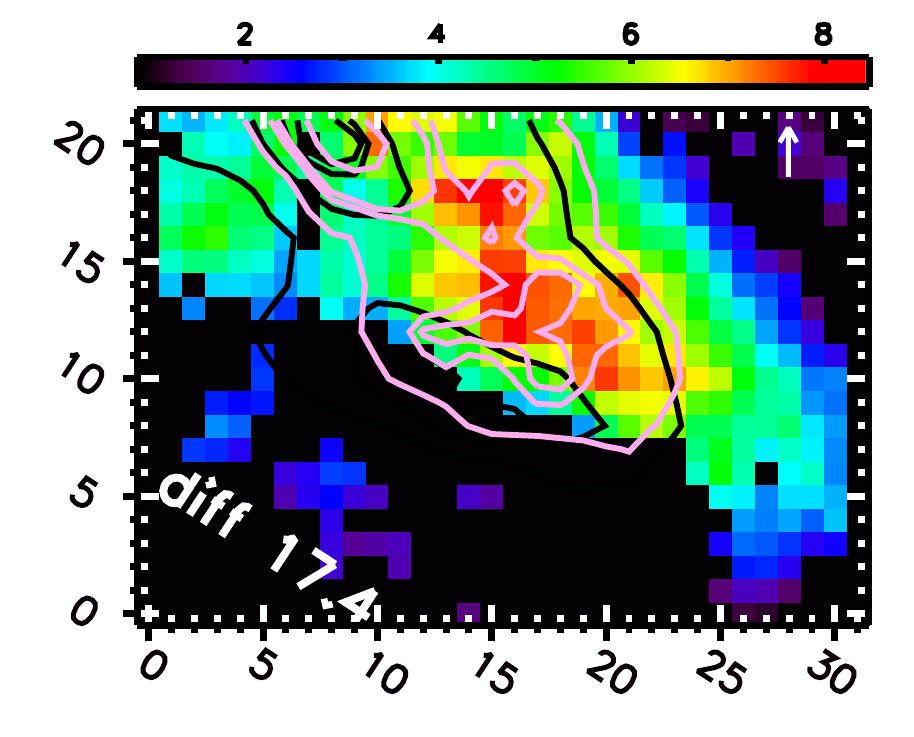}     
\includegraphics[angle=33.4]{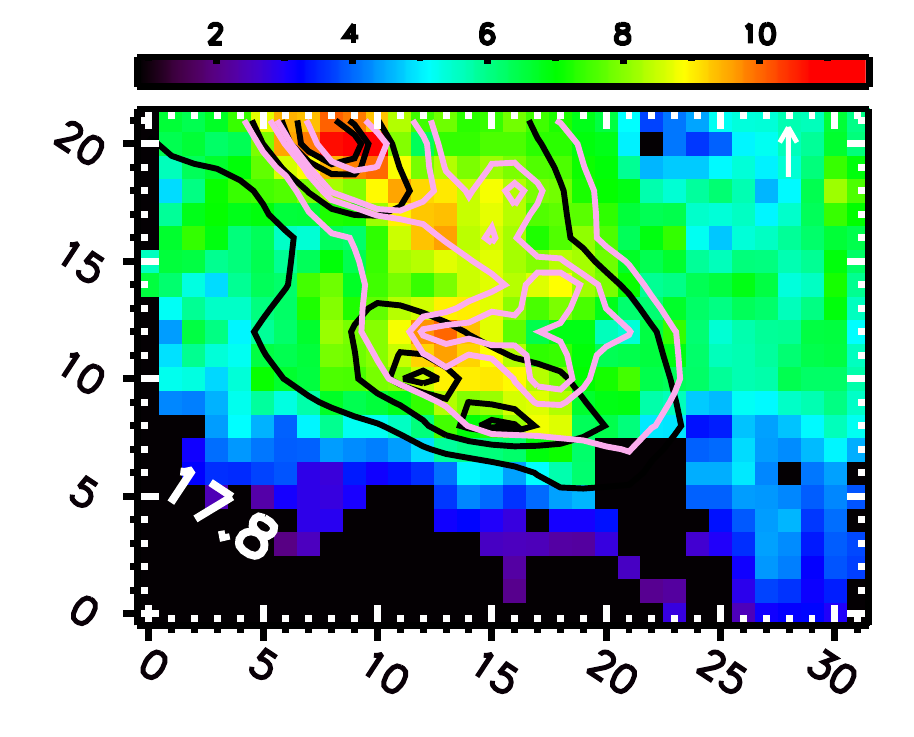}  
\includegraphics[angle=33.4]{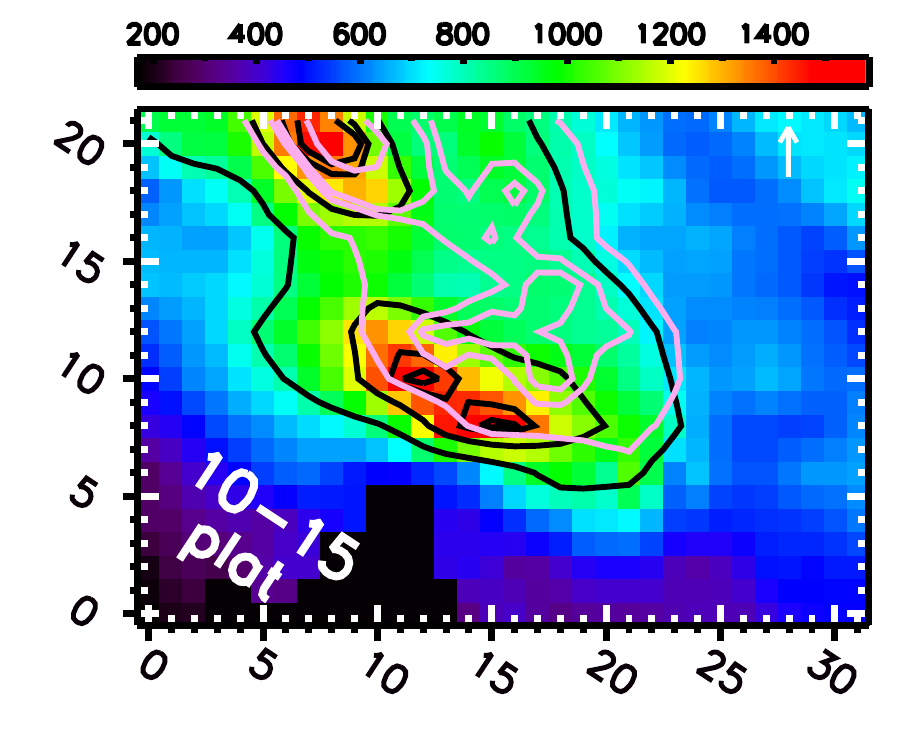}}
\resizebox{13.5cm}{!}{%
\includegraphics[angle=33.4]{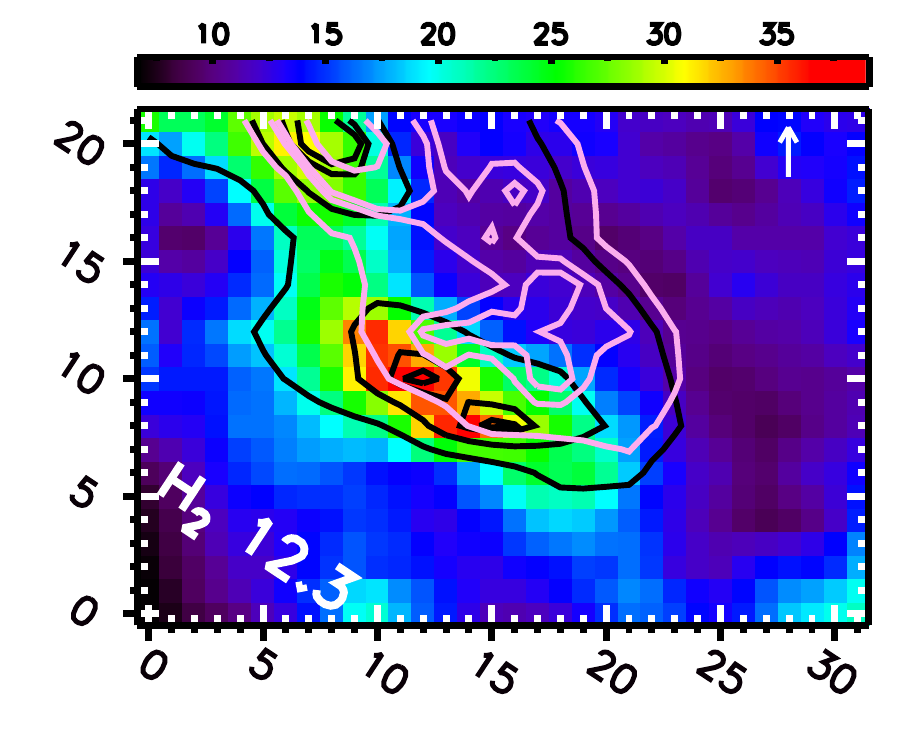}
  \includegraphics[angle=33.4]{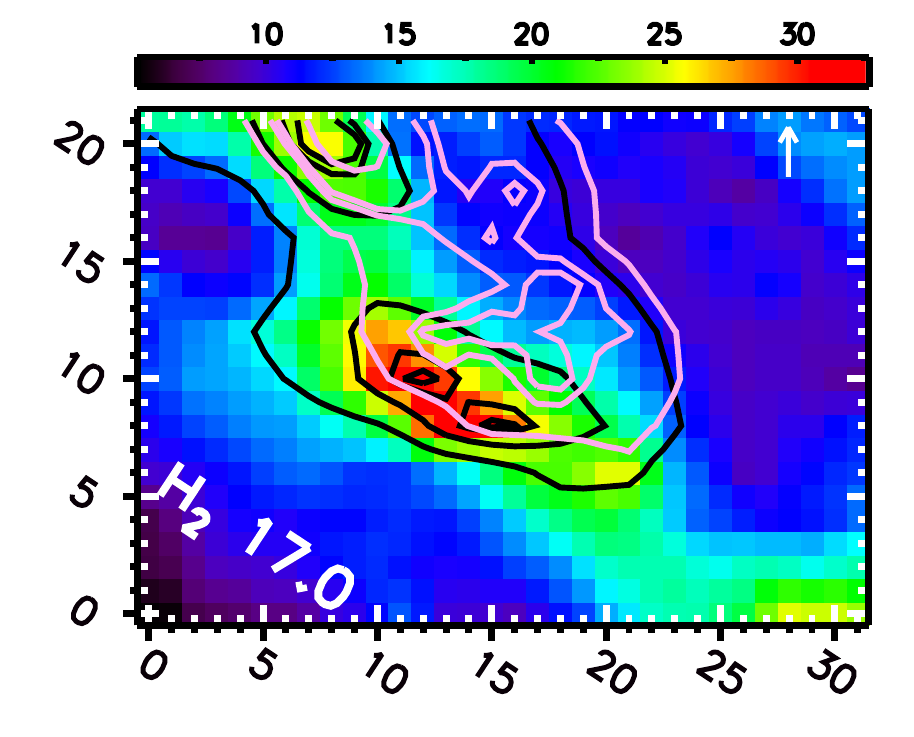}
    \includegraphics[angle=33.4]{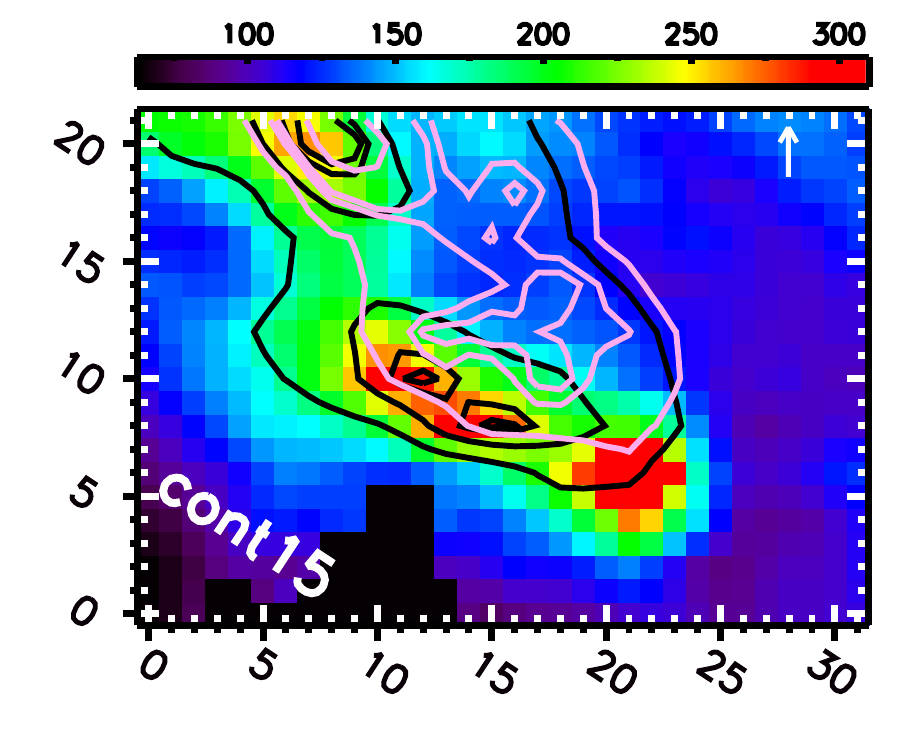}}
\caption{Spatial distribution of the emission features in the 10-20 \mum\, SH data towards NGC~2023 for the south map (applying a local spline continuum and using
a cutoff value of 2 sigma). As a reference, the intensity profiles of the 11.2 and 12.7 \mum\, emission features are shown as contours in black (at 5.0, 7.0, 8.5 and, 9.0 10$^{-6}$ Wm$^{-2}$sr$^{-1}$) and pink (at 1.7, 2.0, 2.97 and, 2.35 10$^{-6}$ Wm$^{-2}$sr$^{-1}$), respectively. Map orientation and units are the same as in Fig.~\ref{fig_slmaps_s}. The white arrow in the top corner indicates the direction towards the central star. The axis labels refer to pixel numbers. The nomenclature is given in the bottom right panel (see also Fig. 1). Note that the 17.4 \mum\, emission is due to both PAH and C$_{60}$ emission while diff 17.4 refers to only the PAH 17.4 \mum\, emission.
}
\label{fig_shmaps_s}
\end{figure*}
%%%%%%%%%%%%%%%%%%%%%%%%%%%%%%%%%%%%%%%%%%%%%%%%%%

%%%%%%%%%%%%%%%%%%%%%%%%%%%%%%%%%%%%%%%%%%%%%%%%%%
\begin{figure*}[tbp]
    \centering
\resizebox{\hsize}{!}{%
  \includegraphics[angle=230.6]{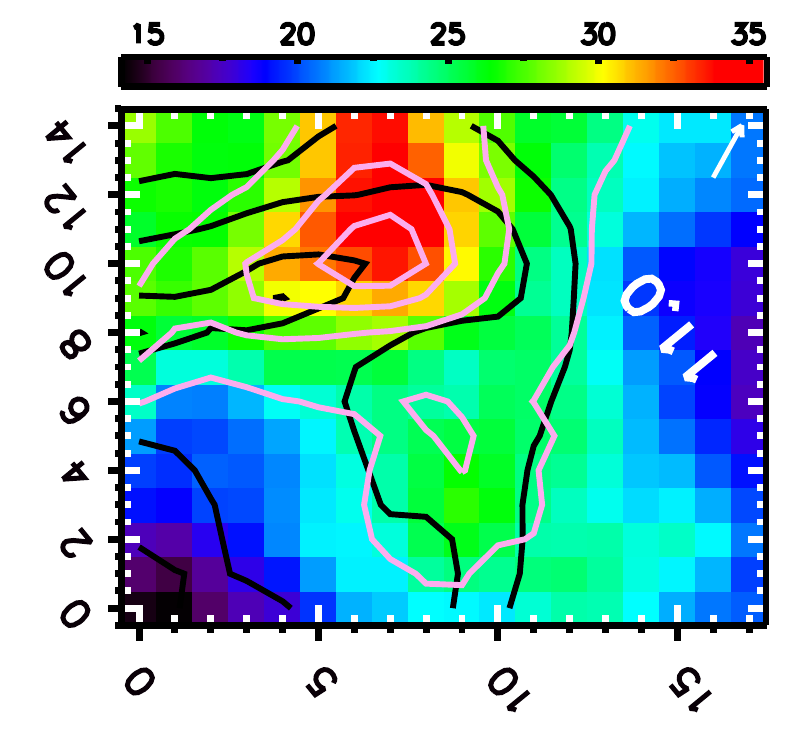}
  \includegraphics[angle=230.6]{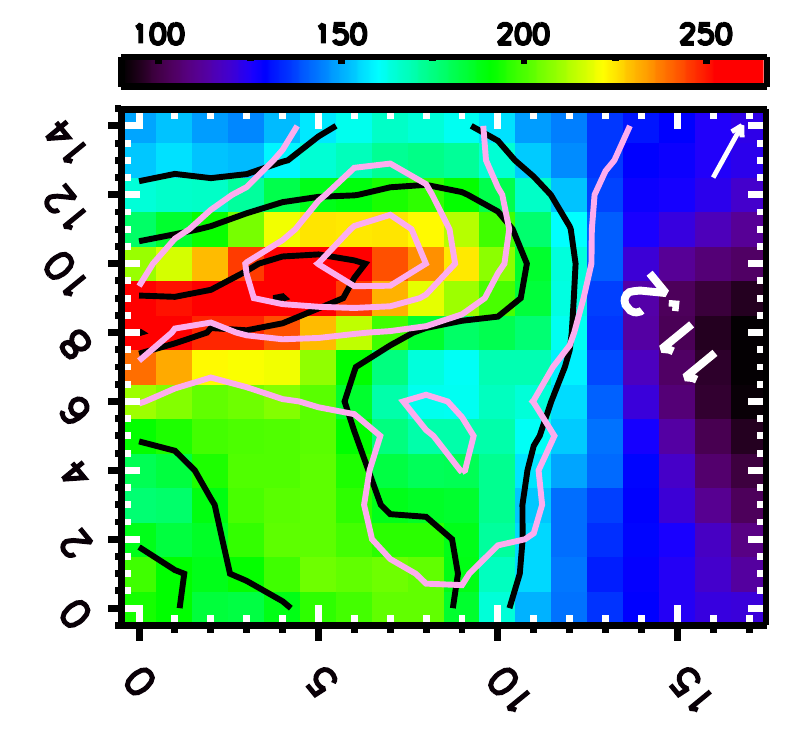}  
  \includegraphics[angle=230.6]{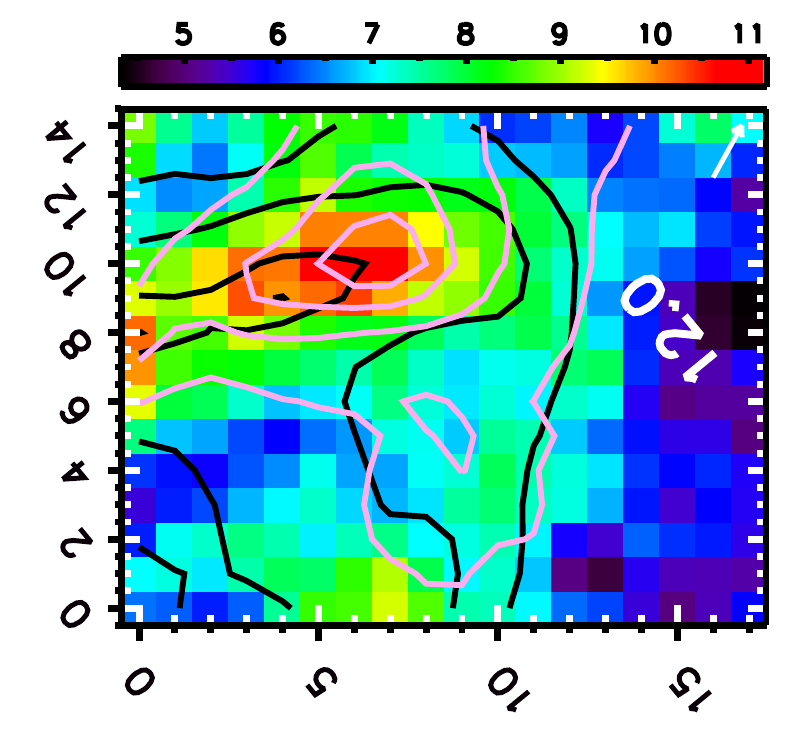}
  \includegraphics[angle=230.6]{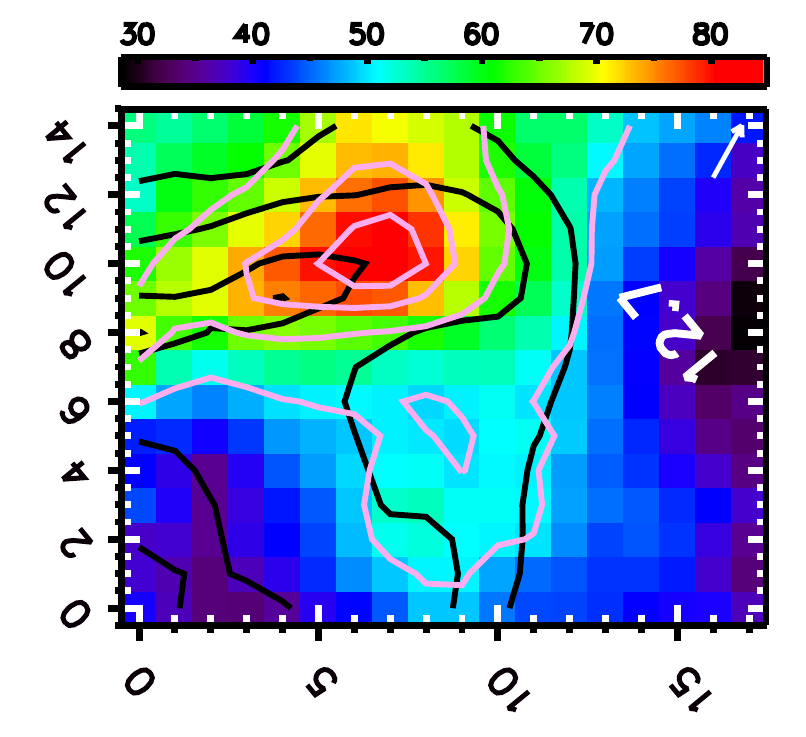}}
  \resizebox{\hsize}{!}{%
  \includegraphics[angle=230.6]{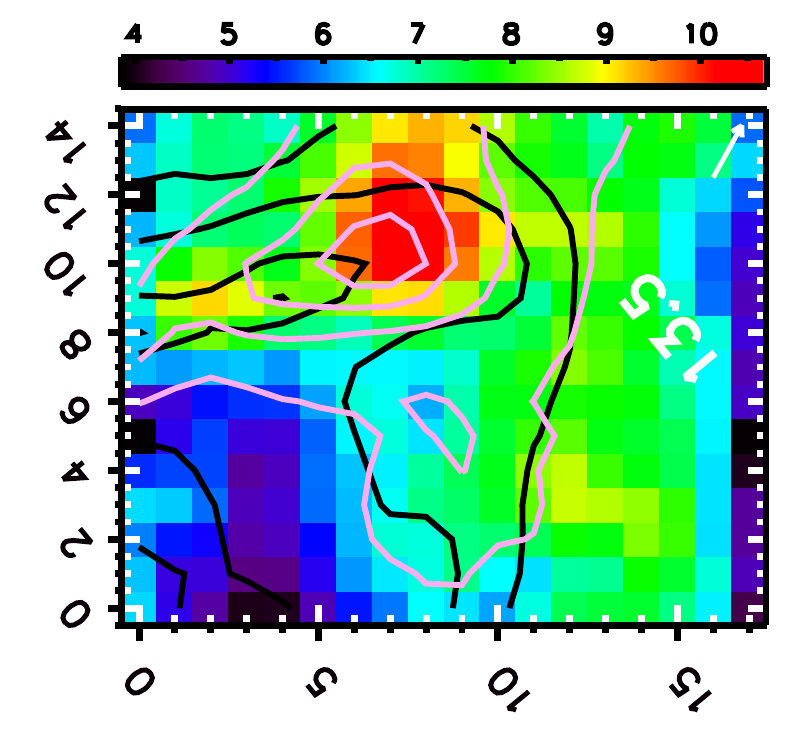}
  \includegraphics[angle=230.6]{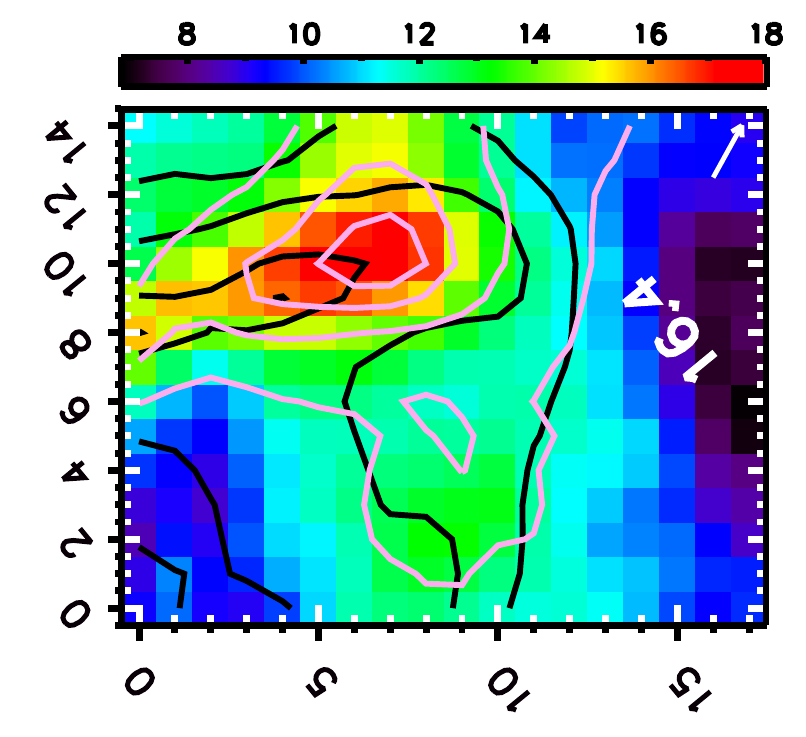}    
  \includegraphics[angle=230.6]{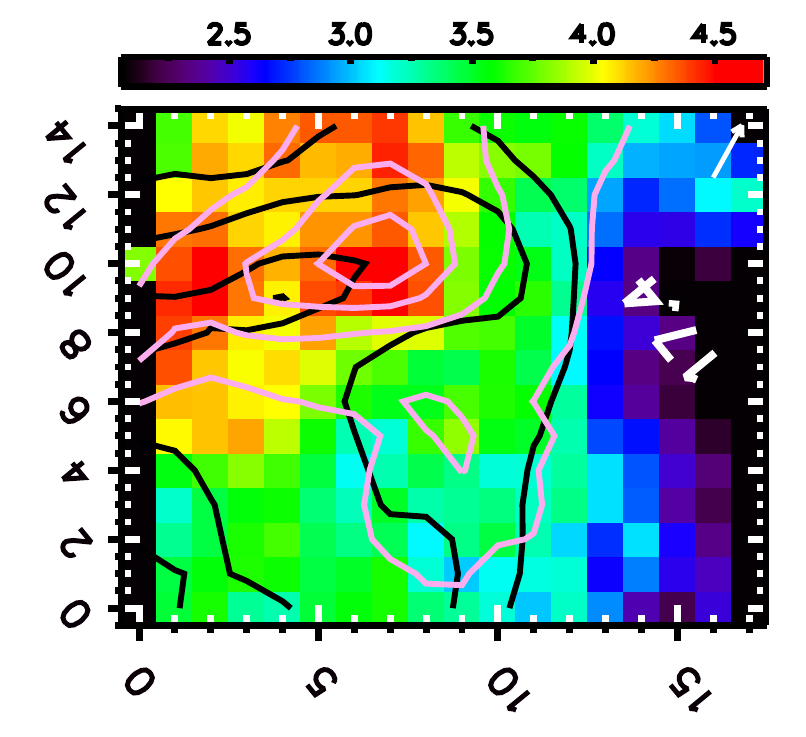}    
  \includegraphics[angle=230.6]{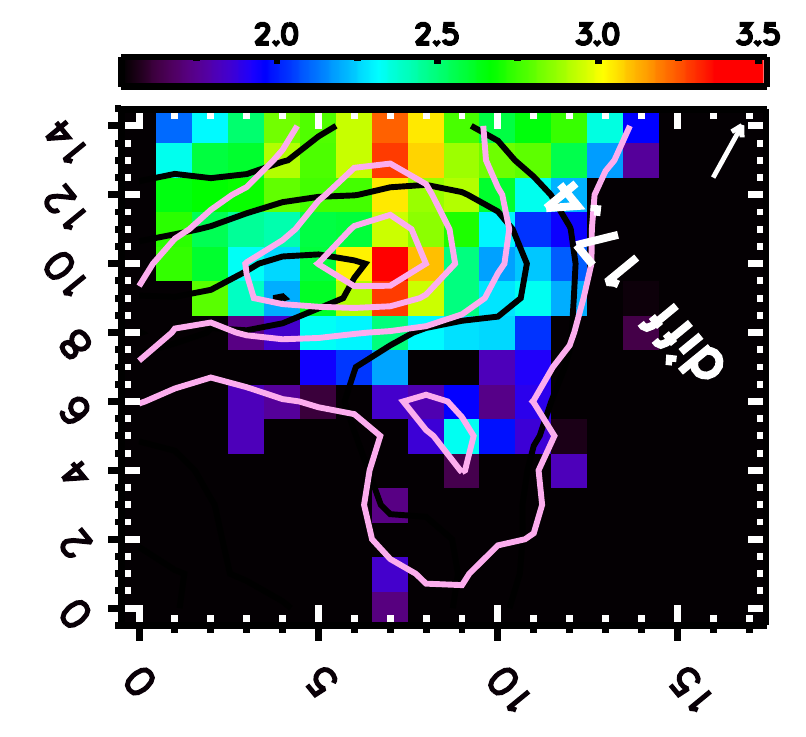} }    
  \resizebox{\hsize}{!}{%
\includegraphics[angle=230.6]{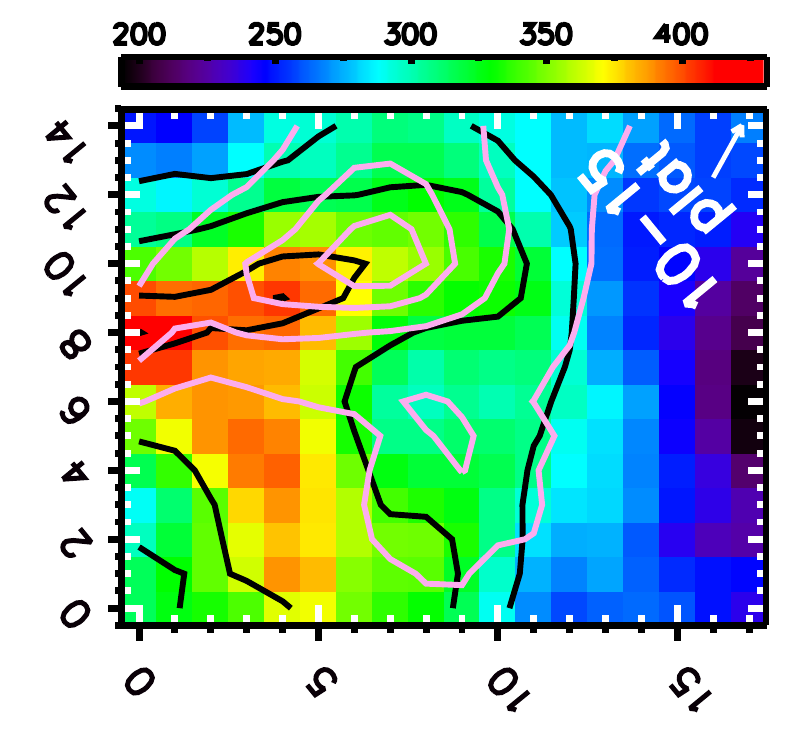}  
\includegraphics[angle=230.6]{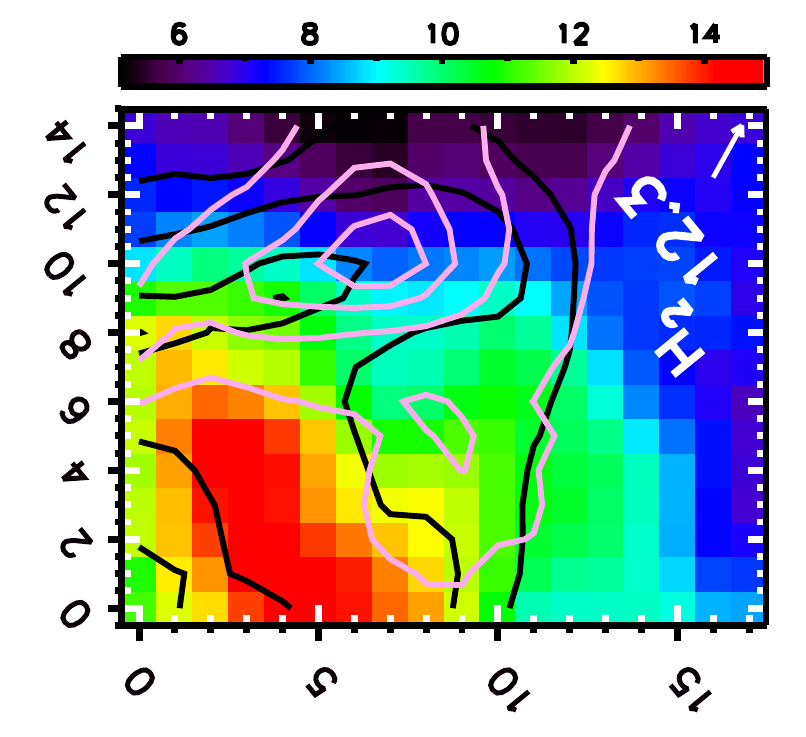}  
\includegraphics[angle=230.6]{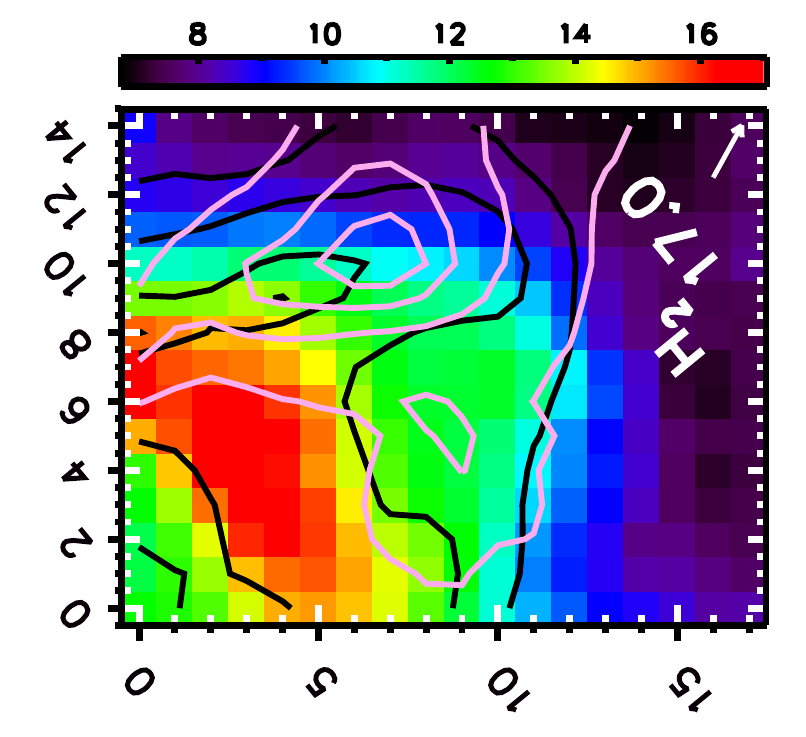}
\includegraphics[angle=230.6]{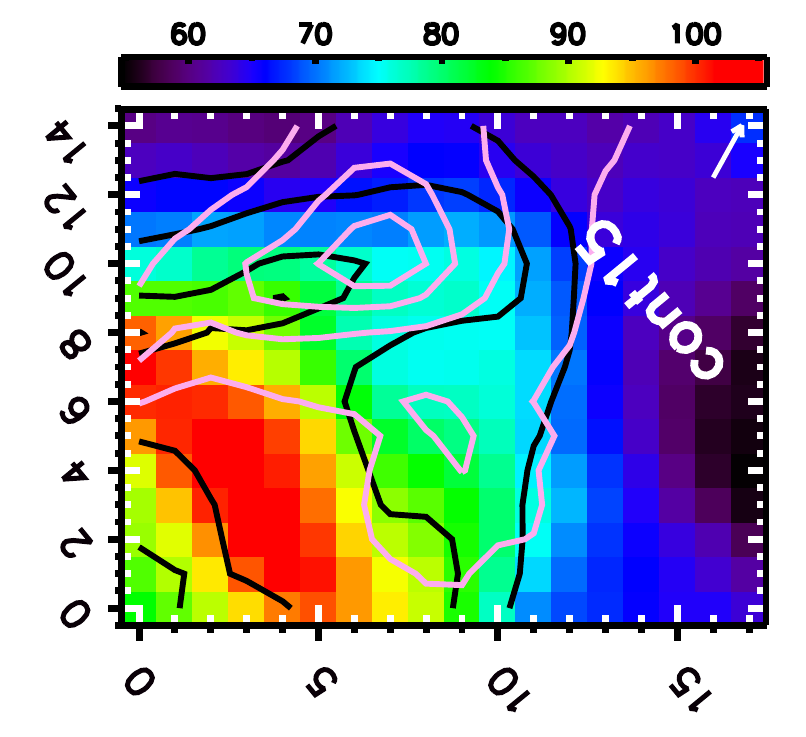}}
 \caption{Spatial distribution of the emission features in the 10-20 \mum\, SH data towards NGC~2023 for the north map (applying a local spline continuum and using
a cutoff value of 2 sigma). As a reference, the intensity profiles of the 11.2 and 12.7 \mum\, emission features are shown as contours in black (at 1.6, 1.9, 2.5 and, 2.66 10$^{-6}$ Wm$^{-2}$sr$^{-1}$) and pink (at 5.0, 6.4, 7.4 and, 8.0 10$^{-7}$ Wm$^{-2}$sr$^{-1}$), respectively. Map orientation and units are the same as in Fig.~\ref{fig_slmaps_n}. The white arrow in the bottom corner indicates the direction towards the central star. The axis
labels refer to pixel numbers. The nomenclature is given in the bottom right panel (see also Fig. 1).  Note that the 17.4 \mum\, emission is due to both PAH and C$_{60}$ emission while diff 17.4 refers to only the PAH 17.4 \mum\, emission.}
\label{fig_shmaps_n}
\end{figure*}
%%%%%%%%%%%%%%%%%%%%%%%%%%%%%%%%%%%%%%%%%%%%%%%%%%                                                                                                 

%%%%%%%%%%%%%%%%%%%%%%%%%%%%%%%%%%%%%%%%%%%%%%%%%%
\begin{figure*}[tbp]
    \centering
\resizebox{18cm}{!}{%
  \includegraphics{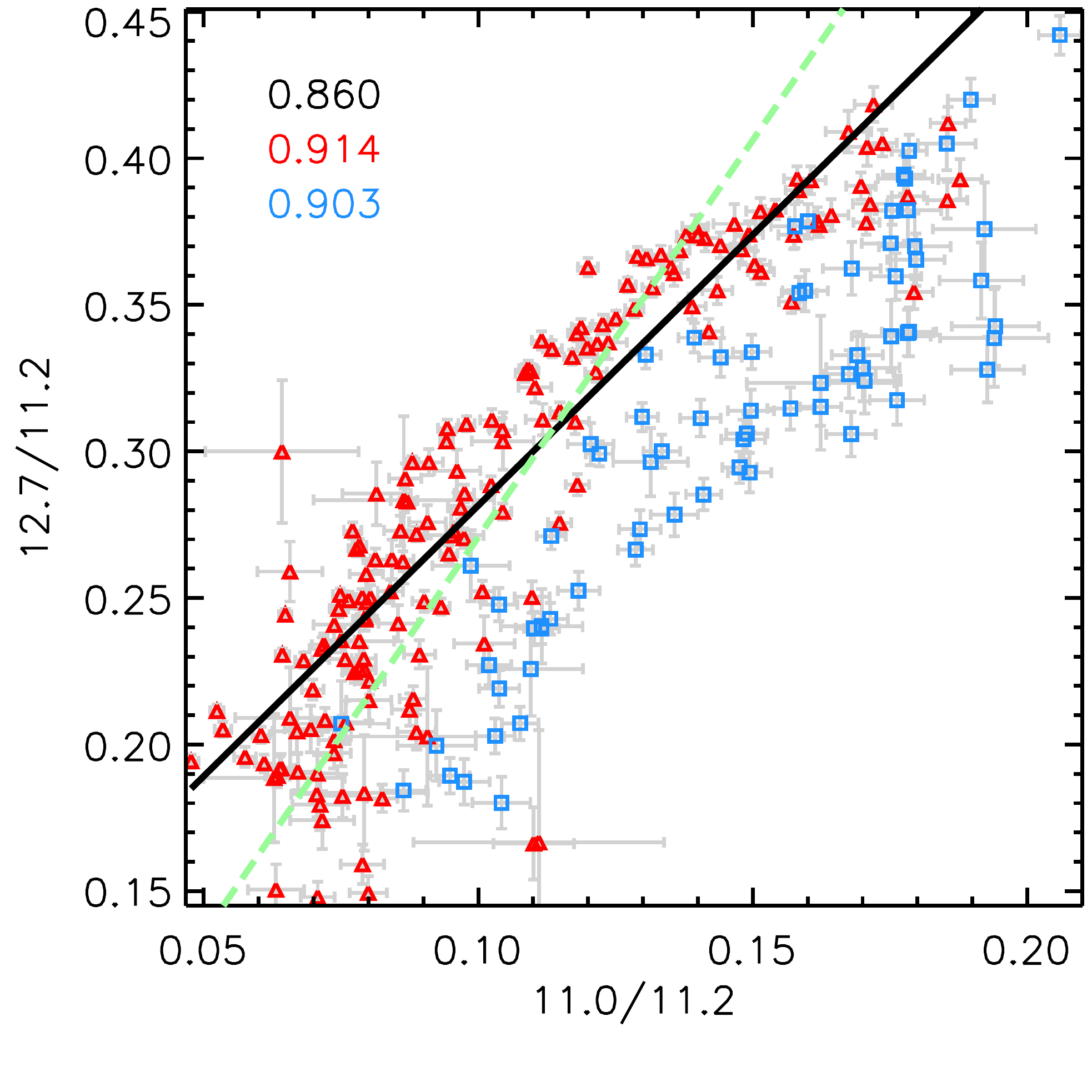}
  \includegraphics{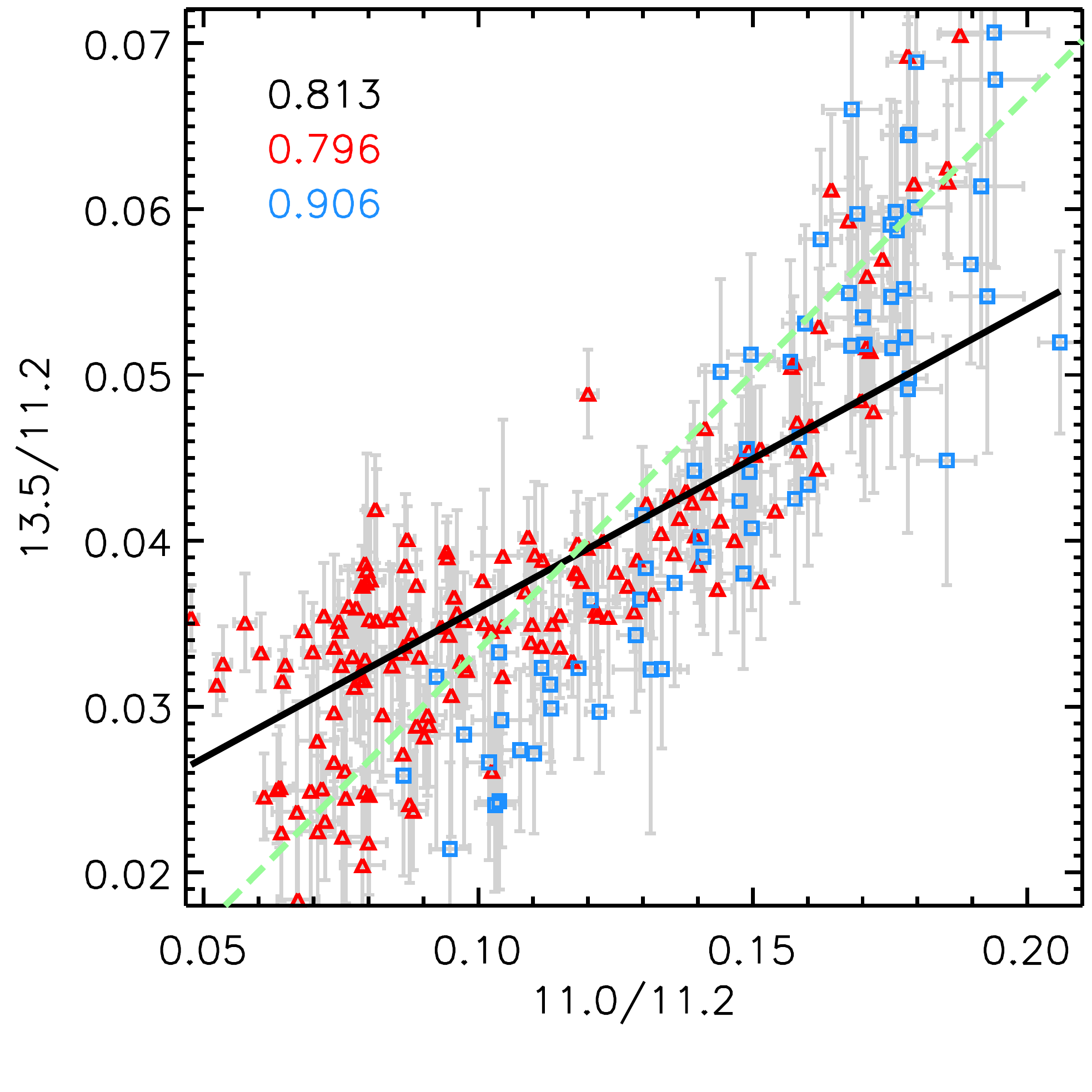}
  \includegraphics{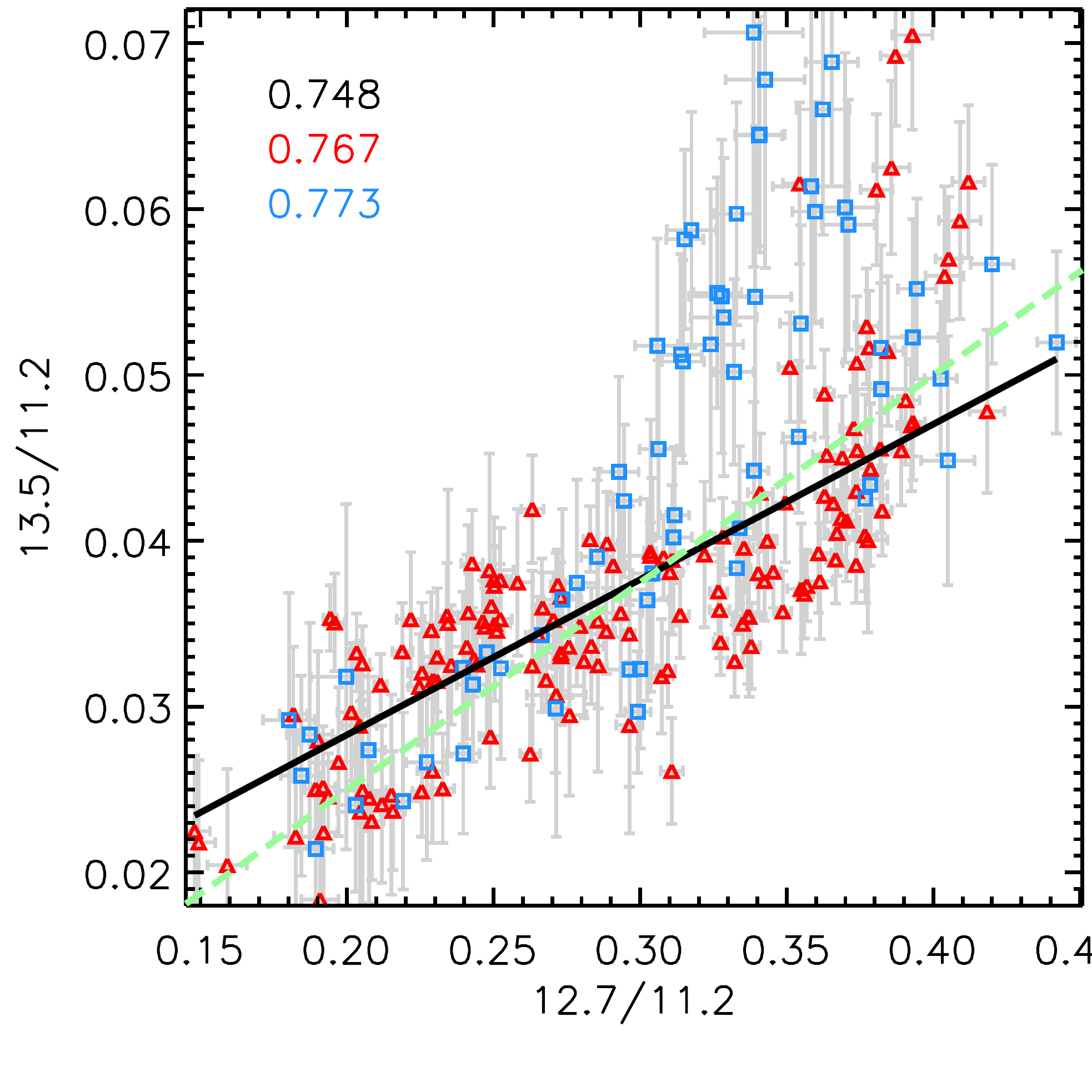}
   \includegraphics{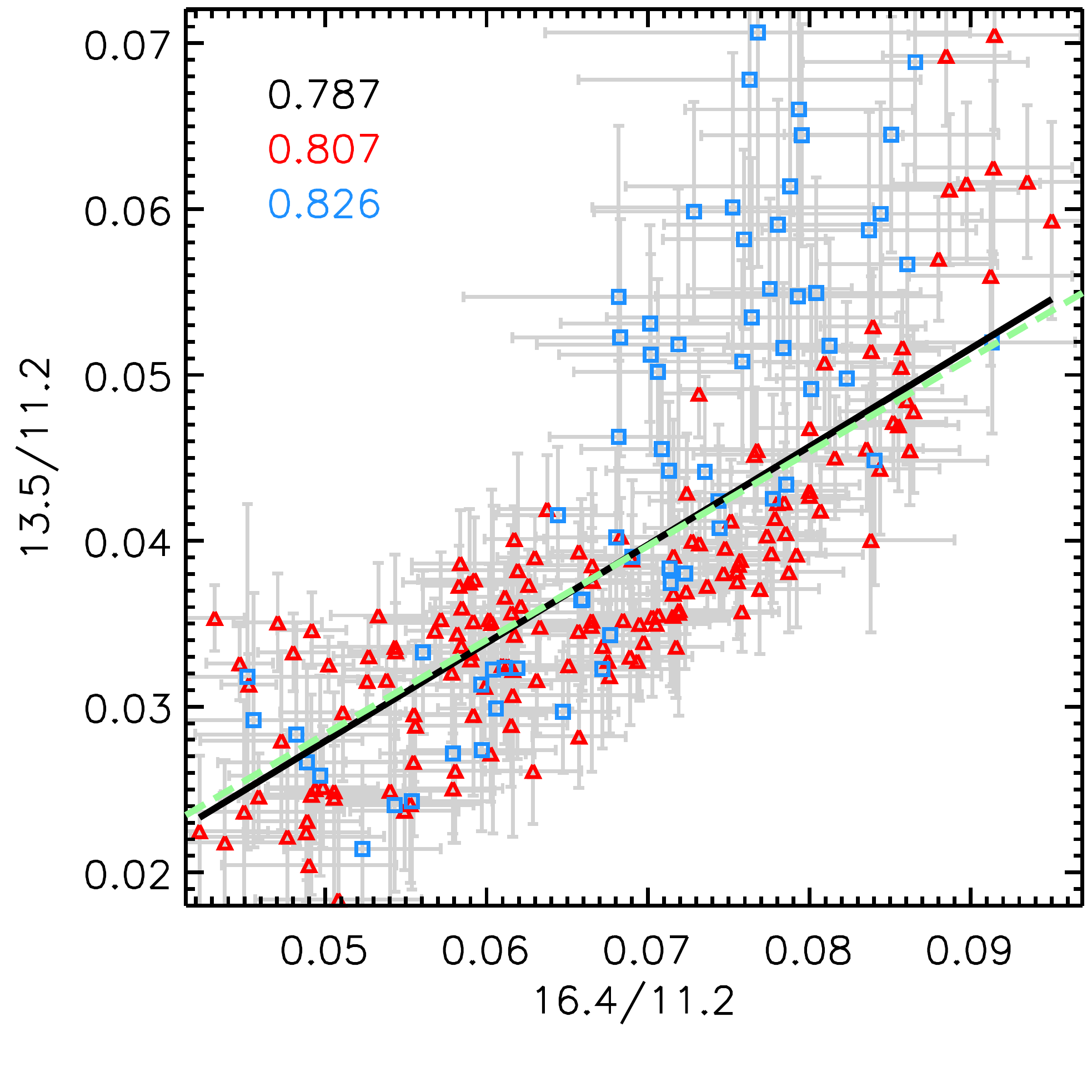}}
\resizebox{18cm}{!}{%
  \includegraphics{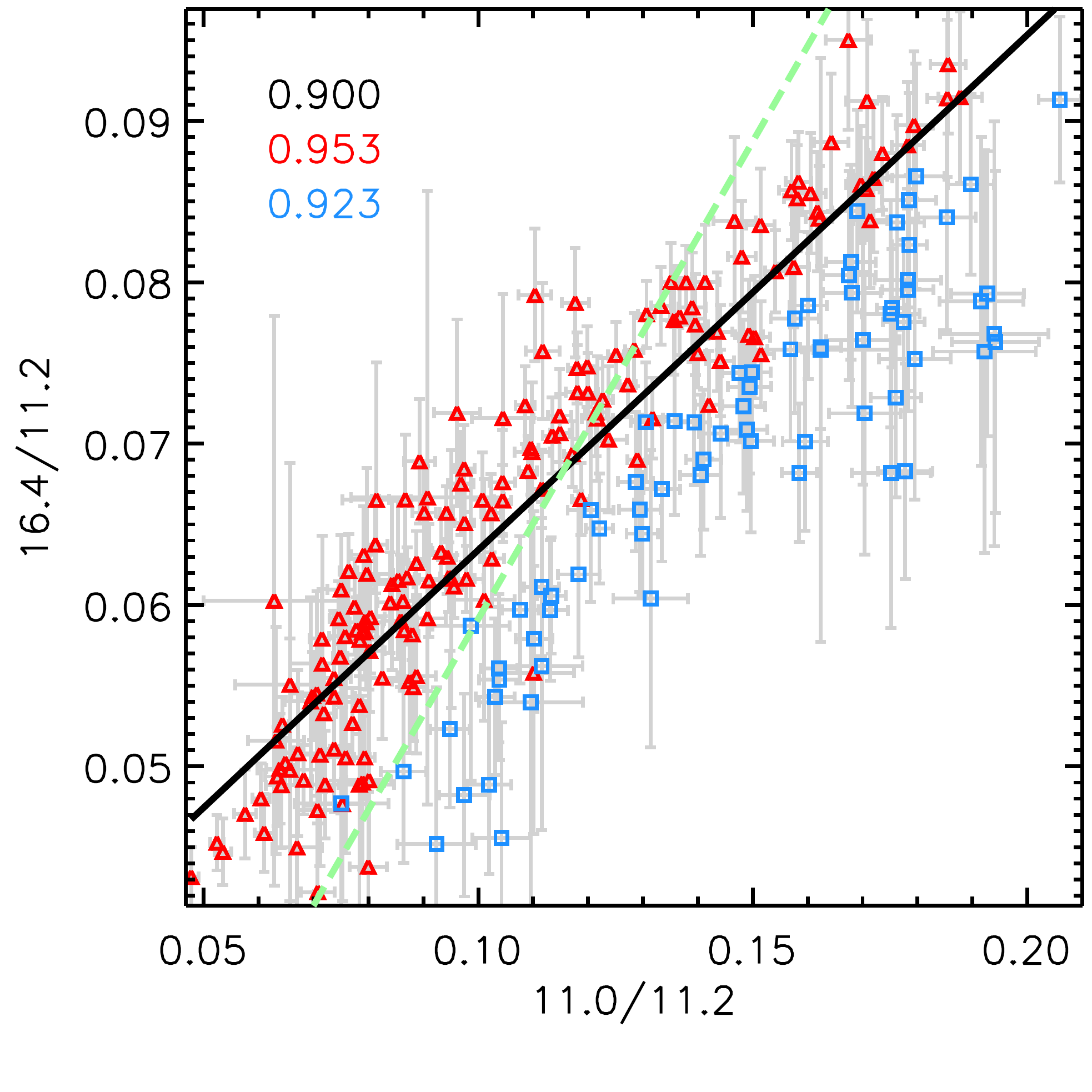}
  \includegraphics{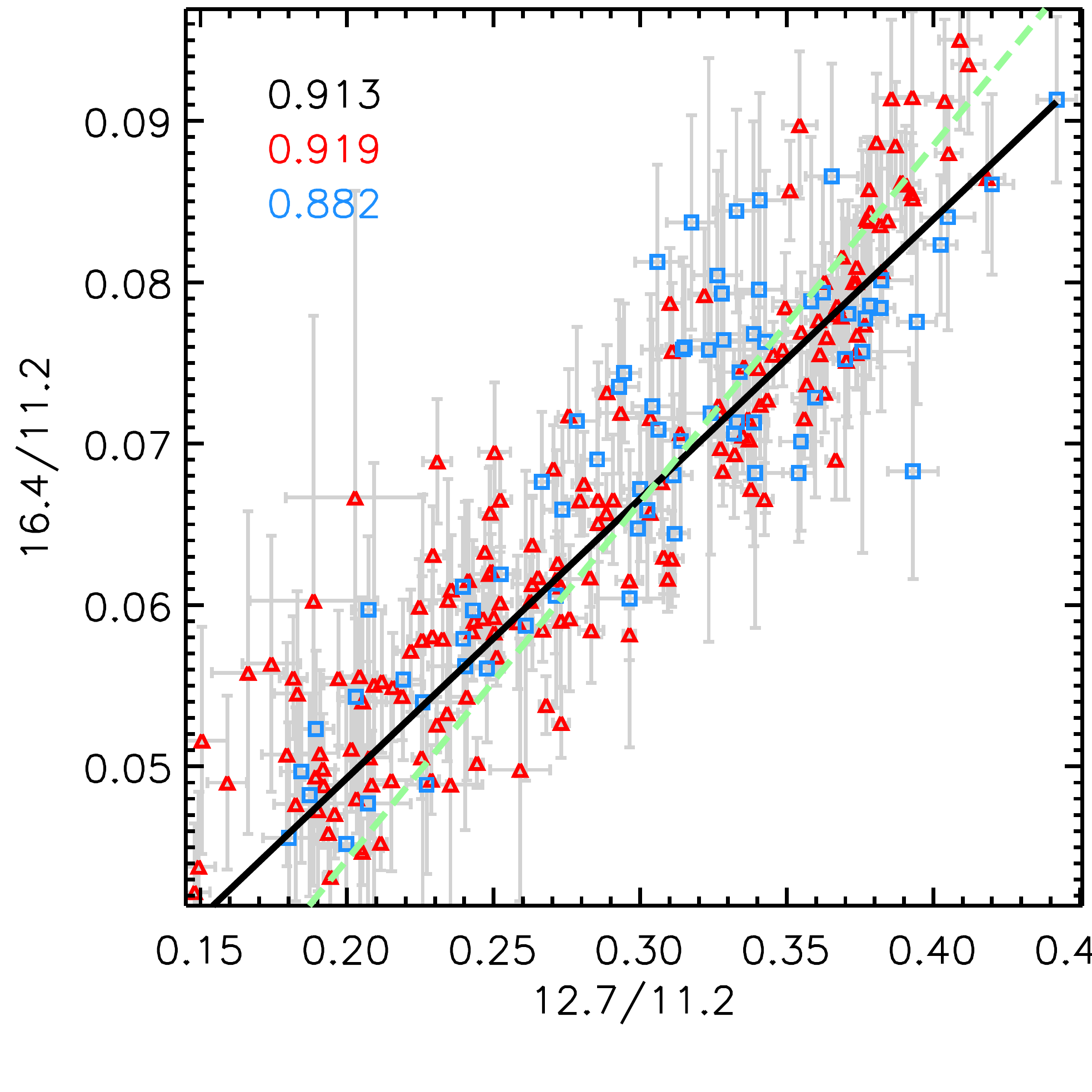}
  \includegraphics{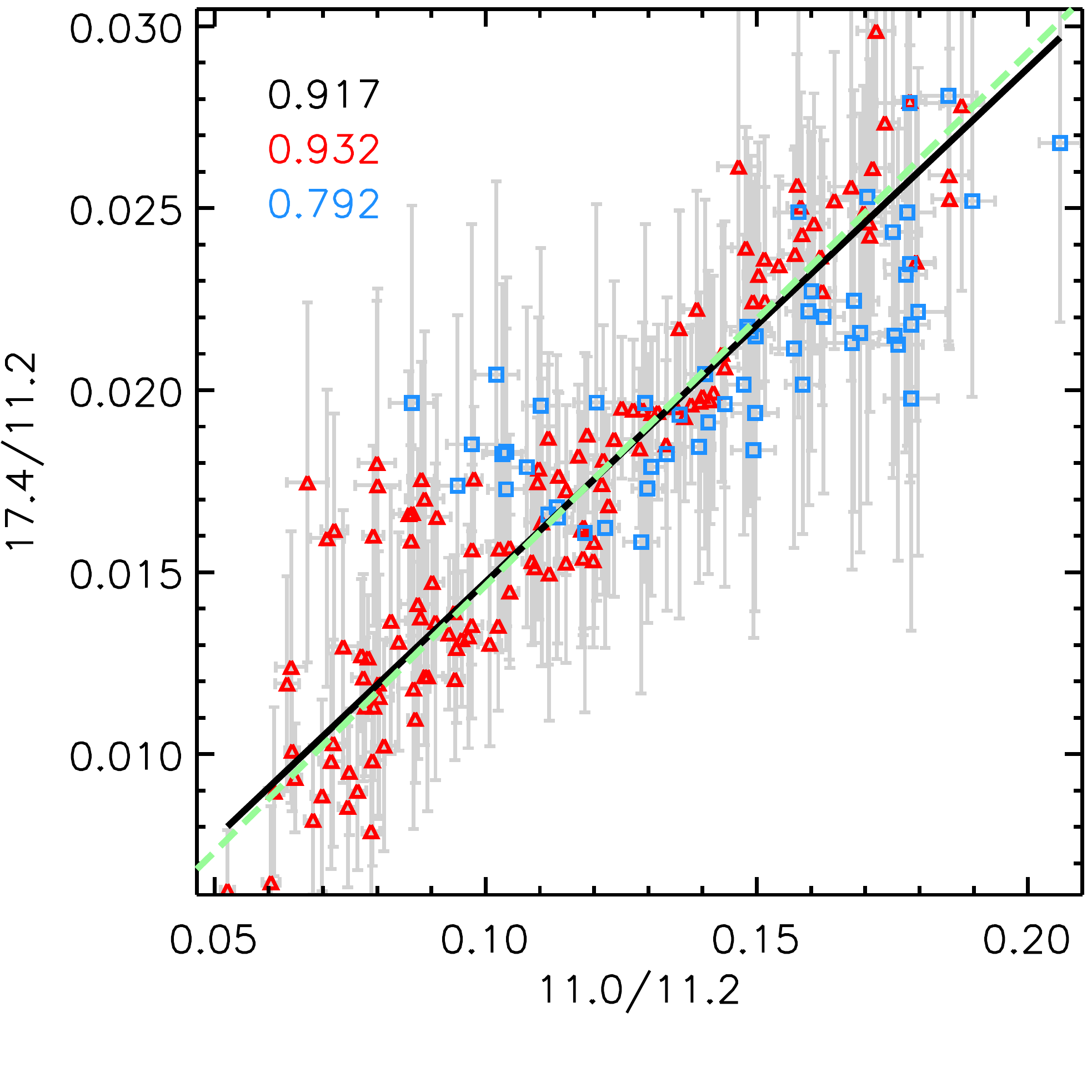}
  \includegraphics{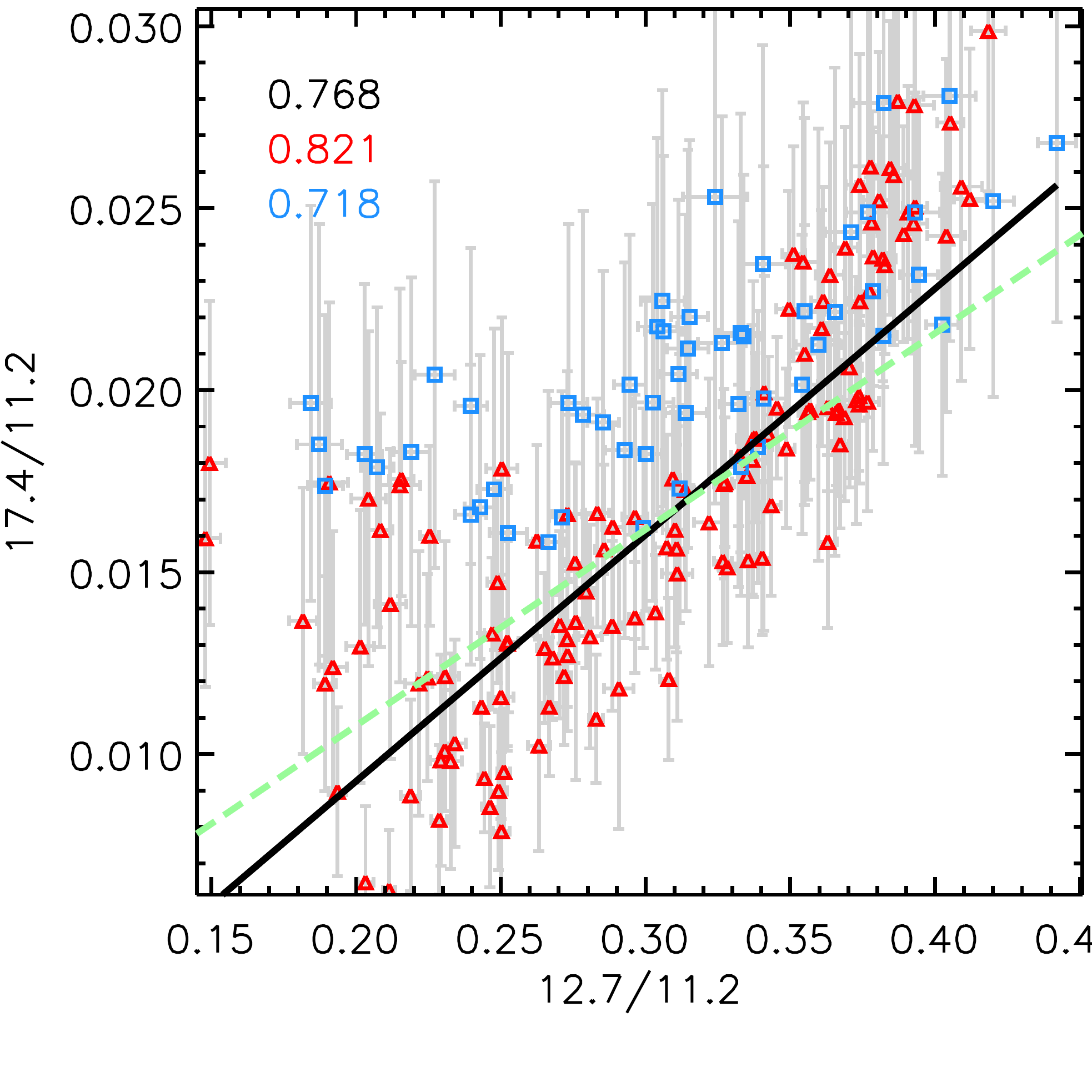}}
\resizebox{18cm}{!}{%
 \includegraphics{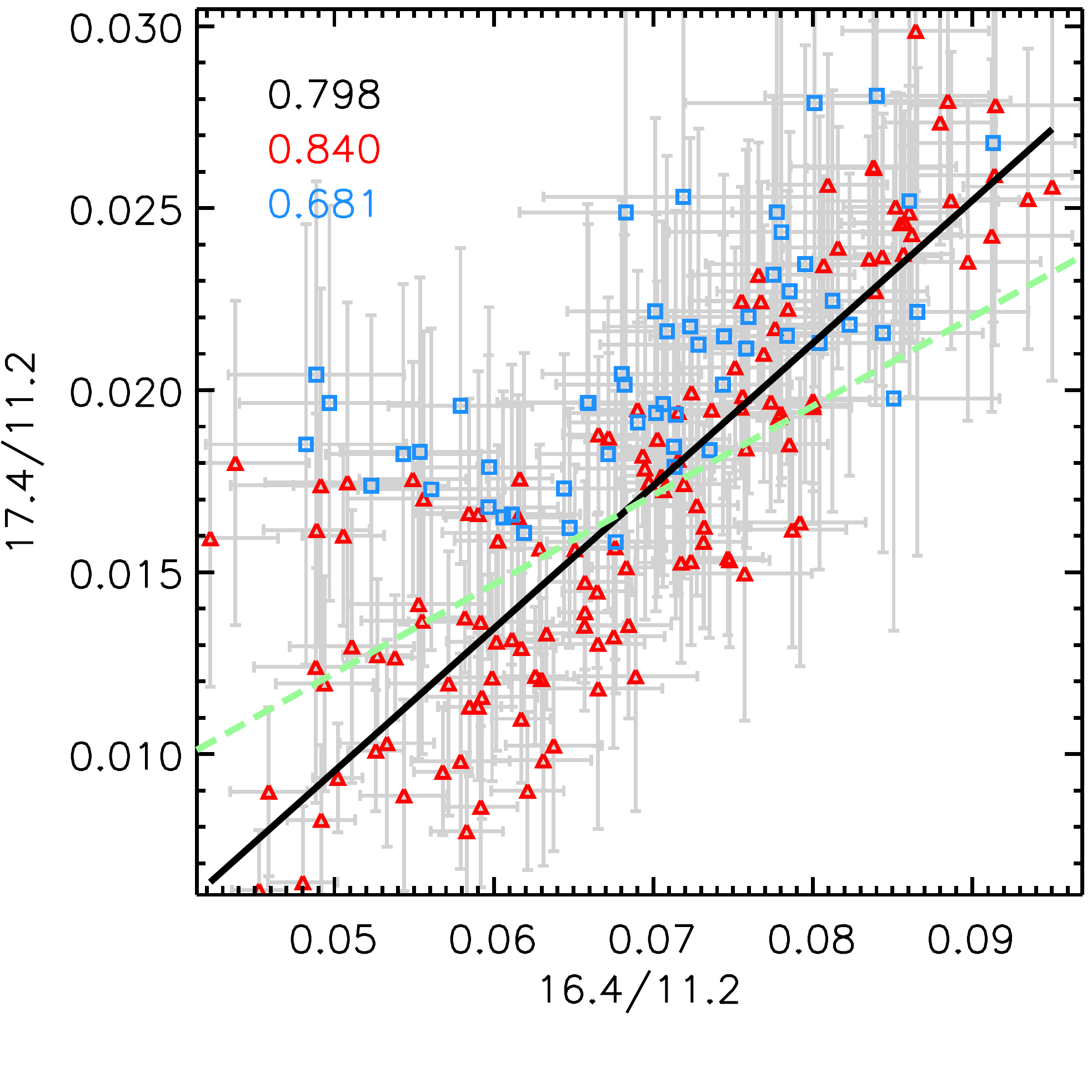}
  \includegraphics{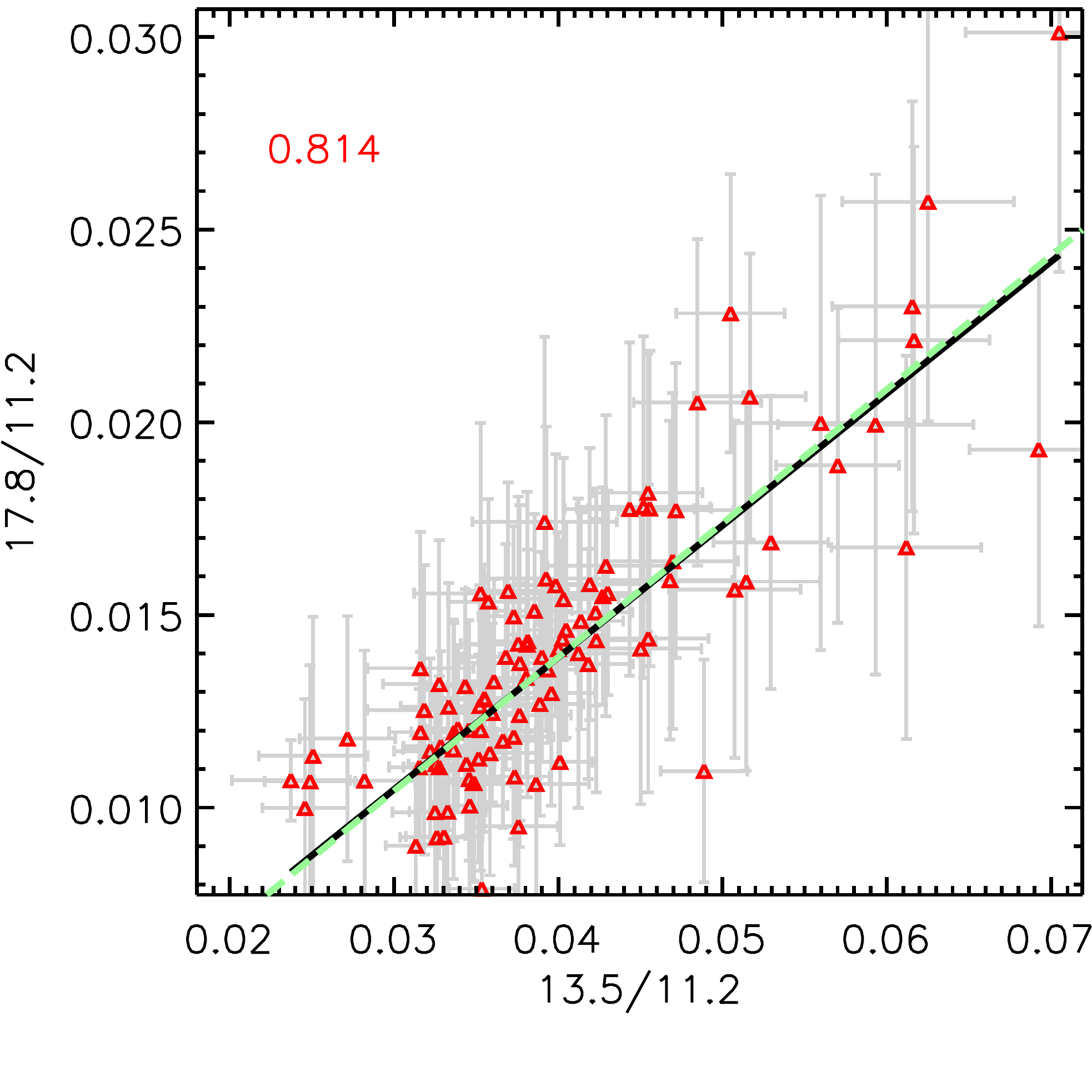}
  \includegraphics{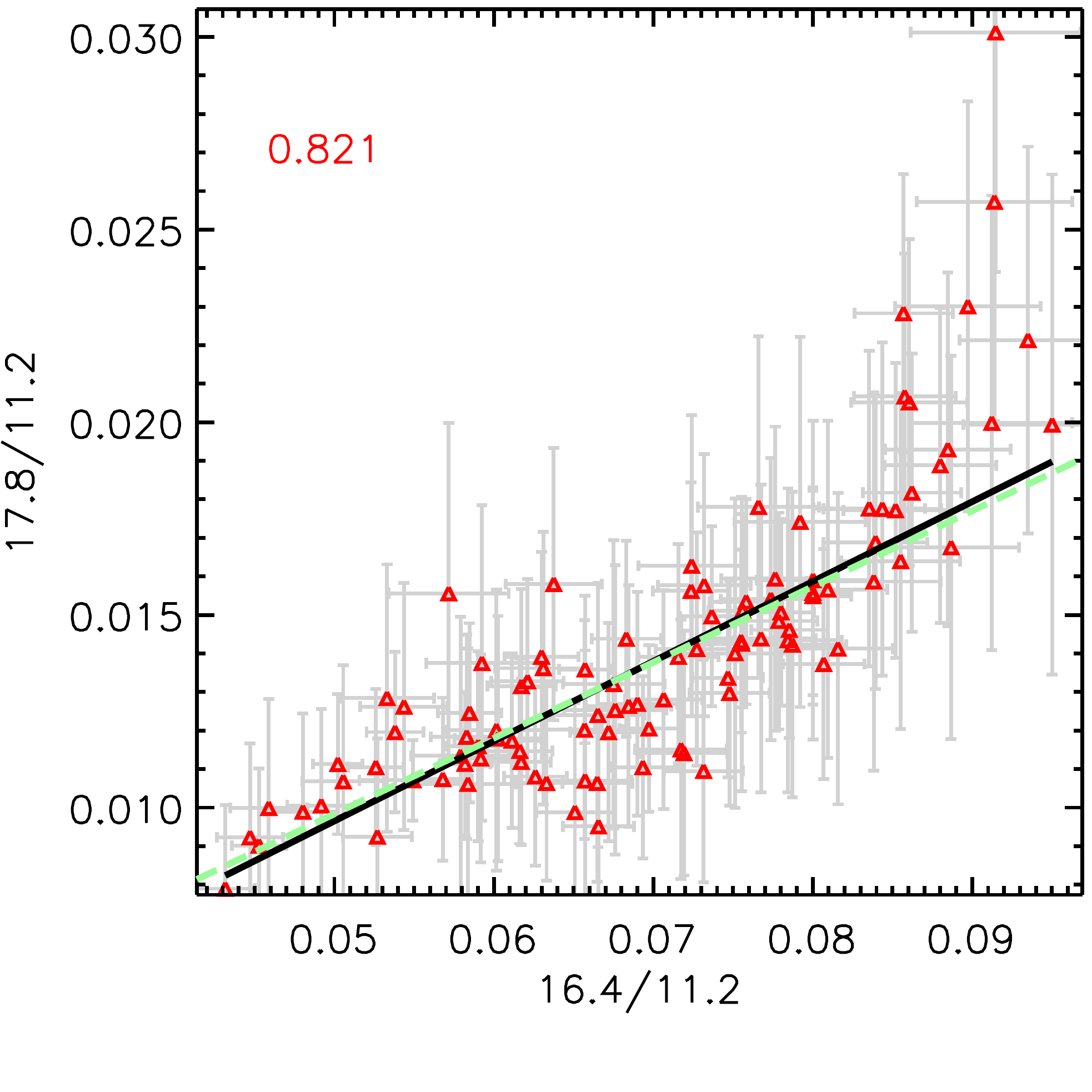}
   \includegraphics{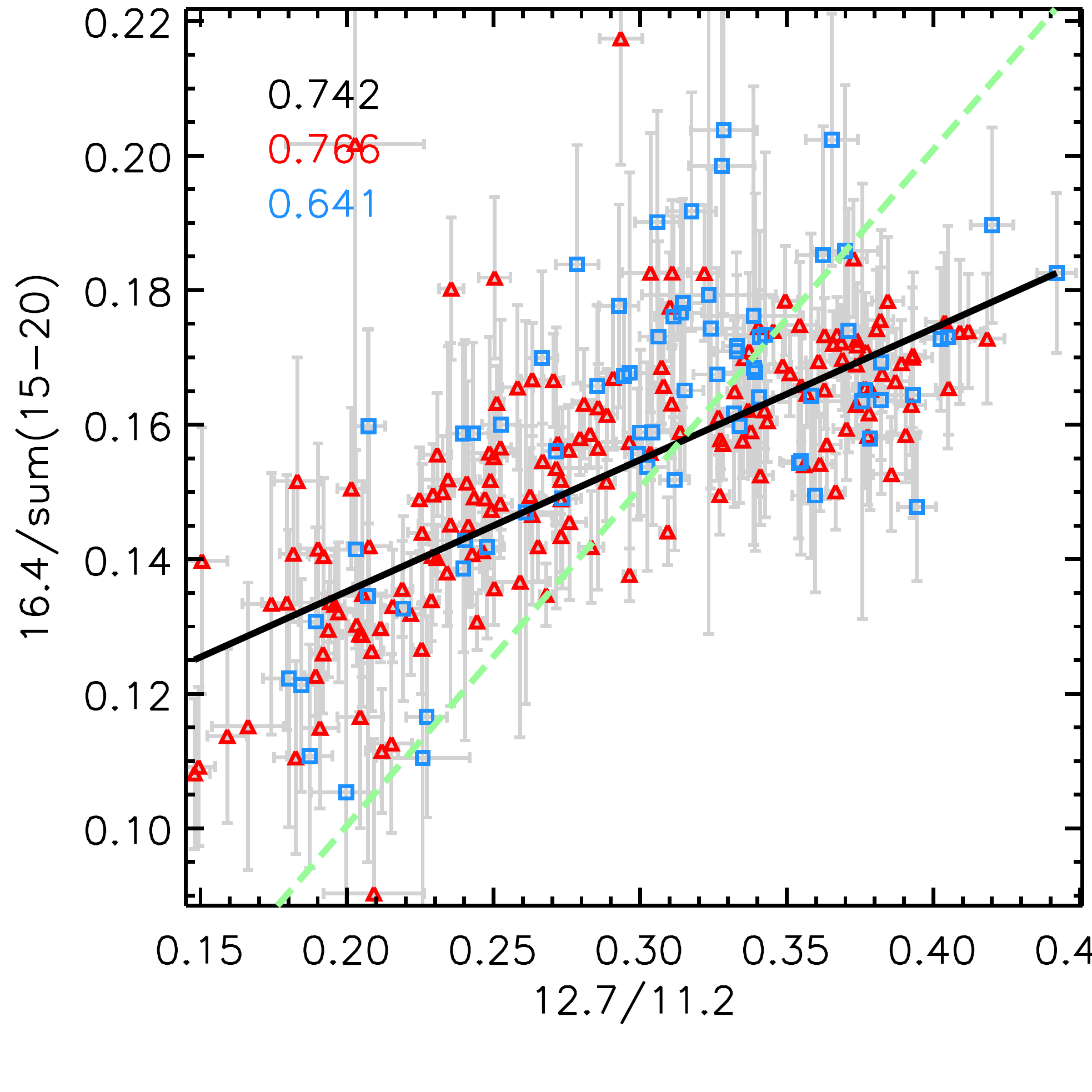}}
\resizebox{18cm}{!}{%
  \includegraphics{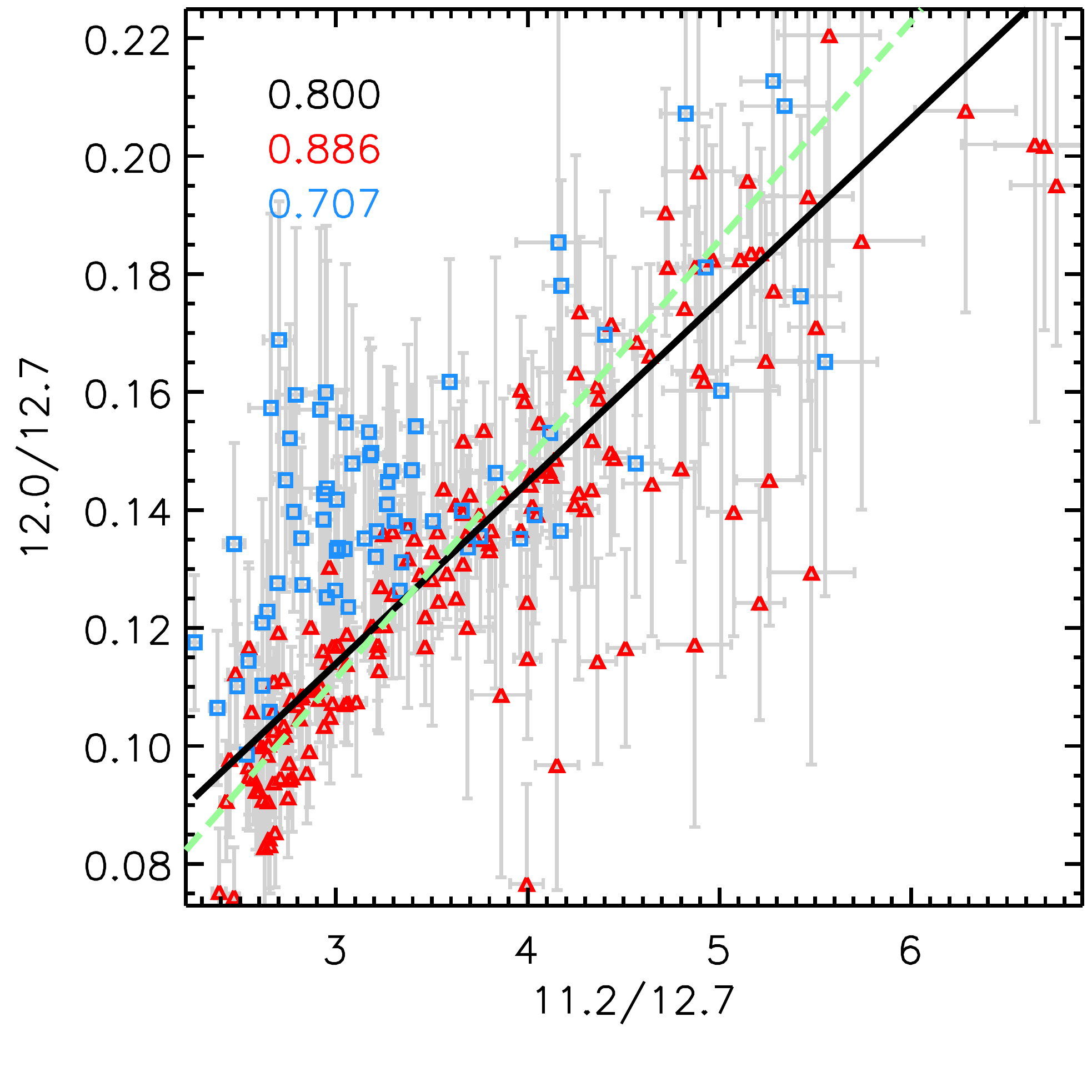}
    \includegraphics{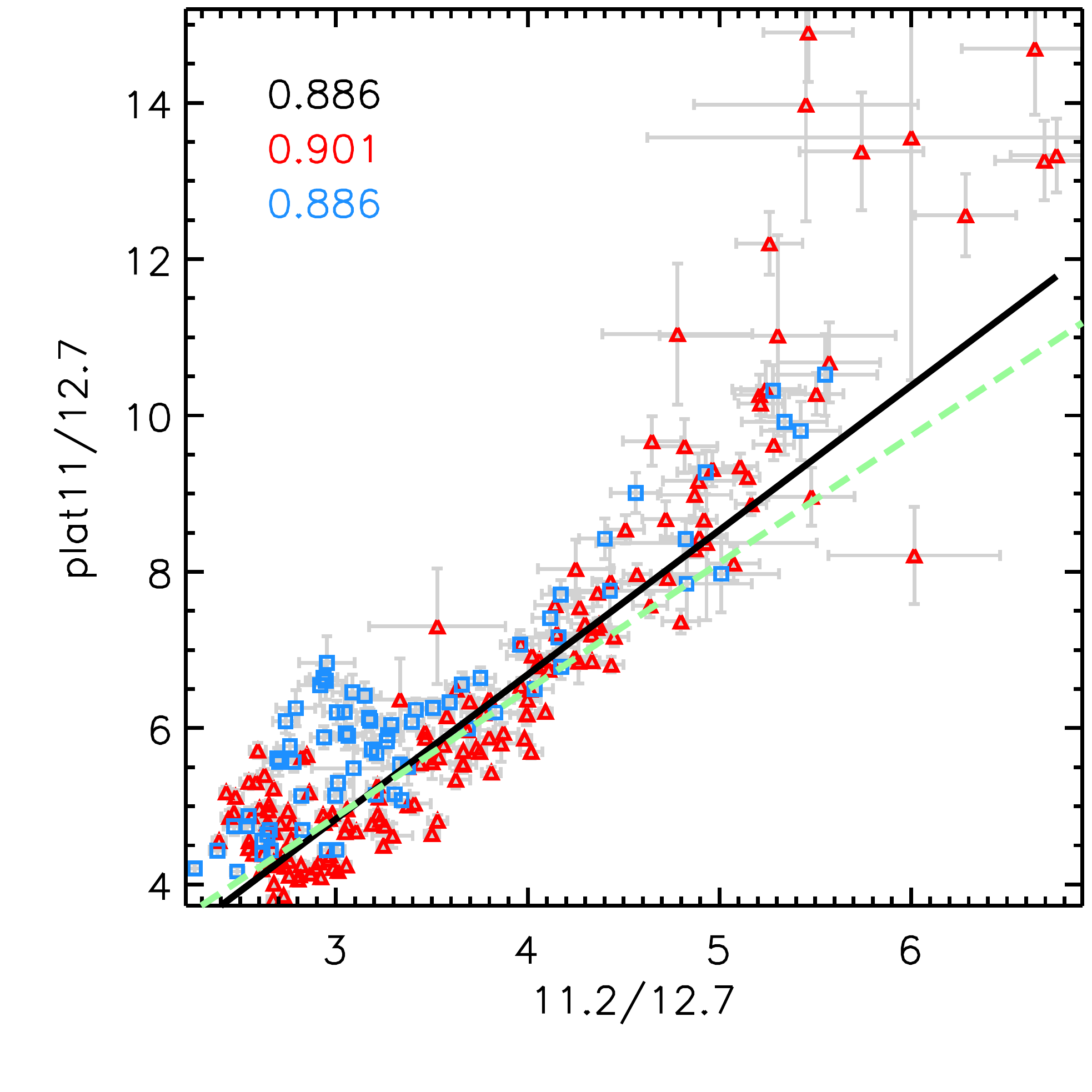}
  \includegraphics{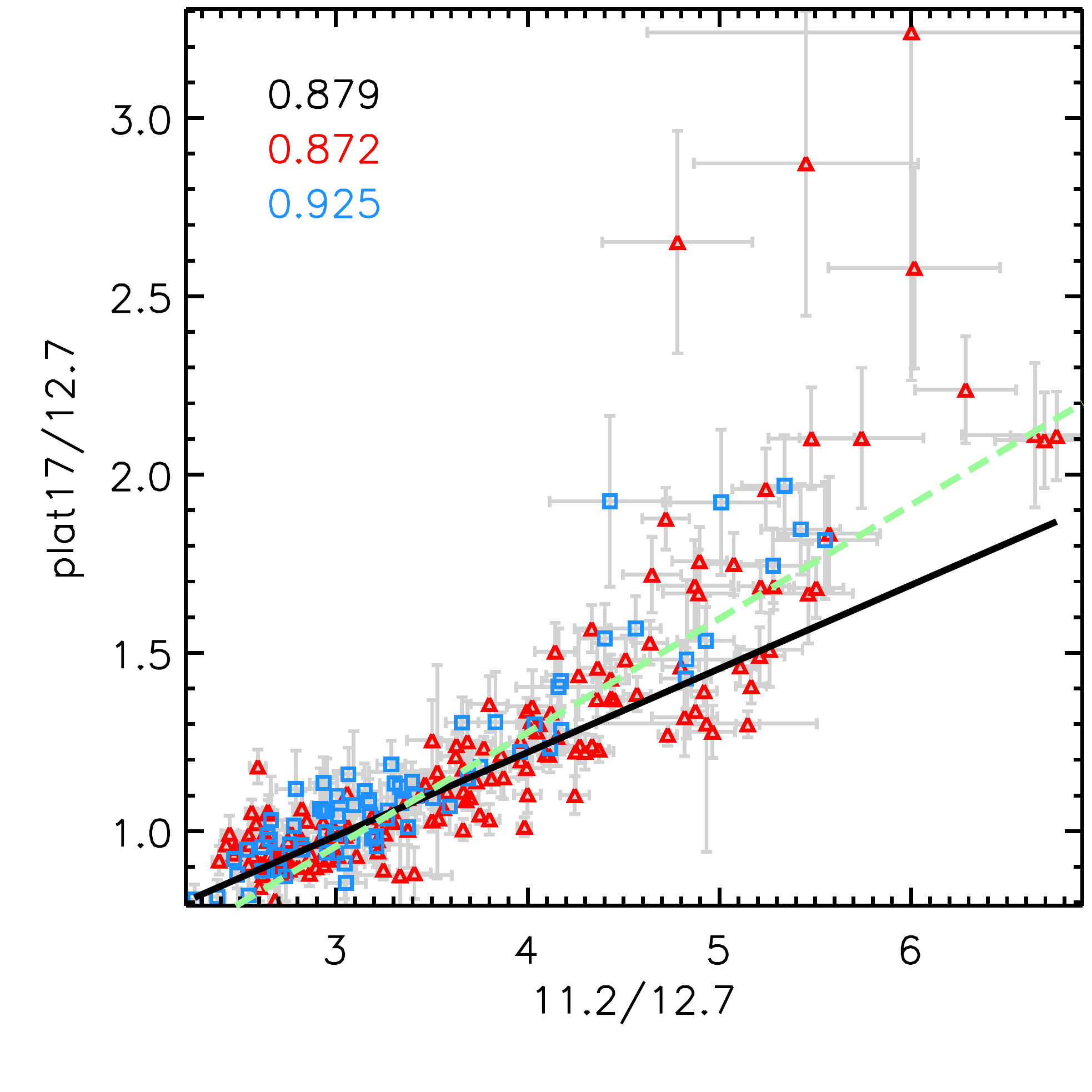}
   \includegraphics{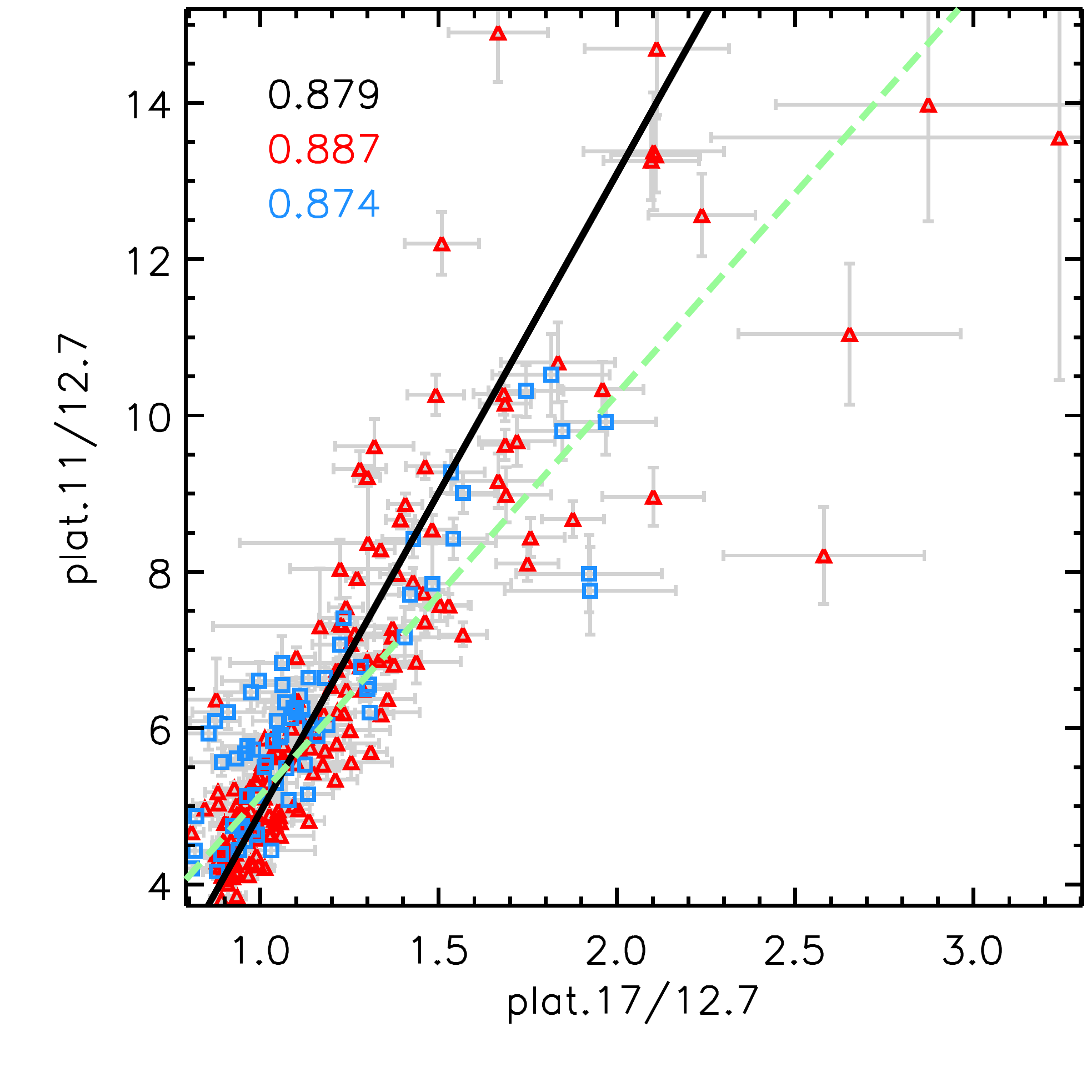}}
  \caption{Correlations in PAH SH intensity ratios across NGC~2023. The red squares are for the south map and the blue triangles for the north map. We applied a
cutoff of 3 sigma. The weighted correlation coefficient is given in the top left corner for both the north and south map (black), for the south map (red) and for the
north map (blue). The Levenberg-Marquardt least-squares minimization fit to the N + S data is shown as a solid line. The dashed line represents a fit to the data forced through (0,0). The fit parameters can be found in Table \ref{fit_parameters}. Plat11 refers to the 10-15 \mum\, plateau, plat17 to the 15-18 \mum\ plateau and 17.4 to the total integrated flux of the 17.4 \mum\, band (including both PAH and C$_{60}$ emission). } 
\label{fig_corr_sh}
\end{figure*}
%%%%%%%%%%%%%%%%%%%%%%%%%%%%%%%%%%%%%%%%%%%%%%%%%%

\subsection{SH data}
\label{shdata}

The spatial distribution of the various emission components in the 10-20 \mum\, SH data are shown in Figs. \ref{fig_shmaps_s} and \ref{fig_shmaps_n} (the range in colours is set by the minimum and maximum intensities present in the map).  Given that paper I gives a detailed discussion of the 15-20 \mum\, emission, we will only briefly summarize these here (see also Figs. 3 and 4 in paper I). The SH data of the south map only covers the S and SSE ridges and the diffuse emission NW of the line connecting both ridges (see Figs.~\ref{fov} and \ref{fig_slmaps_s}). The SH data of the north map covers both the N and NW ridge though the northern part of the NW ridge is missing (see Figs.~\ref{fov} and \ref{fig_slmaps_n}). The feature correlation plots are shown in Fig.~\ref{fig_corr_sh} and Figs. 5 and 6 in paper I. Their fit parameters and correlation coefficients can be found in table~\ref{fittingparameters} and their line cuts in Fig.~\ref{linecuts_ratios}. We normalized the band fluxes to that of another PAH band to exclude the influence of PAH abundance and column density in the correlation plots, as in the SL data analysis.\\

For the south maps a comparison with the SL maps reveals identical maps (in SL and SH) for the 11.2 PAH feature and the H$_2$ emission bands. They clearly peak at the S and SSE positions with the H$_2$ emission, tracing the PDR front, being more concentrated along these two ridges only. The continuum emission differs in the SL and SH data in terms of the concentration on the S and SSE ridge: the SL 14.7 \mum\, emission is more extended than the SH 15 \mum\, emission. The 11.0 PAH feature differs slightly in the SL and SH data: the SH data exhibit peak emission in the SSE ridge while the SL data doesn't. This may be (partially) due to wavelength shifts occurring in the edges of the SL map (see Section \ref{reduction}) and due to different spectral resolution of the SH and SL mode. The morphology of the 12.7 PAH feature in the SH data shows a decreased emission in the S ridge and enhanced diffuse emission NW of the line connecting the S and SSE ridge compared to the 12.7 map based on the SL data, and thus is more similar to e.g. the 7.7 PAH morphology. This discrepancy between the SH and SL data likely originates in the less accurate continuum determination in the SL data due to blending of the emission features (11.2 PAH, 12.0 PAH, H$_2$ and 12.7 PAH).  The spatial distribution of the 12.0 \mum\, PAH emission and the 10-15 \mum\, and 15-18 \mum\, plateau emission is very similar to that of the 11.2 \mum\, PAH emission - although the 12.0 \mum\, band shows weaker peak emission in the S ridge. The 13.5 \mum\, PAH emission shows a morphology in between that of the 11.2 and 16.4 \mum\, PAH emission: compared to the 11.2 \mum\, PAH map, it exhibits a decreased emission in the S ridge and enhanced emission south of the S and SSE ridge and in the north of the map.  In addition, the very weak 14.2 \mum\, PAH band seems to have a spatial distribution similar to the 11.0 \mum\, PAH feature. 
As discussed in paper I, the 16.4 \mum\, PAH emission exhibits a similar spatial variation as does the 12.7 \mum\, PAH emission. The weaker 17.8 \mum\, PAH emission exhibits a morphology between that of the 11.2 PAH and 12.7 PAH emission: its emission peaks in the SSE ridge but it lacks emission in the S ridge and shows slightly enhanced diffuse emission. The 15.8 \mum\, PAH band also shows similarities with the 11.2 \mum\, PAH morphology but is more sharply peaked on the S and SSE ridges. Finally, the 17.4 \mum\, emission (due to both PAH and C$_{60}$) resembles the 11.0 emission most closely. It shows enhanced emission in the north corner of the map but this is mostly due to C$_{60}$ as discussed in paper I. The diffuse emission is dominated by PAH emission and is identical to that seen for the 11.0 emission except that it exhibits decreased emission in the SSE ridge (hence overall, it is identical to the SL 11.0 emission). 

For the north map, the continuum emission, the 11.0 and 11.2 \mum\, PAH bands are the same in both SH and SL data. The SH 12.7 \mum\, emission north map shows small deviations from the SL map, as for the south map. Specifically, for the SH map, the south section of the NW ridge seems to be brighter while the middle section seems to be fainter.  In contrast with the south map, the spatial distribution of the 12.0 and 13.5 \mum\, emission deviates from that of the 11.2 \mum\, emission. Specifically, both bands exhibit a lack of emission in the N ridge while the emission in the NW ridge is more centred towards the lower half of the ridge for the 12.0 \mum\, band and is restricted to the southern part of the ridge only for the 13.5 \mum\, band. The latter is very similar to the 11.0 emission. As discussed in paper I, the 16.4 and 12.7 \mum\, PAH bands exhibit similar morphologies. The 17.4 \mum\, emission peaks in the NW ridge, however, due to its weakness, it's hard to further distinguish its morphology but it seems like it encompasses the entire NW ridge. Finally, the morphology of H$_2$ S(2) emission is the same in both the SH and SL data, though the H$_2$ S(2) SH map is clearly of superb quality with respect to the SL map. Combining the H$_2$ results of both SH and SL observations of the north FOV, we note that the H$_2$ emission peaks in the N ridge for the S(2) and S(1) transitions (as does the continuum emission) while it peaks in the NW ridge for the S(3) and  S(5)\footnote{ as measured using PAHFIT, see Appendix~\ref{pahfit}.} transitions (as does the PAH emission). In contrast, the morphology of all H$_2$ lines in the south map is similar.  A full analysis of this interesting result is beyond the scope of this paper, however the difference might be attributed to different excitation processes for low J versus high J lines. \cite{Sheffer:11} found that in the south map collisions dominate the excitation to the J = 7 level, however in the north the line maps can look quite different if the low J lines are collisionally excited while the high J lines are UV pumped. The higher J lines, whether they are UV pumped or collisionally excited, should arise from regions that are closer to the surface of the PDR where the UV field and temperatures are higher \citep[see e.g.][]{Sheffer:11}.\\

The correlation plots reveal less tight correlations amongst the SH bands and hence the obtained correlation coefficients are lower compared to those obtained with the SL data.  However, the correlation coefficient for the 11.0/11.2 vs. 12.7/11.2 data is similar for both the SL and SH data (see Table \ref{fittingparameters}). Note however that an offset is present between the SH and SL data which is likely due to the less accurate continuum determination in SL resulting in a large underestimation of the weaker band intensities. In addition, an offset is present between the north and south map in both the SL and SH data of which the origin remains unclear. If much weaker emission in the N map were the culprit and thus the accuracy of the 11.0 \mum\, flux, it would effect all correlations involving the 11.0 \mum\, PAH which is not the case. Nevertheless, the fact that we see similar correlation coefficients in the SL and SH data for the 11.0/11.2 vs. 12.7/11.2 data indicates that the PAH features at the longer wavelengths are intrinsically less connected with each other. Keeping this in mind, we can further investigate the behaviour of the longer wavelength bands. 

In paper I, we found correlations between the 11.2 PAH, 15.8 PAH and the 15-18 \mum\, plateau  (correlation coefficients of 0.88--0.89; group 1) and correlations between the 16.4 and 12.7 \mum\, PAH emission (correlation coefficient of 0.91, group 2). In addition to these, we find that the 11.0 \mum\, PAH emission correlates well with the 12.7 and 16.4 PAH emission (correlation coefficients of 0.90--0.95; amongst highest in the SH dataset) and that the 10-15\mum\, plateau correlates well with the features in group 1. Also, as revealed in the south maps, the 17.4 \mum\, emission correlates well with the 11.0 (correlation coefficient of 0.93). Note that this comprises not only PAH emission but also a small contribution of C$_{60}$ emission, in particular in the north corner of the map (see paper I). To a lower degree, the 17.4 \mum\, emission also correlates with the 12.7 and 16.4 \mum\, emission having correlation coefficients of 0.82 and 0.84. 
Finally, the weak 13.5 \mum\, emission band exhibits weaker correlations with the 11.0 and 16.4 \mum\, PAH emission while the weak 12.0 \mum\, PAH emission correlates well with the 11.2 \mum\, PAH emission.

\begin{figure*}[tbp]
\centering
\resizebox{16cm}{!}{%
  \includegraphics{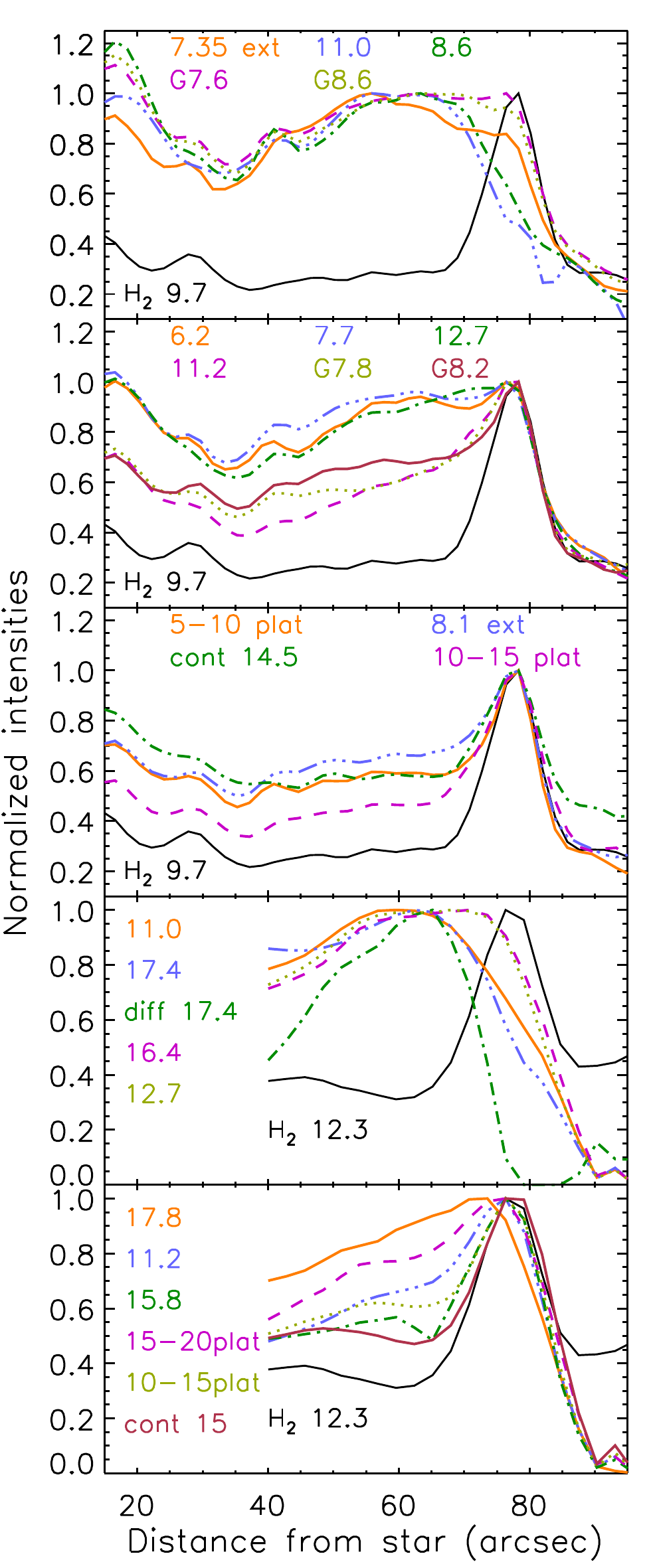}
    \includegraphics{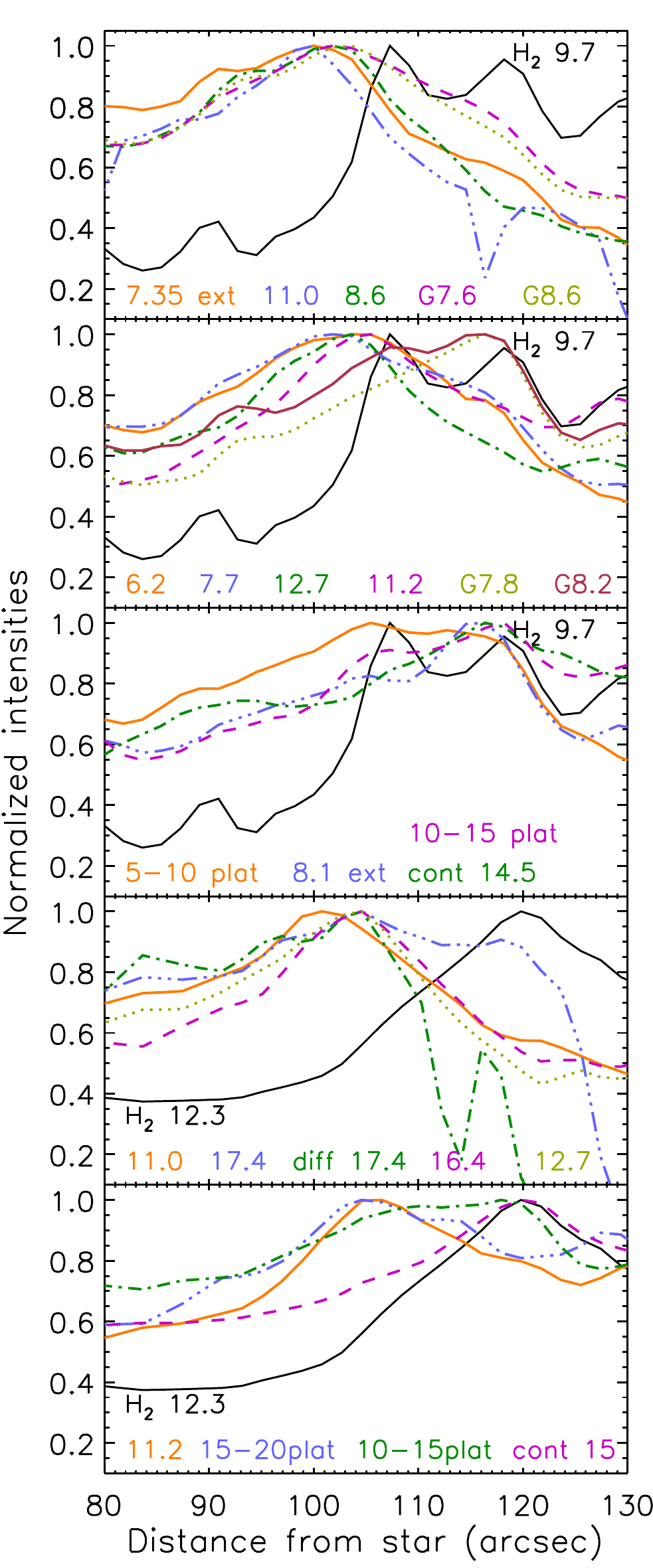}}
\caption{\label{linecuts} Normalized feature/continuum intensity along a projected cut across the south (left) and north (right) FOV directed toward HD37093.  The cut is shown in Figs~\ref{fig_shmaps_s} and \ref{fig_shmaps_n} and is the same for SL and SH data. The top three panels show the SL data and the bottom two the SH data. }
\end{figure*}

\subsection{Spatial sequence}
\label{sequence}
 
The spatial distributions of the individual PAH band intensities (and
continuum emission) show great diversity (Sections \ref{sldata} and \ref{shdata}, and Figs. \,\ref{fig_slmaps_s}, \ref{fig_slmaps_n}, \ref{fig_shmaps_s} and \ref{fig_shmaps_n}).  
They reveal that the (peak) emission for these bands occurs at different distances from the illuminating source.  To further exemplify this, Fig.~\ref{linecuts} presents normalized emission profiles for PAH features, H$_2$ lines and the continuum component for both maps. The following spatial sequence emerges (from furthest to the source to nearest
away from the source):
\begin{itemize}
\item Group 1: 8.1 \mum\, extreme, continuum emission, 10-15 \mum\, plateau 
\item Group 2: G7.8, G8.2, 5-10 \mum\, plateau
\item Group 3: 11.2, 15.8, 15-18 \mum\, plateau, 12.0
\item Group 4: 12.7 (SL)\footnote{A slightly different morphology is present in the SL and SH data (see Section~\ref{shdata}).}, 17.8
\item Group 5: 6.2, 7.7, 16.4, 12.7 (SH)$^{15}$
\item Group 6: G7.6, G8.6, 8.6 (GS)\footnote{For clarity, we make a difference between the 8.6 \mum\, strength when applying a global spline continuum (GS) or local spline continuum (LS).}
\item Group 7: 11.0, 17.4 PAH, 8.6 (LS)$^{16}$
\item Group 8: 7.35 \mum\, extreme
\end{itemize}

The distinction between groups 1, 2, and 3 is made based on the N map where these features peak in either the N or NW ridge or both. And while these features exhibit a slightly different line profile in the S map, they all peak in the S ridge, as does the H$_2$ emission (see Fig.~\ref{linecuts}). Hence, the differentiation between groups 1, 2, and 3 is not clearly present in the south map. Note that the different lines of H$_2$ emission in the N map show different morphologies (see Section~\ref{shdata}). The spatial behaviour of the 15-20 \mum\, bands has already been reported in paper I and \cite{Shannon:15} with respect to the 11.0, 11.2 and 12.7 \mum\, features. Specifically, groups 1 and 2 presented in paper I are part of group 3 and 5 respectively in this spatial sequence. A similar spatial behaviour is found in NGC~7023 for the 15-20 \mum\, bands \citep{Shannon:15}. 

The spatial morphology of individual features, and thus the spatial sequence as given above, is influenced by the applied decomposition of the spectra into individual components. Therefore, we also analyzed the spectral maps with PAHFIT and a detailed comparison with the spline method is given in Appendix \ref{pahfit}. The PAHFIT decomposition also reveals distinct spatial morphology for different components but the (detailed) spatial sequence is different. This originates in the fact that some individual features within PAHFIT trace different emission components than in the spline method: for instance, several PAHFIT features include a fraction of the underlying plateaus (as defined by the spline method). As these plateaus are independently of the features at similar wavelengths, this influences the obtained spatial sequence. 

When considering a specific decomposition method however, rather than being exact, this spatial sequence is indicative of the overall, gradual and continuous (rather than discrete) change in PAH population with distance from the star. Indeed, features in consecutive bins may switch bins and/or consecutive bins may be merged. For example, some features (e.g. G8.2 and continuum emission) exhibit the same spatial distribution in the south map while being distinct in the north map. This reflects the local geometries and different conditions of the maps \citep[the FUV radiation field is $\sim$500\nolinebreak~G$_0$ and $\sim$10$^4$ G$_0$ for respectively the N and S position and the density is $\sim$10$^{4}$ cm$^{-3}$ and $>$10$^{5}$ cm$^{-3}$ respectively;][]{Burton:98, Sheffer:11}. However, the fact that they exhibit different spatial distributions in the north map suggests they are intrinsically different. Hence, towards many other objects, such a detailed sequence is likely not spatially resolved. 

%\clearpage

\section{The infrared emission features and PAHs}
\label{lab}

While these IR emission features at 3.3, 6.2, 7.7, 8.6, 11.3 and 12.7 \mum\, are generally assigned to a family of vibrationally excited PAHs, the identification of the specific molecules has yet to be made. Nevertheless, comparison of the observations with experimental and theoretical studies have put strong constraints on the emitting PAH population. 

First, experimental and theoretical studies of PAHs clearly demonstrate the importance of PAH charge \citep[][and
ref. therein]{Szczepanski:labcations:93, Hudgins:cationsI:94, Langhoff:neutionanion:96,
Kim:gasphasepyrenecation:01, Hudgins:04}. The observed correlations between the main PAH bands (i.e. between the 3.3 and 11.2 \mum\, bands and between the 6.2, 7.7 and 8.6 \mum\, bands) are then interpreted to reflect the charge dominance of the bands: the CH stretching mode at 3.3 \mum\, and the solo CH out-of-plane bending mode at 11.2 \mum\, are dominated by neutral PAHs while the CC stretching modes at 6.2 and 7.7 \mum\, and the CH in-plane bending mode at 8.6 \mum\, are governed by ionized PAHs \citep{Joblin:ngc1333:96, Hony:oops:01, Galliano:08}. At longer wavelengths, the solo CH out-of-plane bending mode at 11.0 \mum\, is assigned to ionized PAHs \citep{Hudgins:tracesionezedpahs:99, Hony:oops:01} while the 15.8 \mum\, band and the broad 15-18 \mum\, plateau \citep[similar to the broad 17 \mum\, band as defined by PAHFIT,][]{SmithJD:07} are assigned to neutrals \citep{Peeters:12, Shannon:15}.

\begin{figure}[tbp]
\centering
  \includegraphics[width=8.5cm]{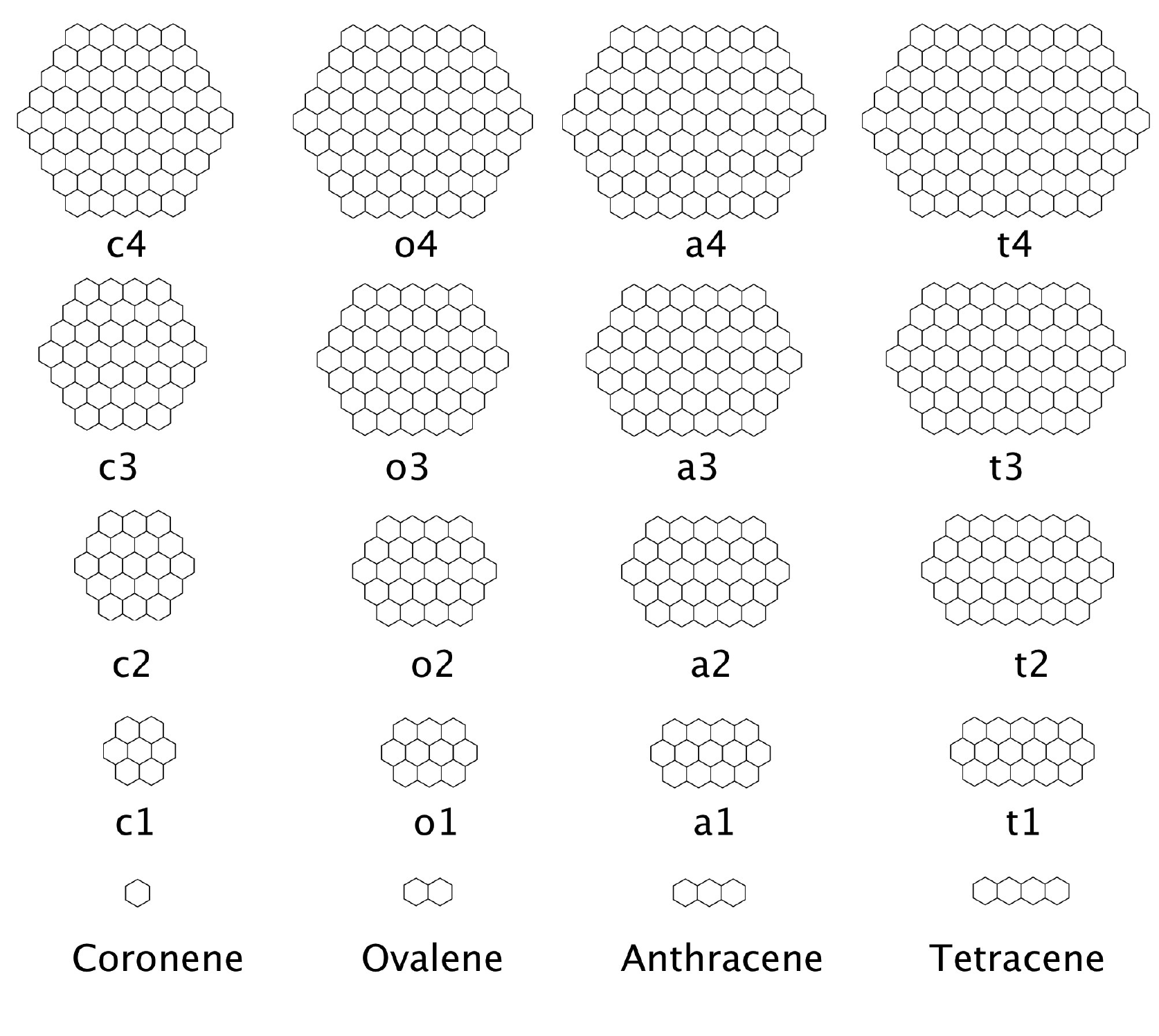}
\caption{\label{molecules} The PAH species, with the number of C-atoms N$_C$ ranging from 24 to 210, and their labels studied in this work. The bottom row shows their central core.  }
\end{figure}

\begin{table}[tbp]
\small
\caption{\label{CHgroups} The number of solo and duo CH groups for the molecules studied.}
\begin{center}
\begin{tabular}{l|cc|l|cc}
molecule & solo & duo & molecule & solo & duo \\
 \hline
 \hline
 & & & & \\[-5pt]
C1 C$_{24}$H$_{12}$ & 0 & 12 & C3 C$_{96}$H$_{24}$   & 12 & 12\\
O1 C$_{32}$H$_{14}$ & 2 & 12 & O3 C$_{112}$H$_{26}$ & 14 & 12 \\
A1 C$_{40}$H$_{16}$ & 4 & 12 & A3 C$_{128}$H$_{28}$ & 16 & 12 \\
T1 C$_{48}$H$_{18}$ & 6 & 12 & T3 C$_{144}$H$_{30}$ & 18 & 12 \\
C2 C$_{54}$H$_{18}$ & 4 & 12 & C4 C$_{150}$H$_{30}$ & 18 & 12 \\
O2 C$_{66}$H$_{20}$ & 8 & 12 & O4 C$_{170}$H$_{32}$ & 20 & 12 \\
A2 C$_{78}$H$_{22}$ & 10 & 12 & A4 C$_{190}$H$_{34}$ & 22 & 12 \\
T2 C$_{90}$H$_{24}$ & 12 & 12 &  T4 C$_{210}$H$_{36}$ & 24 & 12 \\
\hline
\end{tabular}
\end{center}
\end{table}

A secondary effect influencing the intrinsic PAH spectra is the PAH structure. 
The 16.4 \mum\, band seems to be systematically present in PAHs exhibiting pendent rings \citep{Moutou:16.4:00, VanKerckhoven:plat:00, Peeters:plat:04, Boersma:10} and large PAHs with pointy edges \citep{Ricca:12}. However, it is also found to correlate with the 6.2 and 12.7 \mum\, bands \citep{Boersma:10, Peeters:12, Shannon:15}. Hence, both charge and molecular structure likely influences the observed 16.4 \mum\, band \citep[for a detailed discussion, see][]{Peeters:12}.  
Likewise, the 12.7 \mum\, band is assigned to duo and trio CH out-of-plane bending vibrations of large PAHs and hence depends on the PAH edge structure \citep{Hony:oops:01, Bauschlicher:vlpahs1, Bauschlicher:vlpahs2}. It may arise from both neutral and ionized PAHs despite its correlation with the 6.2 and 7.7 \mum\, bands \citep[for a detailed discussion see][]{Hony:oops:01, Boersma:10, Peeters:12}.
In addition, recent theoretical studies have shown that PAH structure also influences the emission at shorter wavelengths, in particular for the 8.6 \mum\, band.   Indeed, symmetric PAHs containing approximately 100 carbon atoms\footnote{Smaller PAHs (less than about 96 carbon atoms) do not show a clear 8.6 \mum\, band \citep{Bauschlicher:vlpahs1, Bauschlicher:vlpahs2} while larger PAHs (more than about 150 carbon atoms up to 384 carbon atoms) still had band {\it positions} consistent with observations but the relative {\it intensities} of the 6.2, 7.7 and 8.6~$\mu$m bands are not consistent with observations \citep{Ricca:12}.} were required to obtain a distinct, sizeable band at 8.6~$\mu$m  \citep{Bauschlicher:vlpahs1, Bauschlicher:vlpahs2}.  Lowering the symmetry by adding more rings or adding irregular edge structures dramatically reduced the size and distinctiveness of the 8.6~$\mu$m band \citep{Bauschlicher:vlpahs2}. 

The reduction in the 8.6~$\mu$m band with lower PAH molecular symmetry and irregular edge structures suggests that the majority of the emitting PAHs have smooth edge structure. Since circular compact PAHs are inherently more thermodynamically stable than non-compact, irregular PAHs, it is not surprising that they dominate the astronomical PAH population.  However, comparison of the computed and observed relative intensities of the 6.2, 7.7, and 8.6~$\mu$m bands suggests the emitting population does not contain only circular, compact species \citep{Ricca:12}. Oval, compact PAHs are also exceptionally stable \citep{Ricca:12} and, until now, have been underrepresented in the collection of computed spectra available to compare with observations. 
Here, we address this deficiency and present the spectra of compact oval PAHs ranging in size from C$_{66}$ to C$_{210}$. The discussion on their astronomical implications will be focussed on the 6 to 9 \mum\, PAH bands and applied to the presented analysis of NGC~2023 (Sect. \ref{discussion}).

\subsection{Model and Methods}
\label{lab_method}

The species studied in this work are shown in Figure~\ref{molecules}.  The bottom line in the figure shows the hexagonal cores around which the rings are systematically added.  They are the carbon skeletons of the molecules benzene, naphthalene, anthracene, and tetracene.  We do not consider these molecules in this work as they have been discussed earlier \citep{Langhoff:neutionanion:96}, but start our study with the second row.  The structures were fully optimized and the harmonic frequencies computed using density functional theory (DFT).  We use the hybrid B3LYP \citep{Becke:93, Stephens:94} functional in conjunction with the 4-31G basis set \citep{Frisch:84}.  All of the DFT calculations are performed using Gaussian~09 \citep{Frisch:09}. The interactive molecular  graphics tool MOLEKEL \citep{Flukiger:00} is used to aid the analysis of the vibrational modes. 

Previous work \citep{Bauschlicher:97} has shown that the computed B3LYP/4-31G harmonic frequencies scaled by a single scale factor of 0.958 are in excellent agreement with the matrix isolation mid-IR fundamental frequencies of the PAH molecules. 

The computed intensities obtained using 14 different functionals, including both pure and hybrids, are in good agreement with each other and with experiment for neutral naphthalene \citep{Bauschlicher:freq:10, Szczepanski:labcations:93}.  As previously noted, the band positions obtained using these functionals also agree well with experiment.  While the same 14 functionals have band positions in good agreement with experiment for naphthalene cation, their intensities are not in particularly good agreement with the two experimental results \citep{Szczepanski:92, Hudgins:cationsI:94, Bauschlicher:freq:10}.  The computed intensities of the ions obtained using the hybrid functionals are in good agreement, as are those obtained using the pure functionals.  However, the pure functional results are somewhat smaller than those obtained using the hybrid functionals.
While there is a systematic difference for the absolute intensities, the relative intensities are more similar for the hybrids and pure functionals.  An inspection of the results in the PAH database shows that the good agreement between theory and experiment for band intensities of naphthalene is quite common for neutrals, as is the qualitative agreement between theory and experiment for the intensities of the cations.  Given the agreement between the different functionals, the agreement of the theory with the experimental data currently available and, the limited number of PAHs that can be studied experimentally, we assume that the theory forms a more consistent set of data than the experiment, and we use the theoretical results to analyze the observed intensities (see Sect. \ref{intensities}).  However, we need to stress that while the computed band positions have been consistently shown to be accurate when compared to available experimental data, the reliability of the computed intensities of ionized PAHs is not yet known.  We suspect that changes in the relative band intensities with shape and size are more reliable than absolute intensities, so we should correctly identify trends in intensity ratios.  Finally we note that any uncertainty in the intensities does not detract from our identification of molecular characteristics that lead to shifts in the band positions (see Sect. \ref{assignments}). 
   
A linewidth of 30 cm$^{-1}$ is taken for the bands shortward of 9~$\mu$m, a linewidth of 10 cm$^{-1}$ is used for the bands between 10 and 15~$\mu$m and a linewidth of 5~cm$^{-1}$ is used between 15-20~$\mu$m; values consistent with current observational and theoretical understanding \citep[see discussion in][]{Ricca:12}. For the 9 to 10~$\mu$m region, the FWHM is scaled in a linear fashion (in wavenumber space) from 30 to 10 cm$^{-1}$. In addition to ignoring any further variations of linewidth as a function of mode, Fermi resonances as well as overtone and combination bands are not taken into account. Despite these limitations, these idealized spectra have proven very useful in better understanding the astronomical spectra.

The computational studies yield integrated band intensities in km/mol, which we broaden in wavenumber space because it is linear in energy.  Thus the units of our synthetic spectra are cm$^{-1}$ for the $x$ axis and km/(mol cm$^{-1}$) for the $y$ axis. The latter is converted to units of a cross section (given in 10$^5 ~$cm$^2$ mol$^{-1}$). The $x$ axis is converted to $\mu$m to compare with observational results, which are commonly reported in $\mu$m.   

Astronomical PAHs are typically observed as the {\it emission} from highly vibrationally excited molecules.  Hence when comparing with observations, our computed 0~K absorption spectra should be shifted to the red to account for the difference between {\it absorption} and {\it emission} from vibrationally excited molecules.  The size of this shift depends on many factors such as the size of the molecule, the anharmonicity of modes, and temperature of the emitting species.  In the past we have redshifted the computed spectra by 15~cm$^{-1}$ to compare with observational spectra \citep[see the discussion in][]{Bauschlicher:vlpahs2}.  {\it For the presentation of the theoretical data of oval compact PAHs (Sect. \ref{lab_results}), we do not apply any shift to the synthetic spectra. However, when comparing to the astronomical observations (Sect. \ref{lab_imp}), we include a 15~cm$^{-1}$ redshift}.

\subsection{IR spectroscopy of the oval, compact PAHs}
\label{lab_results}

The presented theoretical data are part of our ongoing investigations of IR properties of large PAHs \citep{Bauschlicher:vlpahs1, Bauschlicher:vlpahs2, Ricca:12}. In particular, \citet{Ricca:12} reported a detailed description of the spectroscopic properties of the coronene family. Here, we extend this to the ovaline, anthracene, tethracene family. 

The number of vibrational bands in these large molecules is so large that it is not practical to publish them in this manuscript and only plots of the spectra will be given (except for the 3 \mum\, region).  The full list of the band positions and intensities are available on-line in the NASA Ames PAH IR database \citep[][www.astrochem.org\ /pahdb]{Bauschlicher:10, Boersma:14}.

\subsubsection{The CH stretching Vibrations (3.2 - 3.3~$\mu$m)}

The C-H stretching region is very similar for all compact families (coronene, ovalene, anthracene and tethracene), with the aspect ratio having very little effect on this region of the spectra. This 
 thus extends our conclusions for the compact circular PAHs \citep{Ricca:12} to the compact oval PAHs.  A detailed description can be found in Appendix~\ref{lab_results_33}. 

\begin{figure*}
    \centering
\resizebox{15cm}{!}{%
  \includegraphics{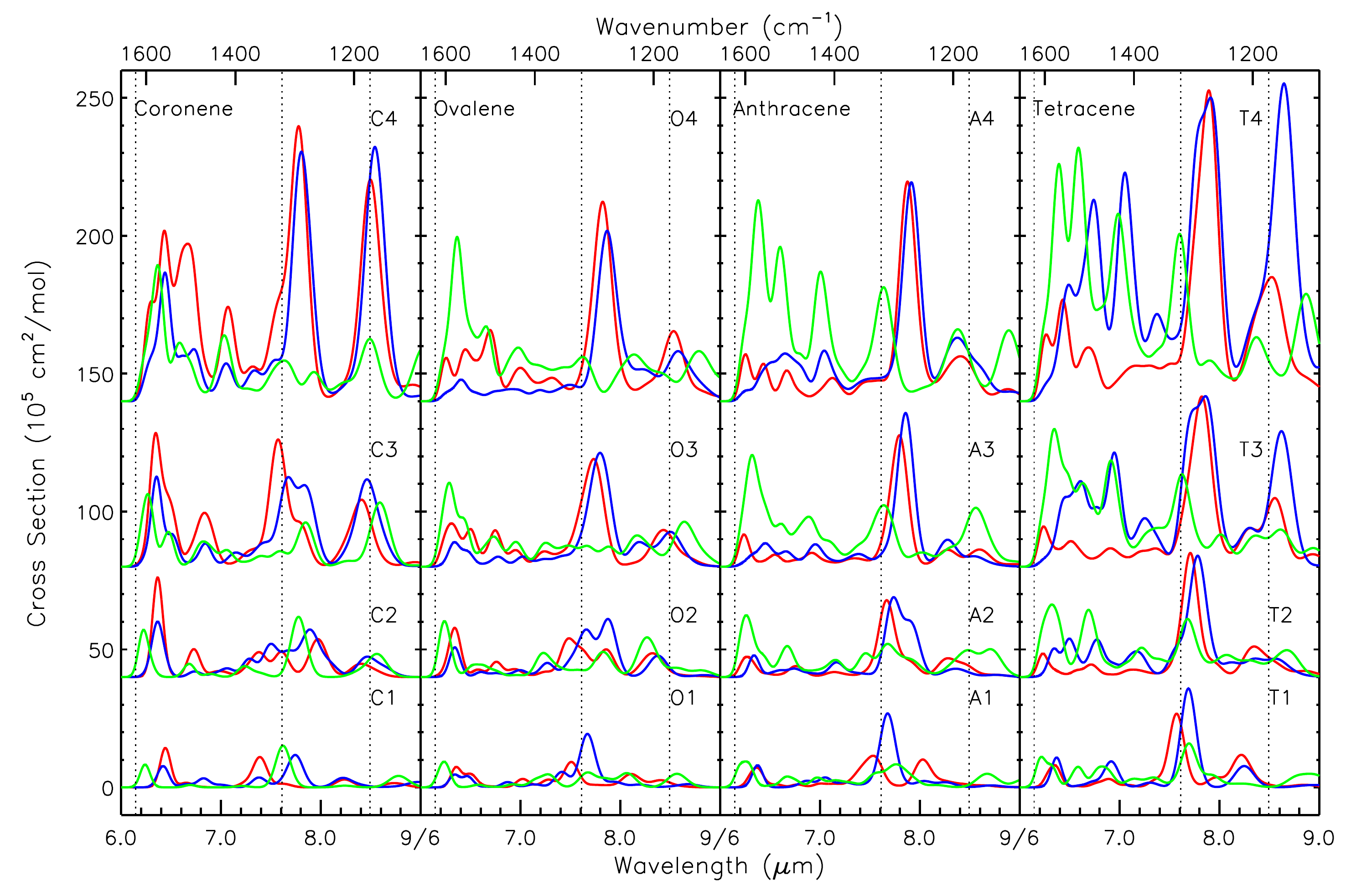}}
\caption{\label{lab6} The 6 to 9~$\mu$m region for the anions (blue), neutrals (x10, green) and cations (red) -- not corrected for redshift. The dotted vertical lines represent the observed positions of the main astronomical PAH bands (at 6.2, 7.7 and 8.6 \mum), blue-shifted by 15 cm$^{-1}$ for comparison with the theoretical data. }
\end{figure*}

\begin{figure*}
    \centering
\resizebox{15cm}{!}{%
  \includegraphics{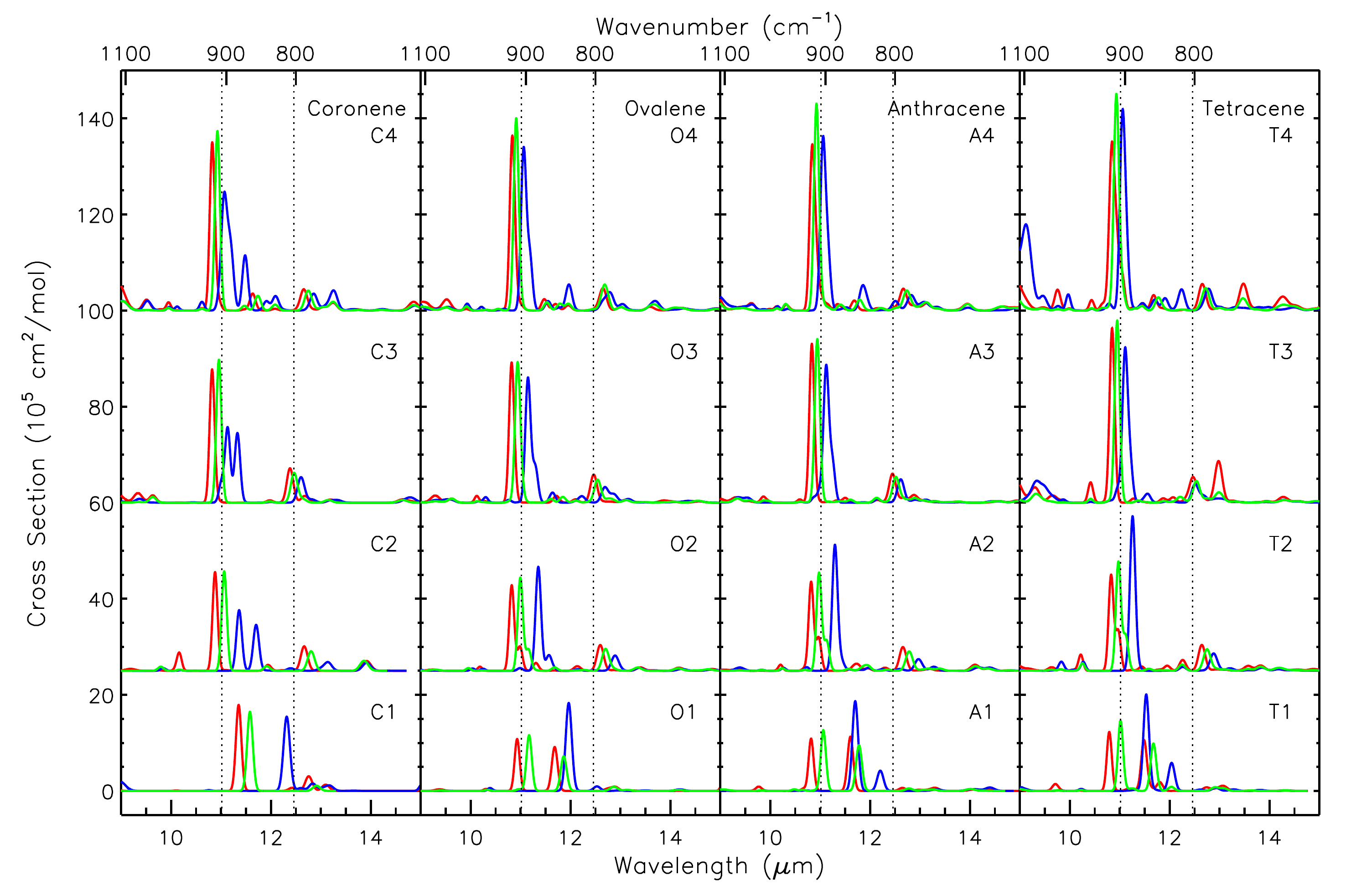}}
\caption{\label{lab9} The 9 to 15~$\mu$m region for the anions (blue), neutrals (green) and cations (red)  -- not corrected for redshift.  The dotted vertical lines represent the observed positions of the main astronomical PAH bands (at 11.2 and 12.7 \mum), blue-shifted by 15 cm$^{-1}$ for comparison with the theoretical data.} 
\end{figure*}

\begin{figure*}
    \centering
\resizebox{15cm}{!}{%
  \includegraphics{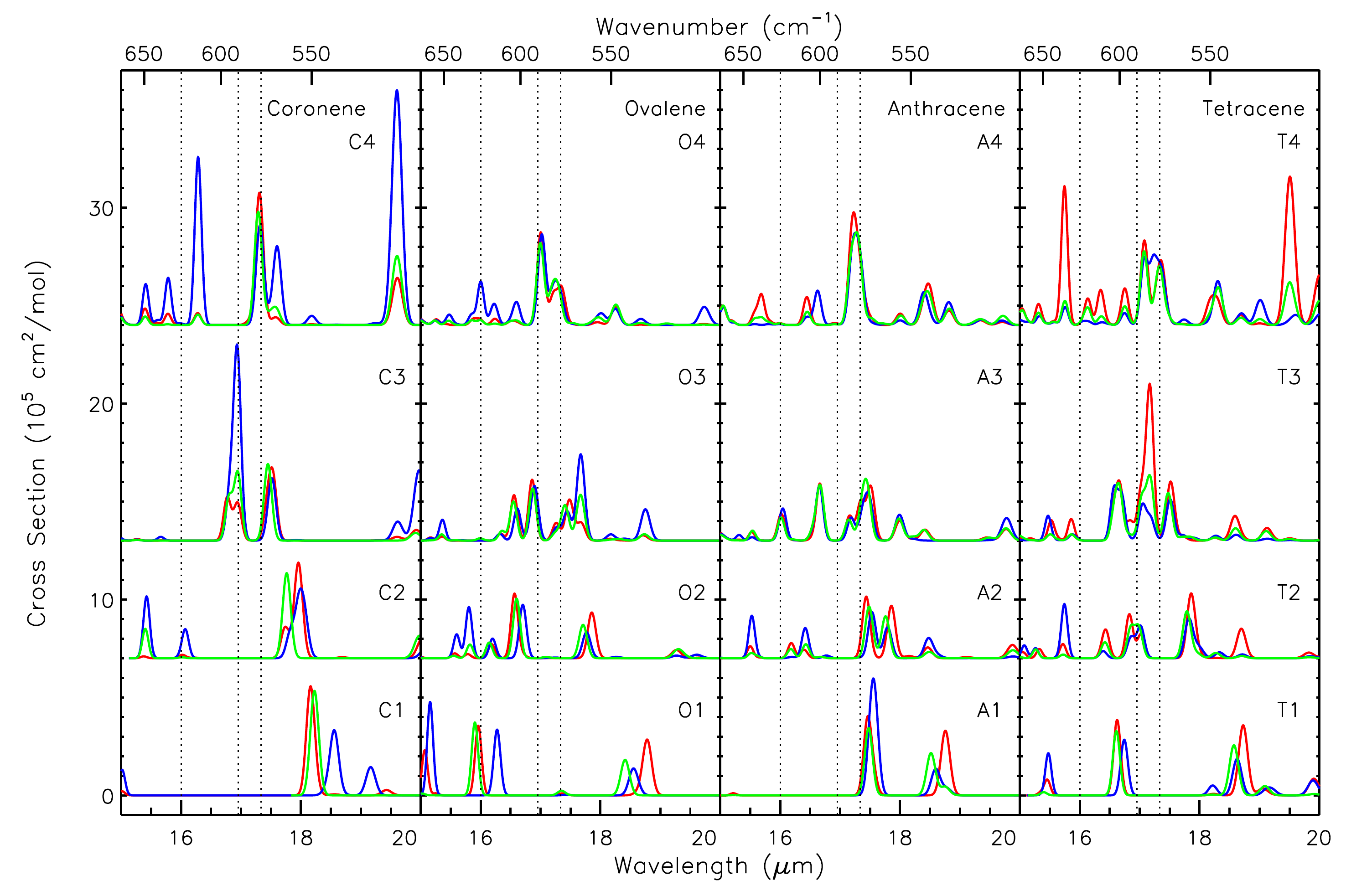}}
\caption{\label{lab15} The 15 to 20~$\mu$m region for the anions (blue), neutrals (green) and cations (red)  -- not corrected for redshift. The dotted vertical lines represent the observed positions of the main astronomical PAH bands (at 16.4, 17.4 and 17.8 \mum), blue-shifted by 15cm$^{-1}$ for comparison with the theoretical data.} 
\end{figure*}

\subsubsection{The CC/CH in-plane stretching Vibrations (6 - 9~$\mu$m)}

The 6-9~$\mu$m region of the neutral, cation, and anion spectra are shown in Figure~\ref{lab6}%-\ref{f5}
. As expected, the intensities of the neutral species are at least 10 times smaller than those of the ions and we do not consider them in detail.  Inspection of the cation spectra show that 6.3, 7.7, and 8.6~$\mu$m bands increase in intensity with increasing size in each family.  However, there are similarities and differences between the families.  The increase in the 7.7~$\mu$m band with size is similar for the different families, while the 6.3 and 8.6~$\mu$m bands in the coronene family grow more rapidly with size than in the others.  The 6.3 and 8.6~$\mu$m band intensities seem to follow a decreasing trend in family from coronene to tetracene, ovalene and finally anthracene.  The 6.3 and 8.6~$\mu$m bands appear to be somewhat coupled, while the 7.7~$\mu$m band seems to vary more independently. 
For the anions, the coronene, ovalene, and anthracene families follow similar trends as the cation, while the tetracene family is clearly different for the anions than the cations, with the intensities of the tetracene anion being the largest of the families considered.

\subsubsection{The CH out-of-plane vibrations (9 - 15~$\mu$m)}

The 9-15~$\mu$m region of the neutral, cation, and anion spectra are shown in Figure~\ref{lab9}. Excluding the smallest member of each family, where the number of duo hydrogens exceeds the number of solo hydrogens, as expected, the neutral and cation spectra look very similar as a function of size and family; the most notable change is the increase in the intensity of the solo band with size and across the families because of the increasing number of solo hydrogens.  The
anions show some differences with family, namely the coalescence of two bands near 11~$\mu$m occurs more slowly for the coronene family than the other families, probably because there are fewer solo hydrogens. 
In addition, the tetracene family shows the growth of a band near 9~$\mu$m with increasing size. It is interesting to note that while the tetracene anion seems to have more intensity in the 6-9~$\mu$m region, it seems more consistent with the other families in the 9-15~$\mu$m range.

\begin{figure*}
    \centering
\resizebox{18cm}{!}{
\includegraphics{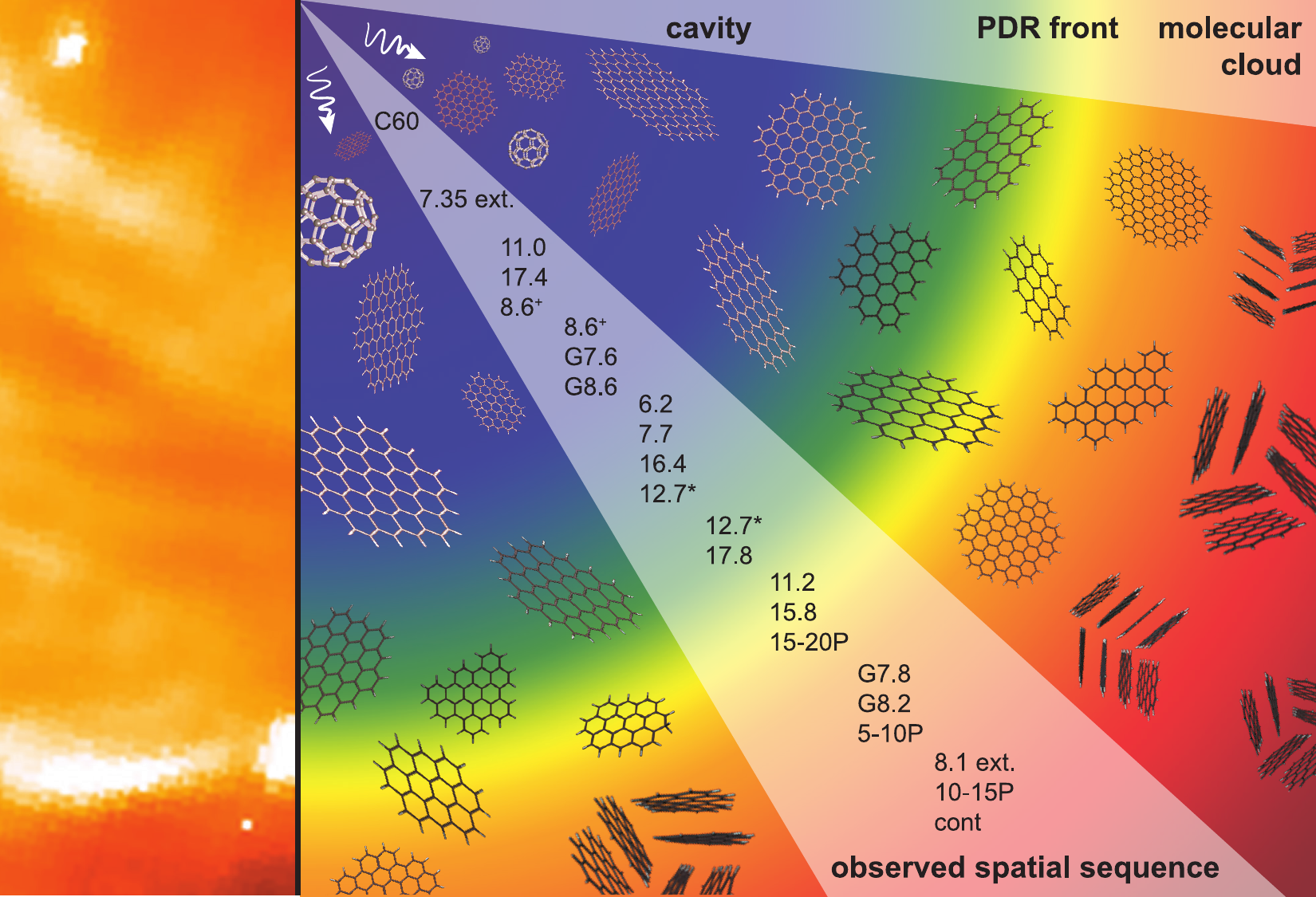}}
\caption{\label{secretdoc} Schematic of the photochemical evolution of the interstellar PAH population in the RN NGC~2023 (modified from \citet{Andrews:15}, their  Fig.~15). 
This figure exemplified the changes in the PAH population as they are more and more exposed to the strong radiation field of the central star in the evaporative flows associated with the PDRs in NGC~2023. With increasing distance from the star, the figure shows the transition from the cavity closer to the star  (lefthand side) to the PDR front$^\#$ and finally to the molecular cloud (righthand side).  Shown are the changing aromatic structures and types of aromatic-rich materials that are consistent with the observed spatial sequence described in Sections~\ref{sequence}.   See Section~\ref{assignments} for details.  }  
$^\#$ as traced by H$_2$ S(3) and S(5) transitions in the north. The H$_2$ S(1) and S(2) transitions are co-located with group 1. See Sections~\ref{shdata} and~\ref{sequence} for details.

$^*$ The SL and SH data indicate a slightly different morphology. $^+$ Group depends on the applied continuum.
\end{figure*}

\begin{figure}
    \centering
\resizebox{7.5cm}{!}{%
  \includegraphics{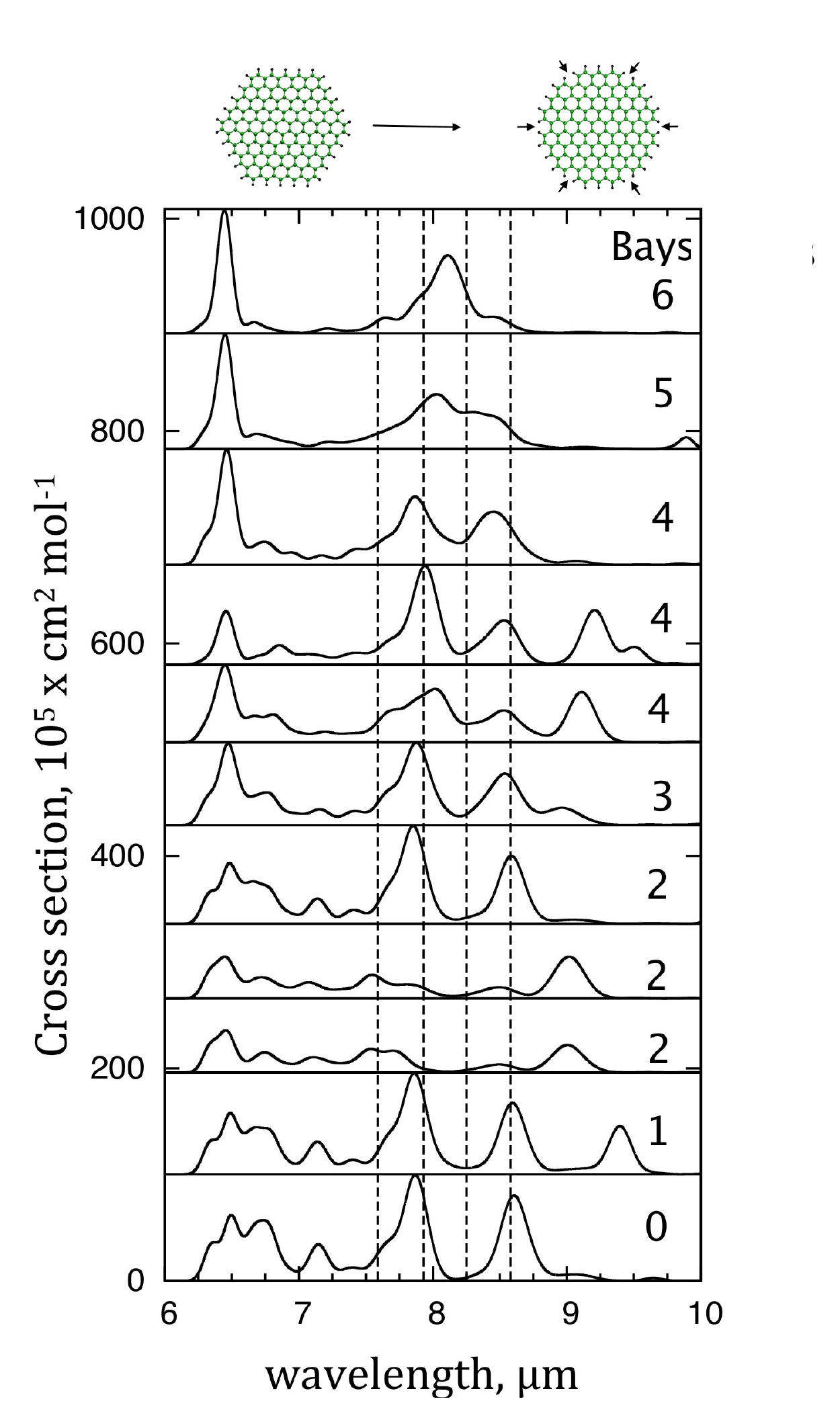}}
\caption{\label{bay-82} Illustration of the effect that bay regions have on the PAH cation spectrum in the 6 to 10 \mum\, range. The parent molecule is C$_{150}$H$_{30}$ and we created bay regions by slightly modifying the parent's structure. This is exemplified at the top of the figure, the black arrows indicate bay regions. The corresponding 6 to 10 \mum\, spectra of the PAH cations are shown with the dotted vertical lines representing the central position of the Gaussian subcomponents at 7.6, 7.8, 8.2 and 8.6 \mum. The number of bay regions for each PAH cation is listed to the right of its spectrum. A redshift of 15 cm$^{-1}$ is applied. }
\end{figure}

\subsubsection{The transition to skeletal vibrations (15 - 20~$\mu$m)}

The 15-20~$\mu$m region of the neutral, cation, and anion spectra are shown in Figure~\ref{lab15}. There tends to be an increase in the number of bands with increasing size, both within a family and across families.   As in our previous studies \citep[e.g.][]{Boersma:10, Ricca:12}, some bands appear to line up with some of the observed bands, however one is unable to make any assignments.

\section{Discussion}
\label{discussion}

\subsection{Theory versus observations}
\label{lab_imp}

The spectroscopic properties of the oval compact PAHs presented
in Section \ref{lab_results} are now combined with earlier studies and analyses of the NASA Ames PAH database \citep[PAHdb,][]{Bauschlicher:10, Boersma:14}, and
applied to the PAH characteristics of NGC~2023 (as discussed in Section \ref{analysis}). In particular, in Sections \ref{assignments} and \ref{intensities}, we will focus on the new decomposition of the 7 to 9 \mum\, PAH emission into the four Gaussian components (G7.6, G7.8, G8.2 and G8.6) in terms of band assignments and relative intensities. We conclude with a discussion on the CH modes in Section \ref{CHmodes}. 

\subsubsection{Band assignments}
\label{assignments}

To summarize the ensuing discussion, we ascribe
the G7.6 \mum\, emission primarily to compact, positively charged PAHs with sizes in the range of 50 to 100 C-atoms,
the G7.8 \mum\, emission to predominantly neutral, very large PAHs (100 $\le \#$ C $<$ 150) / PAH clusters with bay regions or modified duo CH groups (like adding or removing hydrogens or substituting a N for a CH group),
the G8.2 \mum\, emission to PAHs/PAH clusters with multiple bay regions (e.g. PAHs with very irregular structures or corners), and  the G8.6 \mum\, emission to very large compact, symmetric PAHs (96 $\le \#$C $<$ 150).
The interested reader can find a detailed discussion of these assignments in the following paragraphs and of the possible cluster assignment in Section~\ref{82cluster}. Within these assignments and following the discussion in \citet{Andrews:15}, the spatial morphology of the different features and their spatial sequence (as detailed in Section~\ref{sequence}) then reveals the photochemical evolution of the interstellar PAH family as they are increasingly exposed to the strong radiation field of the central star in the evaporative flows associated with the PDRs in NGC 2023. As depicted in Figure~\ref{secretdoc} and summarized in Table~\ref{table_overview}, with decreasing distance from the source of radiation, the PAH family thus evolves evolves from VSG and PAH clusters comprised of PAHs with irregular edge structures and multiple bay regions to very large PAHs with multiple bay regions, irregular edges and modified duo CH groups (cf., G7.8 \& G8.2 components), which are then broken down to a mixture of neutral and positively charged, compact, highly symmetric neutral PAHs (e.g., the circumcoronene family; e.g., 11.2 \mum\, band). Subsequently, these too are photochemically broken down to smaller structures and in this process ``pass" again through the stage of PAHs with irregular edge structures and/or corners (i.e., PAHs with duo/trio CH groups and strong 12.7 \mum\, band; see Section~\ref{CHmodes}). Eventually, only the most stable, compact, and highly symmetric, large ($\ge$ 70 C-atoms) PAH cations and C$_{60}$ fullerenes survive in the region closest to the star (e.g., G7.6 and G8.6 components, and 11.0 \mum\, band). For reference, we note that the grandPAHs as described in \citet{Andrews:15} are located in the PDR front, where the IR continuum and H$_2$ emission peaks.

\begin{table*}[tb]
\scriptsize
\caption{\label{table_overview} List of emission features by groupings presented in Section~\ref{sequence} and schematically illustrated in Figure~\ref{secretdoc}, their aromatic hydrocarbon vibrational assignments, likely carriers, and dominant charge. Note that the carriers of the features are not exclusively due to species in one charge state, but are often dominated by species in a particular charge state; size ranges are quite broad but do show population maxima \citep{Schutte:model:93}; and that the features that comprise the groups deduced here are not confined only to the zones depicted in Figure~\ref{secretdoc}, but do reach their maxima in these zones.}
\begin{center}
\begin{tabular}{lllll}
\multicolumn{1}{c}{Group} & \multicolumn{1}{c}{Feature} & \multicolumn{1}{c}{Vibrational Assignments} & \multicolumn{1}{c}{Carriers$^1$} & \multicolumn{1}{c}{Predominant}\\
& & & & \multicolumn{1}{c}{Charge}\\
\hline 
\hline
 & &  \\[-5pt]
\multicolumn{5}{c}{{\bf {\normalsize Molecular Cloud Edge}}} \\[5pt]
1 & continuum                      &   &  VSG & \\
   & 10--15 \mum\, Plateau   & Blend of aromatic CH$_{oop}$ bends from   & Small PAHs, large PAHs with irregular edges,  & 0 \\
   &                                      & solo, duo, trio and quartet H atoms & PAH clusters, PAH-rich VSGs and small amorphous C particles  & \\
   & 8.1 \mum\,  extreme      &   &   & \\[5pt]
2 & 5--10 \mum\, Plateau     & Blend of weak aromatic CC stretches and CH$_{ip}$ bends & Small PAHs, large PAHs, PAH clusters, PAH-rich VSGs & 0 \\
   & G8.2 \mum\, component & Aromatic CC stretches and CH$_{ip}$ bends & Large PAHs, PAH clusters, PAH-rich VSGs & 0\\
   & G7.8 \mum\, component & Aromatic CC stretches and CH$_{ip}$ bends & Large PAHs, PAH clusters, PAH-rich VSGs & 0\\[5pt]
\multicolumn{5}{c}{{\bf {\normalsize PDR Front$^2$}}}\\[5pt]
3 & 15--18 \mum\, Plateau   & & Large PAHs & 0\\
   & 15.8 \mum\, band          & ? & Large PAHs & 0\\
   & 11.2 \mum\, band          & CH$_{oop}$ bend from solo H atoms & Large PAHs & 0\\[5pt]
\multicolumn{5}{c}{{\bf {\normalsize PDR Front$^2$ \& Cavity}}}\\[5pt]
4 & 17.8 \mum\, band          & ? & Large PAHs & +\\
   & 12.7 \mum\, band$^3$  & Blend of CH$_{oop}$ bends from duo and trio H atoms & Large PAHs & +$^4$\\[5pt]
5 &12.7 \mum\, band$^3$   & Blend of CH$_{oop}$ bends from duo and trio H atoms & Large PAHs & +$^4$\\
   & 16.4 \mum\, band          & ?  & Large PAHs & +\\
   & 7.7 \mum\, band            & CC stretches and CH$_{ip}$ bends & Large PAHs & Roughly 50/50 +,0\\
   & 6.2 \mum\, band            & CC stretches & Large PAHs & +\\[5pt]
\multicolumn{5}{c}{{\bf {\normalsize Cavity}}}\\[5pt]
6 & 8.6 \mum\, band (GS)     &  CH$_{ip}$ bends  & Large PAHs & +\\
   & G7.6 \mum\, component & Aromatic CC stretches and CH$_{ip}$ bends & Large PAHs & +\\
   & G8.6 \mum\, component & CH$_{ip}$ bends & Large PAHs & +\\[5pt]
7 & 8.6 \mum\, band (LS)     &  CH$_{ip}$ bends  & Large PAHs & +\\
   & 11.0 \mum\, band          & CH$_{oop}$ bend from solo H atoms  & Large PAHs & +\\
   & 17.4 \mum\, band          & ?  & Large PAHs & +\\[5pt]
8 & 7.35 \mum\, extreme     &  &  & \\[5pt]
\multicolumn{5}{c}{\bf {\normalsize {Vicinity of Exciting Star$^5$}}}\\[5pt]
    & 17.4 \mum\, band          & CC stretches and bends & C$_{60}$ & 0\\
   & 19.2 \mum\, band          & CC stretches and bends & C$_{60}$ & 0\\
\hline
\end{tabular} 
\end{center}
$^1$Small PAHs have N$_C <$50 and large PAHs N$_C > $50. $^2$ as traced by H$_2$ S(3) and S(5) transitions in the north. The H$_2$ S(1) and S(2) transitions are co-located with group 1. See Sections~\ref{shdata} and~\ref{sequence} for details. $^3$ The SL and SH data show a slightly different morphology, for details see Section~\ref{shdata}. $^4$ The contribution of neutral PAHs varies \citep{Boersma:15, Shannon:16}. $^5$ Note that the PAH and C$_{60}$ morphology is distinct in the south map while co-located in the north map \citep{Sellgren:10, Peeters:12, Castellanos:14}. 
\end{table*}

\paragraph{The G7.6 and G7.8 \mum\, components}
All PAHs with more than $\sim$ 20 carbon atoms have {\it C-C stretching modes} in the 7.6-7.8 \mum\, range.  The frequency of the most intense bands varies with charge, size and structure.  Indeed, \citet{Bauschlicher:vlpahs1, Bauschlicher:vlpahs2} reported that PAH anions emit at slightly longer wavelengths than do the PAH cations. From their comparison of observations to theory, these authors furthermore conclude that the 7.7 \mum\, complex is comprised of a mixture of small and large PAH cations and anions with the {\it``small" species contributing more to the 7.6 \mum\, component and the large species (\#C $\ge$ 100) more to the 7.8 \mum\, component}. In addition, an upper limit to the PAH size of 150 carbon atoms is put forward by \citet{Ricca:12} based on the fact that very large compact PAHs (\#C $\ge$ 150) exhibit broad complex emission between 6 to 7 \mum, which is unlike the astronomical observations. Similarly, \citet{Bauschlicher:vlpahs2} reported that the 7.7 \mum\, complex in irregular PAHs is broader than that of the compact PAHs and is merging with the 8.6 \mum\, band (see their Fig. 8). This suggests that also the molecular edge structure determines the frequency of the most intense bands.
Indeed, a further detailed investigation of the PAHdb suggests that {\it the presence of bays (see Fig.~\ref{bay-82}, top, for examples of bay regions) and modifications to the duo CH groups, like adding or removing hydrogens or substituting a N for a CH group, 
tend to shift the intensity from 7.6 to 7.8 \mum}.  We also find that the broadening and merging of the 7.7 \mum\, complex and the 8.6 \mum\, bands for very large irregular PAHs as reported by \citet{Bauschlicher:vlpahs2} holds in general for PAHs with irregular edge structures. On the other hand, if the molecule is more compact and the shift is caused by adding H or removing H then there is less broadening and hence less merging with the 8.6 \mum\, band.  However, not all such changes in structure shift the intensity in a significant manner.  Hence, multiple factors are likely contributing to the intensity distribution in this range, and we may not have identified and/or quantified all of them.  

\begin{table*}[tbp!]
\small
\caption{\label{t_ratios} The intrinsic intensity ratios$^a$ for the ions for an excitation of 8 eV. }
\begin{center}
\begin{tabular}{lrrrrrrrrr}
Name Formula  & \multicolumn{4}{c}{cation} & & \multicolumn{4}{c}{anion} \\[1pt]
& $I_{8.6}$/$I_{6.2}$& $I_{7.6}$/$I_{6.2}$&  $I_{8.2}$/$I_{6.2}$& $I_{7.8}$/$I_{6.2}$& & $I_{8.6}$/$I_{6.2}$& $I_{7.6}$/$I_{6.2}$&  $I_{8.2}$/$I_{6.2}$& $I_{7.8}$/$I_{6.2}$\\[5pt]
\hline
\hline
\\ [-5pt]
C1 C$_{24}$H$_{12}$        &     0.06&   0.81&   0.12&   0.01& & 0.14&   0.97&   0.28&   0.95\\ 
C2 C$_{54}$H$_{18}$        &     0.16&   0.53&   0.19&   0.28& &  0.40&   0.97&   0.29&   0.90\\
C3 C$_{96}$H$_{24}$        &     0.48&   1.01&   0.10&   0.28& & 1.15&   1.05&   0.14&   1.10\\
C4 C$_{150}$H$_{30}$       &    1.41&   0.94&   0.08&   1.30& & 2.28&   0.84&   0.16&   2.12\\
O1 C$_{36}$H$_{16}$        &     0.39&   0.65&   0.18&   0.31& & 0.08&   2.44&   0.16&   0.95\\ 
O2 C$_{66}$H$_{20}$        &     0.34&   1.13&   0.21&   0.56& &  0.68&   2.05&   0.26&   2.38\\ 
O3 C$_{112}$H$_{26}$       &    0.68&   1.35&   0.10&   1.21& &1.51&   1.85&   0.87&   4.00\\ 
O4 C$_{170}$H$_{32}$       &    1.86&   0.97&   1.07&   3.81&  & 2.83&   1.62&   2.43&   8.03\\ 
A1 C$_{40}$H$_{16}$       &      0.19&   1.79&   0.81&   0.38&  &0.10&   2.61&   0.32&   1.30\\
A2 C$_{78}$H$_{22}$       &      0.72&   2.52&   0.46&   1.50& & 0.42&   1.87&   0.54&   4.37\\
A3 C$_{128}$H$_{28}$       &    0.78&   1.47&   0.38&   4.06&  &  0.82&   1.11&   0.78&   5.62\\ 
A4 C$_{190}$H$_{34}$       &    1.20&   0.79&   0.83&   4.15&  &  2.56&   1.33&   1.89&   5.79\\ 
T1 C$_{48}$H$_{18}$        &      0.35&   2.96&   1.00&   0.35& & 0.25&   2.31&   0.48&   1.40\\ 
T2 C$_{90}$H$_{24}$        &      1.44&   3.21&   0.49&   2.98&  &  0.61&   1.18&   0.40&   2.36\\
T3 C$_{144}$H$_{30}$       &     1.76&   1.45&   0.61&   3.71& & 1.87&   1.67&   0.62&   3.16\\ 
T4 C$_{210}$H$_{36}$       &    1.80&   0.96&   0.64&   3.28&  &3.61&   2.09&   1.23&   4.53\\ 
Avg 1 C1+O1+A1+T1 & 0.21& 1.28& 0.40& 0.21&  &  0.14&   2.02&   0.30&   1.13\\
Avg 2 C2+O2+A2+T2     &      0.43&  1.27&   0.27&   0.82&  &  0.51&   1.38&   0.34&   2.07\\
Avg 3 C3+O3+A3+T3         &  0.75&   1.23&   0.21&   1.54&  &  1.29&   1.33&   0.49&   2.84\\
Avg 4  C4+O4+A4+T4        &  1.51&   0.91&   0.51&   2.73&  &  2.66&   1.28&   1.06&   4.23\\
Avg 5 C3 to T4 &                    1.04&   1.11&   0.32 &  1.99&  &  1.81&   1.31&   0.71&   3.37\\
Ave 6 C2 to T4&                     0.77&   1.18&   0.30&   1.47&  & 1.23&   1.34&   0.54&   2.79\\
Avg 7 Coronene Family  &      0.52 &  0.78&   0.13&   0.48&  &  1.08&   0.97&   0.21&   1.24\\
Avg 8 Ovalene Family   &       0.63&   1.20&   0.26&   1.14&  &  1.34&   1.91&   0.84&   3.92\\
Avg  9 Anthracene Family &     0.88&   1.69&   0.54&   3.06&  & 1.08&   1.47&   0.95&   5.16\\
Avg 10 Tetracene Family   &     1.67&   1.83&   0.58&   3.32&  & 1.56&   1.50&   0.62&   3.02\\[5pt]
\hline
\\[-5pt]
\multicolumn{10}{l}{$^a$ $I_{7.6}$ is the sum of all intensity in the range 7.365 to 7.8~$\mu$m, $I_{7.8}$ in the range 7.8 to 8.08~$\mu$m, $I_{8.2}$ in the}\\
\multicolumn{10}{l}{range 8.095 to 8.365~$\mu$m, $I_{8.6}$ in the range 8.408 to 8.752~$\mu$m, and $I_{6.2}$ in the range 6.2 to 6.6~$\mu$m. A redshift }\\
\multicolumn{10}{l}{of 15 cm$^{-1}$ is  applied before calculating the intensity ratios. See Section \ref{intensities} for details. }\\
\multicolumn{10}{l}{The observations for NGC~2023 fall in the region 1.49 $<I_{G7.6}/I_{6.2}<$ 2.56, 0.35 $<I_{G7.8}/I_{6.2}<$ 1.17, }\\
\multicolumn{10}{l}{0.13 $<I_{G8.2}/I_{6.2}<$ 0.40, and 0.37 $<I_{G8.6}/I_{6.2}<$ 0.72 (Fig.~\ref{ratios-Gcomponents}).}
\end{tabular}
\end{center}
\end{table*}

\paragraph{The G8.2 \mum\, component}
No band assignments exist for the 8.2 \mum\, emission 
as it had previously not been identified as an individual feature in the PAH band family but rather was included in the plateau emission (i.e. 8 \mum\, bump) underlying the 6.2 and 7.7 \mum\, PAH bands when the local spline decomposition is used or in the 7.7 and 8.6 \mum\, features in case of the global spline continuum and Lorentzian decomposition. We therefore investigated all PAHs present in the PAHdb for 8.2 \mum\, emission and found that it seems to originate in C-H in-plane bending modes at bay sites. We further explored this possible origin based on the molecule C$_{150}$H$_{30}$. We have gradually modified the edge structure of this parent molecule to enhance the number of bay regions (see Fig.~\ref{bay-82}, top) and computed the cation spectrum for these new structures. The results are shown in Fig.~\ref{bay-82}\footnote{ These will be added to the PAHdb in the next update.}. The parent structure exhibits two strong emission bands near $\sim$7.8 and $\sim$8.6 \mum. By increasing the number of bay regions, the 7.8 \mum\, band shifts towards 8.2 \mum\, while the 8.6 \mum\, component gets weaker and shifts to shorter wavelengths. Based on this analysis of the PAHdb, we therefore attribute the 8.2 \mum\, emission to {\it C-H in-plane (ip) bending modes in PAHs with multiple bay regions, as for example found in PAHs with very irregular structures or corners}.\footnote{Note that this also holds for neutral PAHs. However, neutral PAHs typically have very weak emission in this wavelength range and hence do not significantly contribute to the emission here.} We do not find a dependence on PAH size: all PAHs with multiple bay regions regardless of size have emission at 8.2 \mum.

\paragraph{The G8.6 \mum\, component}
\citet{Bauschlicher:vlpahs1, Bauschlicher:vlpahs2} reported that the 8.6 \mum\, PAH emission is due to the {\it C-H in-plane bending mode produced by large compact, symmetric PAHs}. Indeed, perusal of the PAHdb indicates that almost any change in symmetry reduces the intensity of the 8.6 \mum\, emission. Typically, PAHs of 96 or more carbon atoms show a significant 8.6 \mum\, band but -- as noted by \citet{Ricca:12} -- PAHs larger than 150 C-atoms show a broad complex of emission between 6 and 7 \mum\, and hence they are not important in the ISM.
For completion, we list here the small and medium sized PAHs in the PAHdb that do exhibit emission at 8.6 \mum: some PAH ions with 20 to 22 carbon atoms, C$_{54}$H$_{18}$ (0, 2+, 3+), and C$_{66}$H$_{19/20}$ anions. 
Keep in mind that PAHs with \#C-atoms $<$ 30 are easily destroyed and not expected to be prevalent in the ISM \citep{Micelotta:shocks:10, Berne:12}.

\begin{figure*}[tbp!]
    \centering
\resizebox{18cm}{!}{%
  \includegraphics{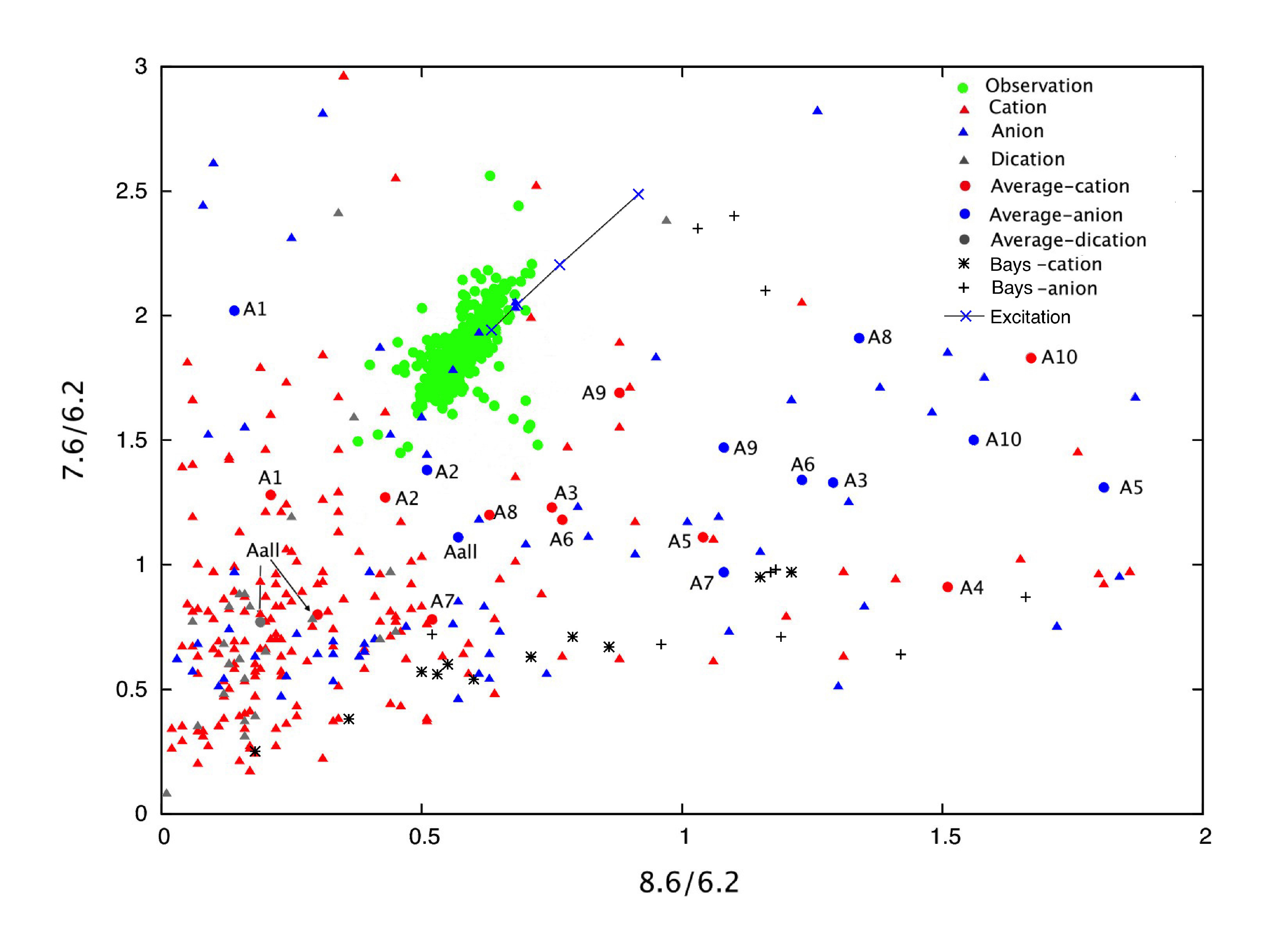}
  \includegraphics{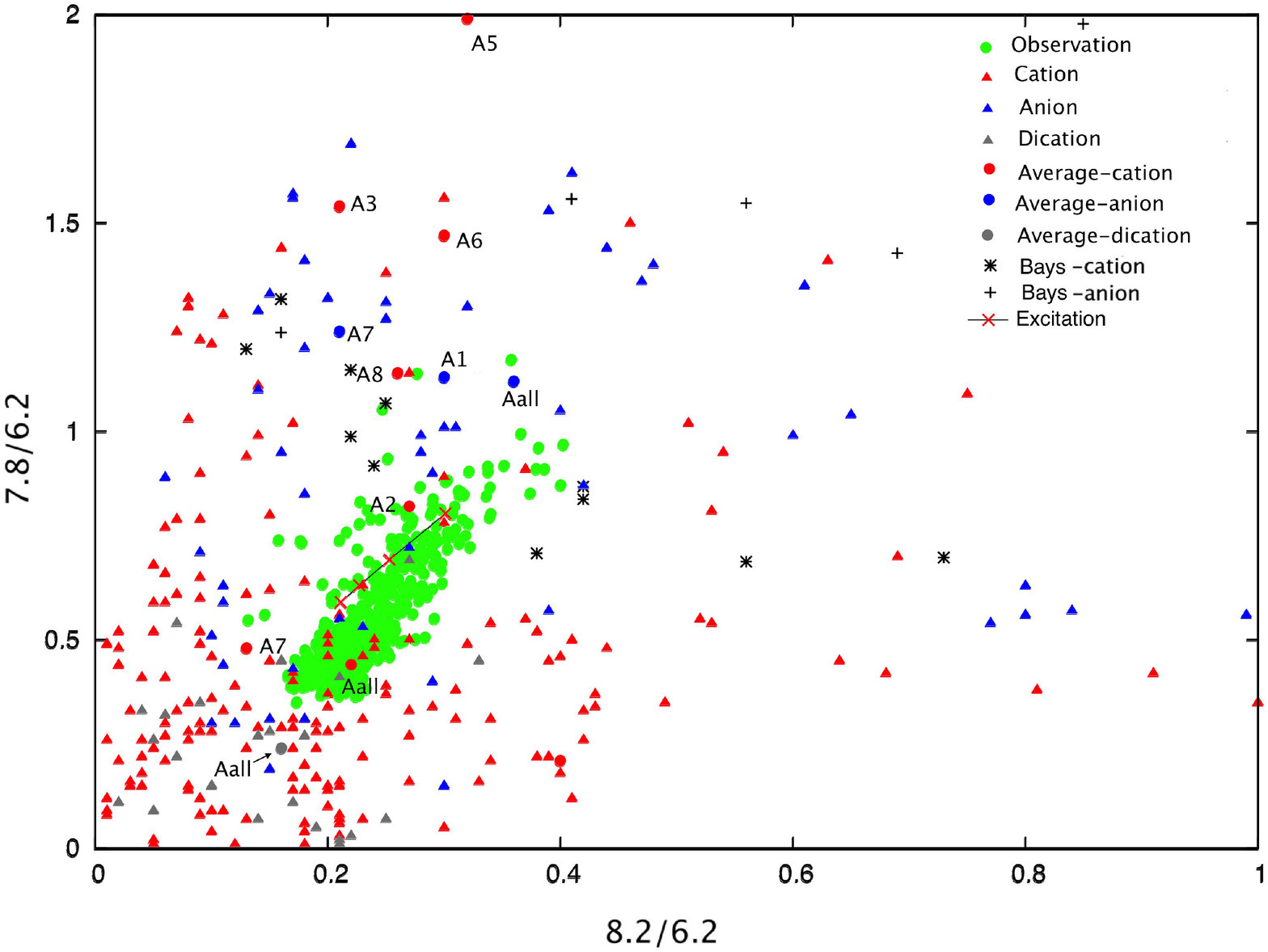}}
\caption{\label{ratios-Gcomponents} Comparison of the intrinsic PAH cations' intensity ratios and the NGC~2023 observations for the 8.6/6.2 versus 7.6/6.2 intensity ratios (left) and for the 8.2/6.2 versus 7.8/6.2 intensity ratios (right) for an excitation of 8 eV.  A redshift of 15 cm$^{-1}$ is  applied before calculating the intensity ratios. The averages are specified in Table \ref{t_ratios} except for A$_{all}$ which refers to the average of all cations/anions respectively with more than 19 carbon atoms. The structures shown in Fig.~\ref{bay-82} are referred to as `Bays'. The effect of  a changing excitation energy for a given molecule is represented by a sold line (for excitation energies of  4, 6, 8 and 10 eV). The spatial distribution of these ratios in NGC~2023 are discussed in Appendix~\ref{ratiomaps}. } %\\[-1pt]}
\end{figure*}

\subsubsection{Relative intensities}
\label{intensities}

The astronomical 8.6/6.2 and 7.7/6.2 intensity ratios are not well reproduced by the PAH intrinsic relative intensities for a collection of PAHs \citep{Ricca:12}. This discrepancy depends on the PAH structure: more elongated PAHs (e.g. oval PAHs) seem to fall closer to the observed values than circular PAHs (i.e. the coronene family). Here we find that the 7-9 \mum\, PAH emission is due to at least two PAH subpopulations. These PAH subpopulations are not traced with the nominal main PAH bands which may explain the lack of overlap between the astronomical and intrinsic intensity ratios. 
Therefore, we calculated the intrinsic intensities in the range of 7.365 to 7.8 \mum\, for the G7.6 component, of 7.8 to 8.08 \mum\, for the G7.8 component, of 8.095 to 8.365 \mum\, for the G8.2 component and of 8.408 to 8.752 \mum\, for the G8.6 component, applying a 15 cm$^{-1}$ redshift and an excitation of 8 eV. This corresponds to the typical photon energy for the illuminating star. We normalized these intensities to the 6.2 \mum\, band intensity. Note that changing the integration range for the 6.2 \mum\, PAH band will change the calculated ratios.  We therefore set this integration range to 6.2--6.6 \mum, corresponding to the range of frequencies of the most intense band in this wavelength range for all PAH ions with \#C $\ge$ 20 (a total of 456 molecules, applying a 15 cm$^{-1}$ redshift). The results for the oval PAH ions are given in Table \ref{t_ratios} and the results for all PAH ions in the PAHdb with \#C $\ge$20 (including the oval PAHs presented here) are shown with the observations in Fig.~\ref{ratios-Gcomponents}. The spatial distribution and the line cuts of these intensity ratios are also shown in Figs.~\ref{linecuts_ratios} and \ref{fig_PAHratios}.

When comparing the theoretical results for the oval PAH molecules discussed in Section \ref{lab_results} (Table \ref{t_ratios}), we note that there is no clear pattern.  The ratios tend to be larger for the anions than the cations.  Within a single family, the $I_{G7.8}/I_{6.2}$ and the $I_{G8.6}/I_{6.2}$ intensity ratios as well as the $I_{G8.2}/I_{6.2}$ anionic intensity ratio do increase with size. A few molecules as well as a few averages are consistent with the observations but do not represent the bulk of the ratios found in NGC~2023. 
Considering the entire database, a large range of values is present for all four intensity ratios, which can not be attributed solely to the smallest PAHs in the PAHdb. The observed range of each intensity ratio individually is well covered by the intrinsic PAH ratios (Fig.~\ref{ratios-Gcomponents}). However, the intrinsic values do not cover well both the range in the observed G7.6/6.2 and in the observed G8.6/6.2 at the same time: indeed, only a handful of molecules (all anions) do exhibit both observed intensity ratios. The match is better for the G7.8/6.2 versus the G8.2/6.2 intensity ratios where several molecules coincide with the observed ratios (including anions, cations and dications). Note that no large PAHs showing strong 8.6 \mum\, emission (\#C $\ge$ 96, see previous section) matches the G7.6/6.2 versus G8.6/6.2 intensity ratios in NGC~2023. The C$_{66}$H$_{19/20}$ anions as well as C$_{20}$H$_{12/14}$ ions do agree\footnote{These include C$_{66}$H$_{19}^-$, PAHdb uid of 712 and 711; C$_{66}$H$_{20}^-$, PAHdb uid of 605 and 116;  C$_{20}$H$_{12}^-$, PAHdb uid of 398; and, to a lesser extent, C$_{20}$H$_{14}^+$, PAHdb uid of 370.}.  Indeed, these large PAHs (\#C $\ge$ 96) exhibit emission near 7.8 \mum\, instead of 7.6 \mum\, in contrast with the observed trend between the 7.6 and 8.6 \mum\, PAH bands. Such an issue does not arise with bay PAH cations responsible for the 8.2 component as they also emit at 7.8 \mum. Here, the complexity occurs for the CH bending modes: the frequency of the solo CH out-of-plane bending mode is longwards of 11.2 \mum, well in the realm of the traditional solo CH out-of-plane bending mode for neutral PAHs. 

Since there are a few matches between the PAHdb and the observations, we can investigate if the observed trend in the G7.6/6.2 versus G8.6/6.2 and G7.8/6.2 versus the G8.2/6.2 intensity ratios (i.e. the slope in the observations) depends on PAH parameters such as molecular structure and PAH excitation. To explore possible dependence on PAH structure, we selected a PAH (C$_{36}$H$_{20}^{+}$, PAHdb uid of 542) that matches well with the G7.8/6.2 versus the G8.2/6.2 observational trend and checked the intensity ratios of related PAHs, i.e. PAHs of the same size but different structure (i.e. C$_{36}$H$_{16}^{+}$, PAHdb uid of 155 and C$_{36}$H$_{16}^{+}$, PAHdb uid of 129).\footnote{While these are not relevant structures as discussed in Section~\ref{assignments}, we were constraint by the limited match between the observations and pure PAH cations.} These however do not match the observations suggesting (although based on a very small sample) that molecular structure is not the dominant driver of the observed trend. Likewise, we explored the effect of different excitation for both trends. We selected C$_{66}$H$_{20}^{-}$ (PAHdb uid of 116) and C$_{66}$H$_{20}^{+}$ (PAHdb uid of 595) for respectively the G7.6/6.2 versus G8.6/6.2 and G7.8/6.2 versus the G8.2/6.2 observational trend and calculated their intensity ratios for excitation energies of 4, 6, 8 and 10 eV.  An increase in excitation energy results in decreased band ratios. While a change in excitation energy mimics both observed trends well (Fig.~\ref{ratios-Gcomponents}), the implied excitation energy variations are much larger than expected in the bright zones of a PDR. Thus, while variations in the excitation energy could be one of the parameters contributing to the observed trends, it is unlikely to be the dominant one.\\

Despite the fact that the PAHdb covers a wide range in relative intensities, only a small number of PAHs have relative intensities that match those observed in NGC~2023. This can be due to (a combination of) various reasons including:
\begin{enumerate}[label=(\roman*)]
\item only a small fractions of PAHs are responsible for these 4 Gaussian components, and thus the 7.7 and 8.6 \mum\, PAH emission, and/or 
\item the PAHdb is incomplete and lacks the necessary PAH molecules to match the observations, and/or 
\item the uncertainty on the relative intensities is considerable and inhibits a meaningful comparison with the observations (see Section \ref{lab_method} for a detailed discussion), and/or 
\item instead of single PAH molecule obeying the observed trends, a selection of PAHs fulfills the observational requirement as a group.  
\end{enumerate}

{\it Point (i)} is consistent with the grandPAH hypothesis: only a set of the most stable PAH species can survive in the harsh conditions of the ISM and are responsible for the observed PAH emission \citep[e.g.][]{Tielens:13, Andrews:15}. 
This also implies however that these (or some of these) grandPAHs are currently not present in the PAHdb as it does not provide a good match with the observations, in particular for the G7.6 and G8.6 subcomponents. However, we note that the PAHdb now does comprise all large, compact, symmetric PAHs (i.e. circular and oval PAHs with 96 $<$ \#C-atoms $<$ 150). 
These are also the most stable ones, an essential aspect to the grandPAH hypothesis. The lack of a good match then may pose a challenge to this hypothesis. 

{\it Point (ii)} is not that surprising as the PAHdb is known to be incomplete and, although biased towards smaller PAHs, is still improving \citep{Bauschlicher:10, Boersma:14}. The current incompleteness and bias may prove to be problematic for the 8.6 \mum\, emission and it's strong connection to the 7.6 \mum\, emission. 
Indeed, the large, compact, symmetric PAHs that exhibit a strong 8.6 \mum\, emission also have emission at 7.8 instead of 7.6 \mum. In addition, any reduction of symmetry decreases the 8.6 \mum\, emission.  This may then indicate that they are not responsible for the 8.6 \mum\, emission but instead (a few) smaller PAHs are (as discussed in Section \ref{assignments}). Within this hypothesis, these structure as well should have been missed in the PAHdb. Indeed, only C$_{66}$H$_{19/20}$ and C$_{20}$H$_{12}$ anions have G7.6 and G8.6 relative intensities well within the realm of the observations while two PAH cations (C$_{20}$H$_{12}^+$;  C$_{20}$H$_{14}^+$) straddle the observed boundaries for these relative intensities. It's interesting though that some of these smaller sized species do emit at 7.6 \mum. 

Finally, we note that excellent fits to astronomical data with the PAHdb have already been obtained \citep[e.g.][]{Cami:11, Boersma:13, Andrews:15}. This may imply that the PAHs may not satisfy the observed trends individually but do so rather as a group. This would however require that all members of the `group' behave spatially the same under a variety of physical conditions as e.g., the G7.6 and G8.6 exhibit the best correlations found between the PAH emission components. As a corollary, photochemistry is very sensitive to excitation and hence size, implying that these PAHs must all have very similar sizes. However, this seemingly contradicts with the assignment of the 7.6 \mum\, emission to PAHs with $\#$ C-atoms $<$ 100 and the 8.6 \mum\, emission to PAHs with $\#$ C-atoms $\ge$ 96.

Hence, it is clear that while none of these arguments undermines the PAH and/or grandPAH hypotheses, it does indicate that further investigations and possibly fine-tuning of the hypotheses are warranted.

\subsubsection{CH modes}
\label{CHmodes}

The spatial distribution of the 8.6, 11.0 and 17.4 \mum\, PAH emission is remarkably similar but quite distinct from that of, for example, the 6.2 and 11.2 \mum\, PAH emission (Section~\ref{analysis}).
Within the framework of the photochemical evolution of PAHs presented in Section~\ref{assignments}, this then implies they arise from similar PAH subpopulations. However, experimental and theoretical studies of PAHs assign these bands to different PAH subpopulations. Indeed, the 8.6 \mum\, PAH emission is due to CHip bending modes in large symmetric PAH ions and/or some smaller PAH ions \citep[Section \ref{assignments}, ][]{Bauschlicher:vlpahs1, Bauschlicher:vlpahs2}. The 11.0 \mum\, PAH emission generally is attributed to solo CH out-of-plane (oop) bending mode in (any) large PAH cation \citep{Hony:oops:01, Bauschlicher:vlpahs1, Bauschlicher:vlpahs2}.  \citet{Candian:15} argue instead for a possible origin in neutral acenes. For NGC~2023, \citet{Shannon:16} report a neutral contribution of roughly 8--16\% to the 11.0 \mum\, band limiting the possible acene contribution in this object. 
Finally, the 17.4 \mum\, PAH emission is systematically seen in large PAHs of all charges \citep{Boersma:10, Ricca:10}. These three ``attributions" represent quite different PAH subpopulations: going to longer wavelengths, the size of the PAH population that can contribute to these bands increases significantly. Yet, they have identical spatial morphologies. We can therefore put further constraints on their assignments based on their observational properties. As argued by \citet{Peeters:12} and \citet{Shannon:15}, the observational characteristics of the 17.4 \mum\, PAH emission points towards a carrier of PAH ions rather than neutrals. In addition, based on their chemical model calculation, \citet{Boschman:15} report that PAH cations are largely dehydrogenated. In this case, these effects combined reduces its PAH sub-population similar to that for the 11.0 \mum\, PAH band. On the other hand, rather few PAHs (present in the PAHdb) have strong 8.6 \mum\, emission and thus the 8.6 \mum\, emission possibly imposes very stringent restrictions to the CH ionic PAH population (and thus the PAH population overall, point (ii) above). Alternatively, while the photochemical evolution of the PAH family in PDRs as described in Section~\ref{assignments} is consistent with the data, other interpretations are possible as well. Indeed, further investigations are needed as the simple assumption that a small collection of 8.6 \mum\, carriers implies a limited set of carriers for the 11.0 and 17.4 \mum\, band as well may not hold. For example, consider large symmetric PAHs that emit strongly at 8.6 \mum.  These PAHs are dominated by solo CH groups with a few duo CH groups present. In addition, their solo CHoop intensity is even stronger than one would expect since the solo CH-groups steal intensity from the duo CH-groups \citep{Bauschlicher:vlpahs1, Ricca:12}. In this case, the 11.0 \mum\, emission is dominated by these large symmetric PAHs minimizing the effect of the larger PAH subpopulation contributing to the 11.0 \mum\, band compared to the PAH subpopulation contributing to the 8.6 \mum\, band. It is yet unknown whether a similar effect is present for the 17.4 \mum\, band. 

Another relatively strong CH band is found at 12.7 \mum. Generally, this band has been ascribed to CHoop bending vibration of duo and trio CH groups in (very) large PAHs \citep{Hony:oops:01, Bauschlicher:vlpahs1, Bauschlicher:vlpahs2, Ricca:12}. In addition, the 12.7 \mum\, intensity is further enhanced by the coupling between the CHoop mode and the C--C ring deformation mode in (elongated) armchair PAHs \citep{Candian:14}. Thus the assignment for the 12.7 \mum\, band, as well as for the G7.8 and G8.2 components, indicate the importance of bay regions, yet their spatial morphologies are very distinct. Indeed, we attribute the G7.8/G8.2 components to large PAHs with bay regions. As the smaller ones of these ($\le$ 70 C-atoms) are broken down, the largest and most stable ones remain abundant the longest (e.g., closest to the star). Theoretical studies have shown that carbon loss from compact PAHs lead to the formation of armchair structures \citep{Bauschlicher:14}. These surviving species are then the carriers of the 12.7 \mum\, band, where we note that, for armchair PAHs, the CHoop mode couples with the C--C skeletal modes, resulting in an enhanced intrinsic strength of the 12.7 \mum\, band \citet{Candian:14}. Clearly, further laboratory studies on the photochemical evolution of PAHs are warranted. 

\subsection{Clusters and VSGs}
\label{vsg}

\subsubsection{PAH features versus plateaus}
We conclude and confirm earlier reports \citep[][paper I]{Bregman:orion:89, Roche:orion:89} that {\it the plateaus are distinct from the features.} As a consequence, we need to fine-tune applied spectral decompositions to obtain the PAH components. 
Several decomposition methods exist in the literature and they differ greatly with respect to the treatment of the broad underlying plateaus. These plateaus are either incorporated in the PAH bands themselves through the wings of Drude or Lorentzian profiles \citep[e.g.][]{Boulanger:lorentz:98, SmithJD:07, Galliano:08} or treated separately \citep[e.g.][this paper]{Hony:oops:01, Peeters:prof6:02, Galliano:08}. Our findings, along with those of \citet{Bregman:orion:89}, \citet{Roche:orion:89} and paper I, strongly imply {\it the plateaus need to be treated independently from the features in spectral decompositions and argue against using Lorentzian and Drude profiles for the PAH features} \citep[see also][]{Tielens:08}.

\subsubsection{Carrier of the plateaus}

The spatial characteristics of the plateaus' emission also reveal further information about their origin.  The morphology of the 10--15 \mum\, plateau in the north map suggests that its carrier may be more closely related to the carrier of the MIR dust continuum than to the PAH molecules responsible for the PAH features.  Likewise, the 5--10 \mum\, plateau emission has a different carrier from that of the 10--15 \mum\, plateau, one that is more closely related to the PAH molecules, for instance by having a size between that of the carriers of the 10--15 \mum\, plateau and that of the PAH bands. In contrast to the 5--10 and 10--15 \mum\, plateau, the morphology of the 15--18 \mum\, plateau resembles more that of the PAH emission. Paper I and \citet{Shannon:15} reported that it likely originates in large neutral PAHs. Indeed, while it has been found to vary independently of the 15--18 \mum\, features overall, it does correlate well with the very weak 15.8 \mum\, band. This is also the only band in this wavelength range that is attributed solely to neutral PAHs because of its correlation with other neutral charge proxies, while the other features have been assigned to either PAH ions or a mixture of neutral and charged PAHs \citep[Paper I;][]{Shannon:15}.

A larger sized carrier for the 5--10 and 10--15 \mum\, plateaus is consistent with previous suggestions. For instance, \citet{ATB} and \citet{Bregman:orion:89} concluded that these broad components likely arise from a semi-continuum produced by a mixture of larger PAHs, PAH clusters and small amorphous carbon particles containing on the order of $\sim$ 400-500 carbon atoms. More recently, blind signal separation (BSS) analysis of spectral maps revealed three spatially distinct components contributing to the `nominal' PAH emission, which are attributed to neutral PAHs, ionized PAHs and evaporating very small grains \citep[PAH$^0$, PAH$^+$ and eVSGs respectively;][]{Boissel:01, Rapacioli:05, Berne:07}. Instead of applying a BSS analysis to our spectral maps, we have used its derivative PAHTAT \citep{Pilleri:12} for comparison with our results (for a detailed discussion on PAHTAT and its application to our data, see Appendix~\ref{PAHTAT}). The spatial morphology of the eVSG component and the plateau emission is very similar though small differences are present (Figures~\ref{fig_PAHTAT_maps}). In hindsight, this is not so surprising as, spectroscopically, the eVSG component is quite distinct and considerably narrower and smaller compared to the plateau emission as defined in this paper (Figure~\ref{fig_PAHTAT_decomp}). 

The similarity between eVSG and 8.1 \mum\, extreme argues for a stronger contribution of the VSG to the 7-9 \mum\, emission than what is found by the PAHTAT analysis and thus the BSS analysis (i.e. the fraction of the eVSG component to the total 7--9 \mum\, emission, in particular near 8.1 \mum). Combined with the fact that we found that the PAH features behave independent of the underlying plateau emission, this suggests that the PAHTAT and BSS ``decomposition" is in need of more components to decompose the PAH emission into its distinct PAH sub-populations. 

\subsubsection{Carrier of the G7.8 and G8.2 component}
\label{82cluster}

As noted, the eVSG emission is spatially most similar to that of the 8.1 \mum\, extreme in the north map (Figure~\ref{fig_PAHTAT_maps}). This suggests that the eVSG component is related to the G8.2 component despite the fact they are spectroscopically two distinct components (see Fig.~\ref{fig_PAHTAT_decomp}). In addition, the spatial distribution seen at 8.1 \mum, the G8.2 component and of the eVSGs is not seen at other wavelengths in the movies regarding the 6.2, 11.2 and 12.7 \mum\, bands (Figure~\ref{fig_movie}) nor is it seen for the G7.6 and G8.6 components (Figure~\ref{fig_maps_decomp}). This is therefore inconsistent with a carrier like the typical PAHs that are responsible for the PAH emission bands. 

Recent results on small neutral PAH clusters \citep{Roser:15} indicate that they have very similar infrared spectral characteristics in the 6-9 \mum\, region (peak position) to those of their respective neutral PAH components. If we assume that this finding also holds for clusters of larger size and/or different charge state, we can extend the earlier proposed assignment for the G8.2 emission based on the PAHdb (see Section~\ref{assignments}) to PAH clusters: PAHs and PAH clusters with multiple bay regions exhibit emission near 8.2 \mum, independent of their charge. In addition, nanograins and/or very small amorphous carbon particles could also possibly contribute to the G8.2 component. However, this cannot be currently confirmed due to the limited amount of infrared data available. An origin in larger sized species (PAH clusters, nanograins and/or very small amorphous carbon particles) is also more consistent with its spatial distribution being more similar to that of the eVSG and continuum emission than that of typical PAH bands. This can be further extended to the G7.8 component which has a similar morphology. While spectroscopically, the eVSG component obtained with the BSS method is quite distinct and considerably broader and weaker compared to the G7.8 emission as defined in this paper (Figure~\ref{fig_PAHTAT_decomp}), it's worth noting that \citet{Rapacioli:05, Berne:07} attributed the 7.8 \mum\, subcomponent to eVSGs. Note that the `traditional'  7.7 \mum\, component has contributions of both the G7.6 and G7.8 component as defined in this paper and thus certainly has a contribution of PAH molecules. 

\subsection{Profile classes}
\citet{Peeters:prof6:02} classified the profiles of the `nominal' 7.7 and 8.6 \mum\, PAH bands in class A, B and C. We applied our 4 Gaussian decomposition to the same sample of ISO-SWS observations as these authors. The class A profiles are well fitted \citep[see also][]{Stock:16b} as well as class B profiles which are not redshifted. However, our decomposition fails to fit the majority of the class B profiles: indeed many class B profiles are redshifted compared to class A profiles and our decomposition (with fixed peak positions and FWHM) does not allow for a red-shifting of the entire band. If we let the peak position of the 4 Gaussian components vary slightly while keeping the same FWHM, we can reproduce all class B profiles. As expected, class B profiles have a lower G7.6/G7.8 intensity ratio compared to class A profiles. Moreover, the G7.6/G7.8 intensity ratio decreases with peak position of the G7.8 component (when allowing for slight red-shifting) albeit with large scatter. Given the assignments discussed earlier, the band profiles thus reflect different PAH populations and size distributions with class B sources.

\subsection{Tracer for PAH charge}
The 6.2, 7.7, 8.6 and 11.0 \mum\, PAH bands have traditionally been used as a tracer for PAH ions and quantitative relations have been deduced to relate the 6.2/11.2, 7.7/11.2 or 8.6/11.2 intensity ratio with the ionization parameter and thus the physical conditions \citep[i.e. the radiation field, G$_0$, the electron density, n$_e$, and the gas temperature, T$_{gas}$; e.g.][]{Galliano:08, Fleming:10, Rosenberg:11, Boersma:15}. Based on a BSS analysis, \citet{Rosenberg:11} argue that the 11.0 \mum\, band may be better suited as a tracer for ions as it is a largely cationic band while the 6.2, 7.7, and 8.6 \mum\, features include a mixture of PAH$^+$ and PAH$^0$ components. The results presented in this paper are partially consistent with this. Clearly, the `nominal' 7.7 \mum\, band traces at least two PAH subpopulations, one of which being PAH cations. By virtue of its strong correlation and spatial resemblance with the `nominal' 7.7 \mum\, band, the 6.2 \mum\, band likely is also contaminated albeit to a lesser degree. To the contrary, the 8.6 and 11.0 \mum\, bands behave similarly on a spatial scale and are located closest to the illuminating star indicating that both bands would work well as a proxy for PAH charge (we present the spatial distribution of the 8.6/11.2 in Appendix~\ref{ratiomaps}). 
The reliability of the 6.2, 7.7 and 8.6 \mum\, bands as a charge proxy was recently investigated by \citet{Boersma:15}. The fitting approach of these authors recognizes that both neutral and charged PAHs contribute to these bands. Nevertheless, these authors find that the 6.2 and 8.6 \mum\, band are indeed a good tracer of PAH charge while the 7.7 \mum\, band is not. This has been attributed to a rapid growth of the 7.8 \mum\, component that is coincident with a stronger red wing for the 11.2 \mum\, band (traditionally assigned to neutral PAHs in the denser regions of the PDR). Comparing the applied decomposition methods, it is clear that their 7.7 \mum\, band includes the G7.6, G7.8, G8.2 and a fraction of the G8.6 components and thus contains a larger contribution of the second PAH subpopulation responsible for the G7.8 and G8.2 components than a `nominal' 7.7 \mum\, band\footnote{These authors extracted the 7.7 \mum\, band in the following way \citep{Boersma:14}: a continuum similar to our global spline continuum is fitted (which we used for the 4 Gaussian decomposition) but the 7.7 \mum\, band intensity includes emission underneath the 8.6 \mum\, band represented, in our decomposition, by the difference between the local and global spline continuum (see Figure \ref{fig_sp_sl}).}. It is therefore not surprising that their 7.7 \mum\, band intensity does not live up to expectations as a charge proxy. It's noteworthy to re-iterate that the 7.7 \mum\, band fails as a tracer for charge in regions that are also characterized by a stronger red wing for the 11.2 \mum\, band. \citet{Rosenberg:11} have attributed this component to eVSGs, another indirect link that PAH clusters and/or VSGs may be the carrier of the G8.2 component.

\subsection{Synopsis}
 
The results presented in this paper, amongst others,  indicate that we are in need to further fine-tune our `simple' picture of the PAH population and its emission. While the charge balance of the PAH population is certainly the major factor driving the observed spatial variations of the major features, other parameters such as chemical structure and size drive more subtle, perhaps second-order, variability that we are now beginning to probe. Hence, our `picture' of the PAH population in space should then reflect the variability of interstellar PAH structure and size, resulting in {\it a spatially and continuously changing PAH population}.  Indeed, at any given location in the nebula, the ionic and neutral PAHs have the same chemical structure because the timescale for ionization and recombination is fast, and charge balance changes on the order of months while the timescale for chemical modifications or processing is much larger \citep[10$^5$ yrs;][]{LePage:03, Berne:12}. However, with for example distance from the illuminating source, not only will the charge balance vary but also the chemical structure of the PAH population. Although the latter may happen on a larger spatial scale than the variation in charge balance. Likewise, the size of the PAHs present have been proven to vary with distance from the illuminating source \citep{Croiset:15}. 

Clearly, this ever changing PAH population is reflected in the observed PAH emission inviting us to go beyond the traditional approach of investigating the `nominal' PAH features. As shown, even within individual PAH emission bands, the spatial morphology
varies with wavelength (Section \ref{decomp7}). One should not think of a PAH band as a single emission feature. Instead, each PAH band is a combination of different components that each respond to the local physical conditions in their own way; they thus each (or their sub-population) have their own spatial distribution. This of course holds for the G7.6 and G8.6 Gaussian components which show a 1:1 correlation and have the highest correlation coefficient (0.987). The animation indicates that even at wavelengths where they are the only Gaussian components contributing (from the four used for the decomposition), the morphology changes continuously. Given that the PAH emission features are due to a collection of PAH molecules, this should not be surprising. In addition, these observations clearly indicate that there are several PAH subpopulations beyond the neutral and charged PAHs.  This then softens the grand-PAH hypothesis in which a few extremely stable PAH molecules make up the PAH emission in space. It would at least imply that instead of one grandPAH set, at each location with unique physical conditions and possibly unique history, a distinct grandPAH collection is present. 
 
\section{Conclusions}
\label{conclusion}
Two \textit{Spitzer}-IRS spectral maps from 5--15 \mum\, of the northern and southern part of the reflection nebula NGC~2023 are presented. The spatial distribution of the individual emission components including the dust continuum emission, the H$_2$ emission and the PAH emission, and relative PAH intensity correlations are investigated. These observational analyses resulted in the following conclusions:

\begin{itemize}
\item PAH emission bands exhibit a variety of different spatial morphologies: their (peak) emission occurs at different distances from the illuminating source revealing a `spatial sequence' within the PAH emission. 

\item Multiple correlations are present between individual PAH bands, including the well-known correlation between the 6.2, 7.7 and 8.6 \mum\, bands. 

\item Despite being well correlated, some PAH bands exhibit distinct spatial morphologies, in particular the traditional charge proxy bands at 6.2, 7.7 and 8.6 \mum. Hence, spectral maps reveal very important subtleties missed by correlation plots. 

\item The 7--9 \mum\, PAH emission is decomposed into 4 Gaussian components (G7.6, G7.8, G8.2 and G8.6). The G7.6 and G8.6 components exhibit the same spatial distribution and this closely resembles that of the 11.0 \mum\, PAH emission. These are the strongest correlated of all PAH components. The G7.8 and G8.2 components have spatial distributions similar to that of the dust continuum emission and not the PAH features emission (in case of the north map) and are not as tightly correlated. Thus, at least two PAH subpopulations with different spatial distributions contribute to the 7--9 \mum\, PAH emission.

\item The morphology of all the major PAH bands varies with wavelength indicating that multiple components contribute to a single feature. In particular, we find that the spatial distribution of the 7 to 9 \mum\, PAH emission continuously vary between two extremes which are bound between $\sim$7.35 and $\sim$8.1 \mum.

\item The PAH features also behave independently from the underlying plateau emission. Existing spectral decompositions clearly need further fine-tuning to treat these components separately. 

\end{itemize}

We compared our observational results with those of other analysis methods such as PAHFIT and PAHTAT. We present spectra of compact oval PAHs ranging in size from C$_{66}$ to C$_{210}$, determined computationally using density functional theory, and investigate trends in the band positions and relative intensities as a function of PAH size, charge and geometry within this family. 

Based on the entire NASA Ames database (PAHdb), we attribute the 7.6 \mum\, emission to compact PAHs with sizes in the range of 50 to 100 C-atoms,
the 7.8 \mum\, emission to very large PAHs (100 $\le \#$ C $<$ 150) with bay regions or modified duo CH groups (like adding or removing hydrogens or substituting a N for a CH group),
the 8.2 \mum\, emission to PAHs/PAH clusters with multiple bay regions (e.g. PAHs with very irregular structures or corners), and  the 8.6 \mum\, emission to very large compact, symmetric PAHs (96 $\le \#$C $<$ 150). Within these framework, the observed spatial sequence then reveals the photochemical evolution of the interstellar PAH family with distance from the illuminating star. 

We compared observed relative intensities with intrinsic relative intensities (as present in the PAHdb) in the 7 to 9 \mum.  The astronomical range in individual relative intensities is well represented by PAHs in the PAHdb but very few are consistent with the observed trends (correlation of G7.6 with G8.6 and of G7.8 with G8.2). We highlight possible reasons for this apparent inconsistency. 

We discuss the assignments of the CH bands within the framework of the presented photochemical evolution of interstellar PAHs. Based on the observed morphologies, we argue that the plateau emission and likely also the G8.2 \mum\, component originate from larger sized species such as larger-than-typical PAHs, PAH clusters and/or very small grains. Finally, we discuss the astronomical implications of our results with regards to PAH charge proxies and the PAH population in space. 

\acknowledgements The authors thank the referee Kris Sellgren for the careful reading of the manuscript and very useful feedback. The authors are extremely grateful to Henry Leparskas for creating Figure~\ref{secretdoc}. EP thanks Dr. D. Stock for fruitful discussions and feedback on the manuscript. We gratefully acknowledge support from the NASA \textit{Spitzer} Space Telescope General Observer Program. EP gratefully acknowledges sustained support from the Natural Sciences and Engineering Research Council of Canada (NSERC: Discovery grant and Accelerator grant). LJA gratefully acknowledges support from NASA's Astrobiology Program and Laboratory Astrophysics Carbon in the Galaxy consortium. AR thanks  NASA's Laboratory Astrophysics program grant NNX11AK09A for its generous support of this work. Studies of interstellar chemistry at Leiden Observatory are supported through advanced-ERC grant 246976 from the European Research Council, through a grant by the Dutch Science Agency, NWO, as part of the Dutch Astrochemistry Network and through the Spinoza premie from the Dutch Science Agency, NWO.

\appendix
\setcounter{figure}{0} \renewcommand{\thefigure}{A.\arabic{figure}}
\setcounter{table}{0} \renewcommand{\thetable}{A.\arabic{table}}

%%%%%%%%%%%%%%%%%%%%%%%%%%%%%%%%%%%%%%%%%%%%%%%%%%
\begin{figure}[tb]
    \centering
\resizebox{9cm}{!}{%
    \includegraphics[width=6.5cm]{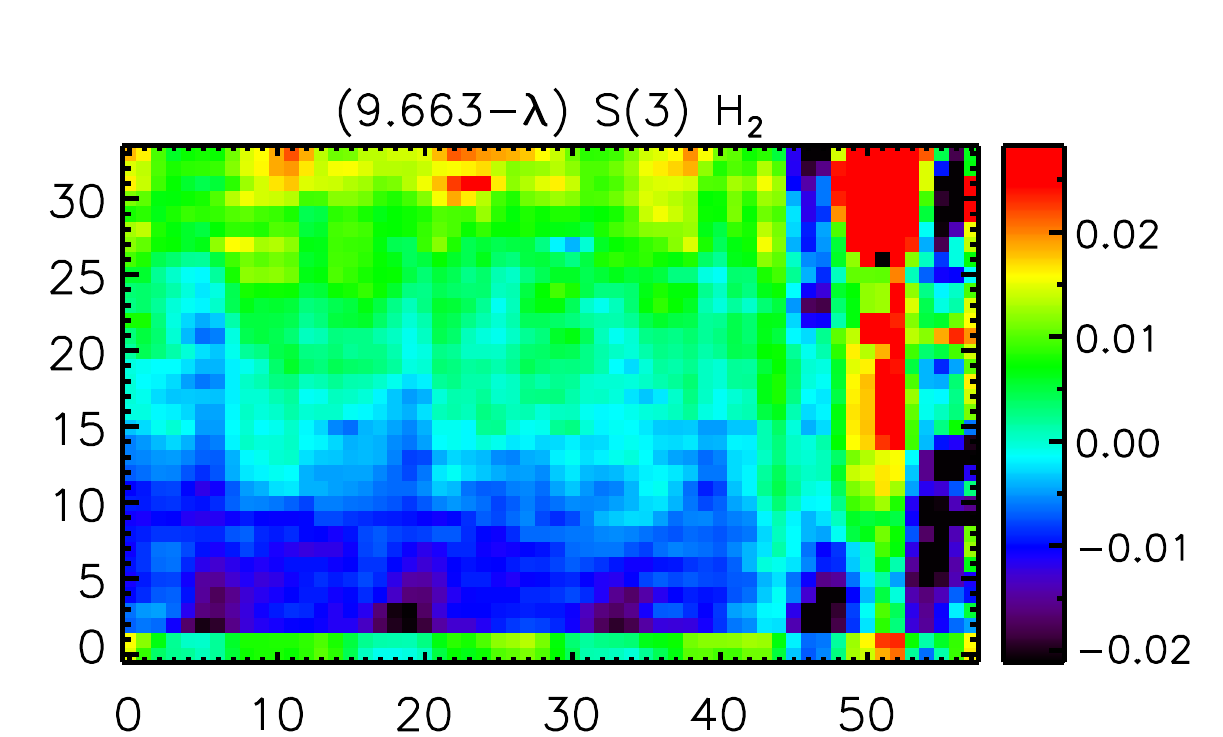}
    \includegraphics[width=5.cm]{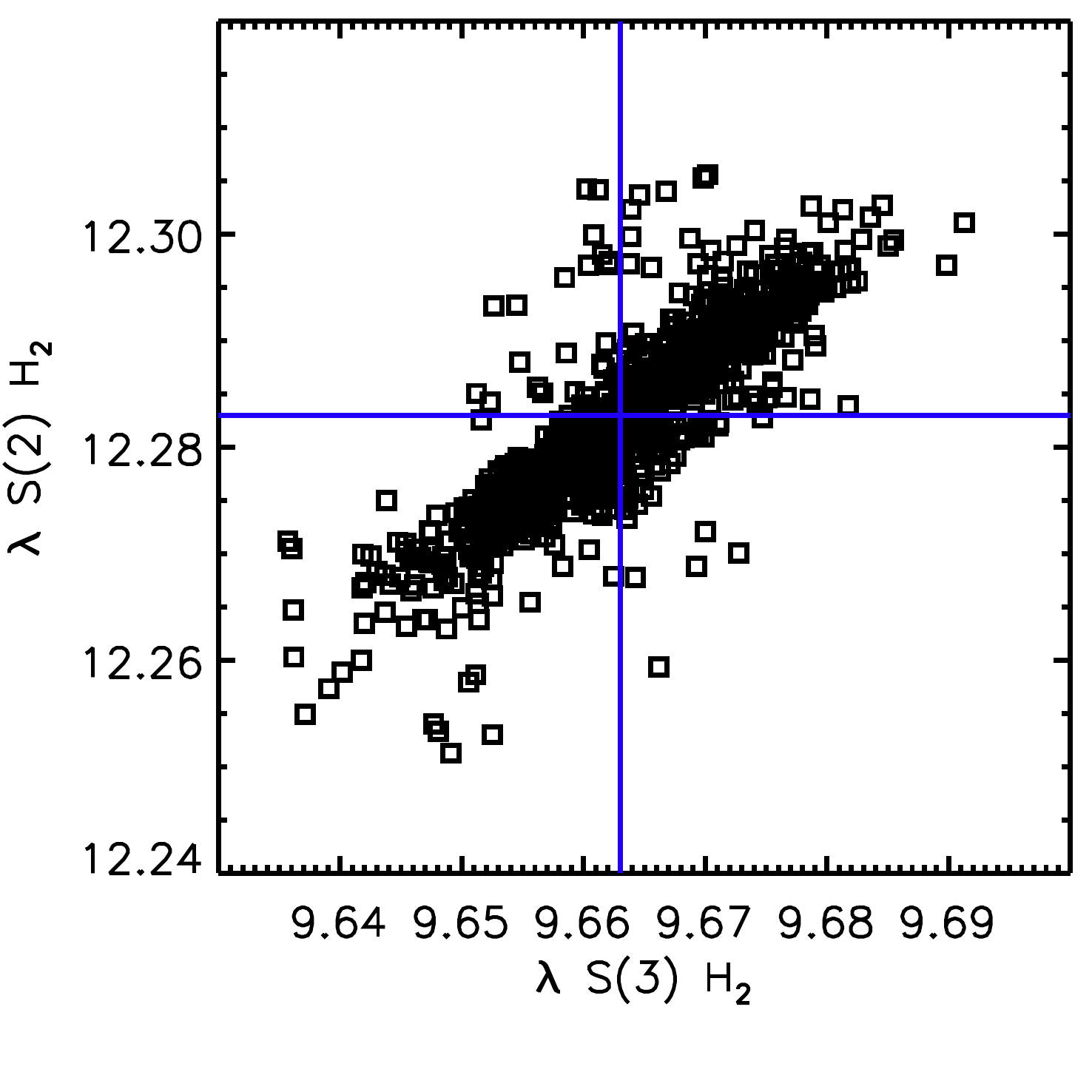}}
\caption{The variation in peak position of H$_2$ emission lines in the 2x2 SL south map: a map of the peak position of the S(3) H$_2$ 9.6 \mum\, emission line (left) and the change in peak positions of the S(3) H$_2$ 9.6 \mum\,  versus the S(2) H$_2$ 12.3 \mum\, emission lines (right).}
\label{fig_waveshift}
\end{figure}
%%%%%%%%%%%%%%%%%%%%%%%%%%%%%%%%%%%%%%%%%%%%%%%%%%

\section{Observed wavelength shifts}
\label{wavelengthshifts}

We've noticed an apparent small (of the order of a few $\times$ 0.01 \mum) wavelength shift in some 2x2 SL spectra, in particular for the south map.   For the north map, only single-pixel spectra from the majority of pixels (not all) in the top row exhibit a wavelength shift. This affects the 2x2 spectra incorporating pixels of the top rows in two ways: 1) these 2x2 spectra will exhibit a shift and 2) the intensity at each wavelength is influenced in particular in wavelength regions with a large intensity gradient. In contrast, for the south map, the single-pixel spectra shift 
from row to row. This is known to occur for maps made with cross-slit step sizes larger than 1 pixel (as is the south SL map) and is a result of an under-sampling of the PSF (Spitzer helpdesk \& Dr. Bregman, private communication).
This effect is generally mitigated by using 2x2 extraction windows as the amplitude of the shift varies approximately between two values (one for even rows and one for uneven rows). However, because the shifts for even and uneven rows are approximately the same and not equal, the 2x2 spectra exhibit residual shifts (of a few $\times$ 0.01 \mum). Moreover, spectra in the second row from the bottom exhibit an opposite shift compared to those in other even rows. This will influence the two bottom rows in the 2x2 spectral maps. The effect of these wavelength shifts can clearly be seen in the peak position of the H$_2$ emission lines (Fig.~\ref{fig_waveshift}). Moreover, it reveals an additional dependence with column\footnote{The red column around x=50 in the left panel of Fig.~\ref{fig_waveshift} is due to a spurious 'spike' in the data on top of the H$_2$ line which we were unable to remove. This features influences the peak position and derived strength of the H$_2$ line (see also Fig.~\ref{fig_slmaps_s}) and leads to a 12.7 \mum\, detection less than 2 sigma (Fig.~\ref{fig_slmaps_s}).}.  
This issue may possibly be resolved by correcting the spectra for these wavelength shifts based on the H$_2$ emission lines on a single pixel scale. However, not all SL2 spectra show H$_2$ emission lines. In addition, the wavelength shift in the SL2 spectra seems to depend on wavelength since all features at the shorter wavelength (including the 6.2 \mum\, PAH band) seem to shift while the wing of the 7.7 \mum\, PAH band does not (although this is tricky to assess as the PAH bands are known to exhibit variations). Hence, we cannot correct this wavelength shift, which is negligible with respect to the width of the PAH features. 
While its effects can be noticed in the top row of the north map and in the bottom two rows of the south map in several feature intensity maps presented in this paper, it does not change the conclusions of this paper.

\section{The spatial distribution of the 8.6 and 11.0 \mum\, emission.}
\label{scaled_maps}

\begin{figure}
    \centering
      \resizebox{6.46cm}{!}{%
     \includegraphics[angle=266.4]{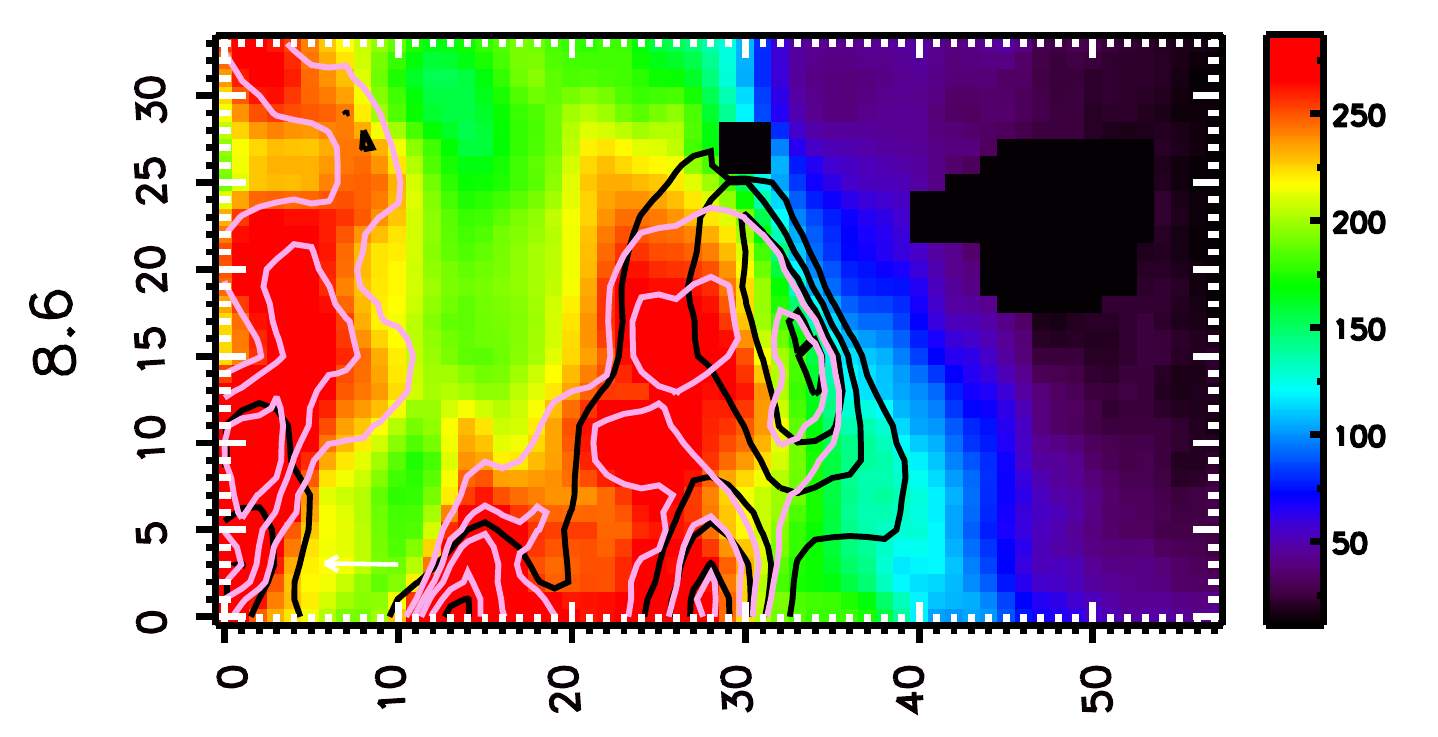}
            \includegraphics[angle=266.4]{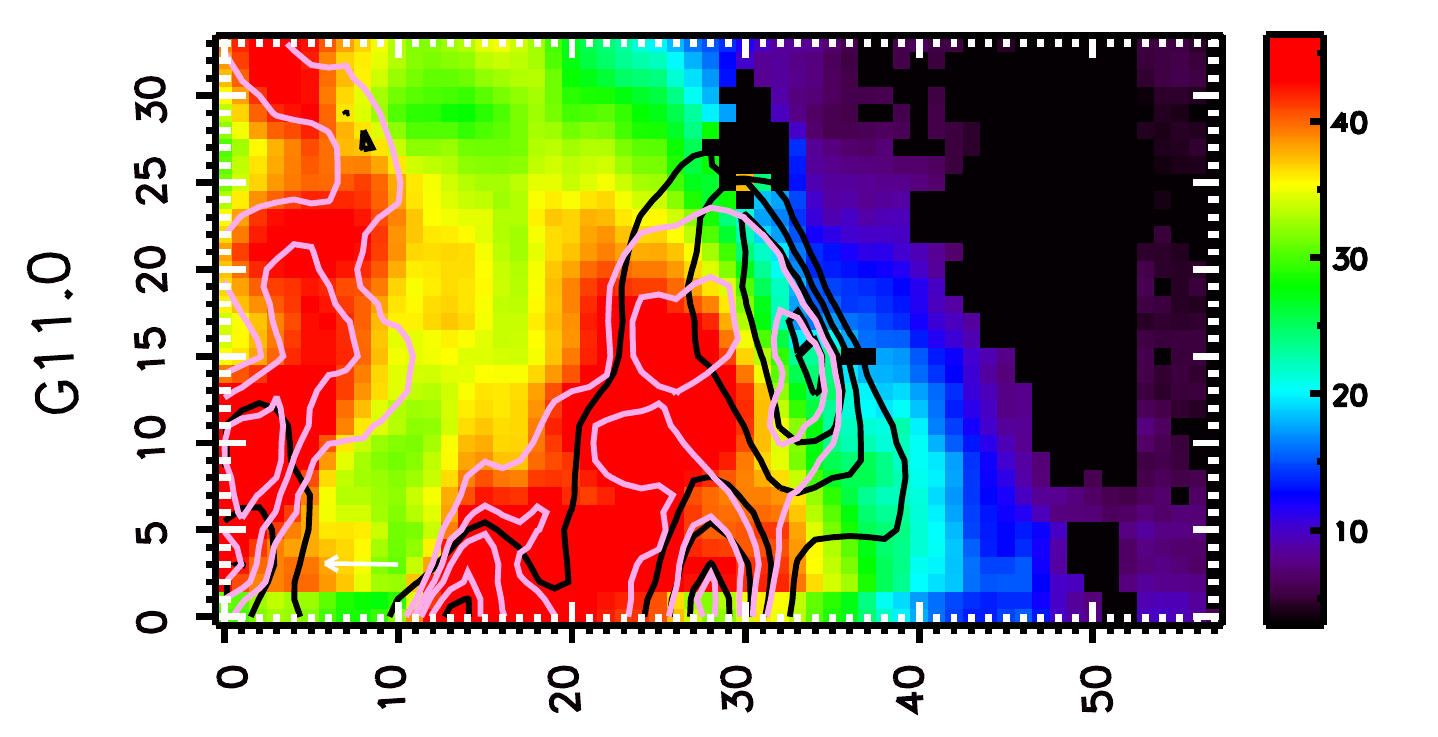} }
\caption{Spatial distribution of the 8.6 and 11.0 \mum\, PAH emission for the south map. The maximum intensity used for the color table is determined on the S' peak emission.  Band intensities are measured in units of 10$^{-8}$ Wm$^{-2}$sr$^{-1}$. The intensity profiles of the 11.2 and 7.7 \mum\, emission features are shown as contours in respectively black (at 3.66, 4.64, 5.64 and, 6.78 10$^{-6}$ Wm$^{-2}$sr$^{-1}$) and pink (at 1.40, 1.56, 1.70 and, 1.90 10$^{-5}$ Wm$^{-2}$sr$^{-1}$). The maps are orientated so N is up and E is left. The white arrow in the top left corners indicates the direction towards the central star. The axis labels refer to pixel numbers. Regions near source C and D excluded from the analysis are set to zero.  The nomenclature and the FOV of the SH map are given in the bottom right panel of Fig.~\ref{fig_slmaps_s}). }
\label{fig_slmaps_scaled}
\end{figure}

For both the 8.6 and 11.0 \mum\, maps, the maximum value for the colour table is determined by the peak emission in the SE ridge, which is located in the bottom two rows (in y-direction) and thus affected by the observed wavelength shift in the data (see discussion in Section~\ref{reduction} and Appendix~\ref{wavelengthshifts}). This in turn affects the visual appearance of these maps.  When the maximum of the intensity range for the colour table is determined from the S' peak emission region (well represented by the pink contours in Fig~\ref{fig_slmaps_scaled}) both PAH maps exhibit an almost identical spatial distribution (see Fig~\ref{fig_slmaps_scaled}),  given rise to their high correlation coefficient and 1:1 dependence.

%%%%%%%%%%%%%%%%%%%%%%%%%%%%%%%%%%%%%%%%%%%%%%%%%%
%%%%%%%%%%%%%%%%%%%%%%%%%%%%%%%%%%%%%%%%%%%%%%%%%%

%%%%%%%%%%%%%%%%%%%%%%%%%%%%%%%%%%%%%%%%%%%%%%%%%%
\begin{figure}
    \centering
\resizebox{8cm}{!}{%
  \includegraphics{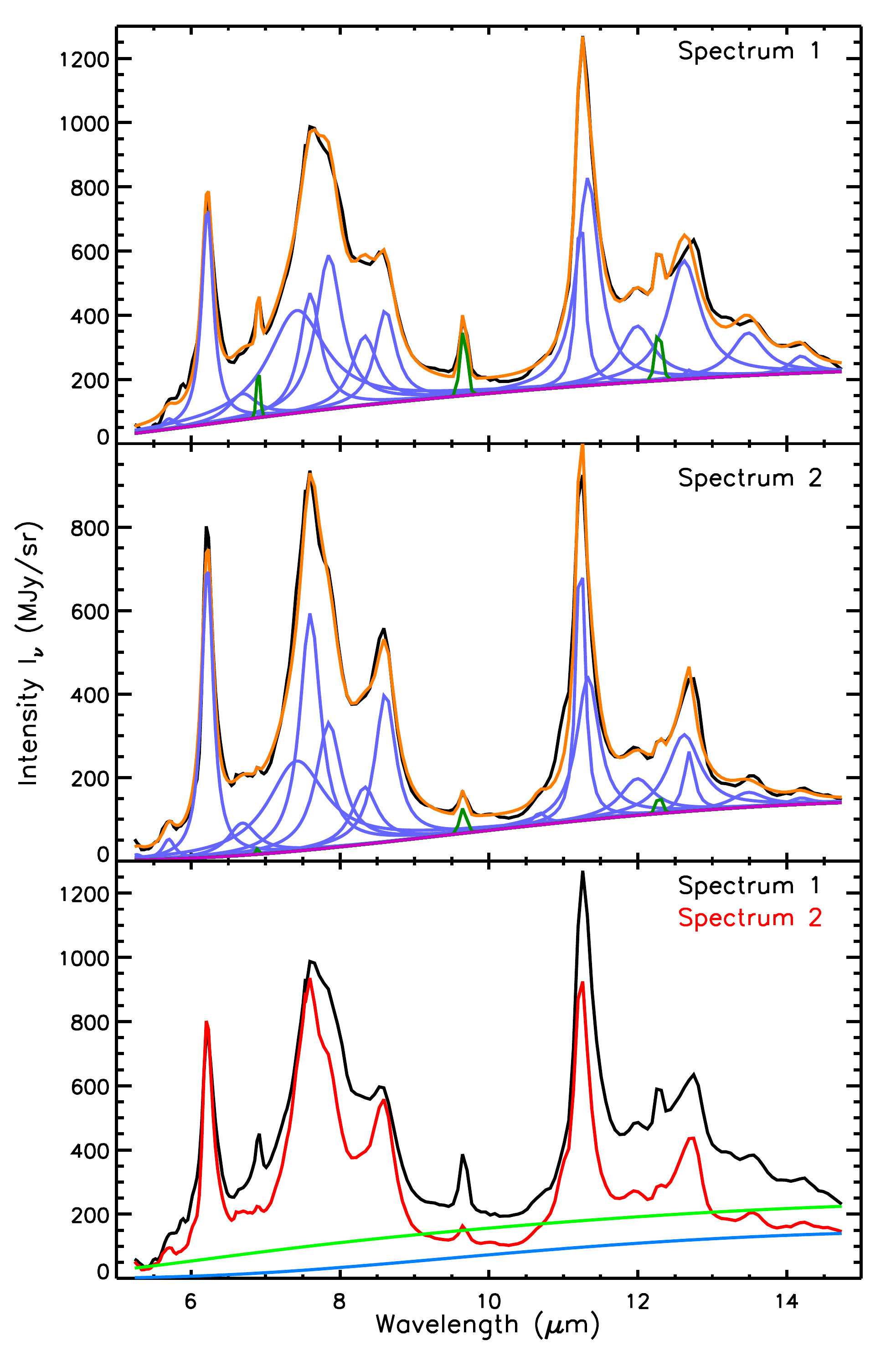}}
\caption{A typical fit to two spectra towards NGC~2023 by PAHFIT (top two panels) with the following colour coding: data (black), fit (red), PAH features (blue), H$_2$ lines (green) and continuum (magenta). The continuum for spectrum 1 is a combination of the 300 and 135K component while the continuum for spectrum 2 constitutes both a 200 and 135K component. The bottom panel compares the two spectra and their continua (in green and blue respectively). }
\label{fig_pahfit_cont}
\end{figure}
%%%%%%%%%%%%%%%%%%%%%%%%%%%%%%%%%%%%%%%%%%%%%%%%%%

\section{Results of PAHFIT decomposition}
\label{pahfit}

To investigate the influence of the decomposition method on the results, we applied PAHFIT \citep{SmithJD:07} to the SL data. PAHFIT models the spectra assuming a combination of starlight continuum, featureless thermal dust continuum, H$_2$ emission lines, fine-structure lines, and dust emission features while taking into account dust extinction. The starlight continuum is represented by blackbody emission, the thermal dust continuum by a set of modified blackbodies, the H$_2$ and fine-structure lines by Gaussian profiles and the dust emission features by a combination of Drude profiles. An example of the resulting fit and its separate components is shown in Fig.~\ref{fig_sp_sl}.  Clearly, this decomposition is quite different from the spline decomposition and hence a direct comparison of the intensities and profiles of the individual emission components derived by the different methods is not possible. A major difference in these two decomposition methods is the fact that the underlying emission plateaus (as defined in the spline decomposition) are accounted for by (the broad wings of) the Drude profiles in the PAHFIT decomposition and hence are part of the individual emission features. Also, several PAH features/complexes are fitted by two or more Drude profiles. In particular, within PAHFIT, the "7.7 \mum\, Complex" is defined as the blend of the Drude profiles with $\lambda$ (fractional FWHM) of 7.42 (0.126), 7.60 (0.044) and 7.85 (0.053) \mum, the "11.2 \mum\, Complex" as the blend of Drude profiles with $\lambda$ (fractional FWHM) of 11.23 (0.012) and 11.33 (0.032) \mum, and the "12.6 \mum\, Complex" as the blend of Drude profiles with $\lambda$ (fractional FWHM) of 12.62 (0.042) and 12.69 (0.013) \mum. 
Finally, the dust continuum emission generally has a lower intensity in the PAHFIT decomposition compared to spline decomposition, influencing the resulting dust features' intensities. Despite these significant differences, it has been shown that the overall conclusions on PAH intensity correlations for a large sample of objects are independent of the chosen decomposition approach \citep[e.g.][]{Uchida:RN:00, SmithJD:07, Galliano:08}. Here, we investigate if this also holds when considering spectral maps of individual sources.

%%%%%%%%%%%%%%%%%%%%%%%%%%%%%%%%%%%%%%%%%%%%%%%%%%
\begin{figure*}
    \centering
\resizebox{16.15cm}{!}{% 17cm
\includegraphics[angle=266.4]{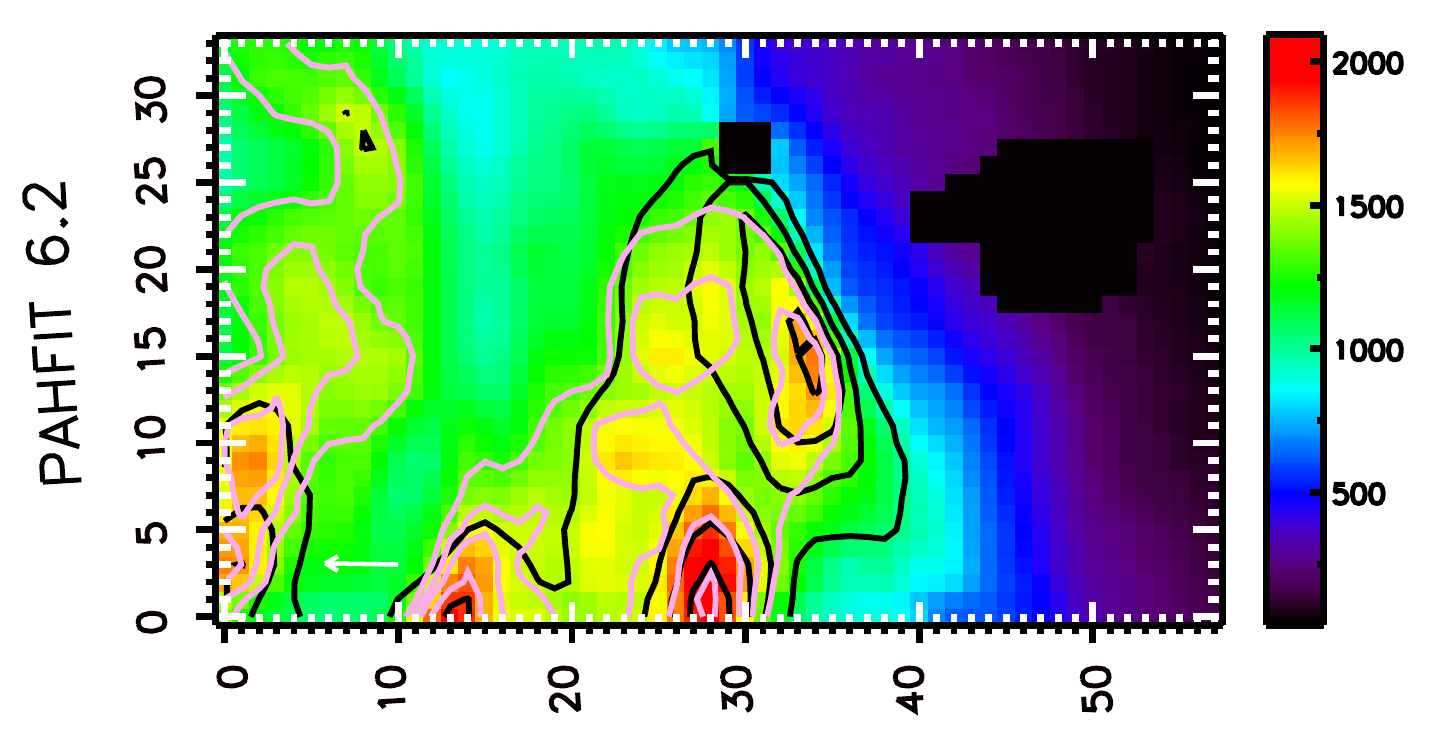}
\includegraphics[angle=266.4]{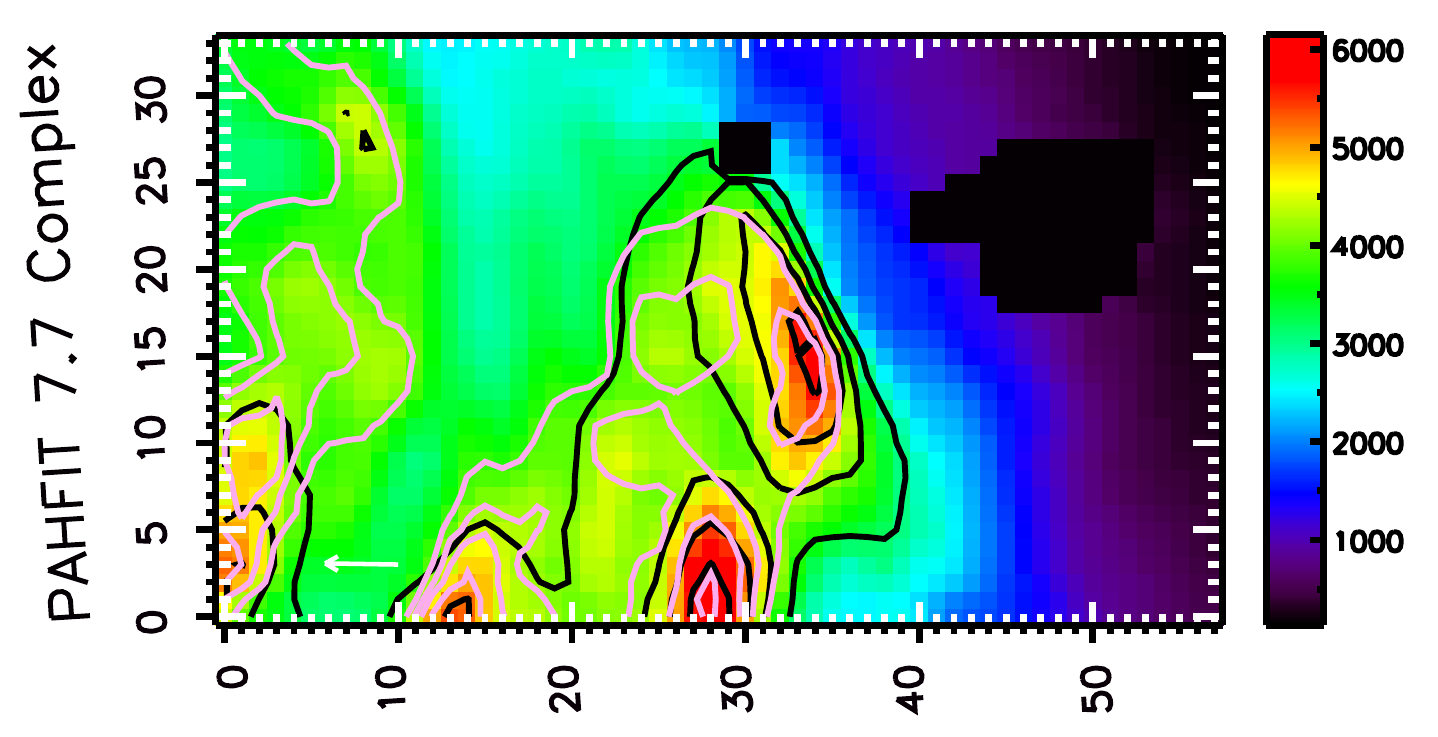}
   \includegraphics[angle=266.4]{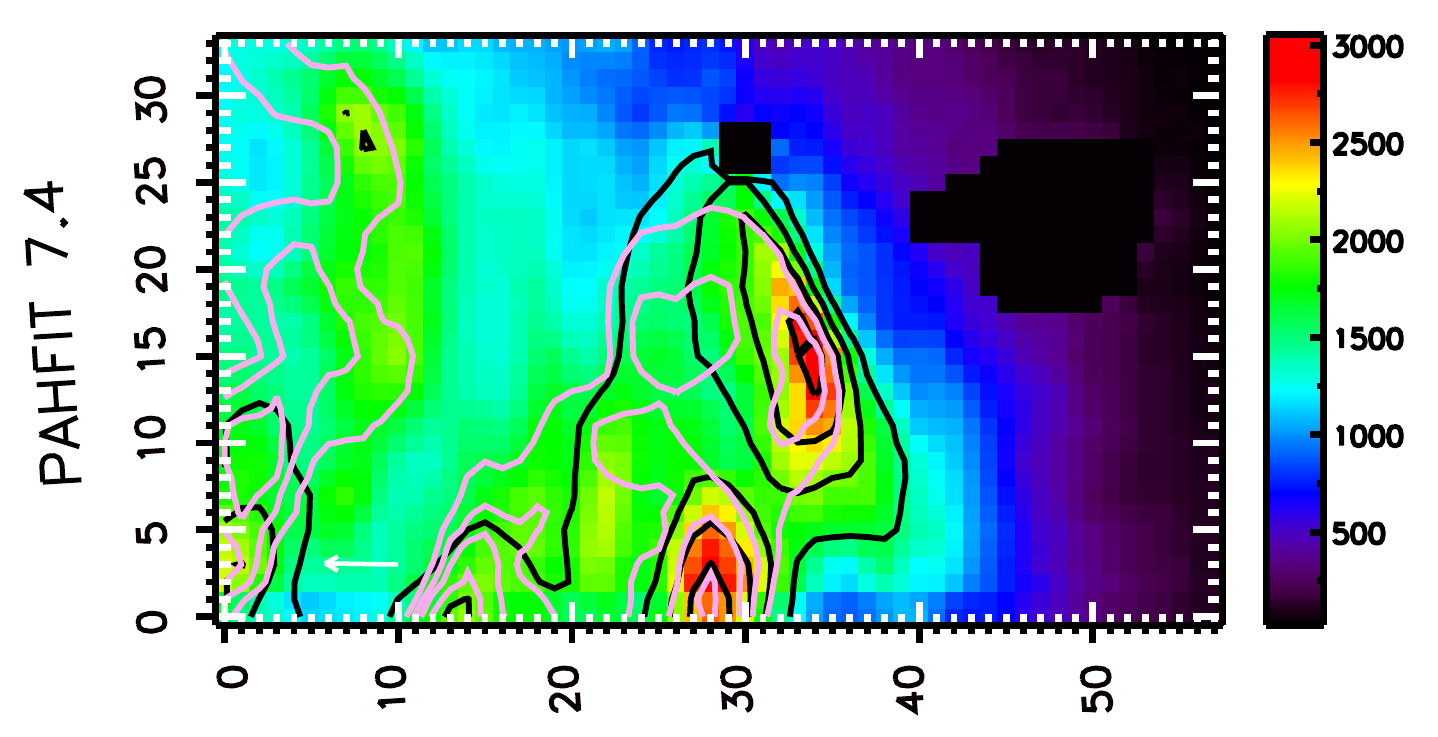}
    \includegraphics[angle=266.4]{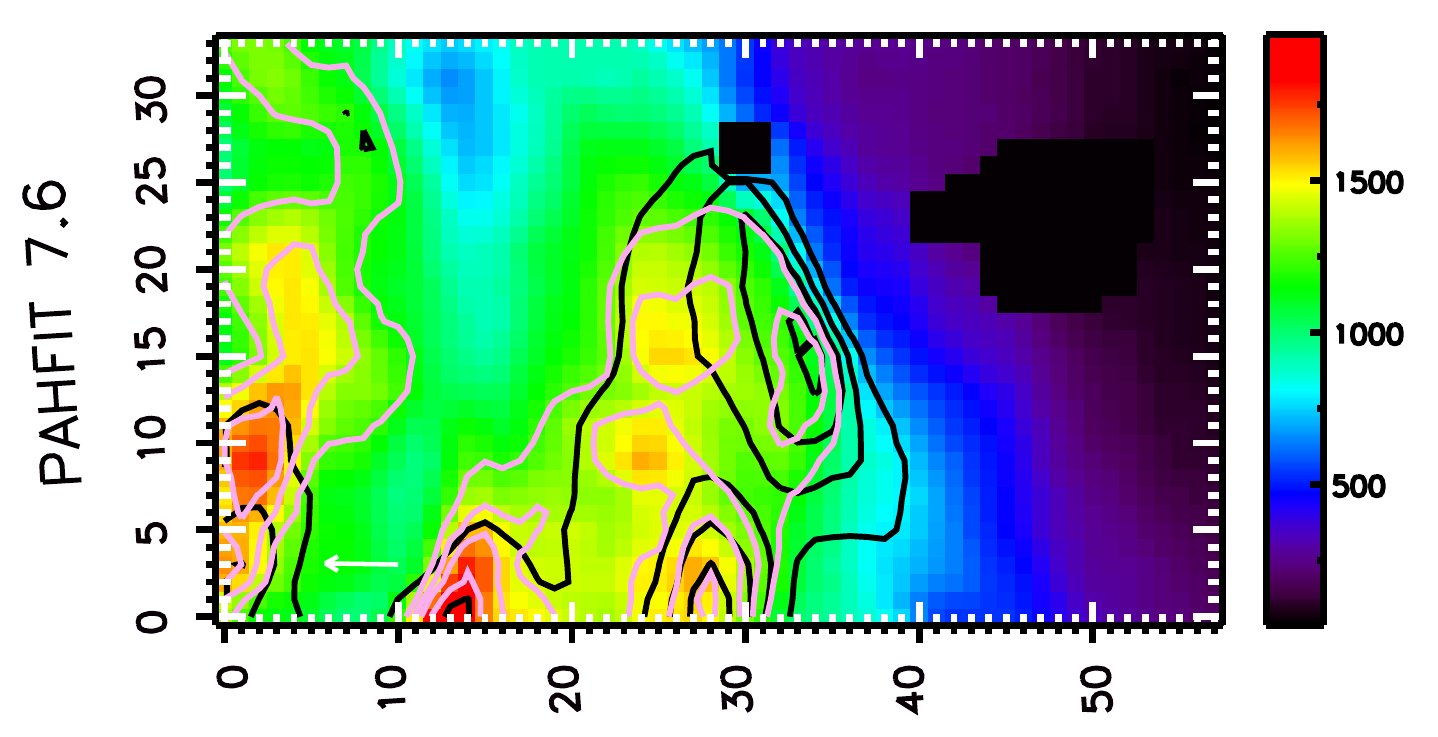}
        \includegraphics[angle=266.4]{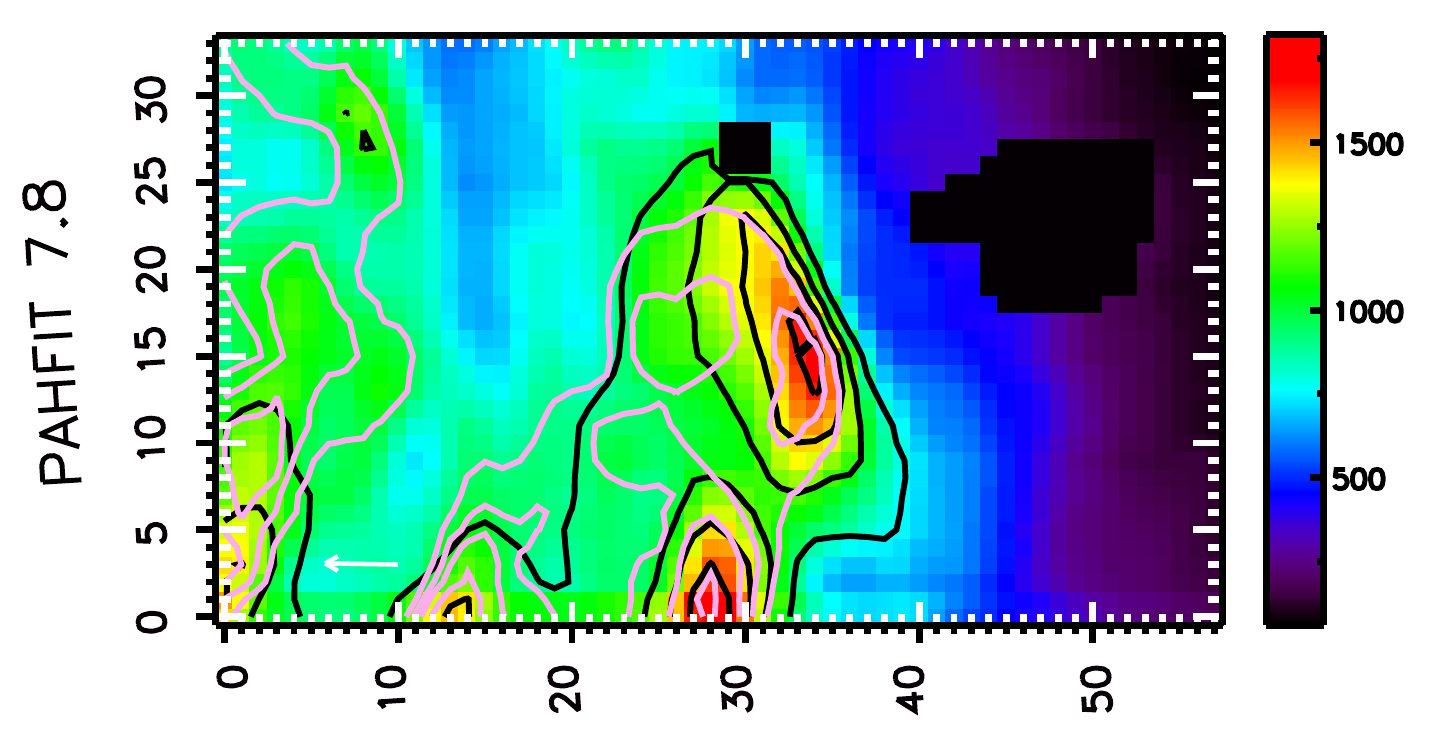}}
   \resizebox{16.15cm}{!}{%
  \includegraphics[angle=266.4]{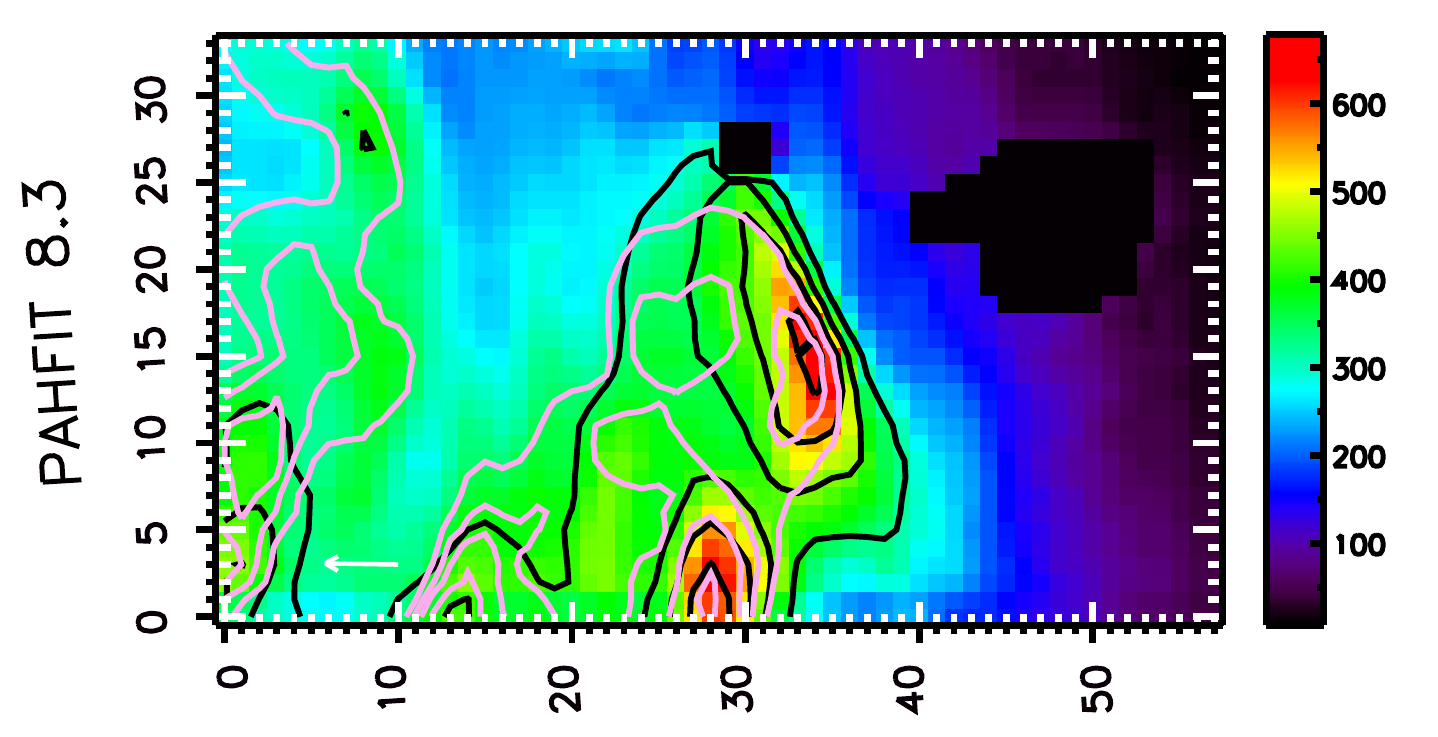}
  \includegraphics[angle=266.4]{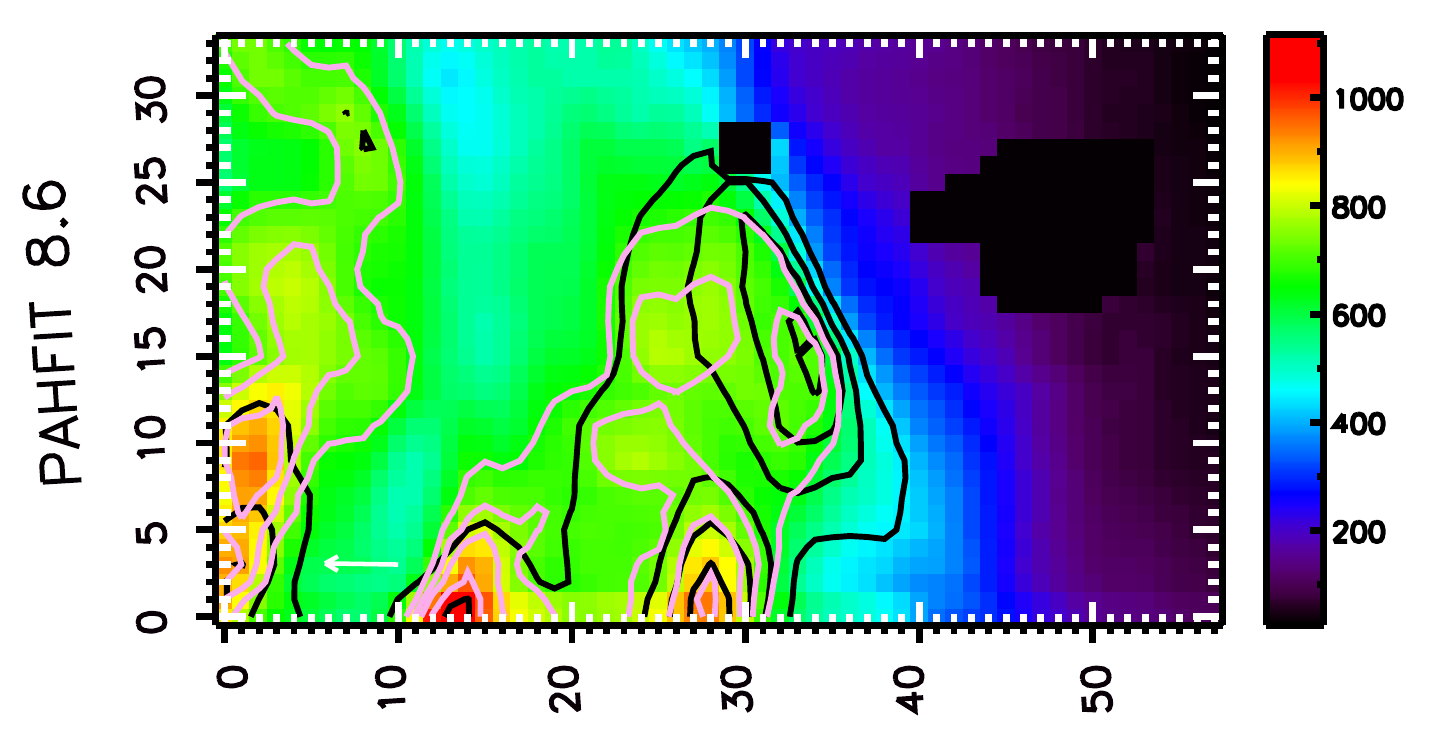}
    \includegraphics[angle=266.4]{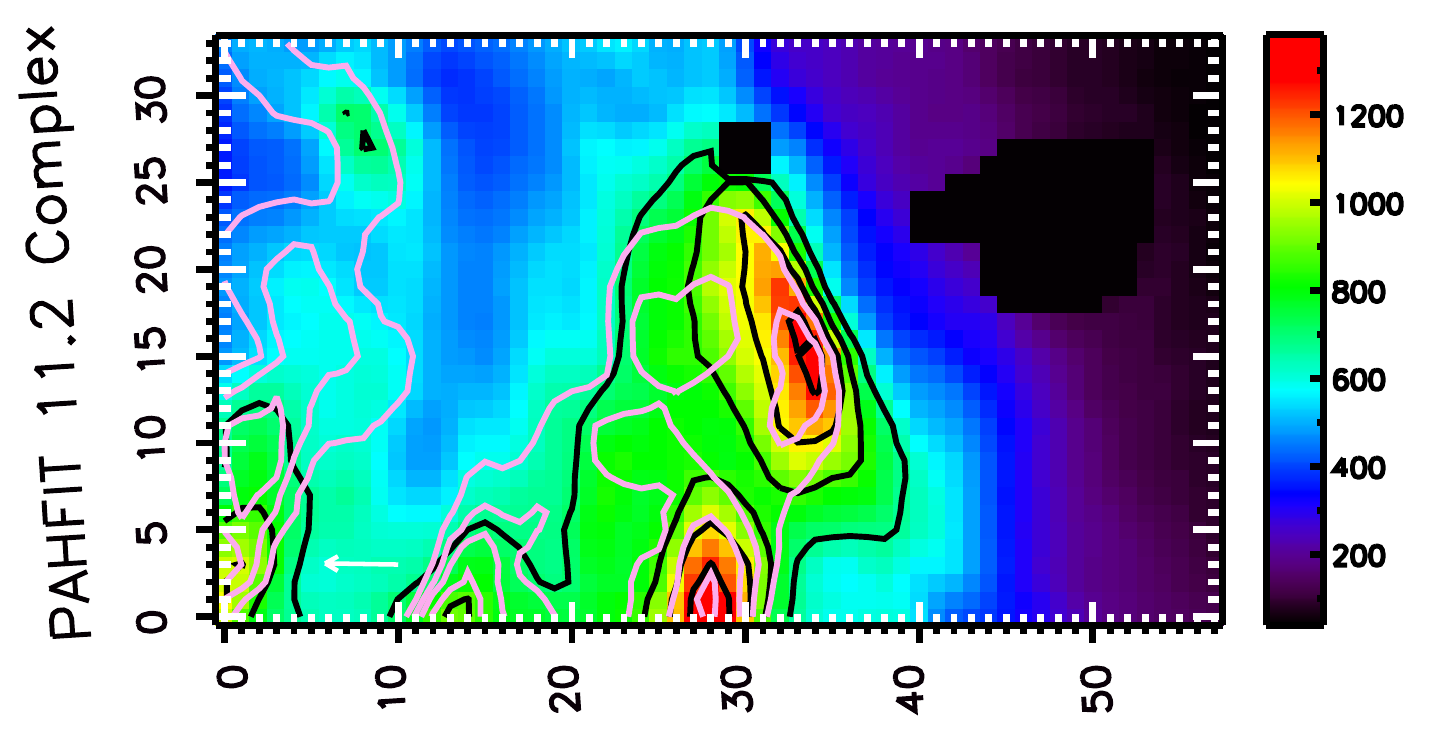}
          \includegraphics[angle=266.4]{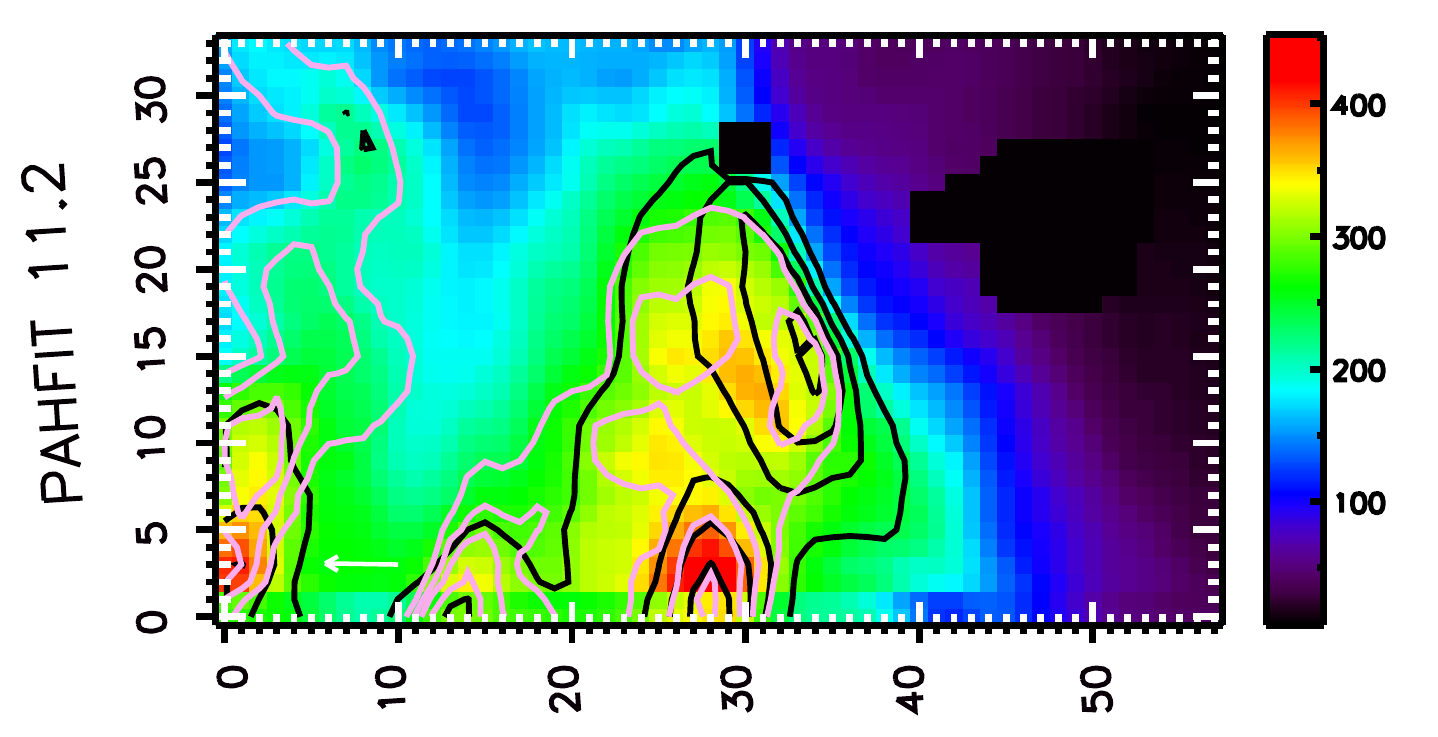}           
          \includegraphics[angle=266.4]{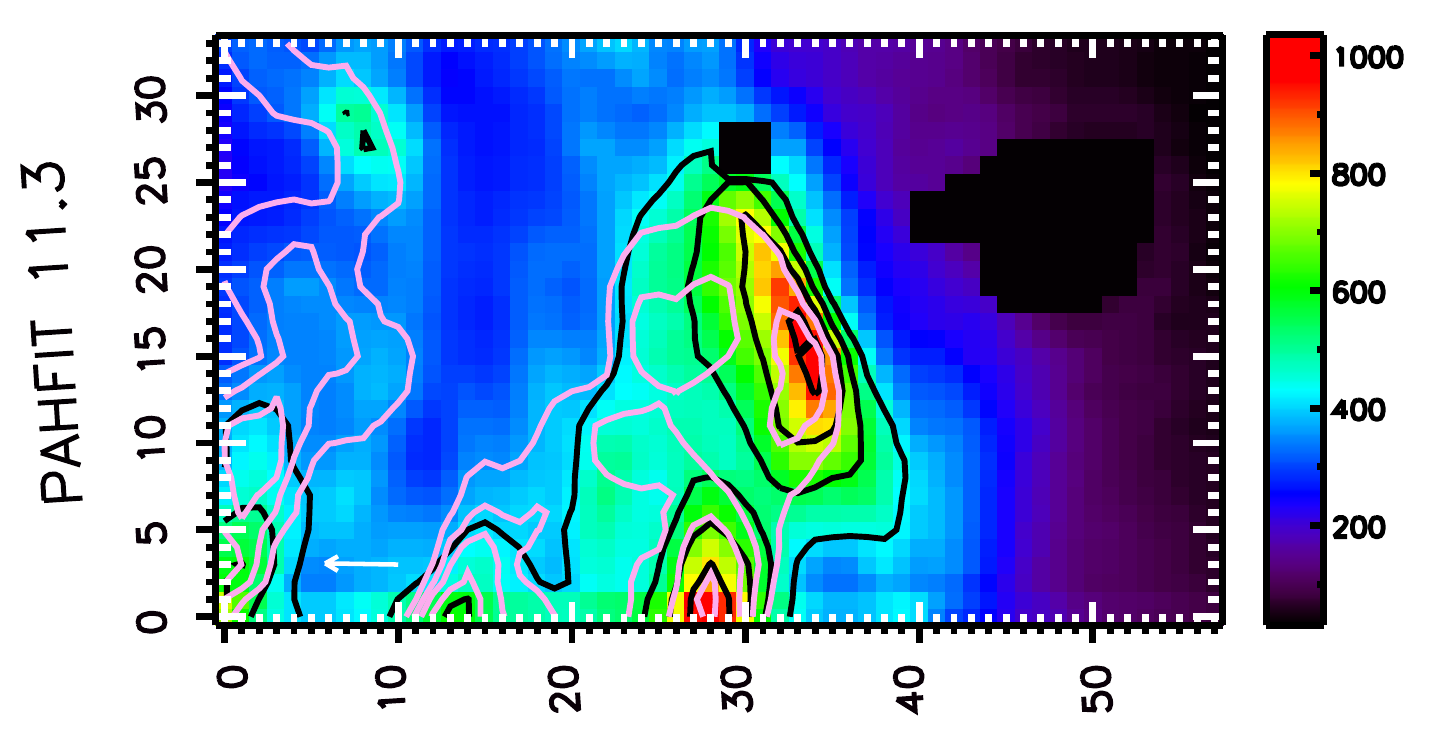}        }
   \resizebox{16.15cm}{!}{%
        \includegraphics[angle=266.4]{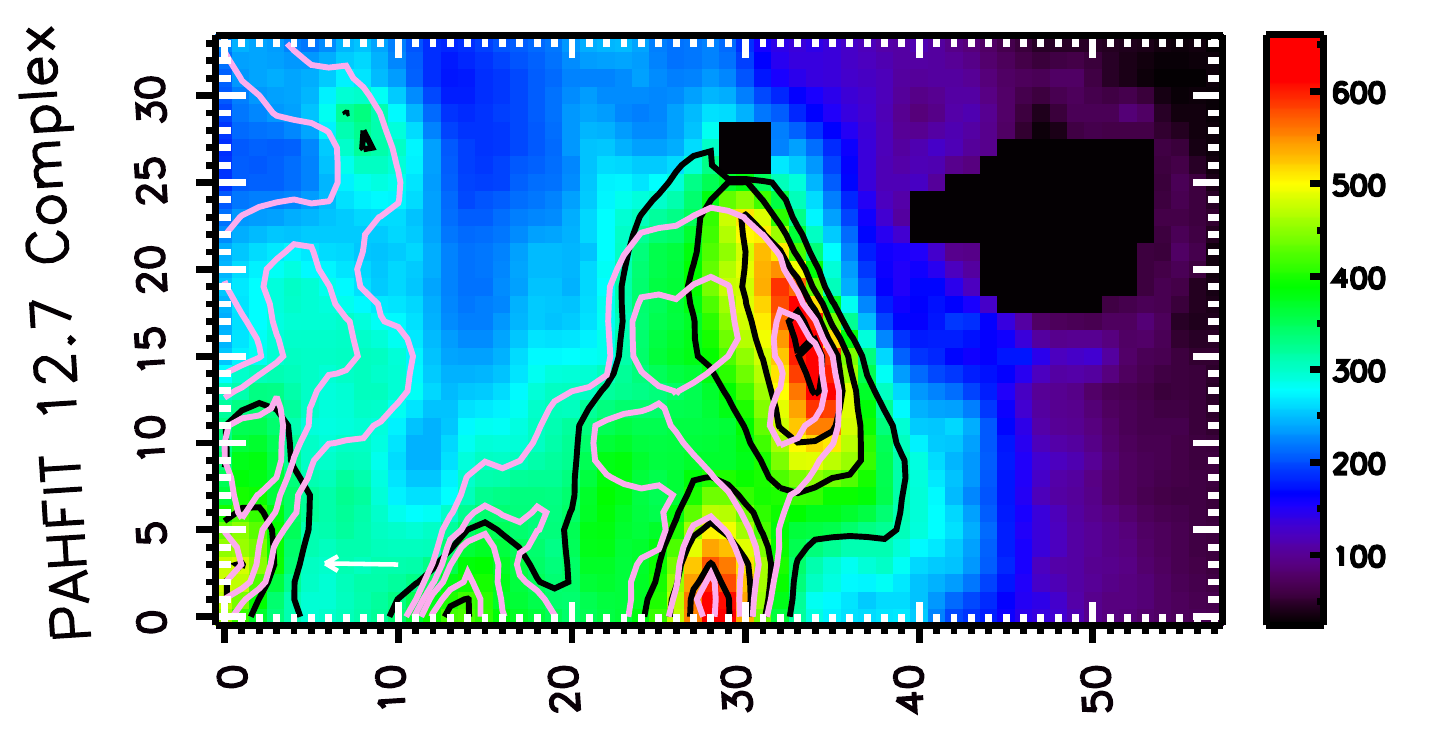}
         \includegraphics[angle=266.4]{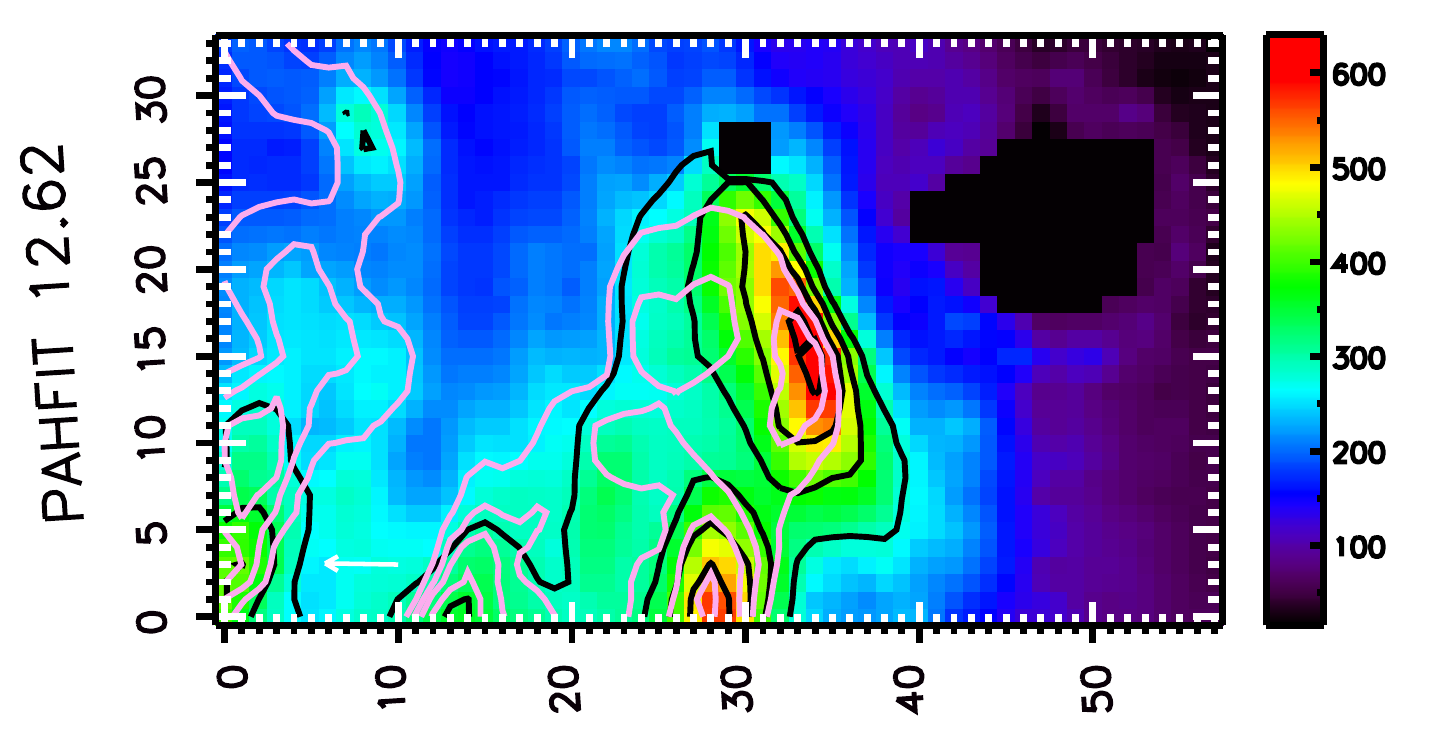}
   \includegraphics[angle=266.4]{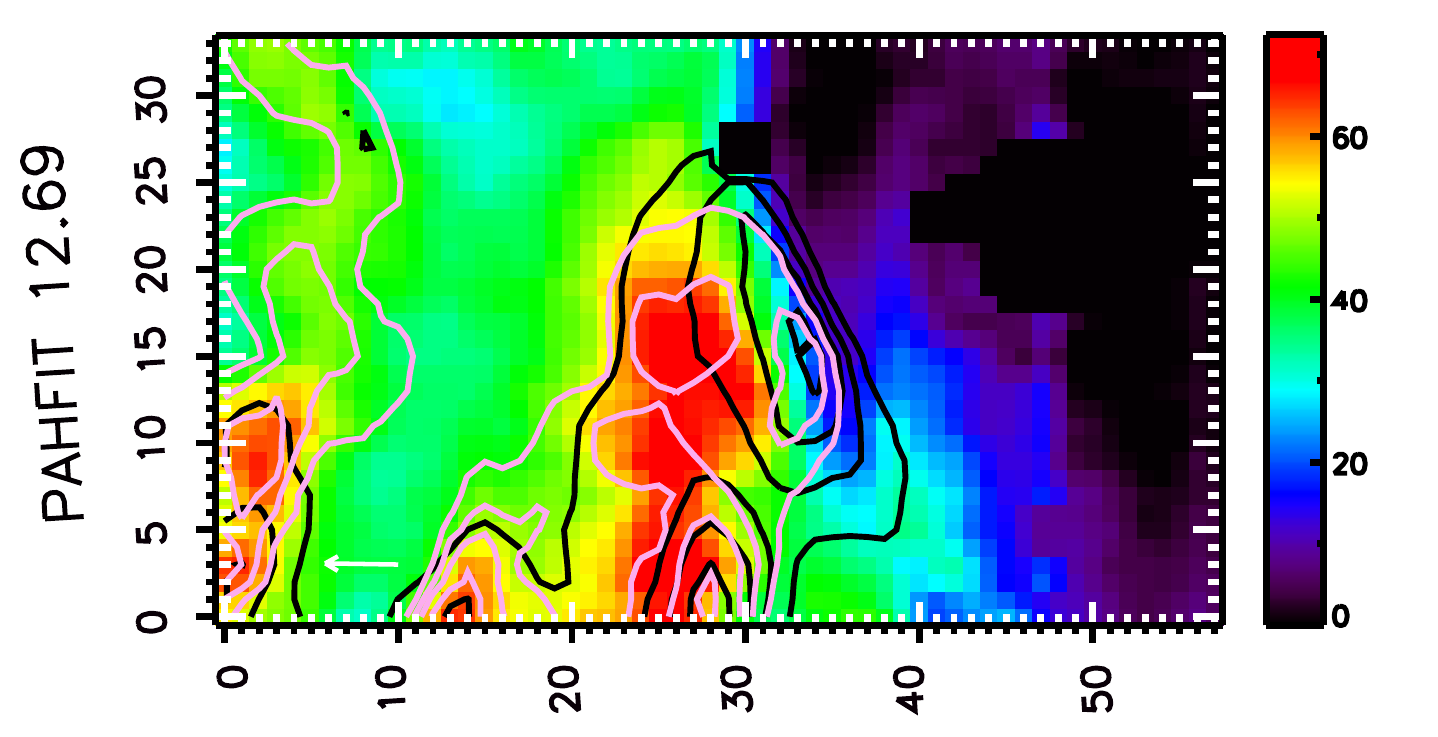}
   \includegraphics[angle=266.4]{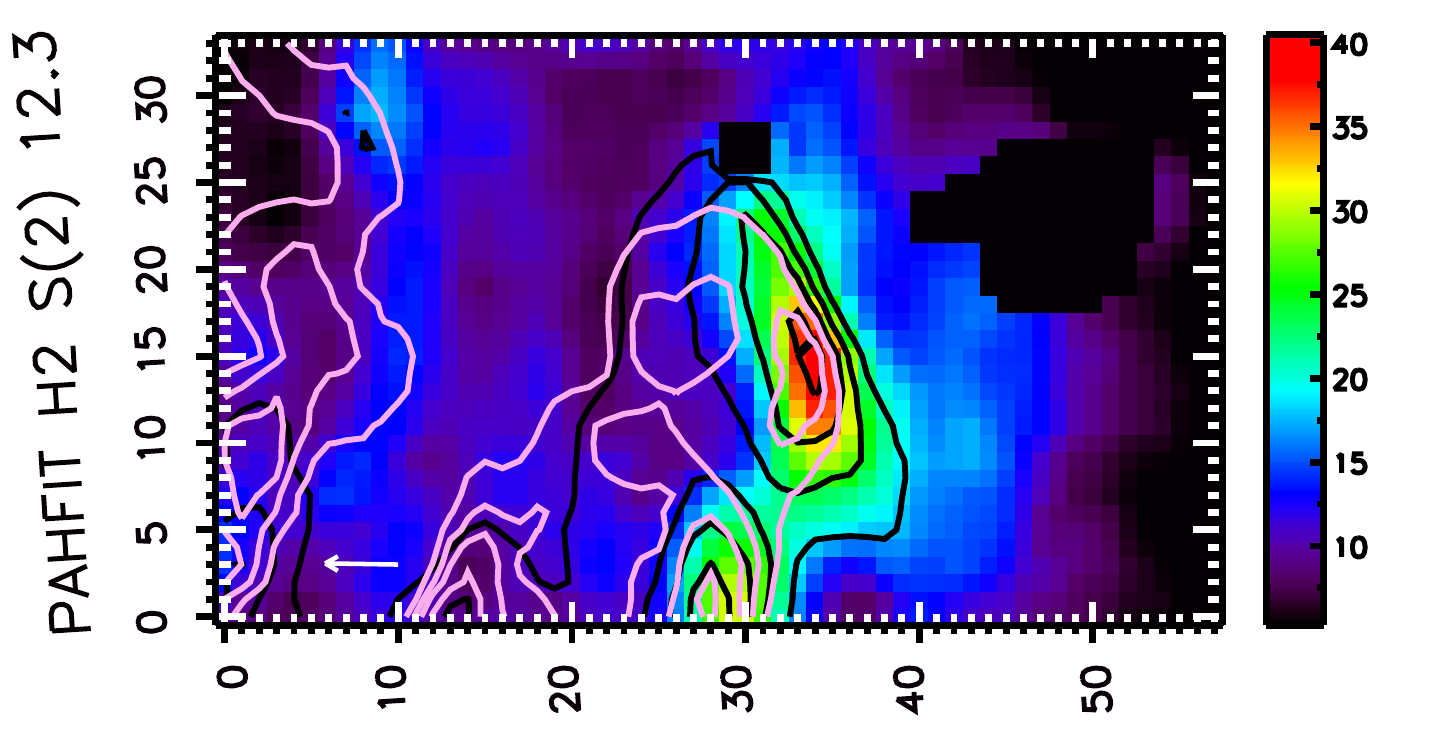}
   \includegraphics[angle=266.4]{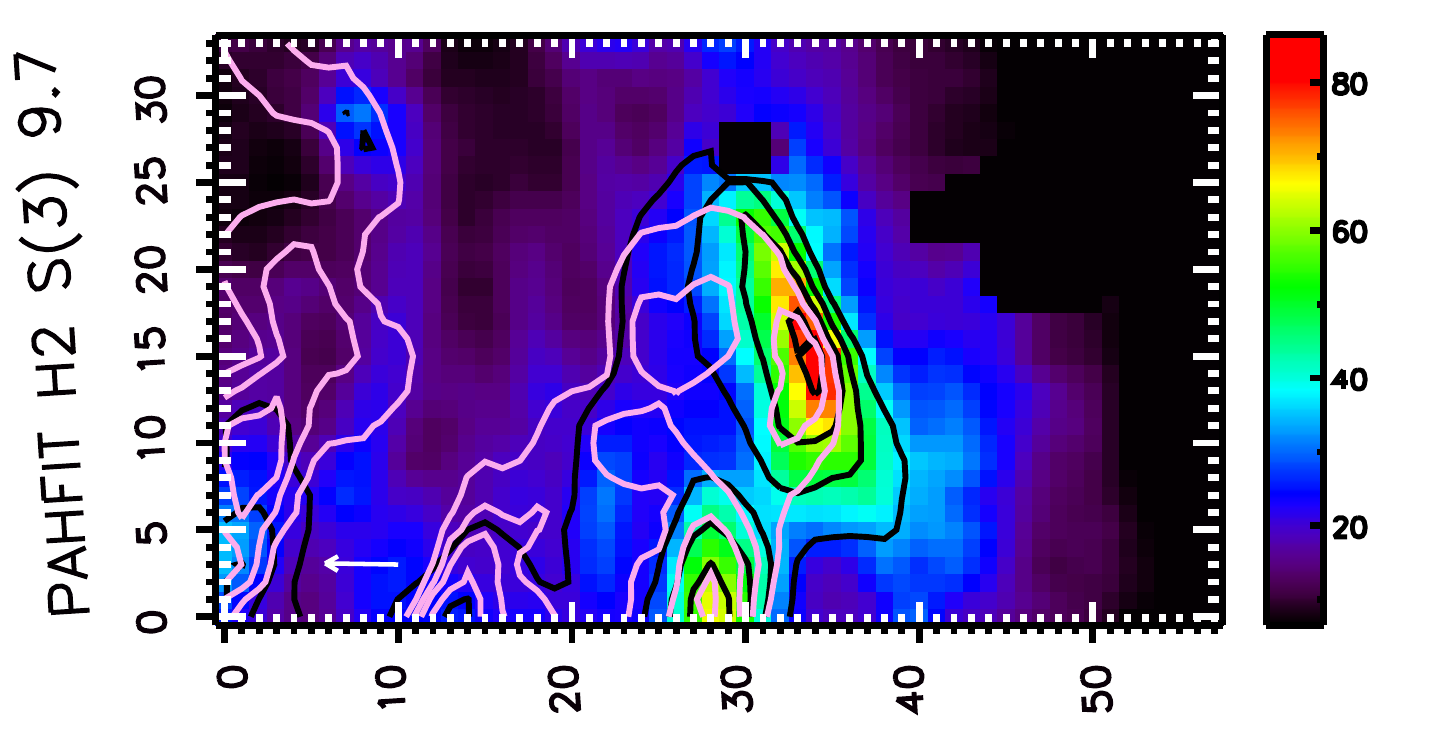}}
              \resizebox{12.92cm}{!}{%
     \includegraphics[angle=266.4]{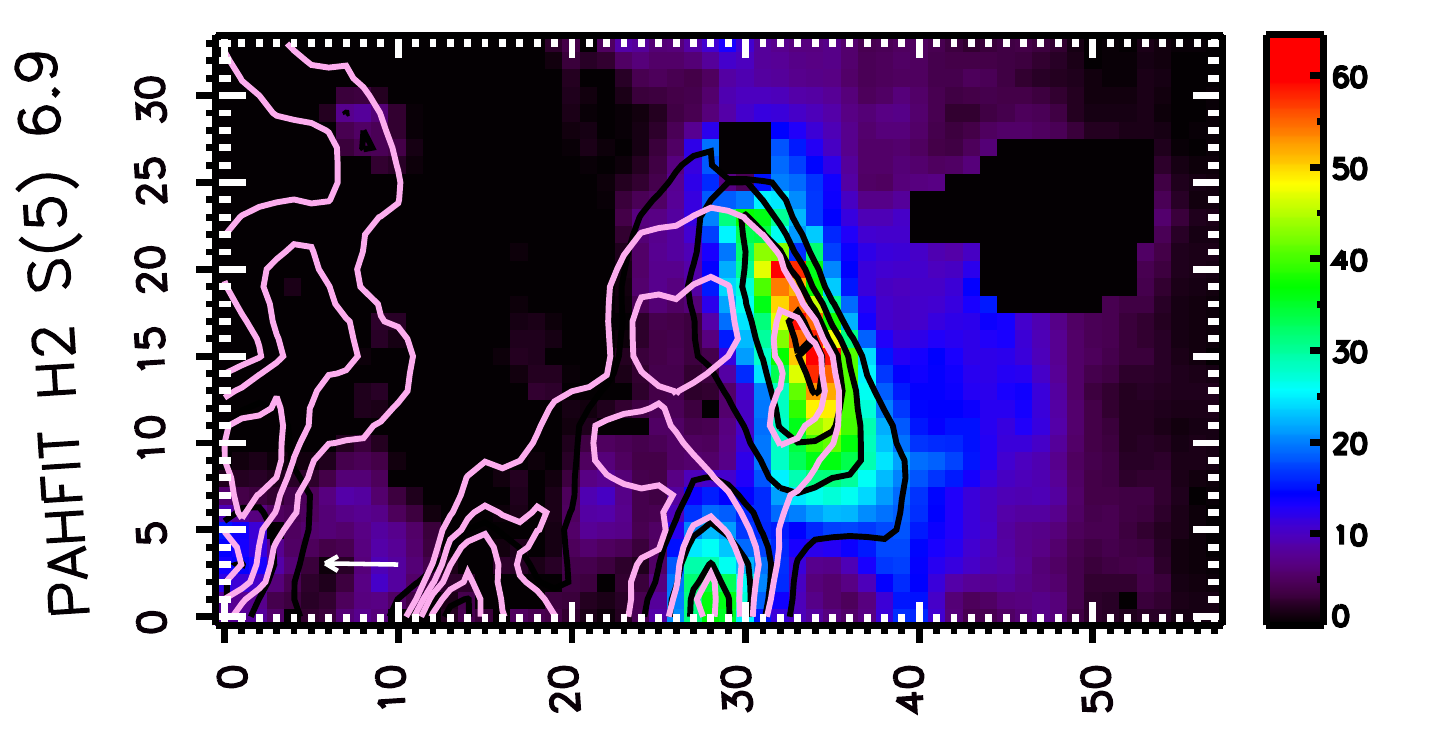}
            \includegraphics[angle=266.4]{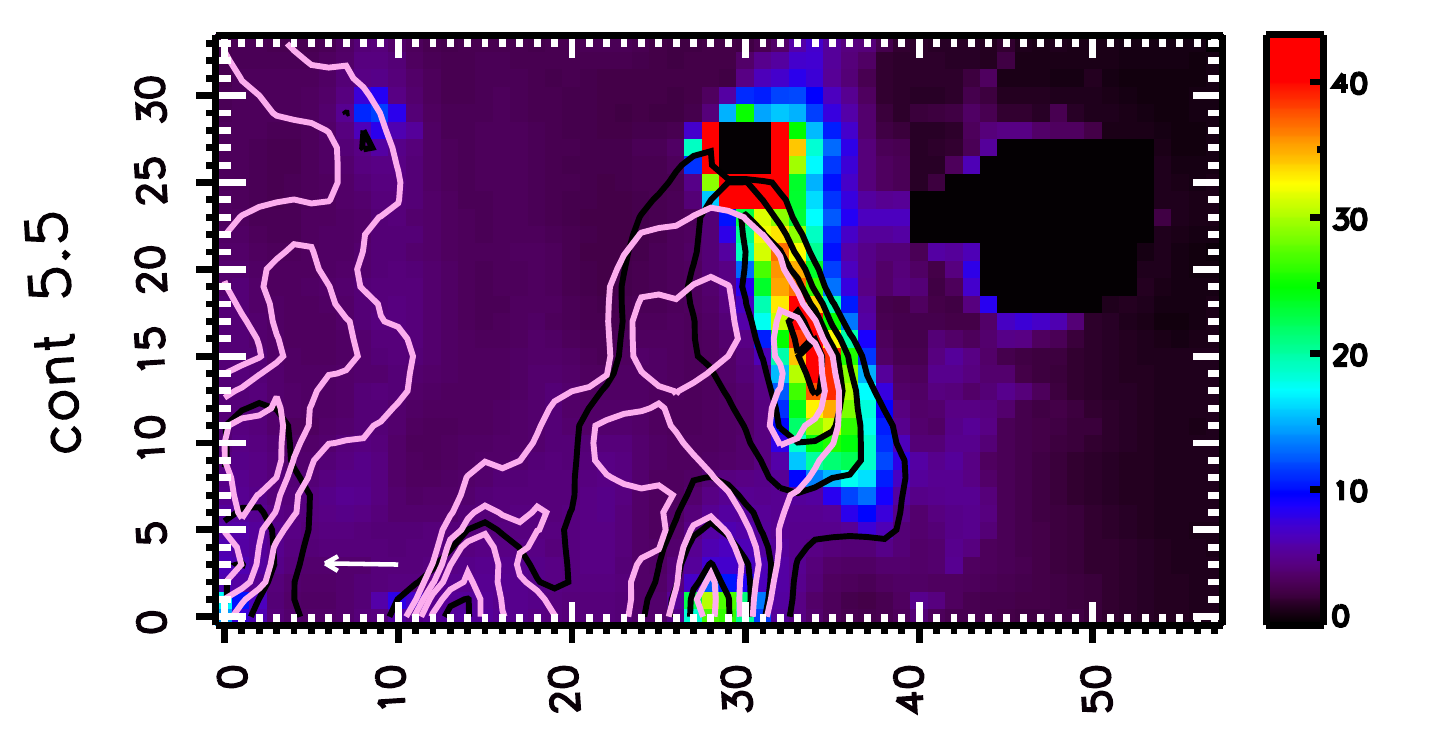}
           \includegraphics[angle=266.4]{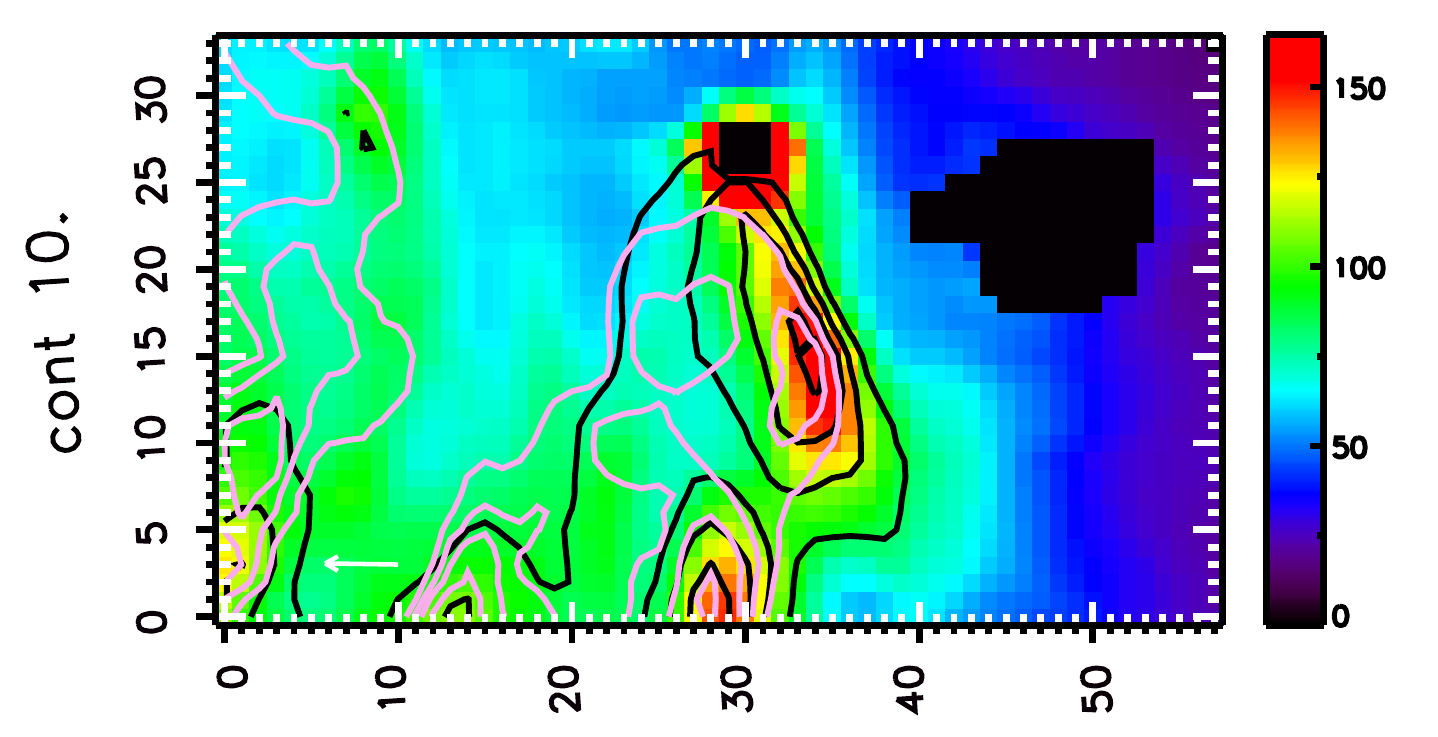}
              \includegraphics[angle=266.4]{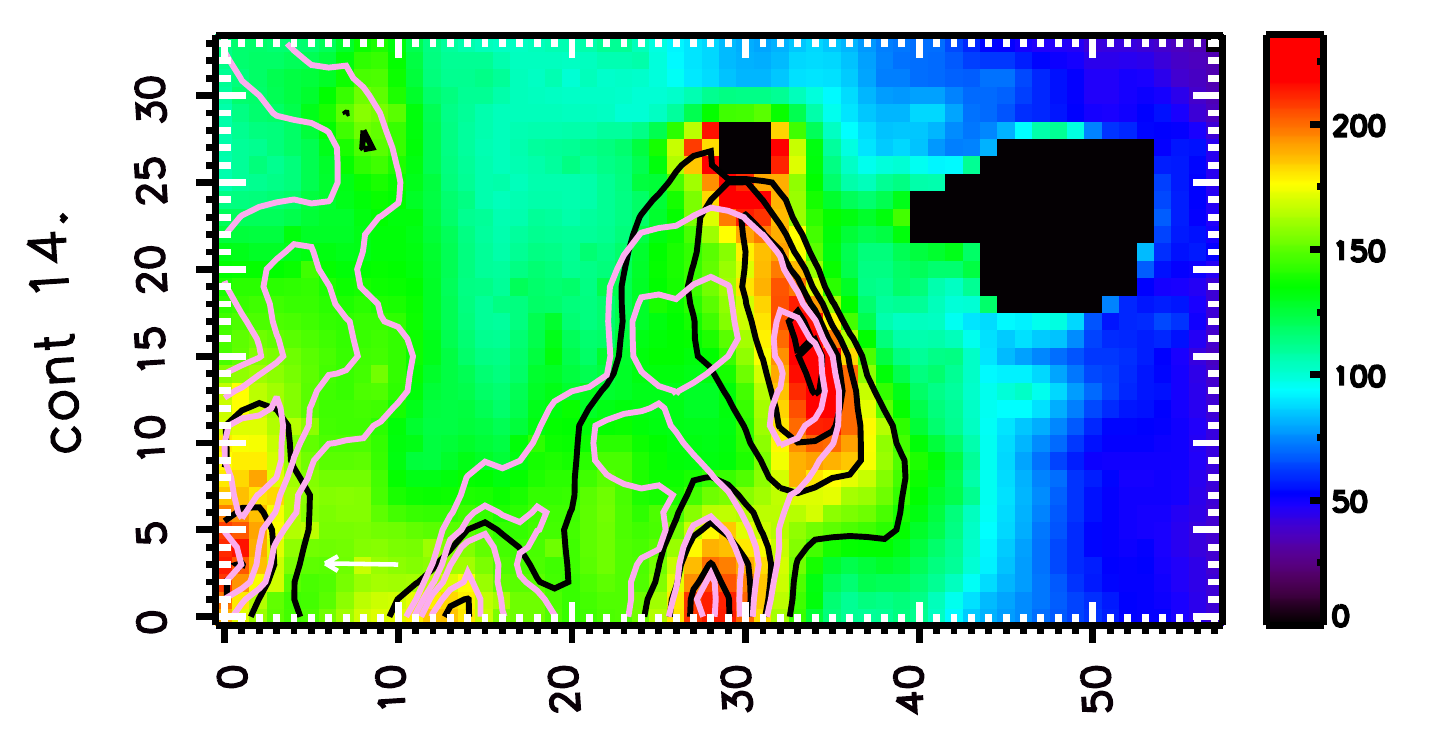} }
\caption{Spatial distribution of the components of PAHFIT for the south map. 
 Band intensities are measured in units of 10$^{-8}$ Wm$^{-2}$sr$^{-1}$ and continuum intensities in units of MJysr$^{-1}$. The intensity profiles of the 11.2 and 7.7 \mum\, emission features are shown as contours in respectively black (at 3.66, 4.64, 5.64 and, 6.78 10$^{-6}$ Wm$^{-2}$sr$^{-1}$) and pink (at 1.40, 1.56, 1.70 and, 1.90 10$^{-5}$ Wm$^{-2}$sr$^{-1}$). The maps are orientated so N is up and E is left. The white arrow in the top left corners indicates the direction towards the central star. The axis labels refer to pixel numbers. Regions near source C and D excluded from the analysis are set to zero.  The range in intensities of the color bar for the south continuum map is determined by excluding the immediate region of source C. The nomenclature and the FOV of the SH map are given in the bottom right panel of Fig.~\ref{fig_slmaps_s}).}  
\label{fig_pahfit_s}
\end{figure*}
%%%%%%%%%%%%%%%%%%%%%%%%%%%%%%%%%%%%%%%%%%%%%%%%%%
\clearpage
%%%%%%%%%%%%%%%%%%%%%%%%%%%%%%%%%%%%%%%%%%%%%%%%%%
\begin{figure*}
    \centering
\resizebox{17.1cm}{!}{% 
\includegraphics[angle=274.1]{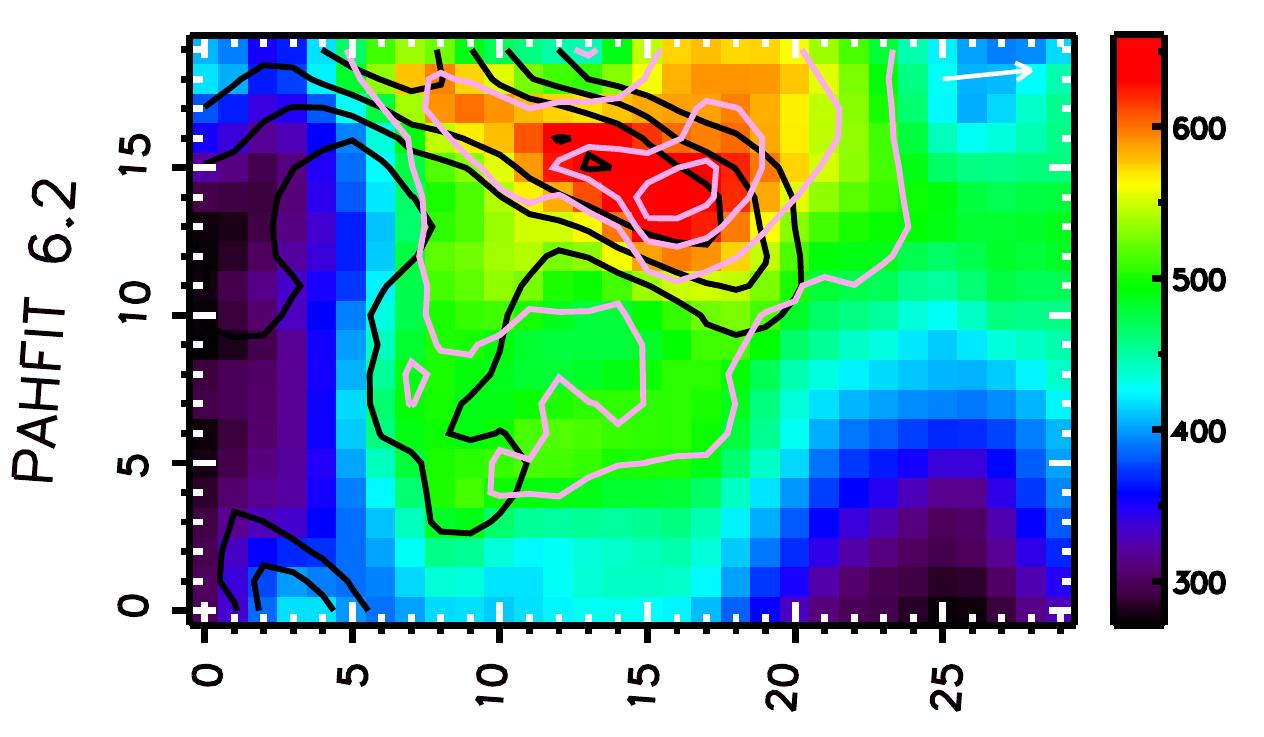}
\includegraphics[angle=274.1]{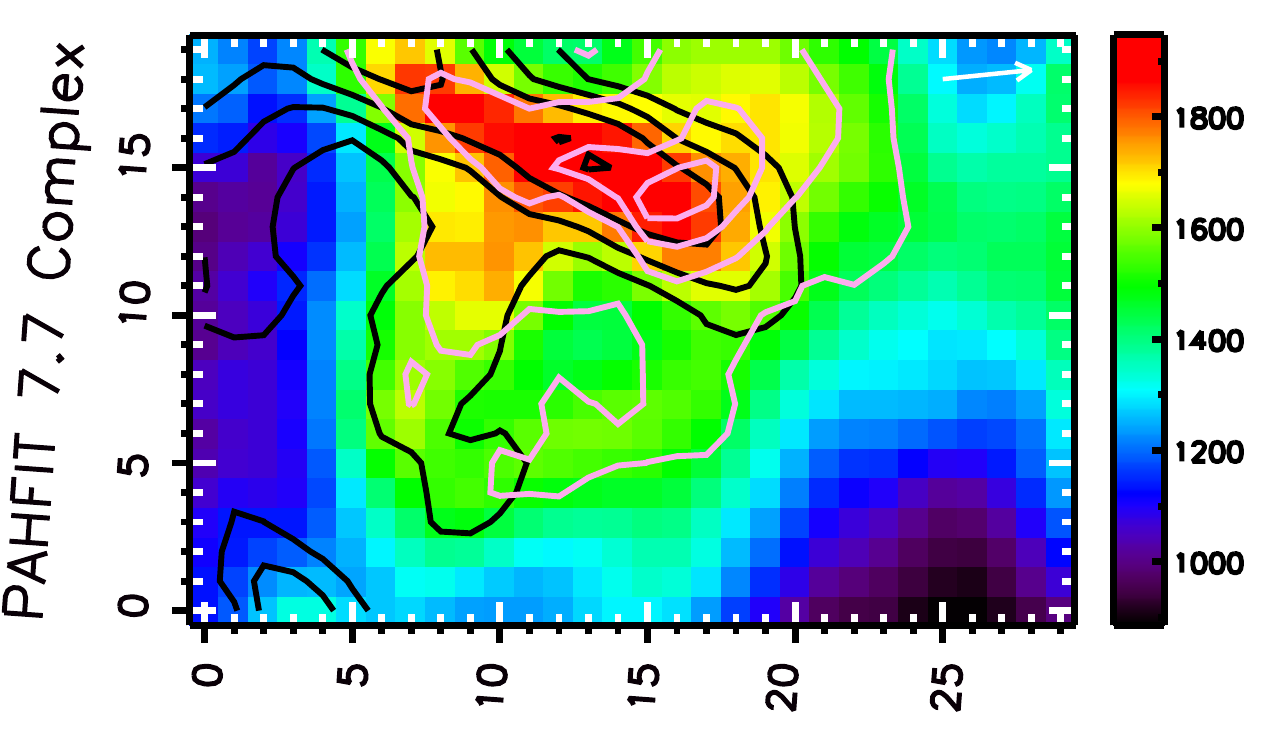}
     \includegraphics[angle=274.1]{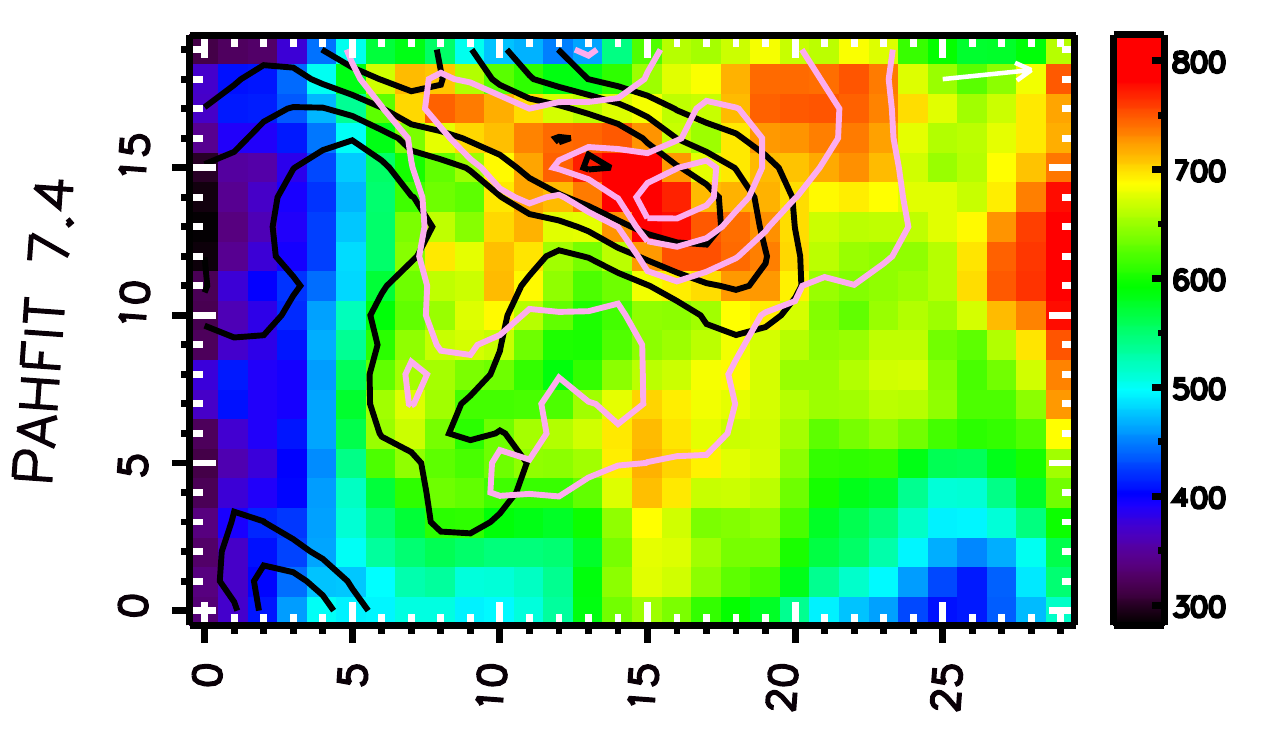}
       \includegraphics[angle=274.1]{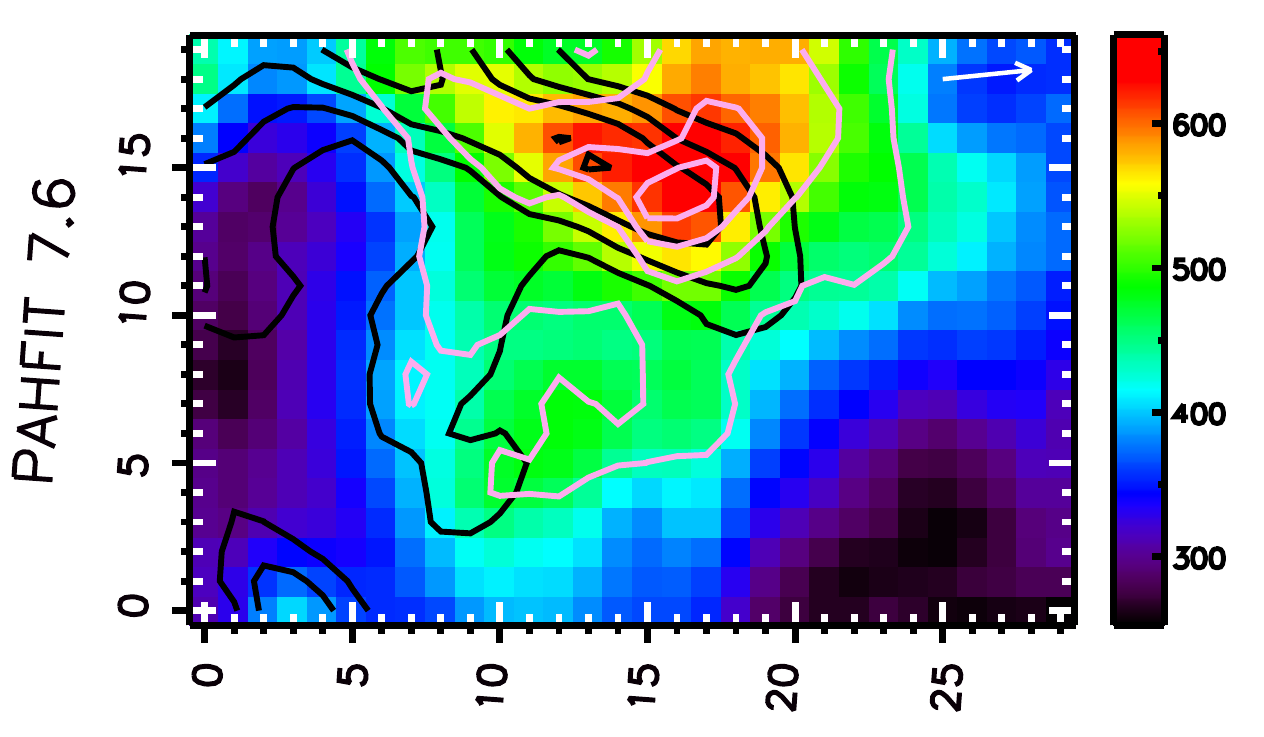}
        \includegraphics[angle=274.1]{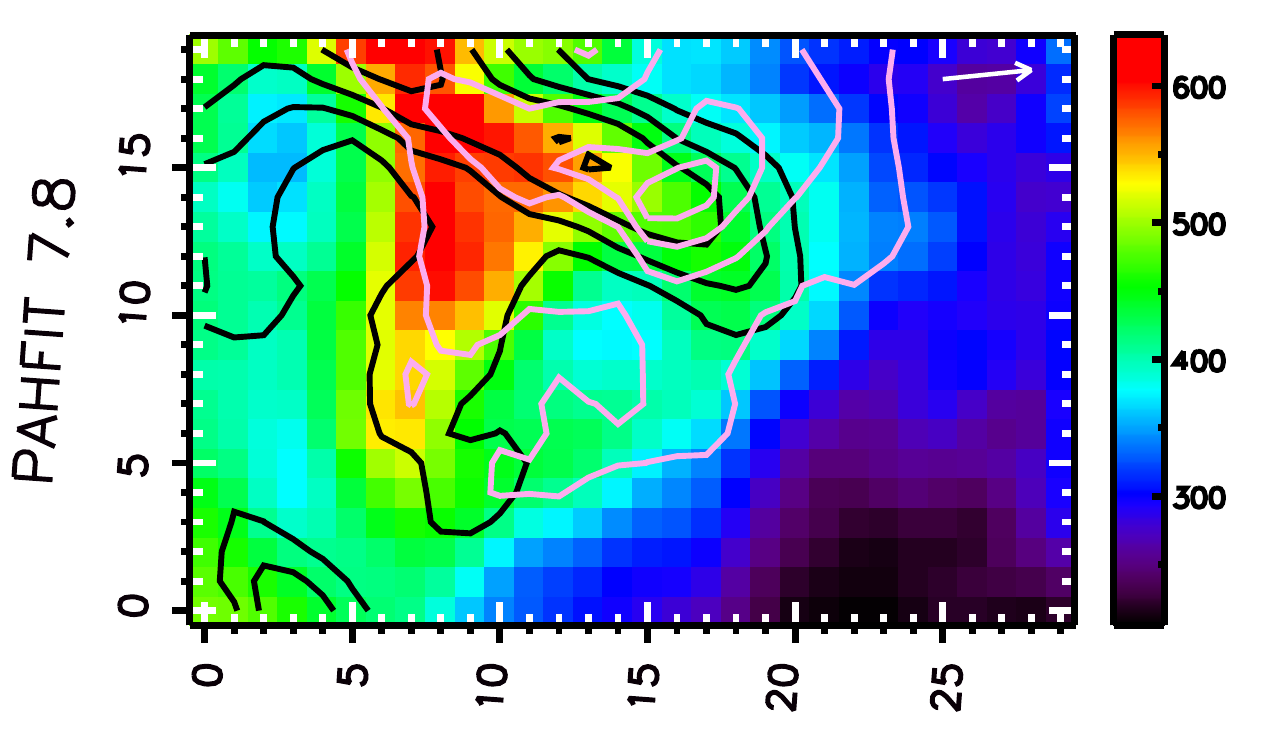}}
   \resizebox{17.1cm}{!}{%
     \includegraphics[angle=274.1]{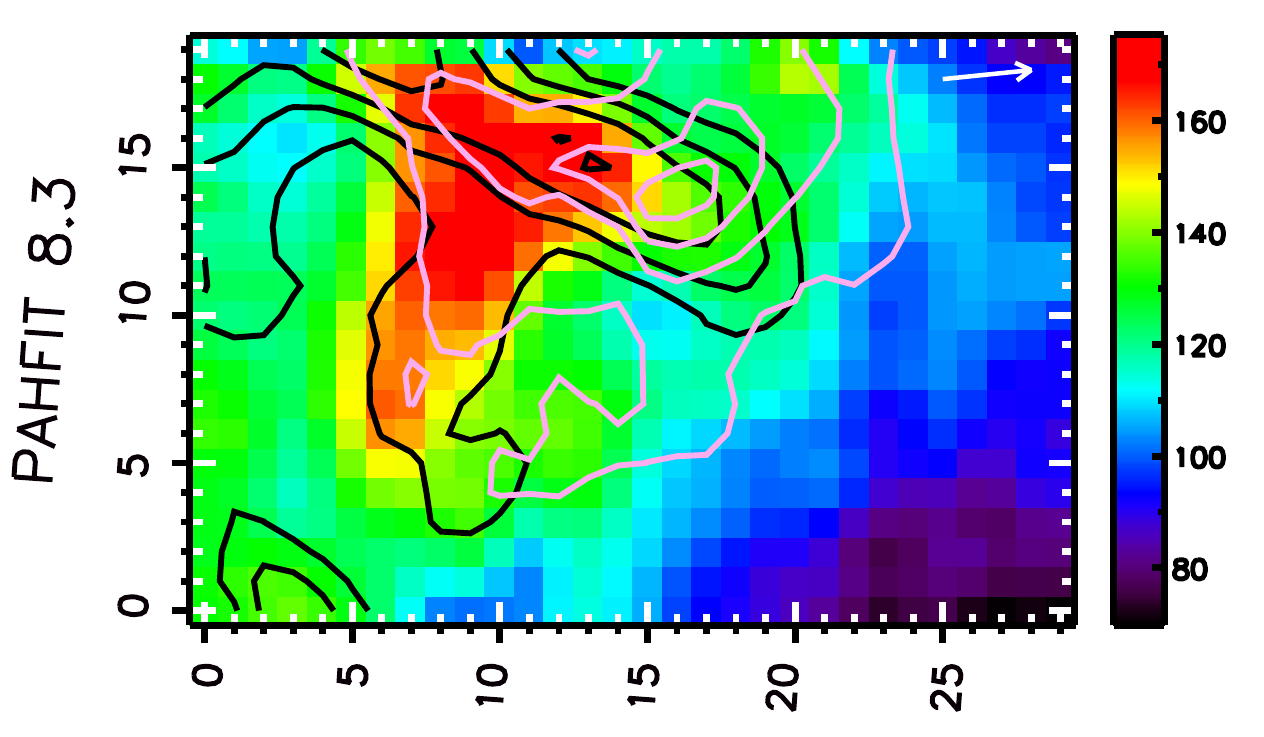}
        \includegraphics[angle=274.1]{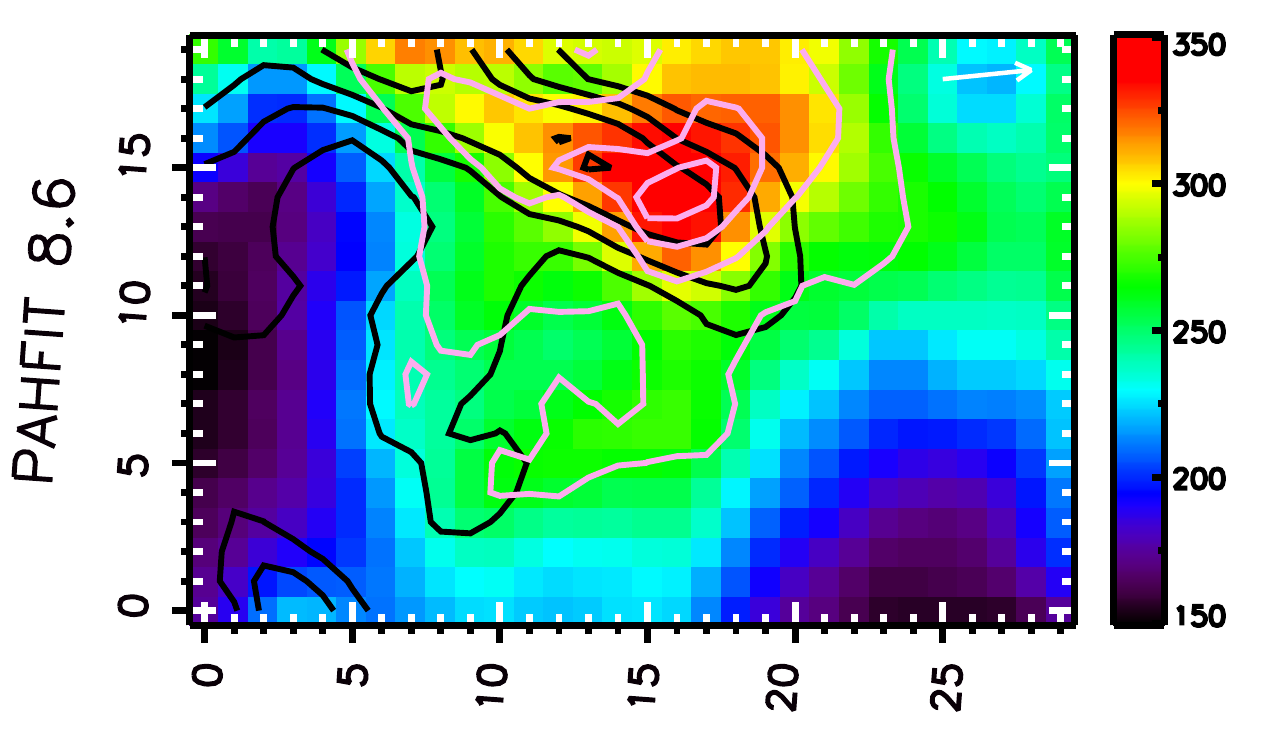}
        \includegraphics[angle=274.1]{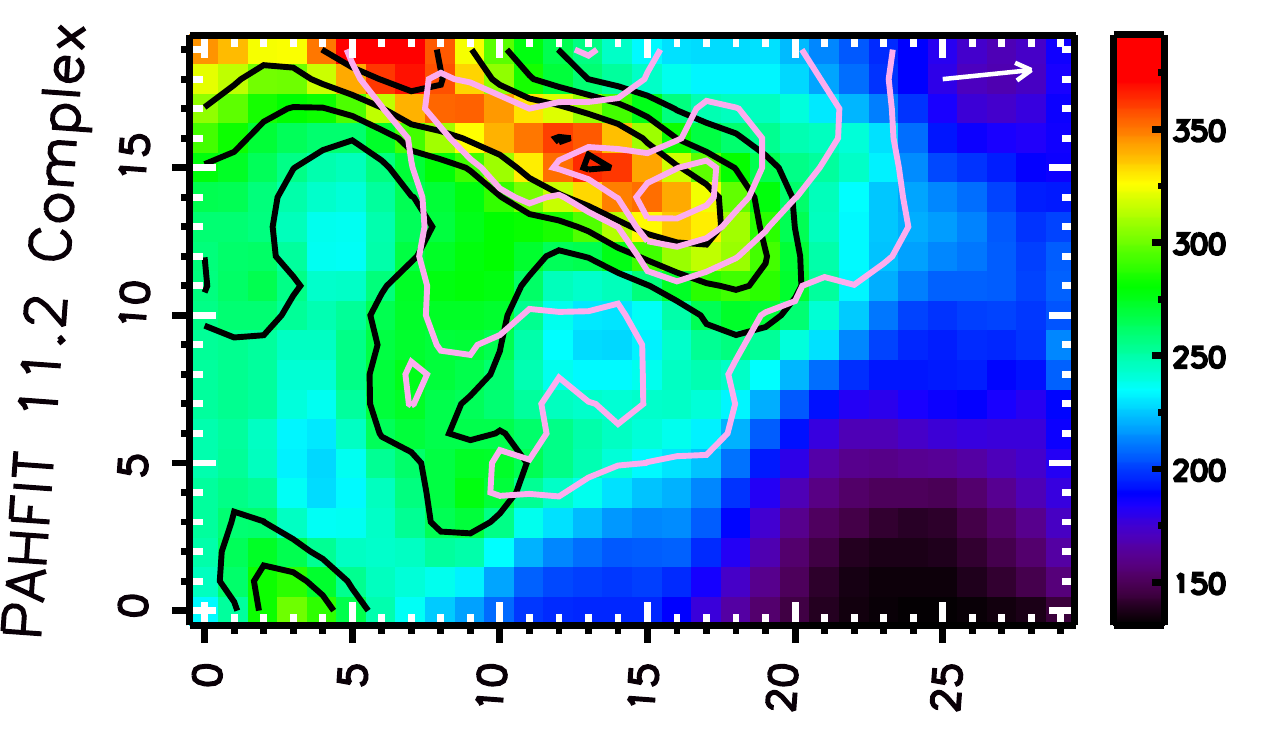}
       \includegraphics[angle=274.1]{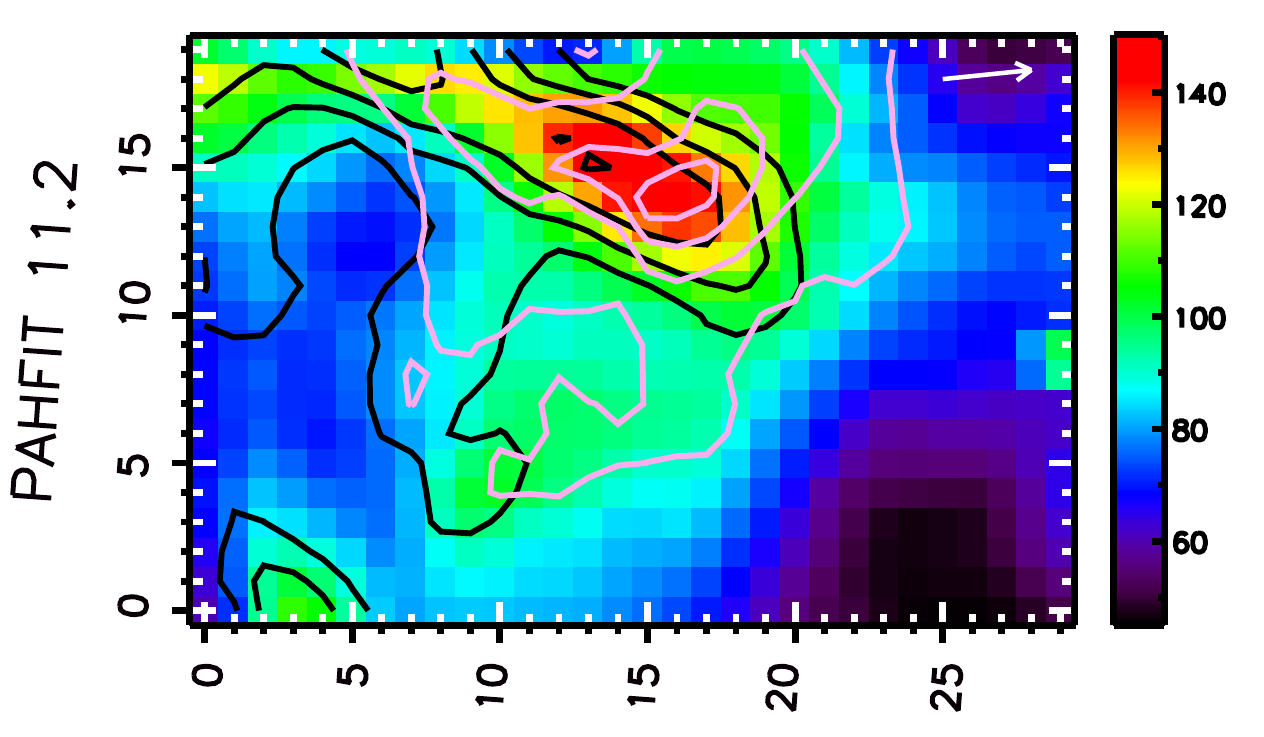}
  \includegraphics[angle=274.1]{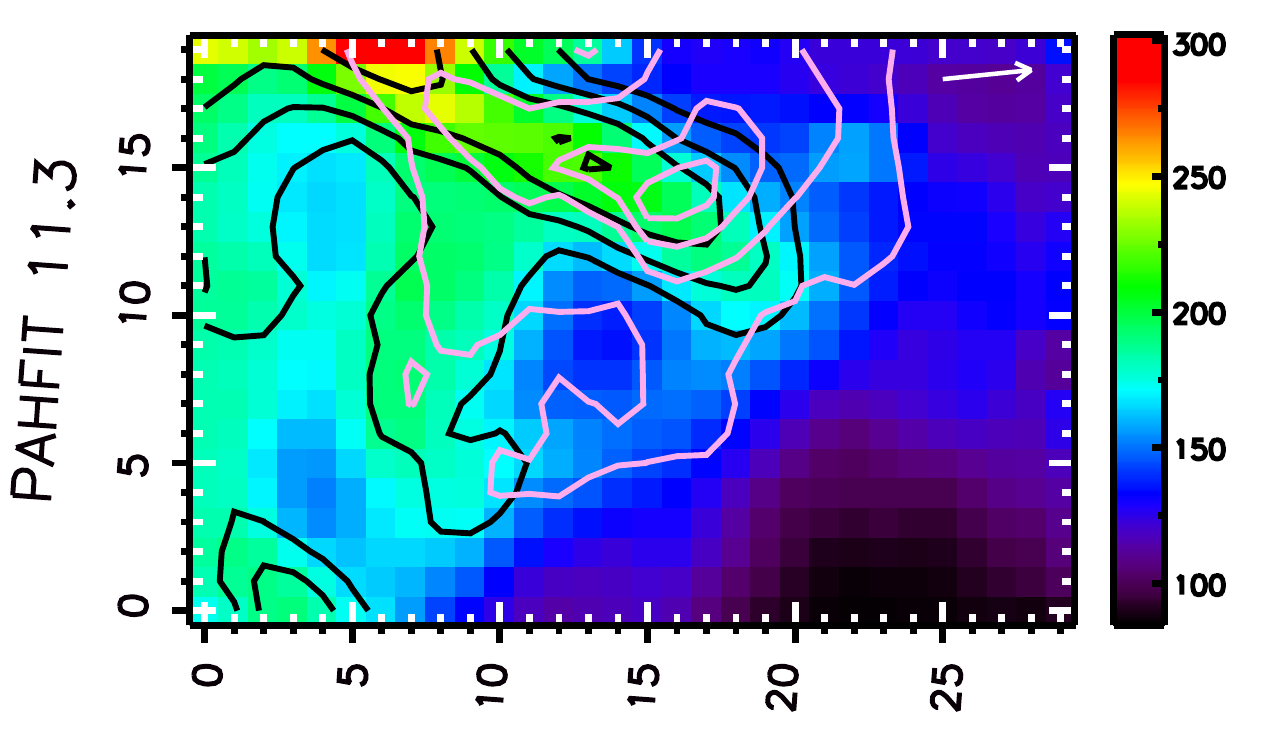}}
   \resizebox{17.1cm}{!}{%
      \includegraphics[angle=274.1]{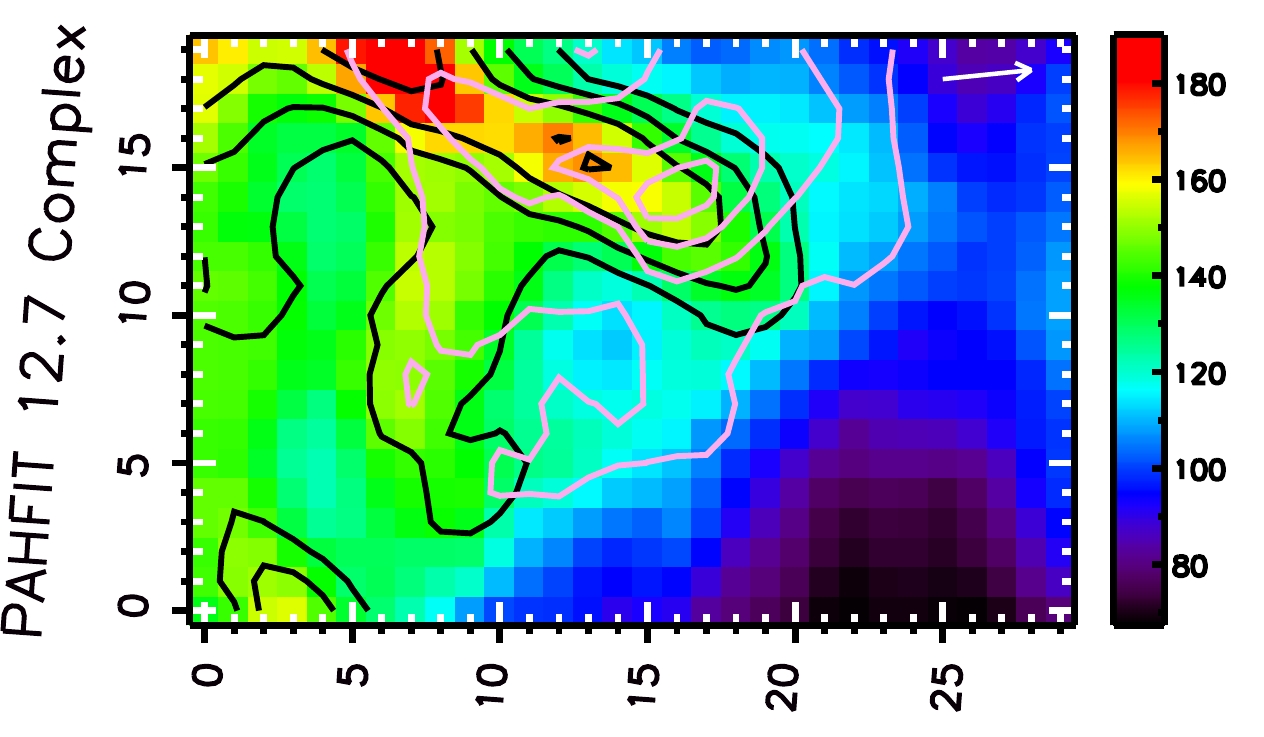}
      \includegraphics[angle=274.1]{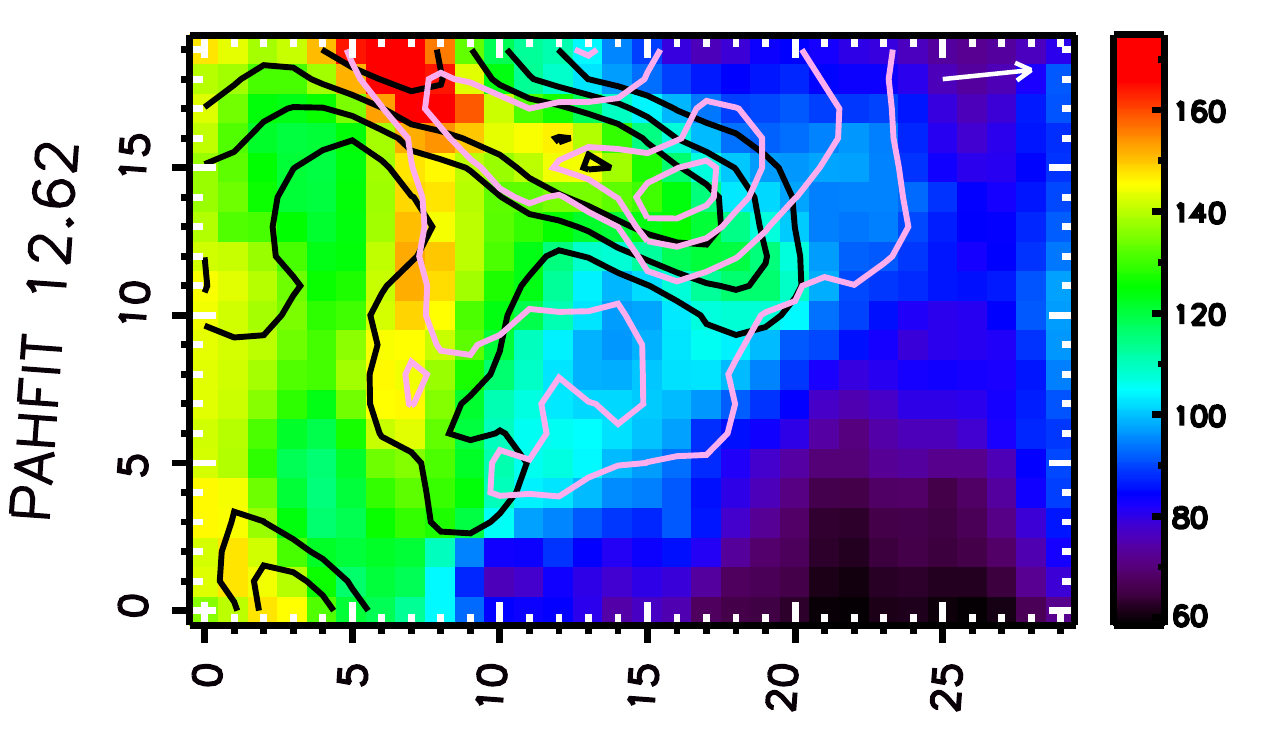}
     \includegraphics[angle=274.1]{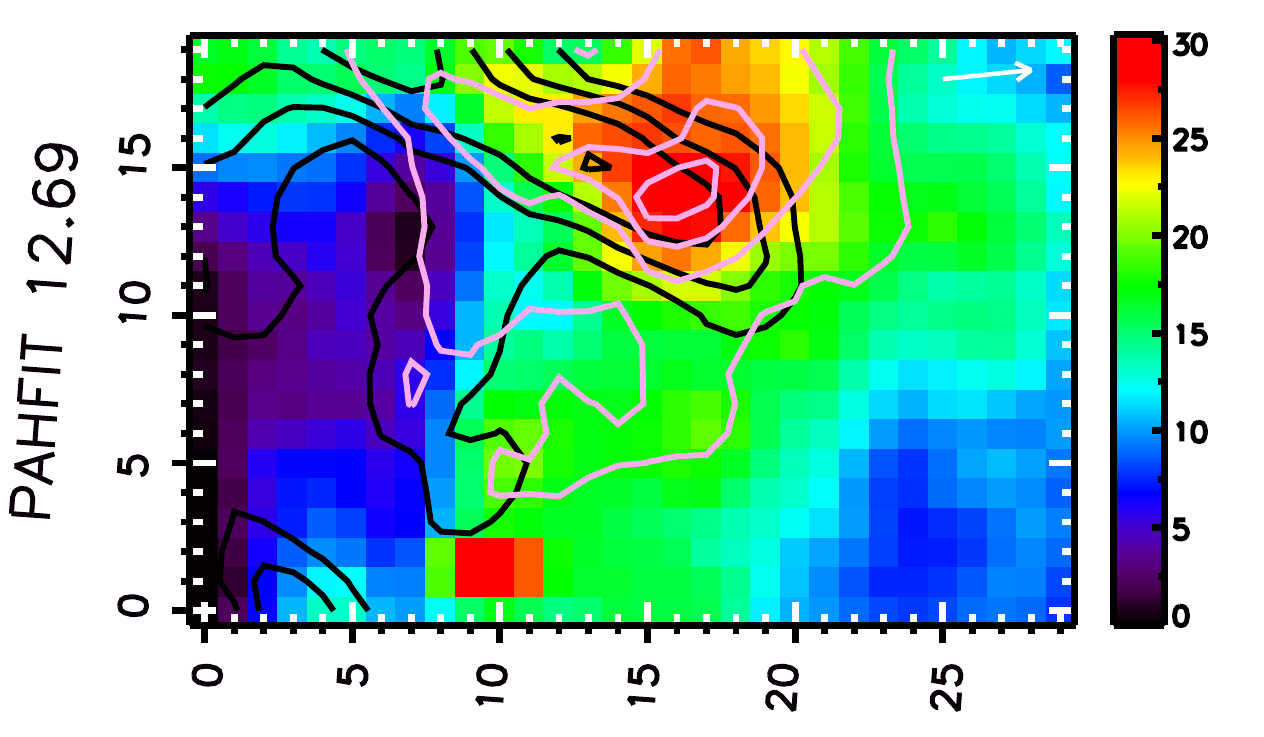}
     \includegraphics[angle=274.1]{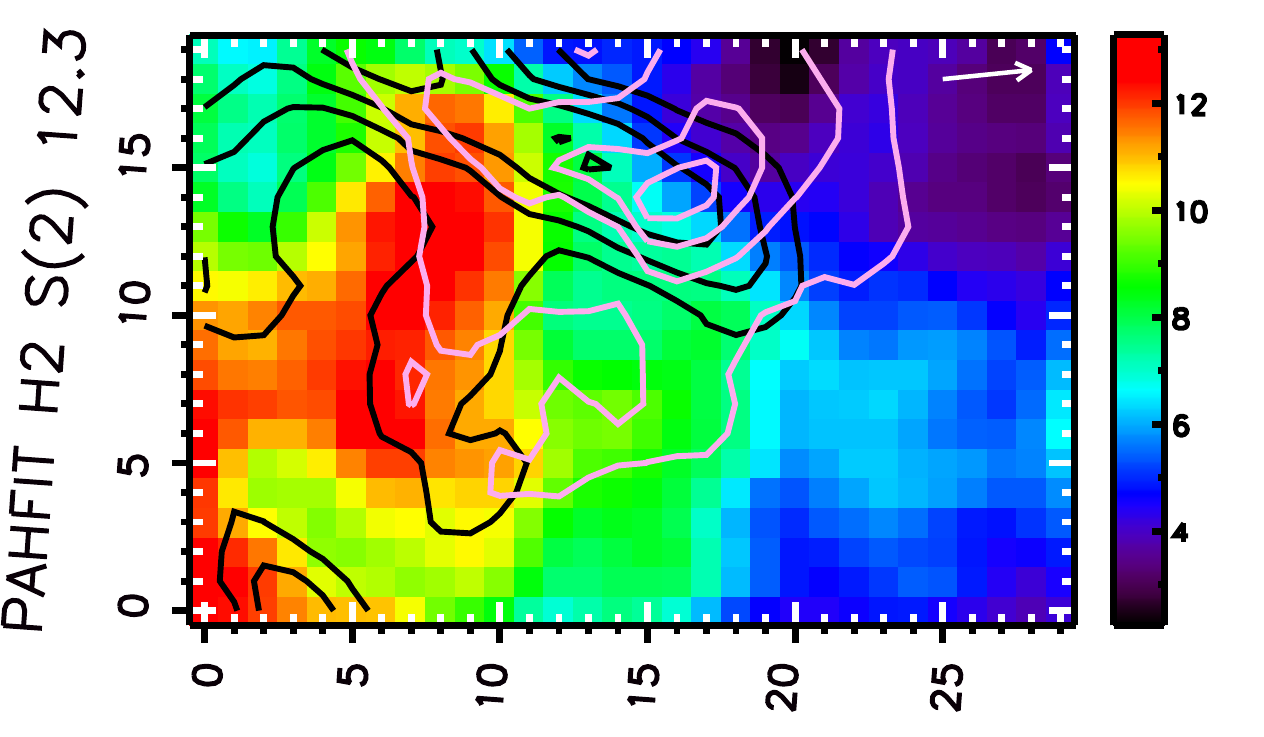}              
     \includegraphics[angle=274.1]{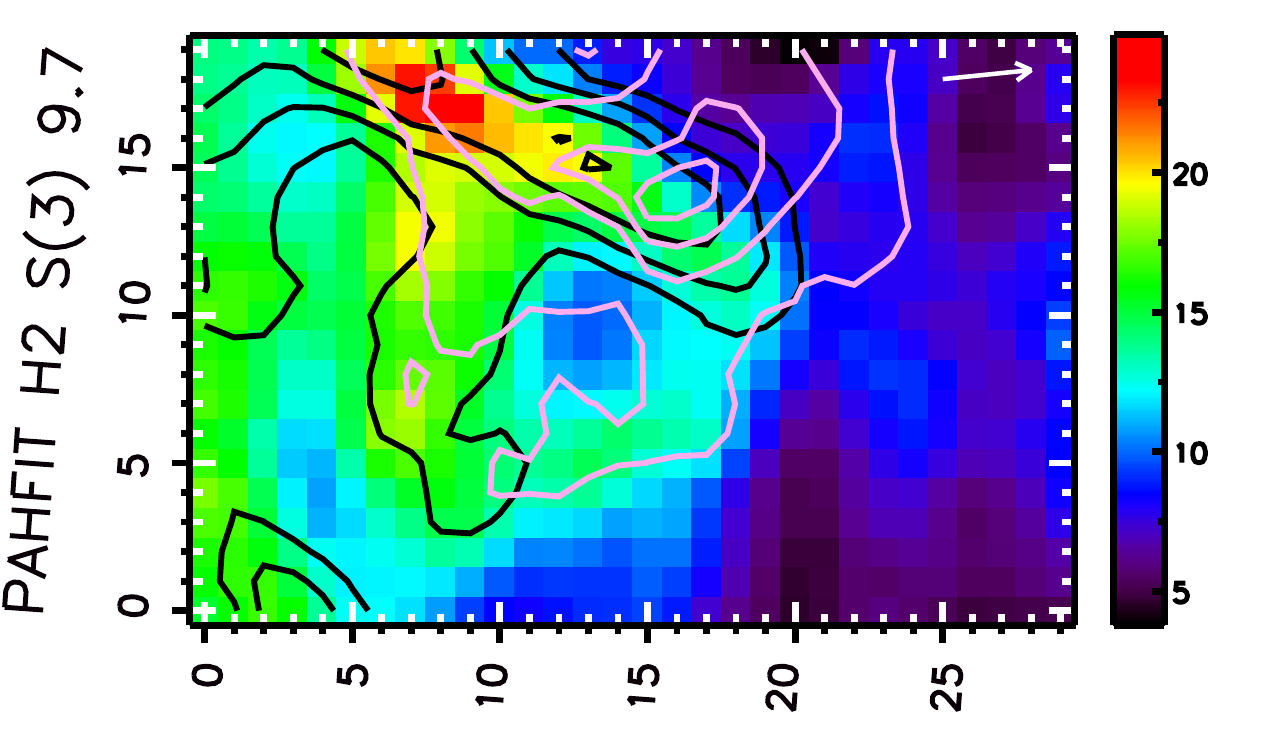}}
               \resizebox{13.68cm}{!}{%
       \includegraphics[angle=274.1]{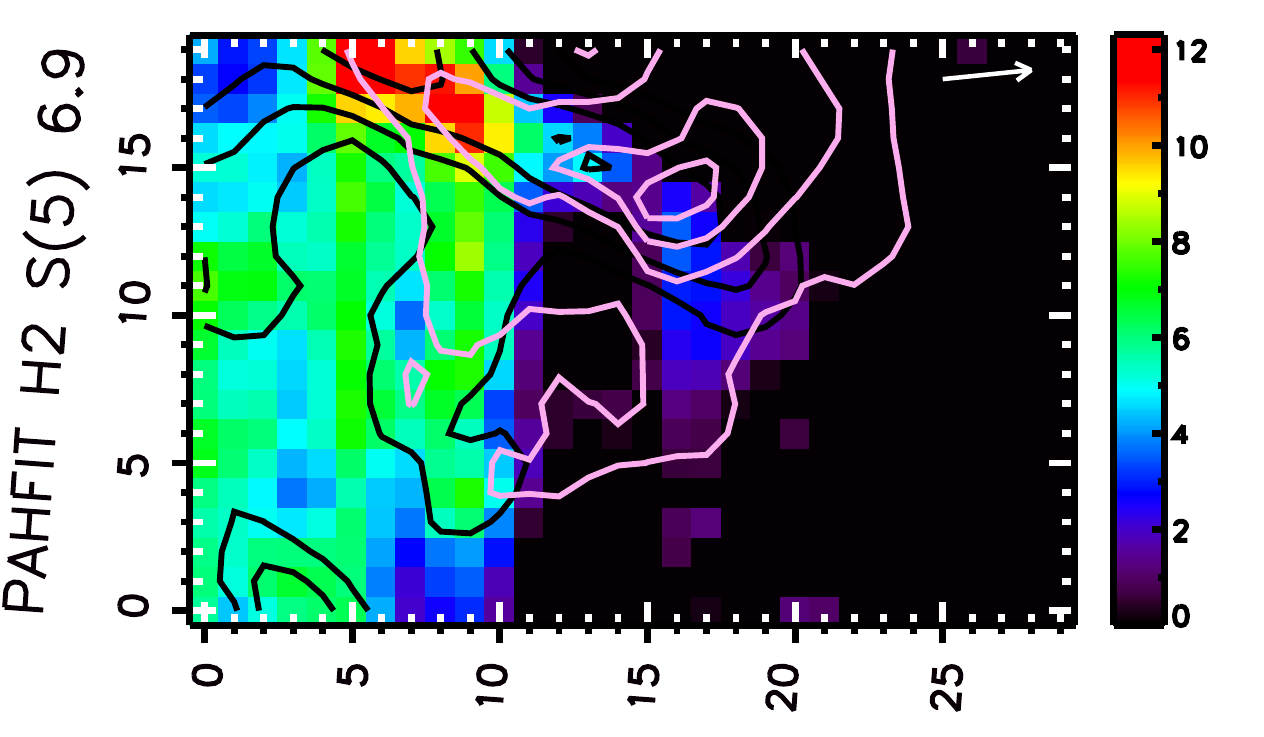}
        \includegraphics[angle=274.1]{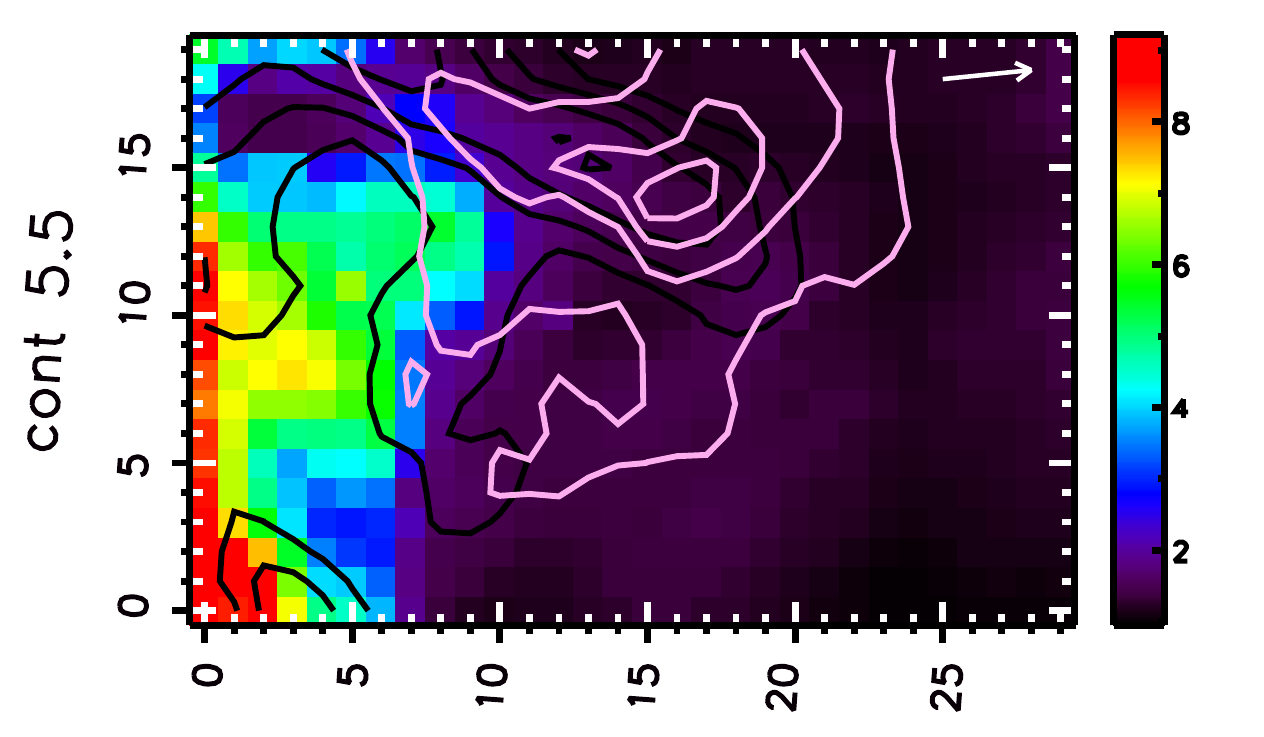}
           \includegraphics[angle=274.1]{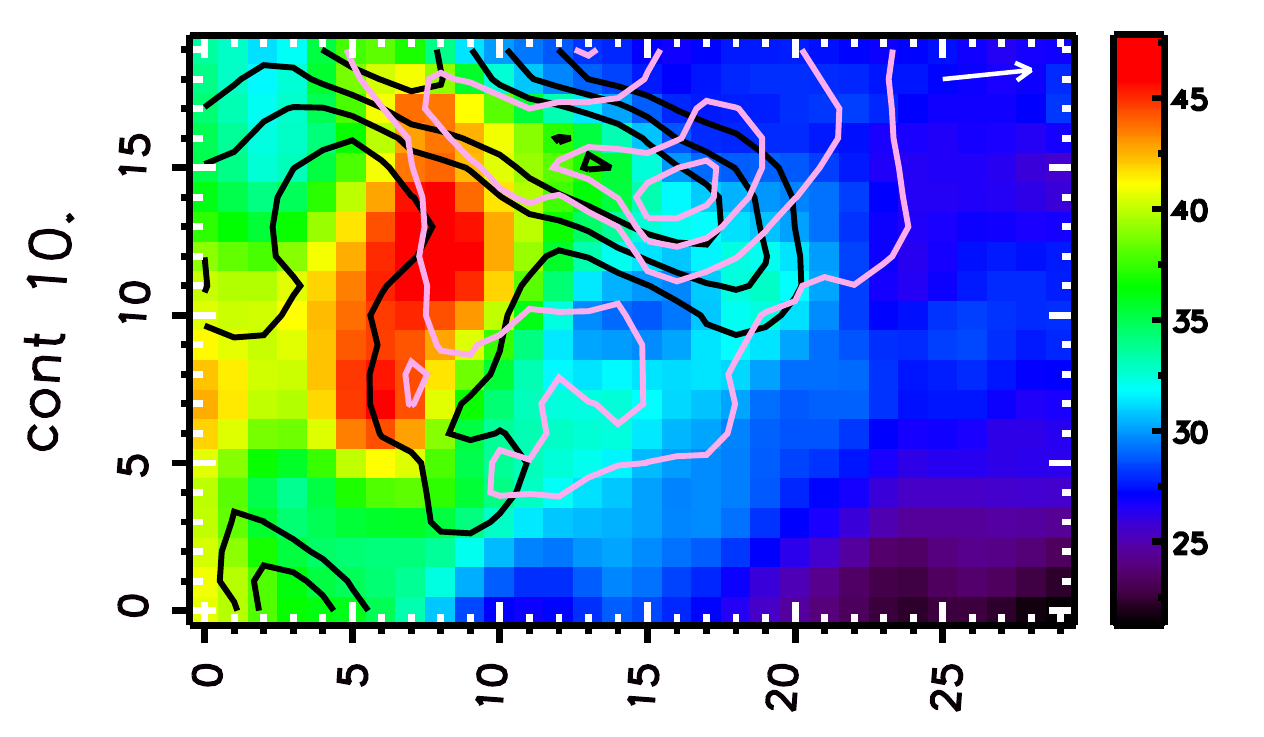}
            \includegraphics[angle=274.1]{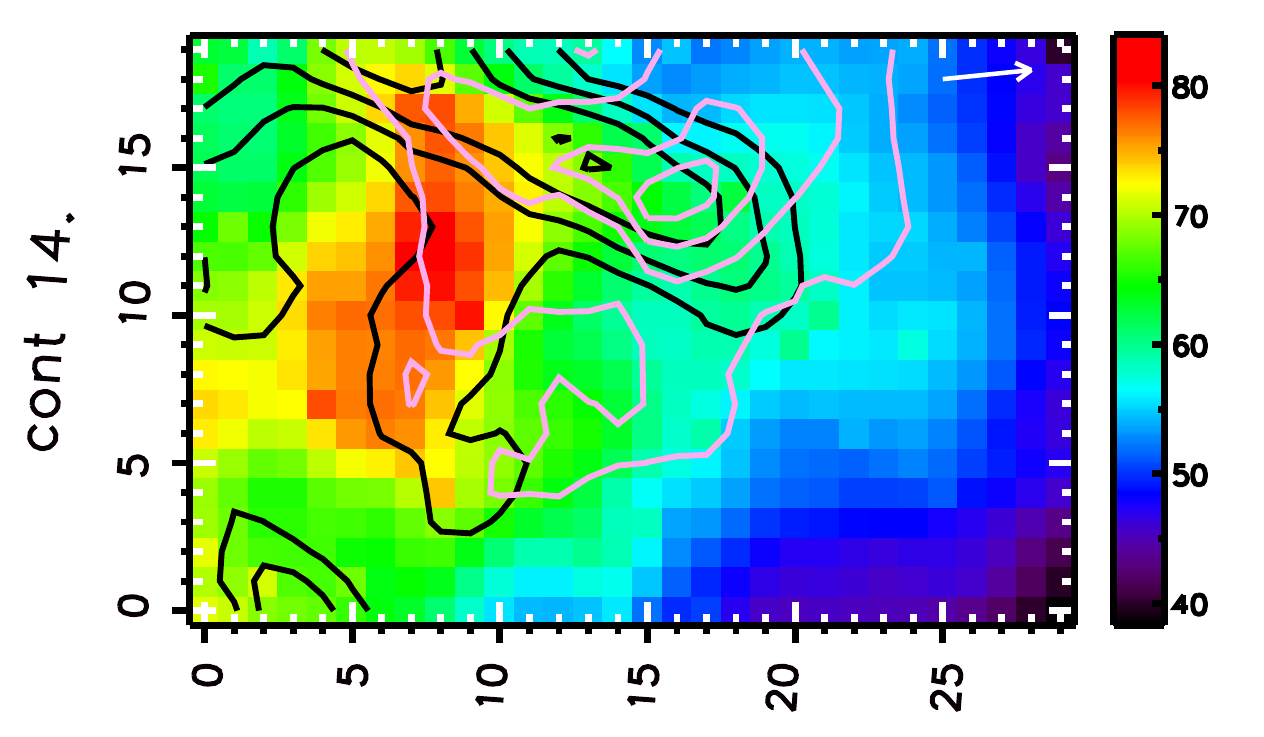}    }
 \caption{Spatial distribution of the components of PAHFIT for the north map. 
  Band intensities are measured in units of 10$^{-8}$ Wm$^{-2}$sr$^{-1}$ and continuum intensities in units of MJysr$^{-1}$. As a reference, the intensity profiles of the 11.2 and 7.7 \mum\, emission features are shown as contours in respectively black (at 1.42, 1.57, 1.75 and, 1.99 10$^{-6}$ Wm$^{-2}$sr$^{-1}$) and pink (at 5.54, 6.30, 6.80 and, 7.20 10$^{-6}$ Wm$^{-2}$sr$^{-1}$). The maps are orientated so N is up and E is left. The white arrow in the bottom right corners indicates the direction towards the central star. The axis labels refer to pixel numbers. The nomenclature and the FOV of the SH map are given in Fig. ~\ref{fig_slmaps_n}.} 
\label{fig_pahfit_n}
\end{figure*}
%%%%%%%%%%%%%%%%%%%%%%%%%%%%%%%%%%%%%%%%%%%%%%%%%%

We set the following conditions when applying PAHFIT.  First, we assumed no extinction is present, similar as for spline decomposition (see Section \ref{cont} for a discussion on extinction). Second, we set the contribution of star light to zero. The effective temperature of the illuminating star is 22,000 K and the resulting fits do not include starlight continuum emission when a contribution is allowed in the fitting routine. Third, we fixed the wavelengths and FWHMs for the dust features and lines (i.e. the default in PAHFIT). Fourth, we set the intensity of the H$_2$ S(7), S(6) and S(4) to zero (at 5.511, 6.109 and 8.026 \mum\, respectively). These H$_2$ lines are not discerned in the spectra. Unfortunately, the H$_2$ S(4) line is blended with the 7.7 \mum\, PAH complex. For the south map, the results from PAHFIT includes emission from this H$_2$ S(4) line when a contribution is allowed in the fitting routine. However, as mentioned, we do not discern its contribution in the spectra, even in the spatial positions where this lines is strongest. Furthermore, its spatial distribution shows emission in the S ridge (as do the H$_2$ S(5) and S(3) lines) but it exhibits additional strong emission in the north-west region of the map where no emission is seen from the H$_2$ S(5) and S(3) lines. Hence, we conclude that this is an artefact of the fitting routine and thus do not allow a contribution of the H$_2$ S(7), S(6) and S(4) lines.

%%%%%%%%%%%%%%%%%%%%%%%%%%%%%%%%%%%%%%%%%%%%%%%%%%
\begin{figure*}[tb]
    \centering
\resizebox{13.6cm}{!}{%
  \includegraphics{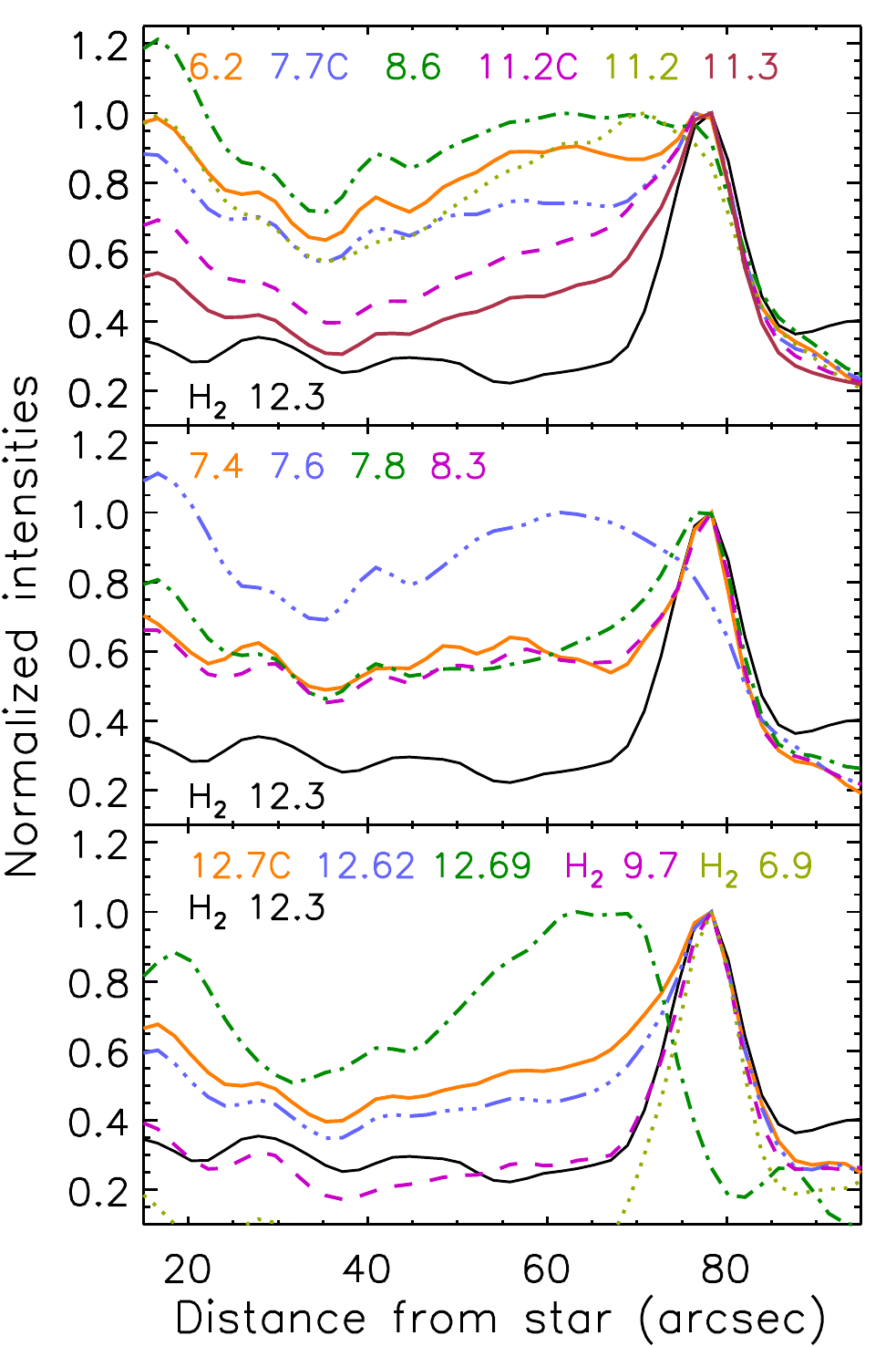}
   \includegraphics{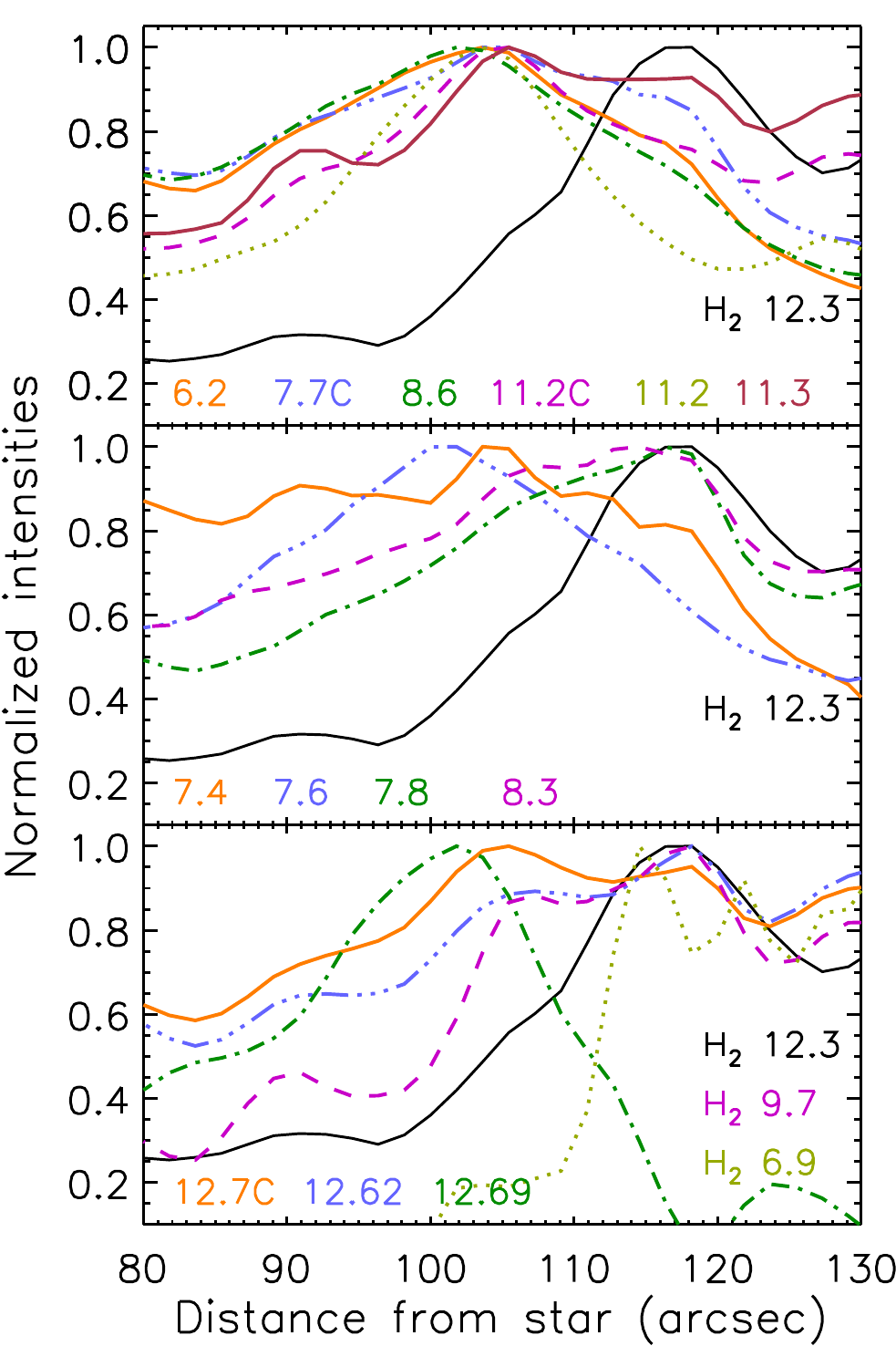}}
\caption{\label{linecuts_PAHFIT} Normalized feature/continuum intensity of the PAHFIT components along the same projected cut across the south (left) and north (right) FOV directed toward HD37093 as in Fig.~\ref{linecuts}. 'C' refers to 'complex'.}
\end{figure*}
%%%%%%%%%%%%%%%%%%%%%%%%%%%%%%%%%%%%%%%%%%%%%%%%%%

The dust continuum emission is fitted by only two and exceptionally three of the maximum allowed eight thermal dust components (with T of 135, 200 and 300 K).  Typically, either the 200K or 300K component is used in combination with the 135K component. The 300K component is used towards the south ridge and sources C and D while the remaining positions require a 200K component. This results however in quite distinct dust continuum intensities at the shorter wavelengths (see Fig. ~\ref{fig_pahfit_cont}): the continua for the spectra fitted with a 300K dust component are very close to the observed intensities at for example 5.5 \mum. In contrast, the continua for the spectra fitted with a 200K dust component have very little intensity at the shorter wavelengths. This distinction in continuum intensity largely reflects the constraints of the fitting routine rather than a physical origin. However, a smaller dust continuum intensity requires a larger contribution of the (wings of the) Drude profiles in order to fit the data. Hence, the strength of the PAH components is influenced by the fitted dust continuum, particularly for the weaker bands thus inhibiting any useful interpretation of these weaker bands. The spatial distribution and line cut of the fitting components as well as these PAH complexes are shown in Figs. \ref{fig_pahfit_s},  \ref{fig_pahfit_n},  and \ref{linecuts_PAHFIT}. A comparison with the results of the spline decomposition (see Sect. \ref{sldata} and Figs. \ref{fig_slmaps_s}, \ref{fig_slmaps_n}, \ref{fig_maps_decomp}, \ref{fig_movie}, and \ref{linecuts}) reveals the following:

\begin{itemize}
\item The spatial distribution of the 6.2 \mum\, PAH emission is very similar for both decompositions. Their small discrepancies between each other can be understood by the different spatial distribution of the LS 6.2 \mum\, PAH band and the LS 5-10 \mum\, plateau which both contribute (with different degrees) to the PAHFIT 6.2 \mum\, PAH band.

\item The morphology of the PAHFIT 7.7 \mum\, Complex is slightly distinct from that of the LS 7.7 \mum\, PAH band arising from the (partial) contribution of the GS 5-10 \mum\, plateau emission to the PAHFIT 7.7 \mum\, Complex and its components. Specifically, the PAHFIT 7.7 \mum\, Complex shows a decreased relative intensity in the horizontal filaments in the northern area of the south map and the broad, diffuse plateau north of the S - SSE ridges. For the north map, it shows an extension towards the N ridge but a decreased extension towards the west of the NW ridge. Likewise, the PAHFIT 7.6, 7.8 and 8.3 \mum\, components resemble the G7.6, G7.8 and G8.2 components respectively. There is slightly more deviation in the south map for the 7.6 and 7.8 \mum\, components: the PAHFIT 7.6 component shows decreased relative intensity in the SSE and S ridges compared to the G7.6 \mum\, emission and the PAHFIT 7.8 \mum\, emission shows a very small increased relative intensity in the S' ridge. The PAHFIT 7.4 \mum\, component follows the PAHFIT 8.3 morphology in the south map. In contrast, in the north map, it shows a unique morphology.

\item The PAHFIT 8.6 \mum\, emission spatially resembles the G8.6 \mum\, emission (GS). Their discrepancies arise from the (partial) contribution of the GS 5-10 \mum\, plateau emission. Nevertheless, the PAHFIT 8.6 \mum\, morphology shows the same trends as the LS 8.6 and GS G8.6 \mum\, bands: 1) lack of intensity in locations where the 11.2 \mum\, emission peaks (S ridge and center/north of NW ridge) and 2) it is distinct from the morphology of the 7.8 and 8.2/8.3 \mum\, components while being similar to the spatial distribution of the 7.6 \mum\, component.

\item Similar to the 6.2 \mum\, PAH band, the LS 11.2 \mum\, band and the PAHFIT 11.2 \mum\, Complex exhibit a similar spatial distribution.  The PAHFIT 11.2 \mum\, Complex includes part of the LS 10-15 \mum\, plateau. The latter's spatial morphology is distinct from that of the LS 11.2 \mum\, PAH band in the north map resulting in small discrepancies between the LS and PAHFIT 11.2 \mum\, band/complex. Within PAHFIT, the 11.2 \mum\, Complex is composed of bands positioned at 11.2 and 11.3 \mum\, which exhibit distinct spatial morphology. Specifically, the 11.3 \mum\, band emission is most similar to the 11.2 \mum\, Complex emission (it contributes the largest fraction to the complex). On the other hand, the morphology of the 11.2 \mum\, band deviates from that of the 11.2 \mum\, Complex and is an intermediate step in a progression towards that of the 7.6 and 8.6 \mum\, bands.

\item The morphology of the PAHFIT 12.7 \mum\, Complex emission is similar to that of the PAHFIT 11.2 \mum\, Complex emission and that of the LS 11.2 \mum\, emission while it is distinct from that of the LS 12.7 \mum\, emission. This distinction arises due to the significant contribution of the LS 10-15 \mum\, plateau to the PAHFIT 12.7 \mum\, Complex. Similar to the 11.2 \mum\, Complex, the two components (positioned at 12.62 and 12.69 \mum) accounting for the 12.7 \mum\, Complex have distinct spatial distributions. Given that the 12.62 \mum\, component contributes on average 11 $\pm$ 6\% and 16 $\pm$ 2\% to the 12.7 \mum\, Complex for the south and north map respectively, the 12.7 \mum\, Complex emission resembles closely the 12.62 \mum\, emission. In contrast, the 12.69 \mum\, emission shows a spatial distribution between that of the LS 12.7 \mum\, emission and that of the LS 11.0 \mum\, emission.
\end{itemize}

%%%%%%%%%%%%%%
\begin{table*}[tb!]
\scriptsize
\caption{\label{fit_parameters} Parameters of the fits to the observed SL (top) and SH (bottom) correlations and their Pearson's correlation coefficients. }
\begin{center}
\begin{tabular}{ll|cccr@{\,}c@{\,}lr@{\,}c@{\,}lr@{\,}c@{\,}l}
 \multicolumn{2}{c}{PAH intensity ratios$^1$} & \multicolumn{3}{c}{weighted correlation coefficient} & \multicolumn{3}{c}{A$^2$} & \multicolumn{3}{c}{B$^2$} & \multicolumn{3}{c}{C$^3$}\\
\multicolumn{2}{c}{y vs. x} & N+S  & N &  S & & &\\
       \hline 
       \hline
 & &  \\[-5pt]
6.2/11.2 & G8.6/11.2 &    0.9721 &    0.9755 &    0.9749 &    0.2583 & $\pm$ &    0.0160 &    1.5154 & $\pm$ &    0.0147 &    1.7476 & $\pm$ &    0.0045\\
6.2/11.2 & 11.0/11.2 &    0.9376 &    0.9320 &    0.9390 &    0.7682 & $\pm$ &    0.0205 &   12.8764 & $\pm$ &    0.2044 &   20.3914 & $\pm$ &    0.1560\\
6.2/11.2 & 12.7/11.2 &    0.8884 &    0.8462 &    0.9164 &   -0.5619 & $\pm$ &    0.0489 &    7.7560 & $\pm$ &    0.1480 &    6.0840 & $\pm$ &    0.0230\\
7.7 (LS)/11.2 & 6.2/11.2 &    0.9713 &    0.9828 &    0.9729 &   -0.4170 & $\pm$ &    0.0357 &    2.0201 & $\pm$ &    0.0191 &    1.8020 & $\pm$ &    0.0042\\
7.7 (LS)/11.2 & 8.6 (LS)/11.2 &    0.9363 &    0.9346 &    0.9392 &    1.1322 & $\pm$ &    0.0351 &    4.2934 & $\pm$ &    0.0636 &    6.2698 & $\pm$ &    0.0389\\
7.7 (LS)/11.2 & 11.0/11.2 &    0.9216 &    0.9328 &    0.9208 &    1.0326 & $\pm$ &    0.0495 &   26.9987 & $\pm$ &    0.4939 &   37.0259 & $\pm$ &    0.2652\\
7.7 (LS)/11.2 & 12.7/11.2 &    0.8809 &    0.8218 &    0.9139 &   -1.6614 & $\pm$ &    0.1059 &   15.9992 & $\pm$ &    0.3206 &   11.0683 & $\pm$ &    0.0512\\
G7.6/11.2 & G8.6/11.2 &    0.9872 &    0.9887 &    0.9870 &    0.1574 & $\pm$ &    0.0189 &    3.0693 & $\pm$ &    0.0174 &    3.2105 & $\pm$ &    0.0045\\
G7.6/11.2 & 11.0/11.2 &    0.9317 &    0.9342 &    0.9322 &    1.1601 & $\pm$ &    0.0425 &   26.3498 & $\pm$ &    0.4240 &   37.6225 & $\pm$ &    0.2621\\
G7.6/11.2 & 12.7/11.2 &    0.9197 &    0.8647 &    0.9449 &   -0.4904 & $\pm$ &    0.0251 &    4.9616 & $\pm$ &    0.0759 &    3.5100 & $\pm$ &    0.0143\\
(G7.6+G7.8)/11.2 & 6.2/11.2 &    0.9391 &    0.9539 &    0.9447 &    1.1117 & $\pm$ &    0.0466 &    1.7208 & $\pm$ &    0.0246 &    2.3025 & $\pm$ &    0.0077\\
(G7.6+G7.8)/11.2 & G86/11.2 &    0.9502 &    0.9450 &    0.9525 &    1.5320 & $\pm$ &    0.0369 &    2.6292 & $\pm$ &    0.0337 &    4.0092 & $\pm$ &    0.0170\\
(G7.6+G7.8)/11.2 & 11.0/11.2 &    0.8751 &    0.8676 &    0.8800 &    2.2136 & $\pm$ &    0.0578 &   24.3539 & $\pm$ &    0.5796 &   46.3829 & $\pm$ &    0.4653\\
(G7.6+G7.8)/11.2 & 12.7/11.2 &    0.8363 &    0.7307 &    0.8753 &   -0.1693 & $\pm$ &    0.1147 &   14.3072 & $\pm$ &    0.3477 &   13.8005 & $\pm$ &    0.0536\\
G8.2/11.2 & G7.8/11.2 &    0.7462 &    0.8220 &    0.7519 &    0.1082 & $\pm$ &    0.0107 &    0.3508 & $\pm$ &    0.0111 &    0.4630 & $\pm$ &    0.0022\\
G8.6/11.2 & 11.0/11.2 &    0.9465 &    0.9507 &    0.9468 &    0.3393 & $\pm$ &    0.0125 &    8.4567 & $\pm$ &    0.1245 &   11.7402 & $\pm$ &    0.0761\\
G8.6/11.2 & 12.7/11.2 &    0.9197 &    0.8647 &    0.9449 &   -0.4904 & $\pm$ &    0.0251 &    4.9616 & $\pm$ &    0.0759 &    3.5100 & $\pm$ &    0.0143\\
8.6 (LS)/11.2 & 6.2/11.2 &    0.9370 &    0.9369 &    0.9391 &   -0.3666 & $\pm$ &    0.0132 &    0.4735 & $\pm$ &    0.0069 &    0.2883 & $\pm$ &    0.0020\\
8.6 (LS)/11.2 & 11.0/11.2 &    0.9507 &    0.9518 &    0.9519 &    0.0055 & $\pm$ &    0.0087 &    5.9965 & $\pm$ &    0.0862 &    6.0484 & $\pm$ &    0.0290\\
8.6 (LS)/11.2 & 12.7/11.2 &    0.9411 &    0.9145 &    0.9588 &   -0.5709 & $\pm$ &    0.0145 &    3.4803 & $\pm$ &    0.0439 &    1.8109 & $\pm$ &    0.0135\\
12.7/11.2 & 11.0/11.2 &    0.9025 &    0.8622 &    0.9363 &    0.1676 & $\pm$ &    0.0031 &    1.7087 & $\pm$ &    0.0306 &    3.3512 & $\pm$ &    0.0305\\
plat.7/11.2 & 6.2/11.2 &    0.8702 &    0.9107 &    0.8608 &    2.5193 & $\pm$ &    0.1218 &    2.6167 & $\pm$ &    0.0638 &    3.9329 & $\pm$ &    0.0192\\
plat.7/11.2 & 7.7 (LS)/11.2 &    0.8343 &    0.9153 &    0.8210 &    3.1482 & $\pm$ &    0.1156 &    1.2678 & $\pm$ &    0.0335 &    2.1812 & $\pm$ &    0.0129\\
plat.7/11.2 & plat11/11.2 &   -0.1112 &    0.3343 &   -0.1735 &  -89.0386 & $\pm$ &   41.8972 &   60.5135 & $\pm$ &   26.1939 &    4.8298 & $\pm$ &    0.0389\\
7.7 (LS)/6.2 & 8.6 (LS)/6.2 &    0.3374 &   -0.1060 &    0.3976 &    0.4450 & $\pm$ &    0.1193 &    4.8382 & $\pm$ &    0.4239 &    6.4170 & $\pm$ &    0.0434\\
(G7.6+G7.8)/6.2 & G86/6.2 &    0.4472 &    0.5021 &    0.4378 &   -1.3032 & $\pm$ &    0.2145 &    6.2744 & $\pm$ &    0.3716 &    4.0188 & $\pm$ &    0.0093\\
G8.2/G8.6 & G7.8/G8.6 &    0.8326 &    0.9646 &    0.8165 &    0.1398 & $\pm$ &    0.0065 &    0.2944 & $\pm$ &    0.0080 &    0.4646 & $\pm$ &    0.0021\\
plat.7/G8.6 & 7.7 (LS)/G8.6 &    0.4106 &    0.6013 &    0.4023 &  -40.7359 & $\pm$ &    3.6317 &   15.0708 & $\pm$ &    1.1572 &    2.0776 & $\pm$ &    0.0108\\
11.2/G8.6 & 12.7/G8.6 &    0.7307 &    0.8088 &    0.7544 &   -1.9183 & $\pm$ &    0.0959 &    9.6982 & $\pm$ &    0.3385 &    3.0312 & $\pm$ &    0.0231\\
plat. 7/plat.11 & 6.2/plat.11 &    0.9469 &    0.9600 &    0.9470 &    1.0956 & $\pm$ &    0.0493 &    3.0234 & $\pm$ &    0.0456 &    4.0115 & $\pm$ &    0.0207\\
plat. 7/plat.11 & 7.7 (LS)/plat.11 &    0.9174 &    0.9521 &    0.9153 &    1.2655 & $\pm$ &    0.0566 &    1.5907 & $\pm$ &    0.0289 &    2.2246 & $\pm$ &    0.0137\\
      \hline 
  & &  \\[-5pt]
12.7/11.2 & 11.0/11.2 &    0.8598 &    0.9026 &    0.9142 &    0.0969 & $\pm$ &    0.0067 &    1.8465 & $\pm$ &    0.0605 &    2.7079 & $\pm$ &    0.0279\\
13.5/11.2 & 11.0/11.2 &    0.8129 &    0.9064 &    0.7955 &    0.0179 & $\pm$ &    0.0011 &    0.1804 & $\pm$ &    0.0100 &    0.3340 & $\pm$ &    0.0045\\
13.5/11.2 & 12.7/11.2 &    0.7474 &    0.7725 &    0.7658 &    0.0096 & $\pm$ &    0.0018 &    0.0935 & $\pm$ &    0.0060 &    0.1250 & $\pm$ &    0.0013\\
13.5/11.2 & 16.4/11.2 &    0.7874 &    0.8264 &    0.8071 &   -0.0017 & $\pm$ &    0.0021 &    0.5920 & $\pm$ &    0.0309 &    0.5670 & $\pm$ &    0.0052\\
13.5/11.2 & 17.4/11.2 &    0.5938 &    0.4945 &    0.6156 &    0.0165 & $\pm$ &    0.0015 &    1.3549 & $\pm$ &    0.0957 &    2.4152 & $\pm$ &    0.0534\\
16.4/11.2 & 11.0/11.2 &    0.9002 &    0.9228 &    0.9534 &    0.0315 & $\pm$ &    0.0010 &    0.3191 & $\pm$ &    0.0086 &    0.5915 & $\pm$ &    0.0062\\
16.4/11.2 & 12.7/11.2 &    0.9133 &    0.8818 &    0.9192 &    0.0147 & $\pm$ &    0.0015 &    0.1730 & $\pm$ &    0.0050 &    0.2212 & $\pm$ &    0.0012\\
17.4/11.2 & 11.0/11.2 &    0.9167 &    0.7917 &    0.9317 &    0.0006 & $\pm$ &    0.0005 &    0.1412 & $\pm$ &    0.0042 &    0.1464 & $\pm$ &    0.0012\\
17.4/11.2 & 12.7/11.2 &    0.7676 &    0.7181 &    0.8211 &   -0.0043 & $\pm$ &    0.0011 &    0.0677 & $\pm$ &    0.0037 &    0.0539 & $\pm$ &    0.0007\\
17.4/11.2 & 16.4/11.2 &    0.7982 &    0.6805 &    0.8398 &   -0.0101 & $\pm$ &    0.0013 &    0.3918 & $\pm$ &    0.0188 &    0.2445 & $\pm$ &    0.0034\\
diffuse 17.4/11.2 & 11.0/11.2 &    0.6812 &    0.8953 &    0.5791 &    0.0035 & $\pm$ &    0.0012 &    0.0621 & $\pm$ &    0.0096 &    0.0895 & $\pm$ &    0.0021\\
diffuse 17.4/11.2 & 12.7/11.2 &    0.7397 &    0.9237 &    0.7000 &   -0.0052 & $\pm$ &    0.0020 &    0.0483 & $\pm$ &    0.0060 &    0.0332 & $\pm$ &    0.0007\\
diffuse 17.4/11.2 & 16.4/11.2 &    0.5825 &    0.8311 &    0.5471 &   -0.0019 & $\pm$ &    0.0023 &    0.1817 & $\pm$ &    0.0317 &    0.1560 & $\pm$ &    0.0036\\
17.8/11.2 & 13.5/11.2 &    0.8143 &    0.0000 &    0.8143 &    0.0002 & $\pm$ &    0.0010 &    0.3425 & $\pm$ &    0.0269 &    0.3476 & $\pm$ &    0.0048\\
17.8/11.2 & 16.4/11.2 &    0.8210 &    0.0000 &    0.8210 &   -0.0007 & $\pm$ &    0.0009 &    0.2072 & $\pm$ &    0.0139 &    0.1967 & $\pm$ &    0.0024\\
19.0/11.2 & 11.0/11.2 &    0.6486 &   -0.1943 &    0.6552 &   -0.0009 & $\pm$ &    0.0022 &    0.1647 & $\pm$ &    0.0198 &    0.1569 & $\pm$ &    0.0058\\
19.0/11.2 & 12.7/11.2 &    0.3716 &   -0.5228 &    0.4091 &    0.0009 & $\pm$ &    0.0039 &    0.0523 & $\pm$ &    0.0128 &    0.0554 & $\pm$ &    0.0025\\
19.0/11.2 & 16.4/11.2 &    0.5476 &   -0.2527 &    0.5702 &   -0.0117 & $\pm$ &    0.0040 &    0.4295 & $\pm$ &    0.0594 &    0.2577 & $\pm$ &    0.0108\\
11.2/12.7 & 13.5/12.7 &    0.3424 &    0.0000 &    0.3424 &   -6.9288 & $\pm$ &    1.5988 &   82.9106 & $\pm$ &   12.9525 &   27.3724 & $\pm$ &    0.4232\\
15.8/12.7 & 11.2/12.7 &    0.8878 &    0.0000 &    0.8878 &   -0.0200 & $\pm$ &    0.0039 &    0.0231 & $\pm$ &    0.0012 &    0.0173 & $\pm$ &    0.0003\\
16.4/12.7 & 11.2/12.7 &    0.6118 &    0.6826 &    0.6001 &    0.1714 & $\pm$ &    0.0050 &    0.0151 & $\pm$ &    0.0015 &    0.0659 & $\pm$ &    0.0008\\
16.4/12.7 & plat.11/12.7 &    0.7017 &    0.7446 &    0.6987 &    0.1703 & $\pm$ &    0.0038 &    0.0095 & $\pm$ &    0.0007 &    0.0401 & $\pm$ &    0.0005\\
16.4/12.7 & plat.17/12.7 &    0.6980 &    0.6305 &    0.7156 &    0.1445 & $\pm$ &    0.0055 &    0.0727 & $\pm$ &    0.0052 &    0.2079 & $\pm$ &    0.0018\\
plat.11/12.7 & 11.2/12.7 &    0.8862 &    0.8865 &    0.9012 &   -0.7051 & $\pm$ &    0.2186 &    1.8476 & $\pm$ &    0.0703 &    1.6229 & $\pm$ &    0.0114\\
plat.11/12.7 & plat.17/12.7 &    0.8794 &    0.8744 &    0.8866 &   -3.2396 & $\pm$ &    0.2889 &    8.1680 & $\pm$ &    0.2745 &    5.1377 & $\pm$ &    0.0391\\
plat.17/12.7 & 11.2/12.7 &    0.8785 &    0.9253 &    0.8717 &    0.2831 & $\pm$ &    0.0266 &    0.2345 & $\pm$ &    0.0081 &    0.3192 & $\pm$ &    0.0020\\
12.7/16.4 & 11.2/16.4 &   -0.2435 &   -0.2654 &   -0.2428 &    5.1265 & $\pm$ &    0.1473 &   -0.0468 & $\pm$ &    0.0097 &    0.3068 & $\pm$ &    0.0041\\
12.7/15-18plat. & 11.2/15-18plat. &    0.0481 &    0.0714 &    0.0490   &  &  &  &  &  &  &  & \\
16.4/sum(15-20) & 12.7/11.2 &    0.7424 &    0.6412 &    0.7663 &    0.0961 & $\pm$ &    0.0033 &    0.1955 & $\pm$ &    0.0107 &    0.5022 & $\pm$ &    0.0047\\
\hline
\end{tabular} 
\end{center}
$^1$ (LS) refers to the PAH band strength when using the local spline continuum, plat. 7 to the 5--10 \mum\, plateau, plat. 11 to the 10-15 \mum\, plateau, 15-18 plat. to the\\
15-18 \mum\ plateau and diffuse 17.4 to the estimated fraction of the 17.4 \mum\, band attributed to PAHs (i.e. we subtracted the C$_{60}$ component which is estimated to \\
be 0.48 * C$_{60}$ at 19 \mum\, and assumes a
constant ratio between the 17.4 and 18.9\mum\, C$_{60}$ emission).  \\
$^2$ Parameters of the fit: y = A + Bx\\
$^3$ Slope of the fit through (0,0) 
\end{table*}
%%%%%%%%%%%%%%

\section{Fitting results}
\label{fittingparameters}
The parameters of the fits to the observed correlations and their Pearson's correlation coefficients are listed in Table \ref{fit_parameters}.

\section{PAH ratio maps}
\label{ratiomaps}
\begin{figure*}[tbp]
\centering
\resizebox{16cm}{!}{%
  \includegraphics{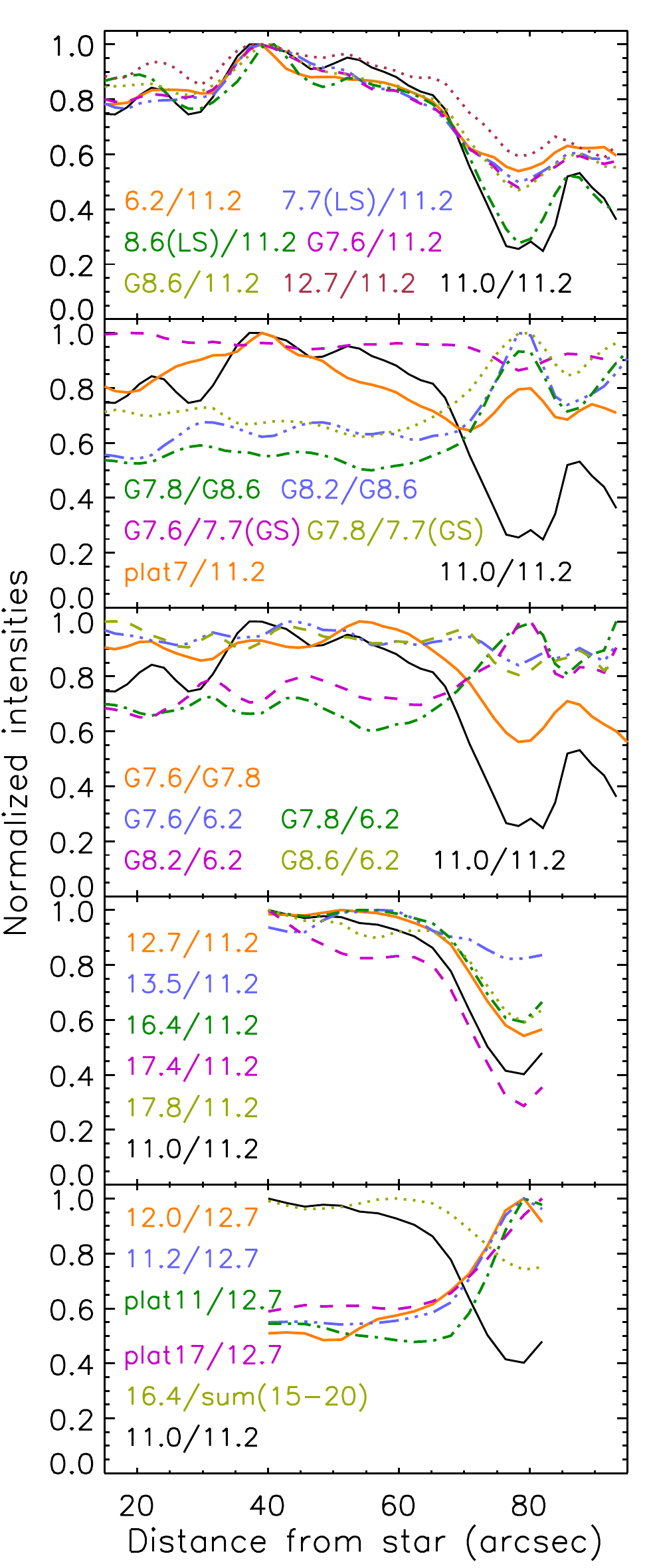}
    \includegraphics{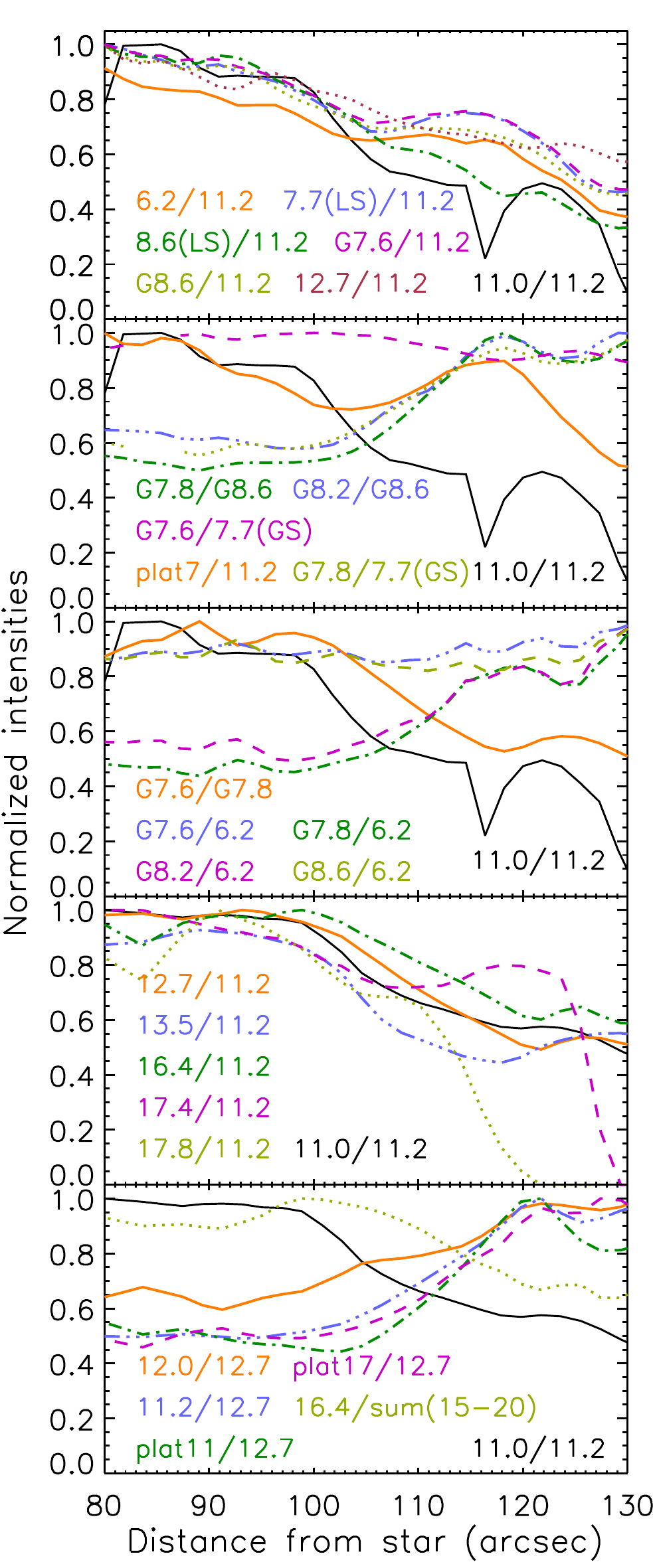}}
\caption{\label{linecuts_ratios} Normalized features intensity ratios along a projected cut across the south (left) and north (right) FOV directed toward HD37093. The cut is shown in Figs~\ref{fig_shmaps_s} and \ref{fig_shmaps_n} and is the same for SL and SH data. The top three panels show the SL data and the bottom two the SH data. }
\end{figure*}

\begin{figure*}[tb]
    \centering
\resizebox{13.6cm}{!}{%
  \includegraphics[angle=266.4]{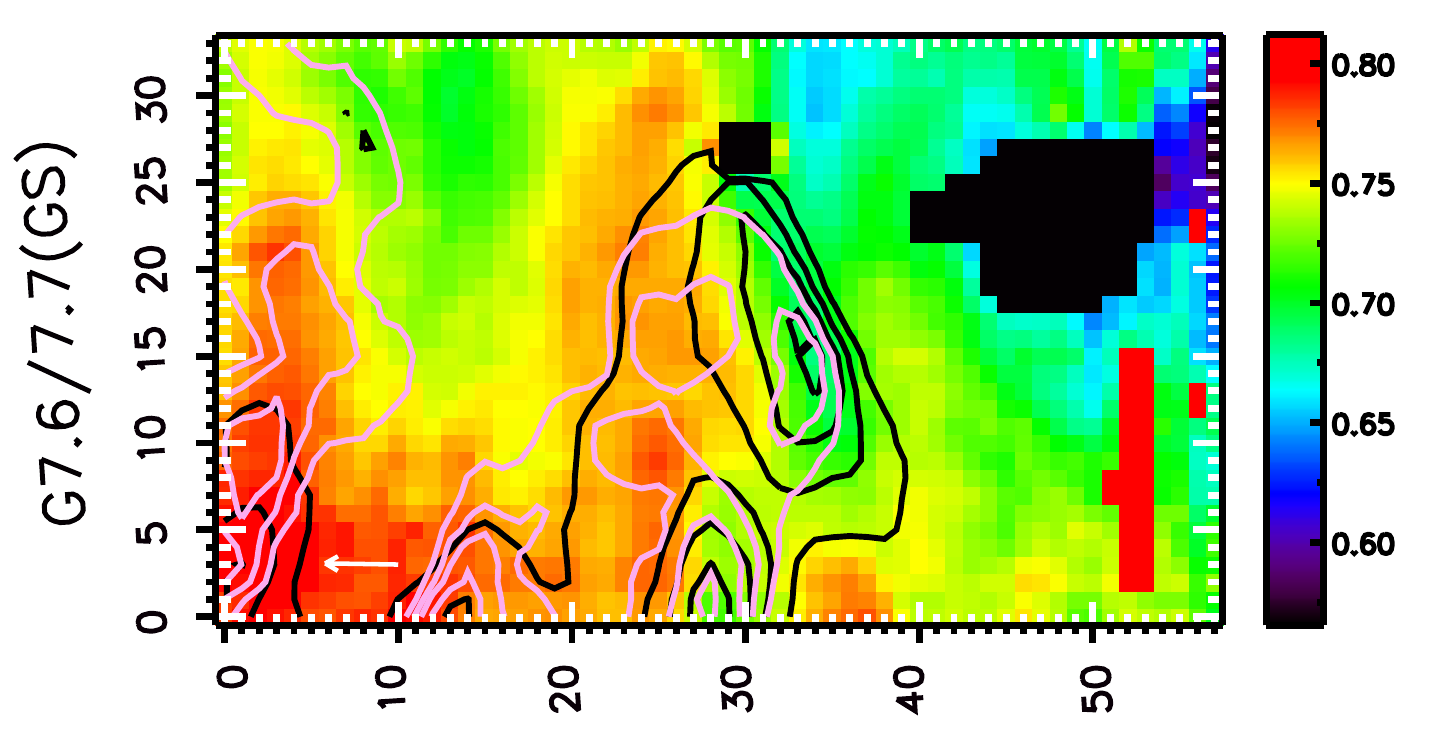}
  \includegraphics[angle=266.4]{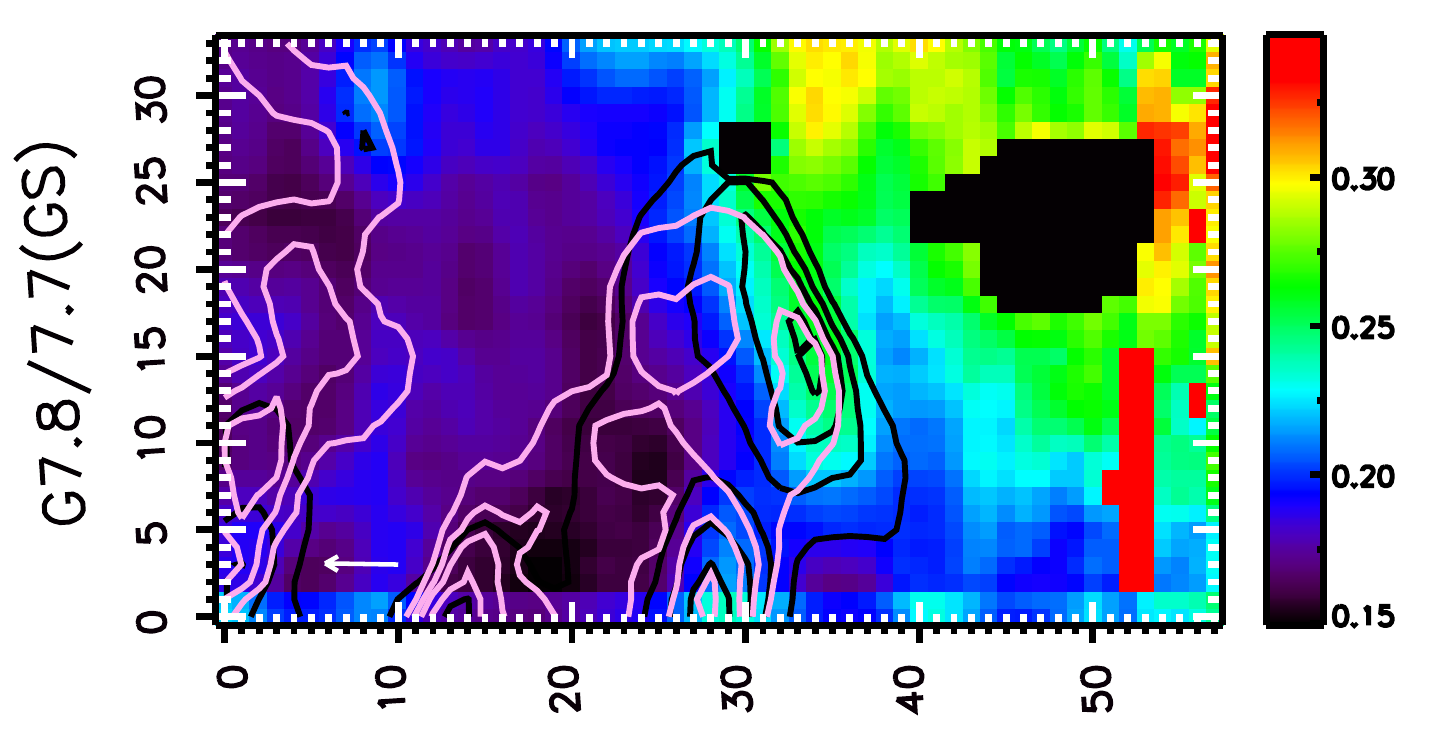}
    \includegraphics[angle=266.4]{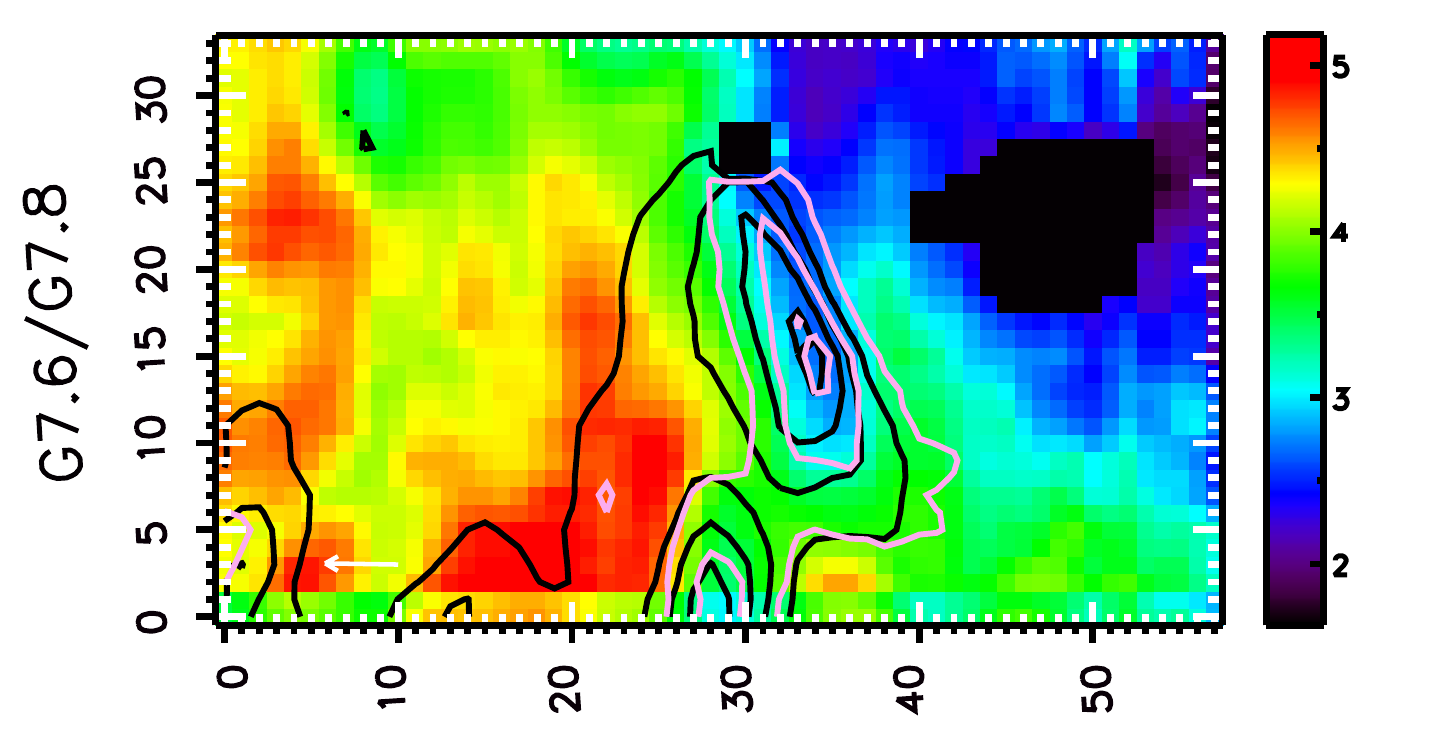}
    \includegraphics[angle=266.4]{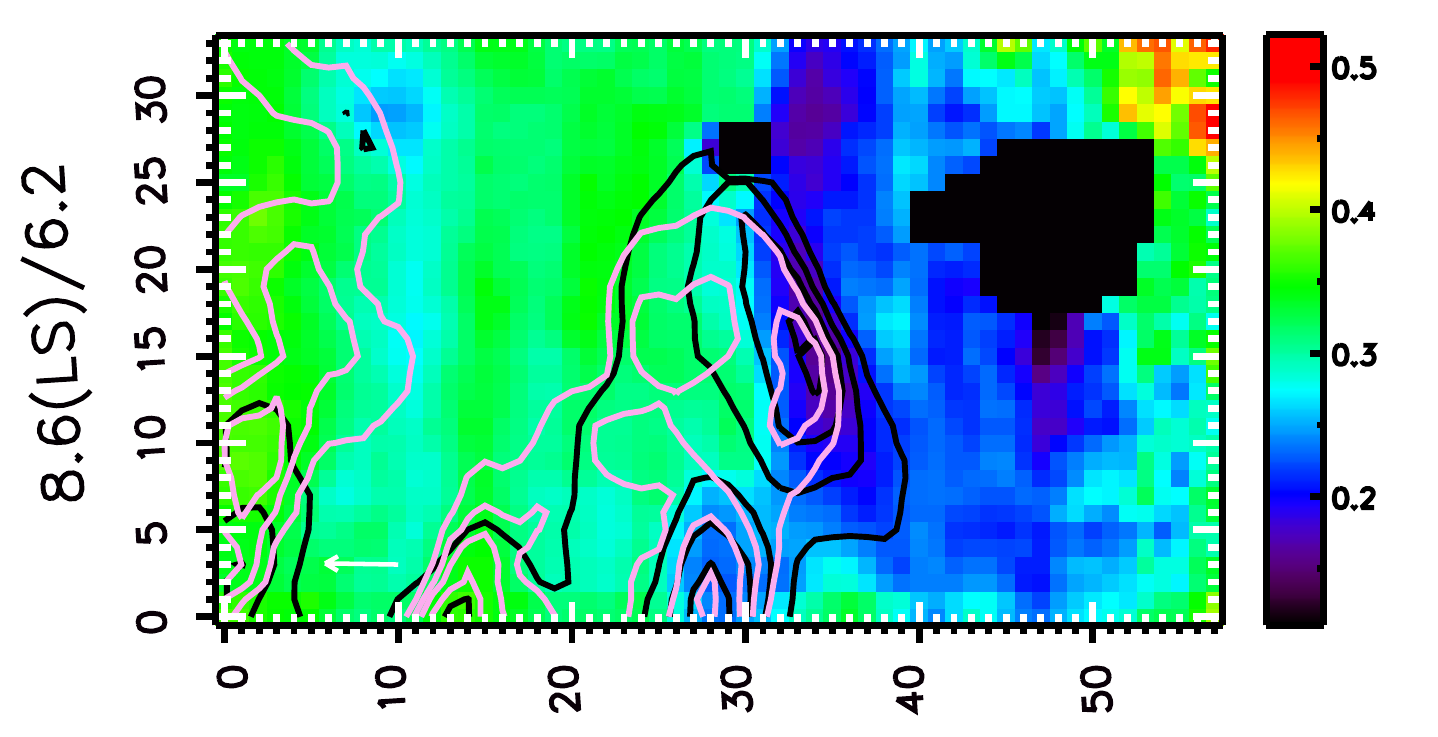}}
\resizebox{13.6cm}{!}{%
  \includegraphics[angle=266.4]{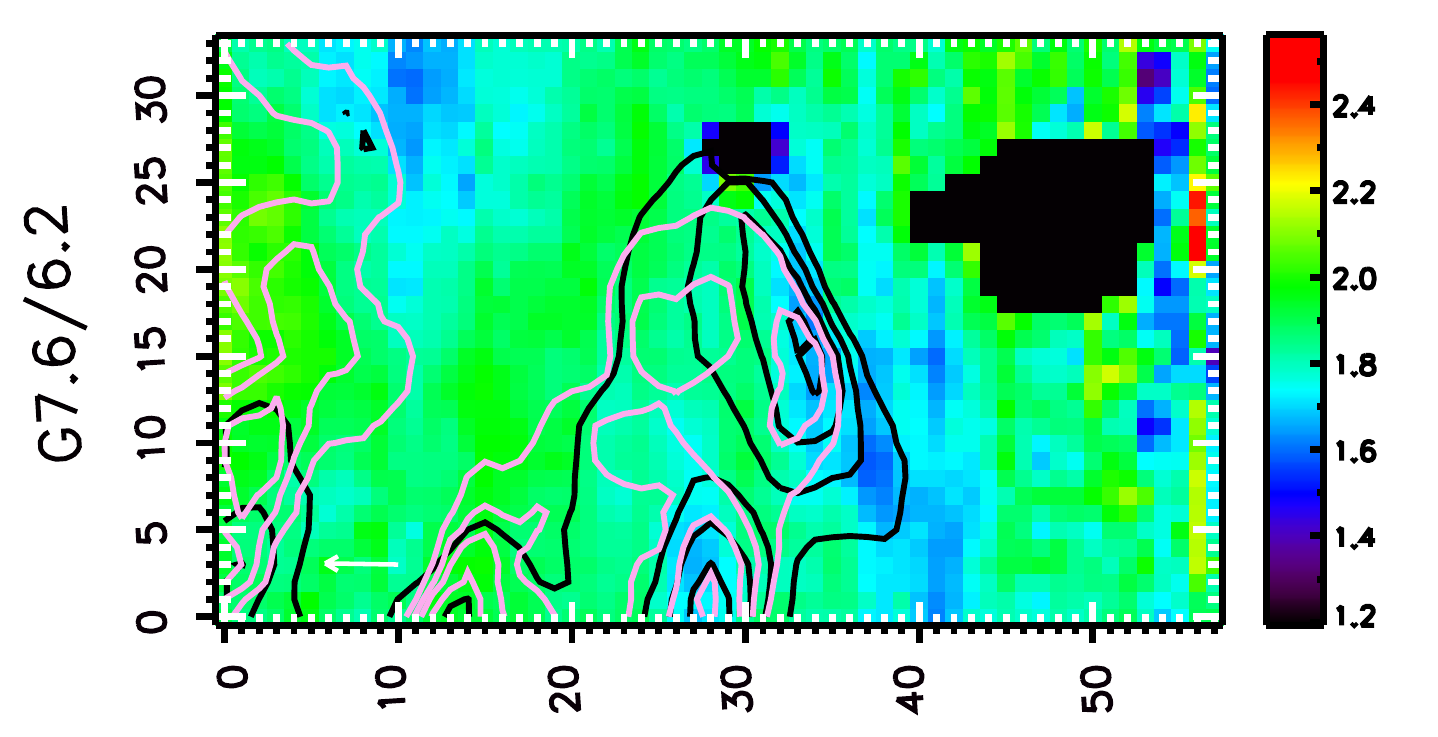}
  \includegraphics[angle=266.4]{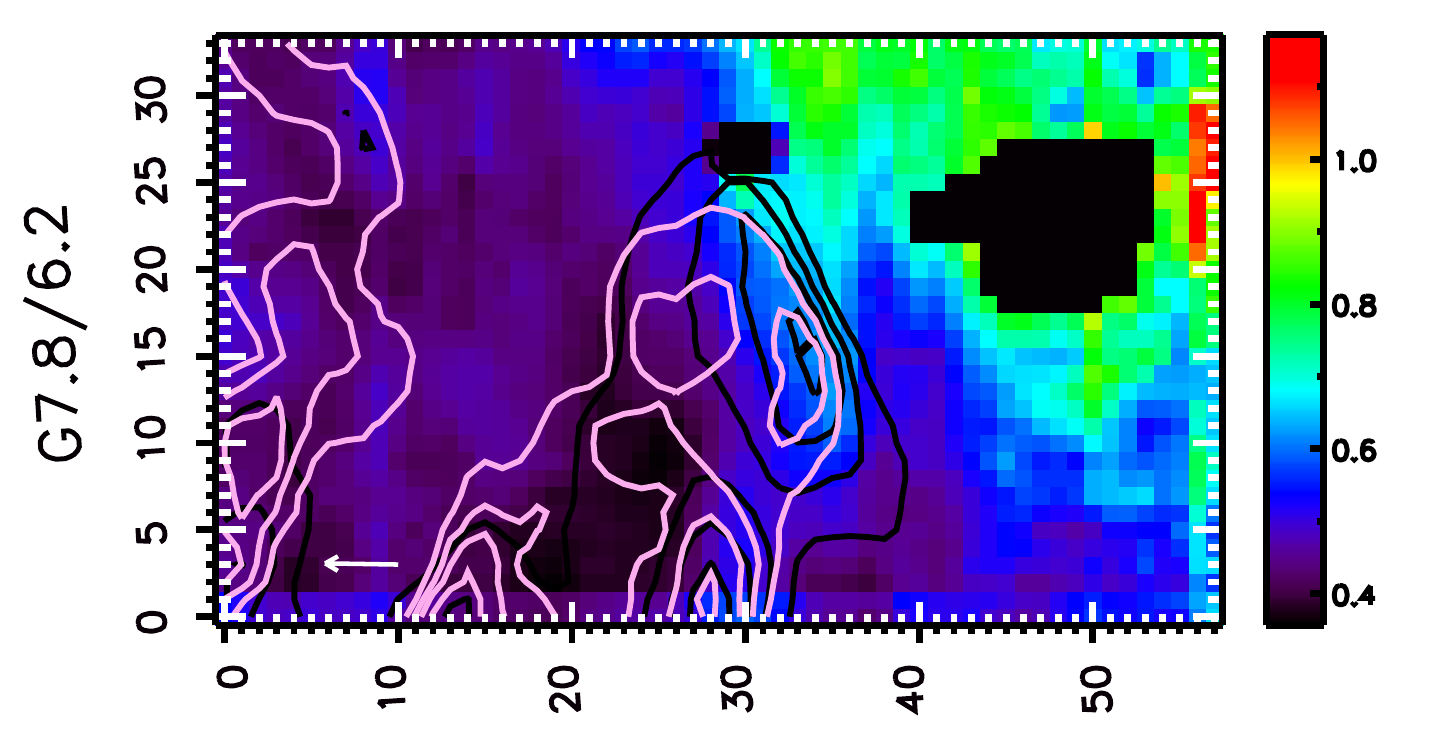}
    \includegraphics[angle=266.4]{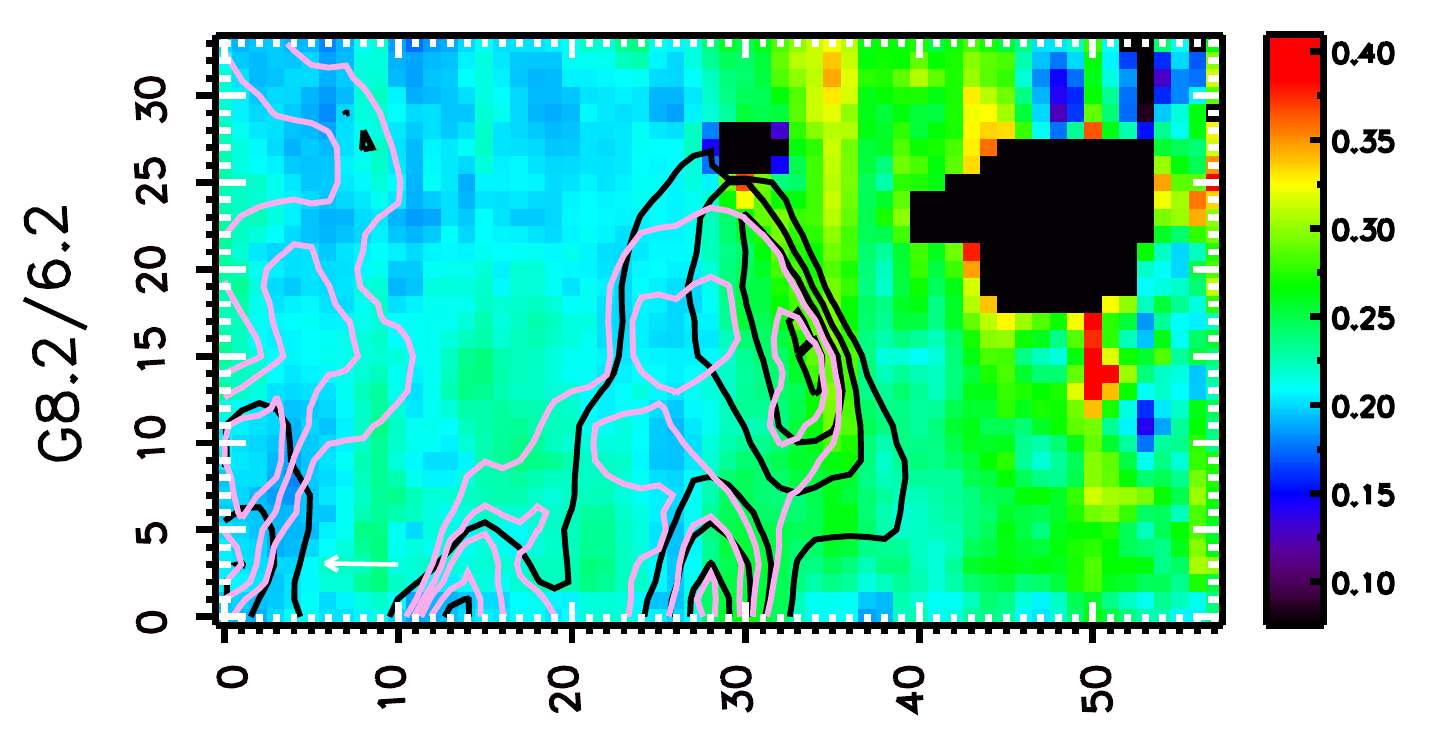}
    \includegraphics[angle=266.4]{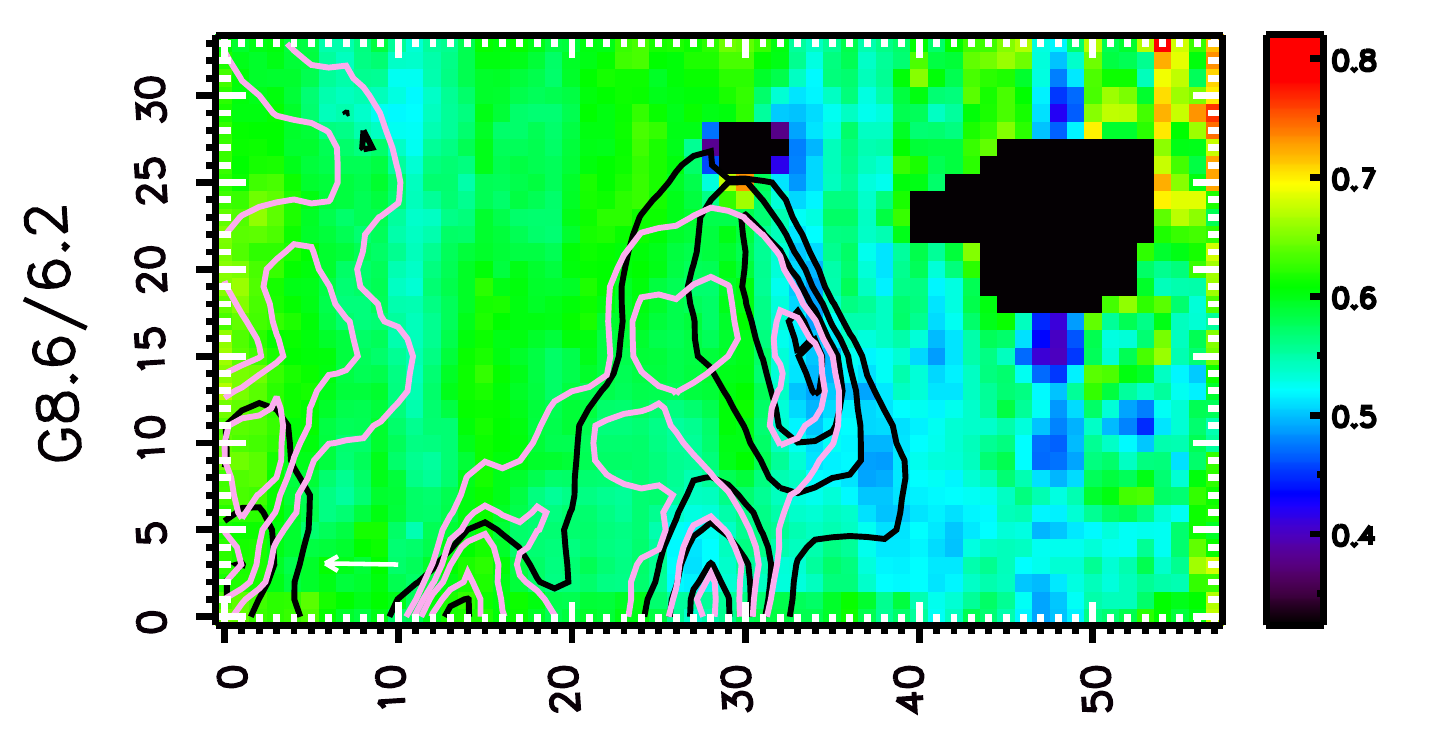}}
\resizebox{14.4cm}{!}{%
   \includegraphics[angle=274.1]{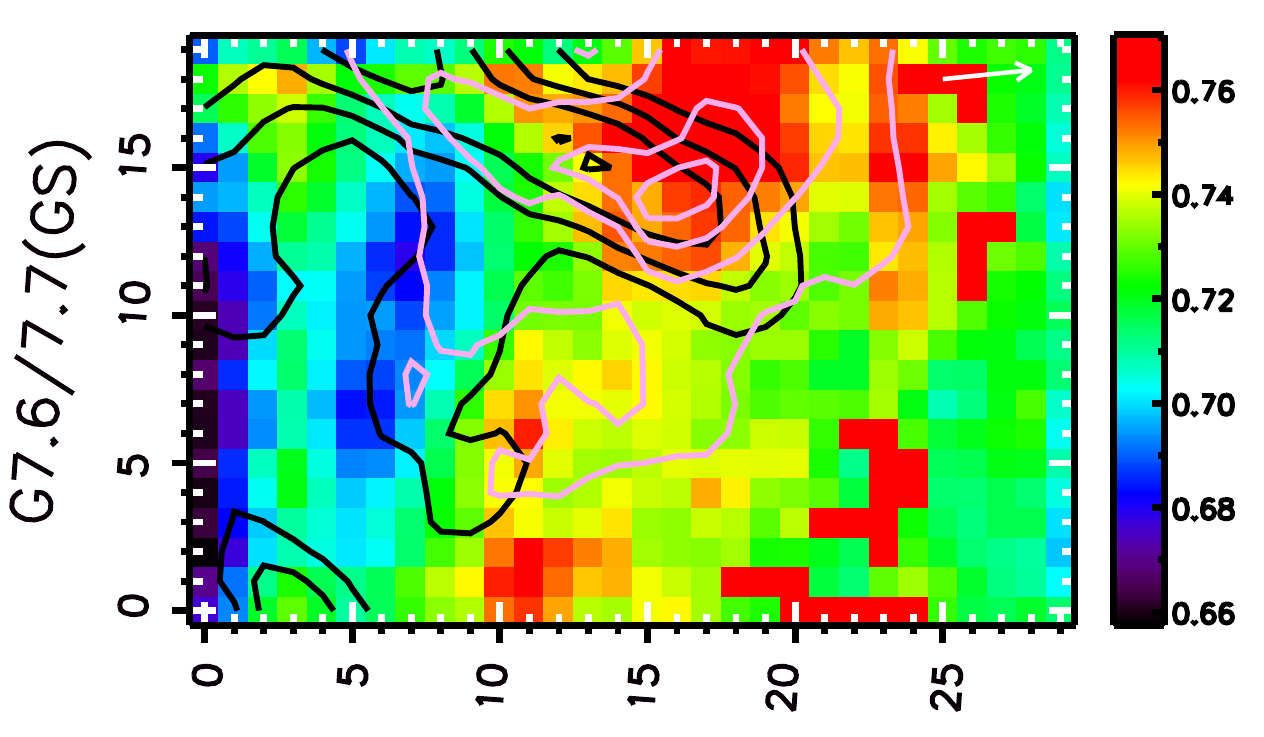}
  \includegraphics[angle=274.1]{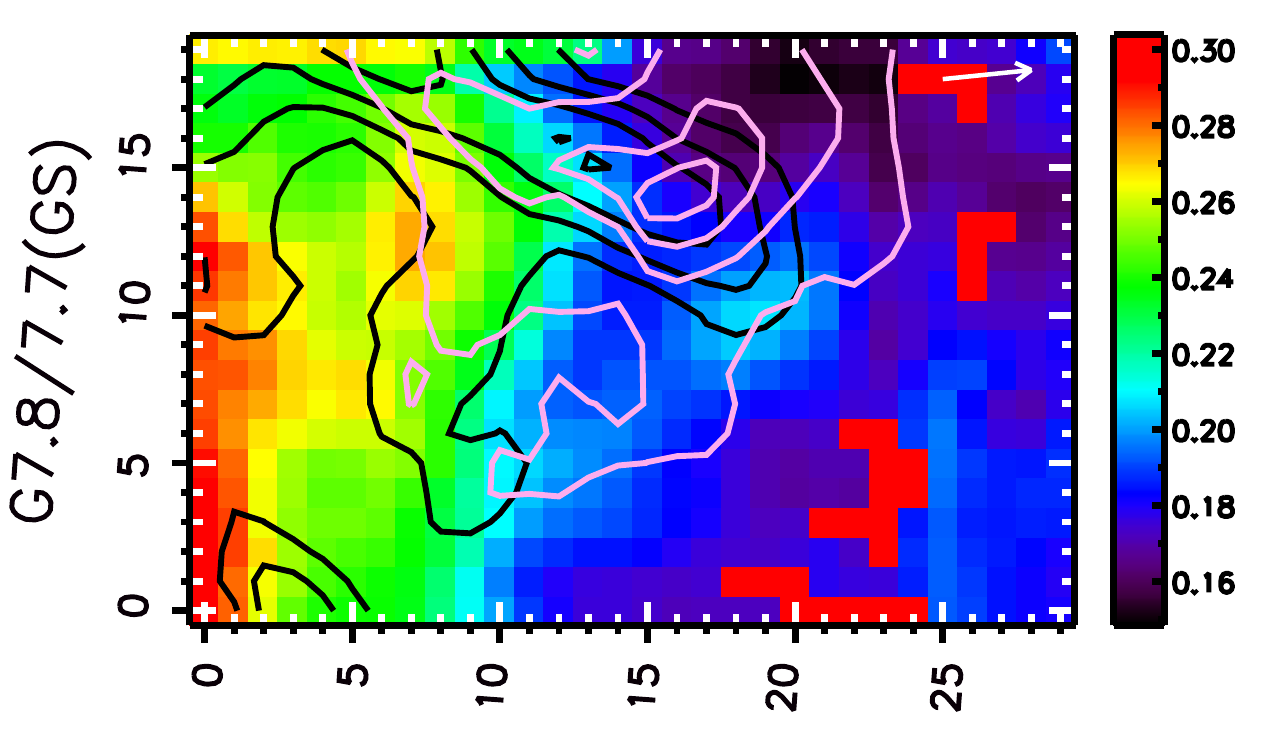}
     \includegraphics[angle=274.1]{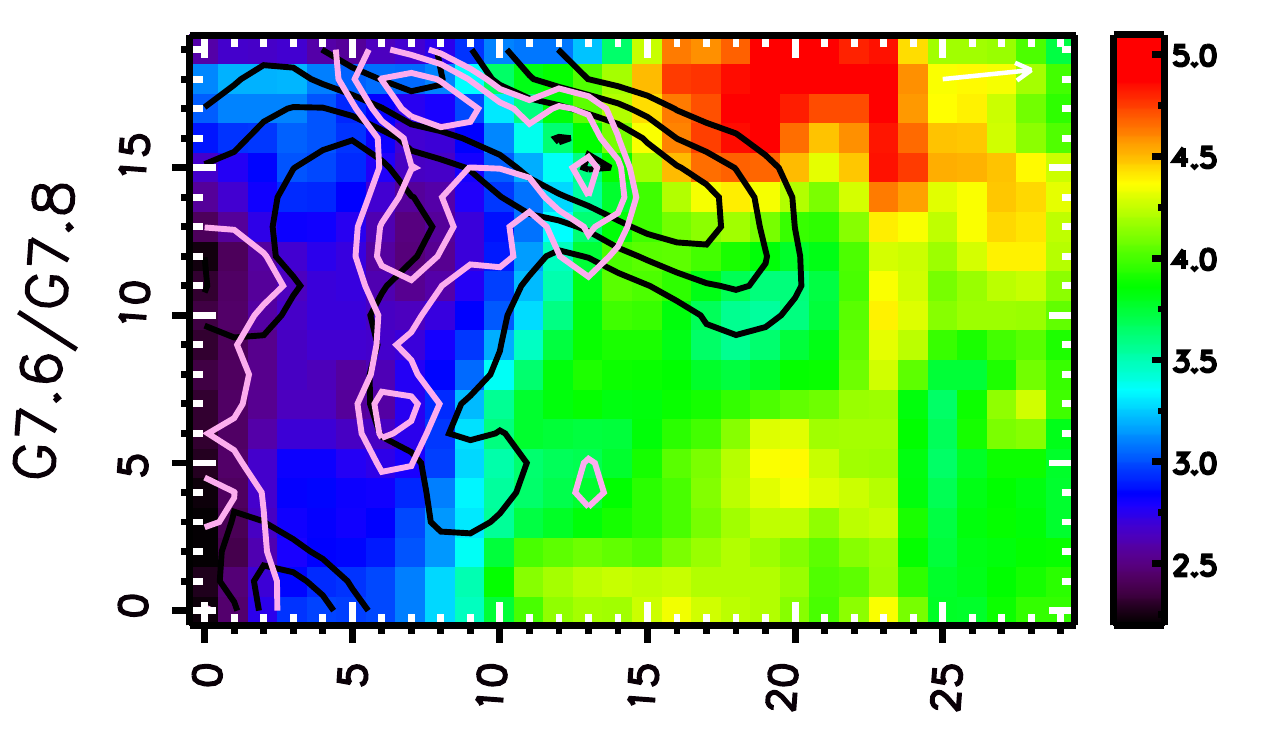}
      \includegraphics[angle=274.1]{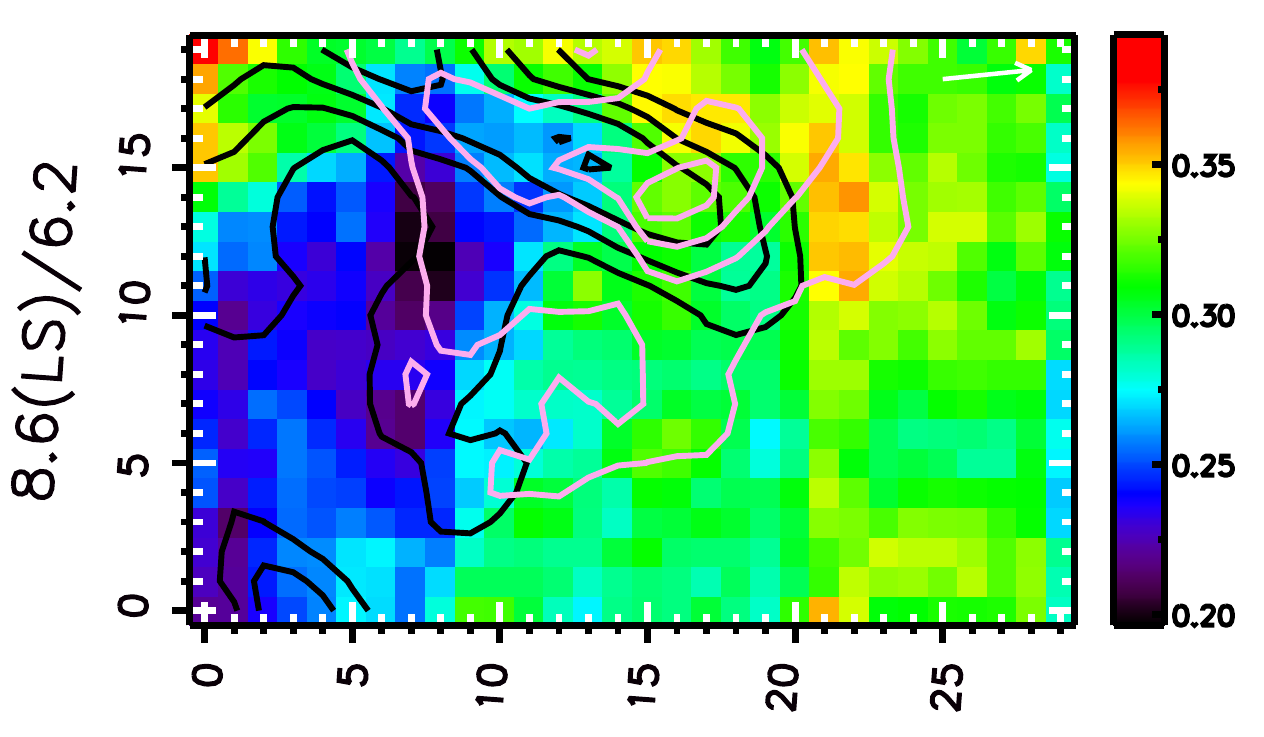}}
\resizebox{14.4cm}{!}{%
   \includegraphics[angle=274.1]{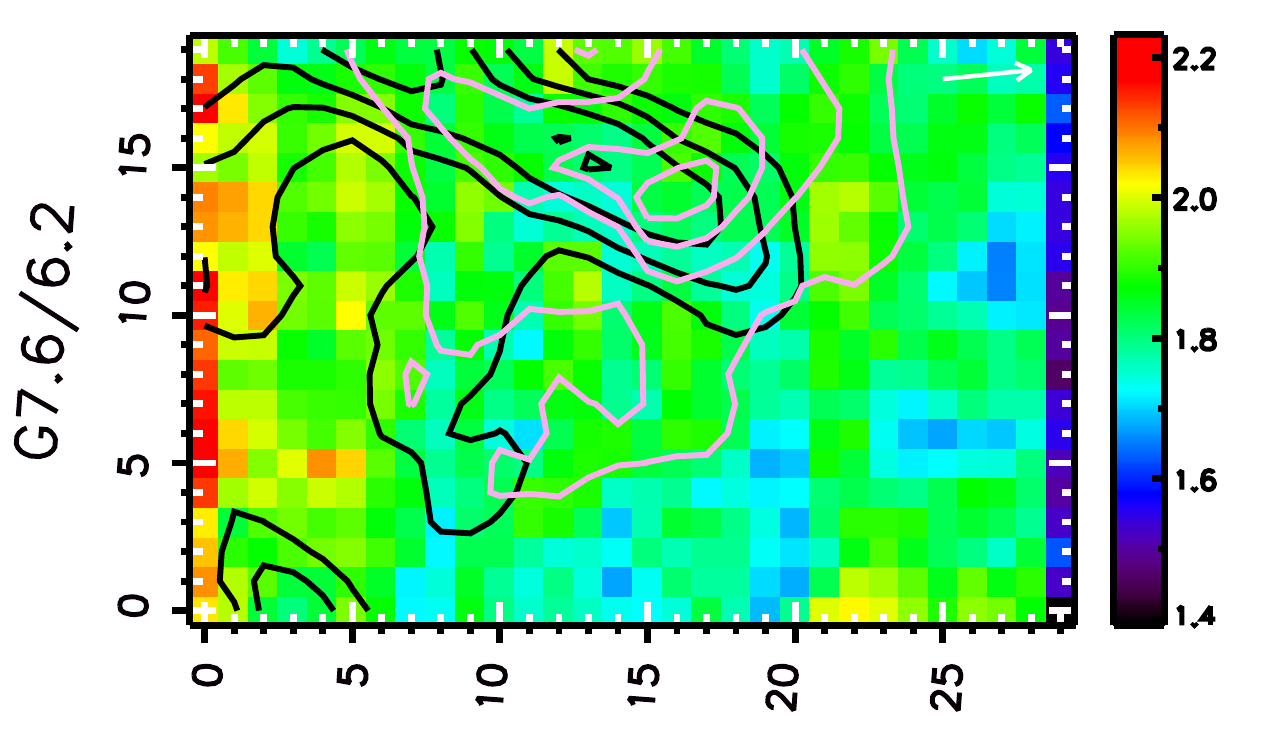}
  \includegraphics[angle=274.1]{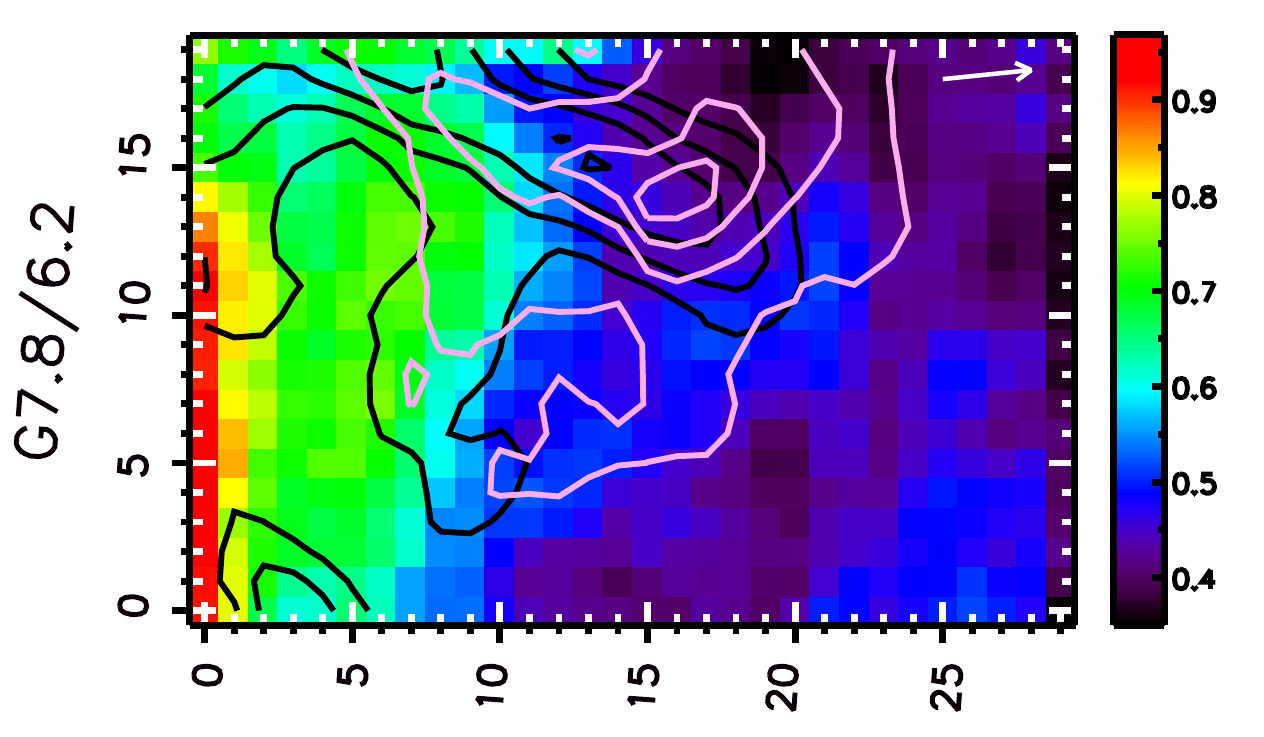}
     \includegraphics[angle=274.1]{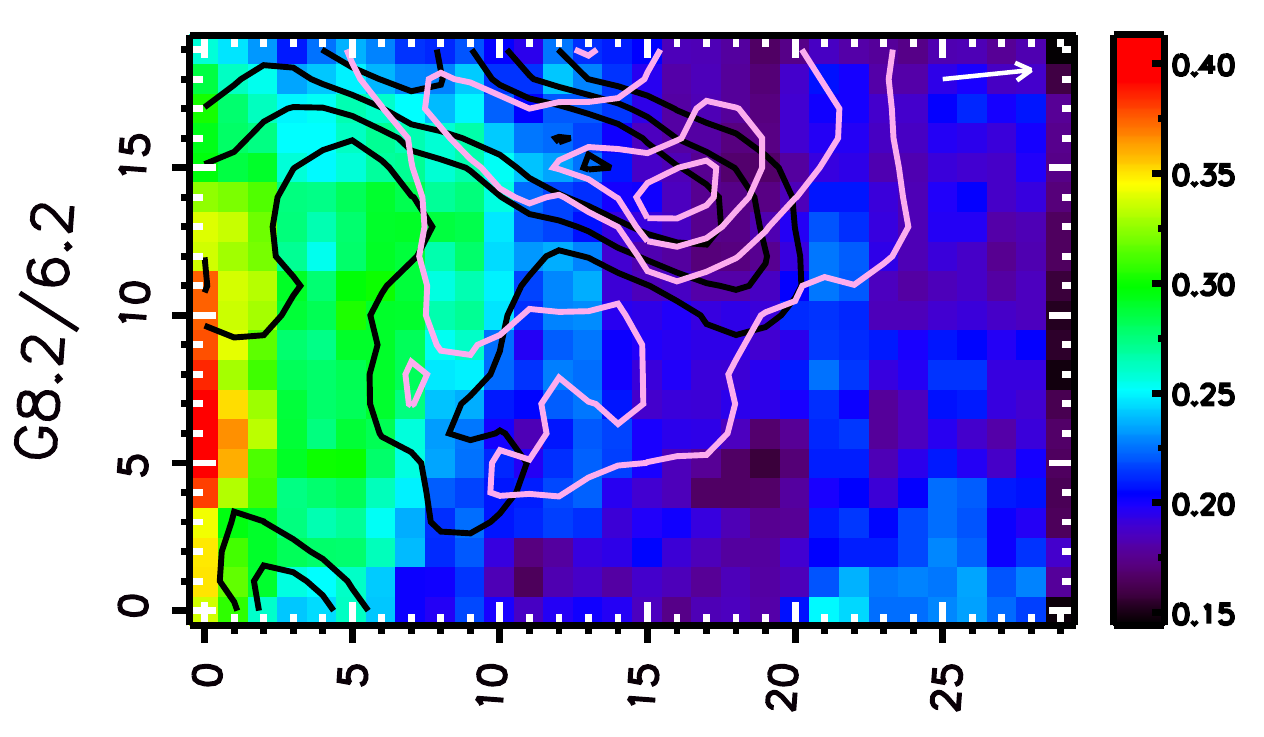}
      \includegraphics[angle=274.1]{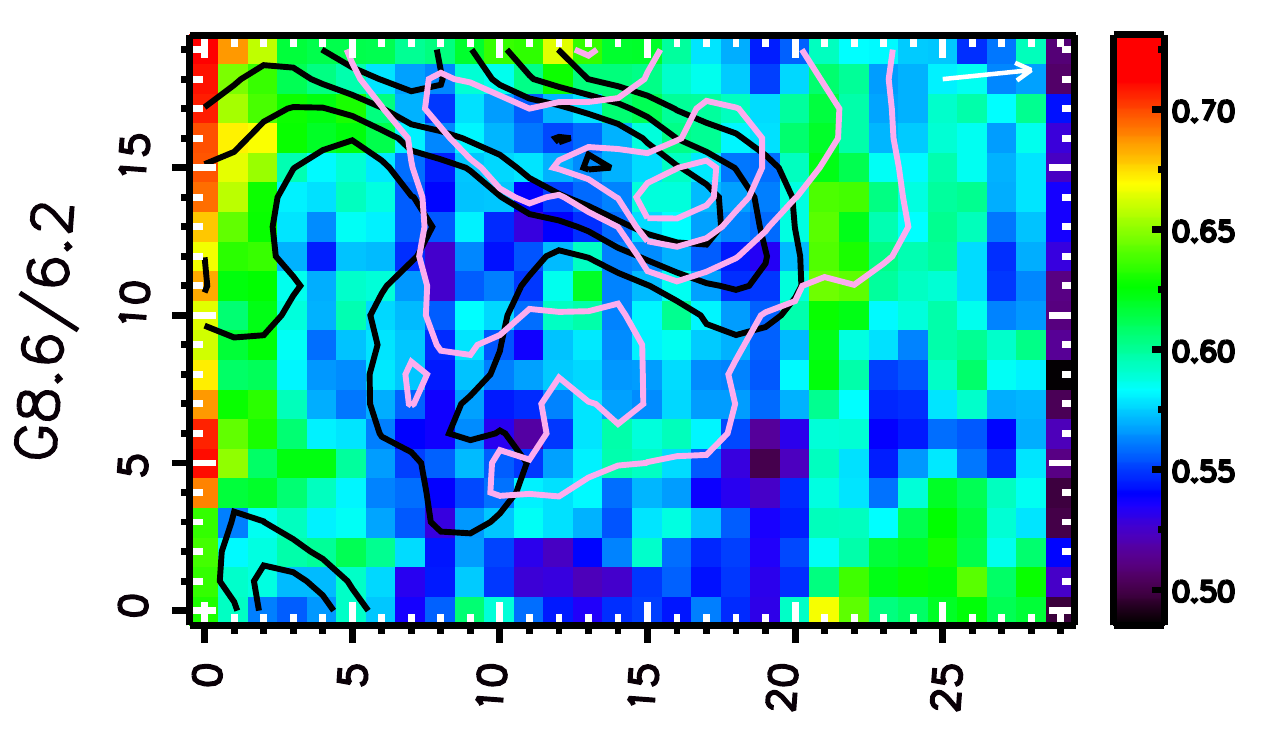}}
\caption{Spatial distribution of selected PAH ratios for the south map (top panels) and north map (bottom panels). Map orientation, contours, units and symbols are the same as in Figs. \ref{fig_slmaps_s} and \ref{fig_slmaps_n}. For the G7.6/G7.8 map, we show the H$_2$ S(3) 9.7 \mum\, instead of the 12.7 \mum\, PAH bands (in pink contours).   }
\label{fig_PAHratios}
\end{figure*}

Line cuts of PAH ratios shown in Figs.~\ref{fig_corr}, \ref{fig_corr_sh}, and \ref{ratios-Gcomponents} are presented in Fig.\ref{linecuts_ratios}.  A selected number of PAH ratio maps are presented in Fig.~\ref{fig_PAHratios}. 

The first set shows the fraction of the 7.7 \mum\, feature due to either the G7.6 or G7.8 component.  As we use the global spline (GS) continuum before applying the four Gaussian decomposition, we also use this continuum for the 7.7 \mum\, feature in these maps. Clearly, the G7.6 component makes up the dominant fraction of the 7.7 \mum\, complex. The spatial distribution of the G7.6/7.7(GS) is closest to, but not similar, that of the 11.0 \mum\, emission.  

%%%%%%%
%%%%%%%

 \citet{Boersma:14} showed that peak emission of the ratio of the 7.6 and 7.8 \mum\, subcomponents is parallel with the PDR front and is located between the PDR front and the star in NGC~7023. We report a similar result for NGC~2023: G7.6/G7.8 clearly peaks in regions where ionized PAHs reign (as traced by the 8.6 and 11.0 \mum\, PAH bands) and shows minimal emission on the PDR front (as traced by H$_2$) and in the molecular cloud. 

Given the results of this paper, the ion-to-neutral PAH fraction is best traced by the 8.6/11.2 or 11.0/11.2 PAH ratio.  Since both ratios have similar morphologies, Fig.~\ref{fig_PAHratios} only shows that of the 8.6/11.2 PAH ratio. This ion-to-neutral PAH fraction is smallest in all ridges in both the N and S maps and thus also in the PDR fronts, and peaks closest to the illuminating star. 

Finally, we show the spatial distributions of the ratios presented in Fig.~\ref{ratios-Gcomponents}: the four Gaussian components normalized on the 6.2 \mum\, PAH band. Given the charge attribution of the G7.6, G8.6 and 6.2 PAH features, the G7.6/6.2 and G8.6/6.2 shows little variation across the FOV. This is less clear in the N map, originating in the fact that the range of the color table is set by the edge columns (in x coordinates). Nevertheless, the weak variation in pixels not located at the edges is present in both maps. Note the clear decrease in emission on the PDR fronts of the S and SSE ridges in the south map, another indication that the 6.2 \mum\, PAH band is not solely due to ions.

%%%%%%%%%%%%%%%%%%%%%%%%%%%%%%%%%%%%%%%%%%%%%%%%%%
%%%%%%%%%%%%%%%%%%%%%%%%%%%%%%%%%%%%%%%%%%%%%%%%%%

\section{Influence of the decomposition on the plateaus' behavior}
\label{influence_decomp_plat}

While the applied decomposition may well be flawed to some extent, the observed spatial differences of the plateaus and the PAH features nevertheless imply that the bulk of the emission in the plateaus behaves independently from that of the PAH features (Section~\ref{vsg}). Indeed, we applied a linear fit to the dust continuum (cf. plateau continuum) which is consistent with the observed dust continuum emission in PDRs \citep[e.g.][]{Berne:07, Compiegne:08}. 
The uncertainty on the dust continuum emission is therefore such that it will only have a relatively small influence on the strength of the plateau emission and hence does not change its spatial distribution significantly. The disentangling of the plateau emission and the features is more challenging. In this regard, the `in-between' morphology of the 5--10 \mum\, plateau may possibly indicate that we have not selected the `true' plateau emission, though there is no way to tell based on this dataset. But we can turn the problem around: if both the features and the underlying plateau arise from the same PAH population, a similar spatial behavior would occur, independently on how the division is made between these individual components. Clearly, within the PAH population several subpopulations exist (the most obvious one is neutral vs charged PAHs) and so spatial differences are present within the PAH features themselves \citep[as discussed in this paper and many others, e.g. ][]{Joblin:ngc1333:96, Sloan:99, Hony:oops:01, Galliano:08}. However, none of the `nominal' PAH bands have their strongest emission in the NW ridge in the north map as do the 5--10 and 10--15 \mum\, plateaus,  indicating that the plateaus have a different carrier. A caveat is in place here for the 5--10 \mum\, plateau. In this paper, we showed that the `nominal' 7--9 \mum\, PAH emission is comprised of at least two sub-populations, one of which does show strong emission in the NW ridge (as the continuum emission; cf. G7.8 and G8.2, Fig.~\ref{fig_maps_decomp}). Even more so, the morphology of the 7--9 \mum\, PAH emission changes with each wavelength and varies between two extremes, one peaking in the NW ridge and one peaking in the N ridge (as the `nominal' PAH emission; Fig.~\ref{fig_movie}). Fig.~\ref{5to10plat} clearly shows though that even in the $\sim$ 7.7--8.3 \mum\, wavelength range, the features vary independently of the plateau emission. This `complication' does not hold for the other 'nominal' PAH bands (e.g. 6.2 and 11.2 \mum\, PAH bands) and thus does not affect the 10--15 \mum\, plateau. We thus conclude and confirm earlier results \citep[][paper I]{Bregman:orion:89, Roche:orion:89} concluding that the plateaus are distinct from the features.

\begin{deluxetable}{l@{\hspace{8pt}}r@{\hspace{6pt}}r@{\hspace{6pt}}r@{\hspace{8pt}}r@{\hspace{6pt}}r@{\hspace{6pt}}r@{\hspace{8pt}}r@{\hspace{6pt}}r@{\hspace{6pt}}rrrrrr}
  \tabletypesize{\small}
   \tablecolumns{12} 
  \tablecaption{\label{t1} C-H band position maximum (in~$\mu$m), total intensity (in km/mol), and intensity per C-H (in km/mol/H).  No redshift is applied.}
  \startdata 
    Molecule          & \multispan3 \hfil Cation \hfil  & \multispan3 \hfil  Neutral \hfil & \multispan3 \hfil  Anion \hfil \\
    \noalign{\vskip 4pt}
                      & \multicolumn{3}{c}{\hrulefill}  & \multicolumn{3}{c}{\hrulefill}   & \multicolumn{3}{c}{\hrulefill} \\[2pt] 
                   & \multispan3 \hfil $\lambda$ \hspace{10pt} $I$\hspace{5pt} $I$(C\rlap{H)} \hfil & \multispan3 \hfil $\lambda$ \hspace{13pt} $I$ \hspace{4pt} $I$(C\rlap{H)} \hfil & \multispan3 \hfil $\lambda$ \hspace{12pt} $I$  \hspace{4pt}$I$(C\rlap{H)} \hspace{6pt}\hfil  \\   [2pt]      

                      & \\ 
                      \hline
                      \hline
                      \\ [5pt]         
C1 C$_{24}$H$_{12}$  & 3.240&    1.8&    0.1&\ \ \   3.263&    8.9&    0.7& \ \ \  3.298&   20.7&    1.7\\[2pt]
O1 C$_{32}$H$_{14}$  &  3.242&    3.3&    0.2&  3.263&   10.8&    0.8&  3.291&   24.3&    1.7\\[2pt]
A1 C$_{40}$H$_{16}$   &  3.245&    4.9&    0.3&  3.264&   13.0&    0.8&  3.288&   27.4&    1.7\\[2pt]
T1 C$_{48}$H$_{18}$   &  3.247&    6.8&    0.4&  3.264&   14.7&    0.8&  3.284&   30.3&    1.7\\[5pt]
C2 C$_{54}$H$_{18}$   &  3.249&    7.6&    0.4&  3.264&   15.5&    0.9&  3.284&   28.7&    1.6\\[2pt]
O2 C$_{66}$H$_{20}$   &  3.250&    9.6&    0.5&  3.264&   18.2&    0.9&  3.280&   32.5&    1.6\\[2pt]
A2 C$_{78}$H$_{22}$  &  3.252&   11.8&    0.5&  3.265&   21.0&    1.0&  3.281&   36.3&    1.7\\[2pt]
T2 C$_{90}$H$_{24}$   &  3.254&   14.6&    0.6&  3.266&   23.5&    1.0&  3.279&   40.8&    1.7\\[5pt]
C3 C$_{96}$H$_{24}$ &  3.256&   16.0&    0.7&  3.268&   24.9&    1.0&  3.279&   39.5&    1.6\\[2pt]
O3 C$_{112}$H$_{26}$   &  3.256&   18.2&    0.7&  3.266&   27.4&    1.1&  3.276&   43.1&    1.7\\[2pt]
A3 C$_{128}$H$_{28}$    &  3.258&   20.6&    0.7&  3.268&   30.6&    1.1&  3.277&   47.1&    1.7\\[2pt]
T3 C$_{144}$H$_{30}$    &  3.259&   24.5&    0.8&  3.268&   34.3&    1.1&  3.277&   53.6&    1.8\\[5pt]
C4 C$_{150}$H$_{30}$  &  3.261&   27.1&    0.9&  3.268&   34.4&    1.1&  3.276&   52.8&    1.8\\[2pt]
O4 C$_{170}$H$_{32}$    &  3.261&   29.0&    0.9&  3.272&   38.6&    1.2&  3.276&   56.8&    1.8\\[2pt]
A4 C$_{190}$H$_{34}$   &  3.262&   32.2&    0.9&  3.269&   43.1&    1.3&  3.277&   62.4&    1.8\\[2pt]
T4 C$_{210}$H$_{36}$    &  3.263&   37.6&    1.0&  3.270&   47.6&    1.3&  3.276&   69.8&    1.9\\[5pt]
& \\[2pt]
 \enddata 
\end{deluxetable}

\begin{figure}
    \centering
\resizebox{10cm}{!}{%
  \includegraphics{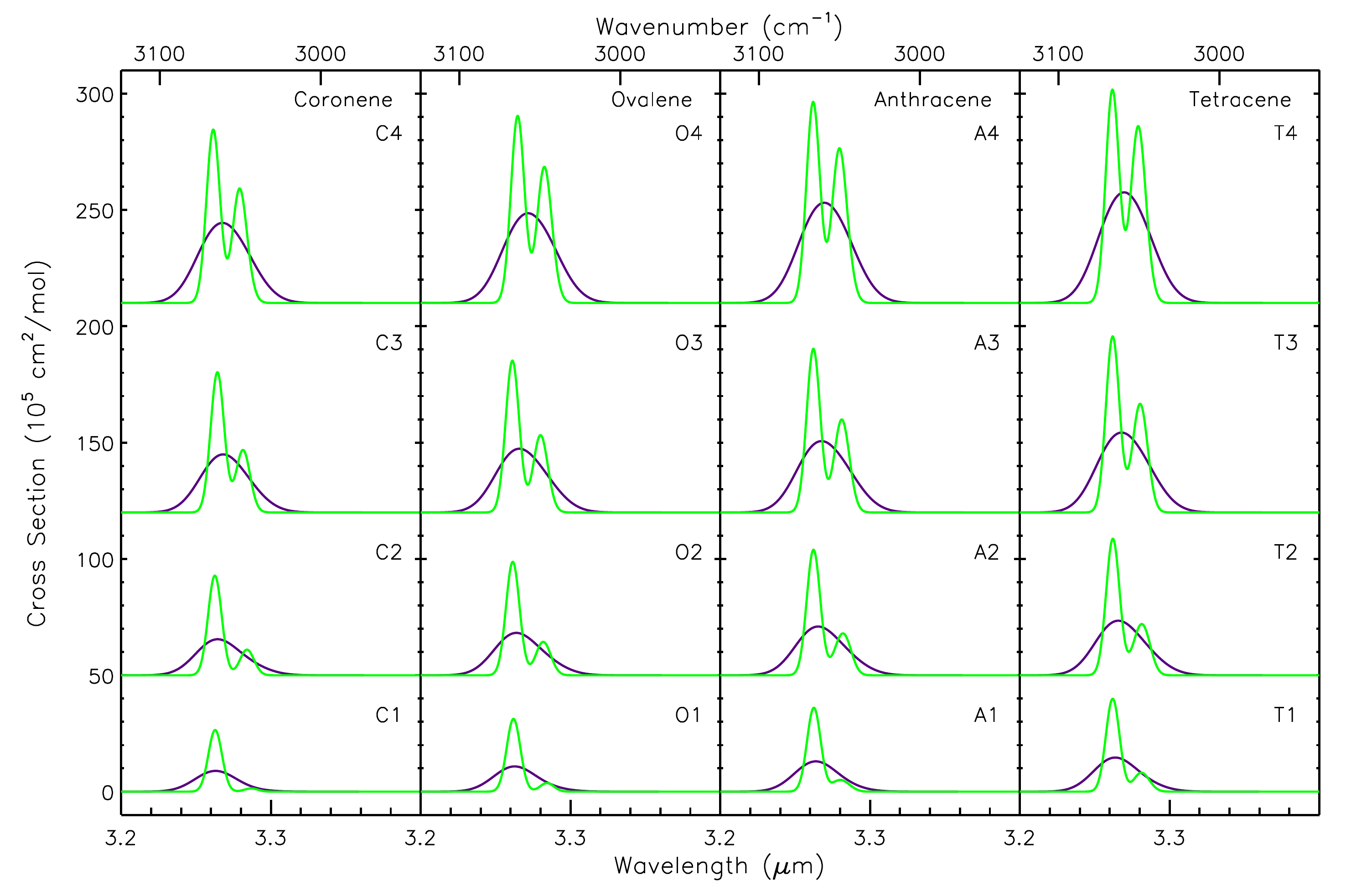}}
\caption{\label{lab3} The C-H stretching region for the neutrals.  The green curves are for 15~cm$^{-1}$ FWHM while the black line is for 30~cm$^{-1}$.  For each family, the molecules are ordered from smallest on the bottom to largest on the top. The duo and solo C-H stretch, at shorter and longer wavelength respectively, are discernible in green curves.} 
\end{figure}

\section{The CH stretching Vibrations (3.2 - 3.3~$\mu$m) of oval, compact PAHs}
\label{lab_results_33}

The band position maxima for the C-H stretch of the molecules shown in Figure~\ref{molecules} is summarized in Table~\ref{t1}.  We first note that the trends for all four families are similar and follow those noted previously \citep{Bauschlicher:vlpahs1, Bauschlicher:vlpahs2, Ricca:12}.  In summary, there is a slow increase in band position with increasing size for the neutral and cation and a slow decrease for the anions. For all sizes, the band position of the cations is always at the shortest wavelengths, followed by that of the neutrals and subsequently that of the anions. As with other large PAHs, the intensity of the cations is slightly weaker than that of the neutrals while the anion intensities are about twice as large. The intensity per H increases with size for all charge states, with the increase being the smallest for the anion and largest for the cation.   As a result, the cation intensity becomes rather sizeable for the largest species.

The C-H stretching bands, plotted using a FWHM of 15 and 30 cm$^{-1}$, are shown in Figure~\ref{lab3}.  All of the species studied have 12 duo hydrogens, with the number of solos varying with size and with family class (see Table~\ref{CHgroups}). Since the duo C-H stretches and the solo C-H stretches fall at slightly different wavelengths, the constancy of the C-H stretching frequency, when plotted with a 30 cm$^{-1}$ bandwidth, might appear to be a bit odd. To investigate this in more detail, we also plot the C-H stretching region using a FWHM of 15~cm$^{-1}$.  The important factor to note is that while the number of duo hydrogens is constant, the intensity of the shorter wavelength duo C-H stretch increases as the intensity of the longer wavelength solo C-H component increases.  That is, there is a
coupling of the solo and duo C-H stretching modes so that the intensities of both increase with increasing molecular size. This results in a much smaller change in the 30~cm$^{-1}$ FWHM band maximum than would have been expected if there was no coupling of the solo and duo C-H stretches.

%%%%%%%%%%%%%%%%%%%%%%%%%%%%%%%%%%%%%%%%%%%%%%%%%%
\begin{figure}[tb!]
    \centering
\resizebox{11cm}{!}{%
  \includegraphics{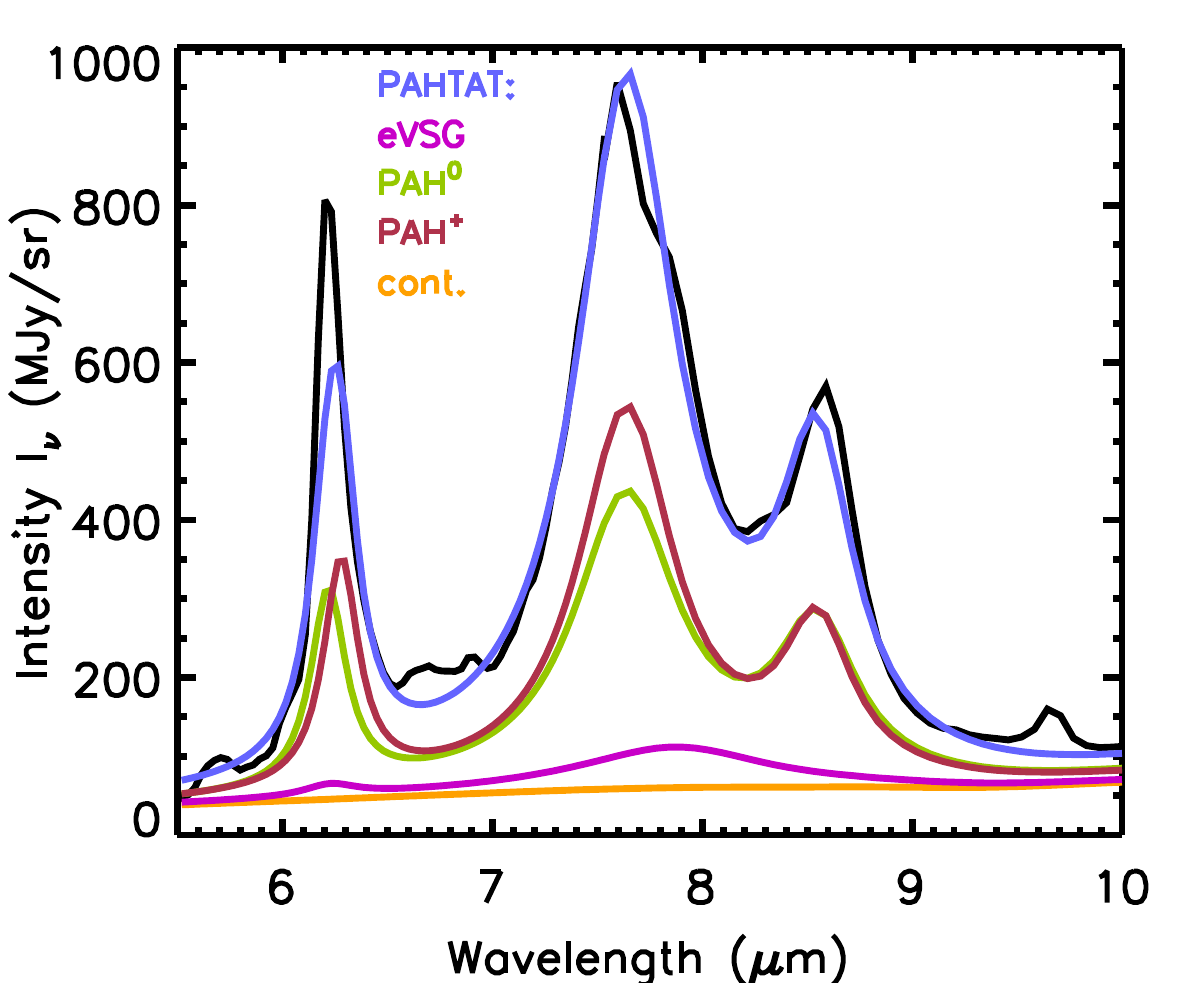}
  \includegraphics{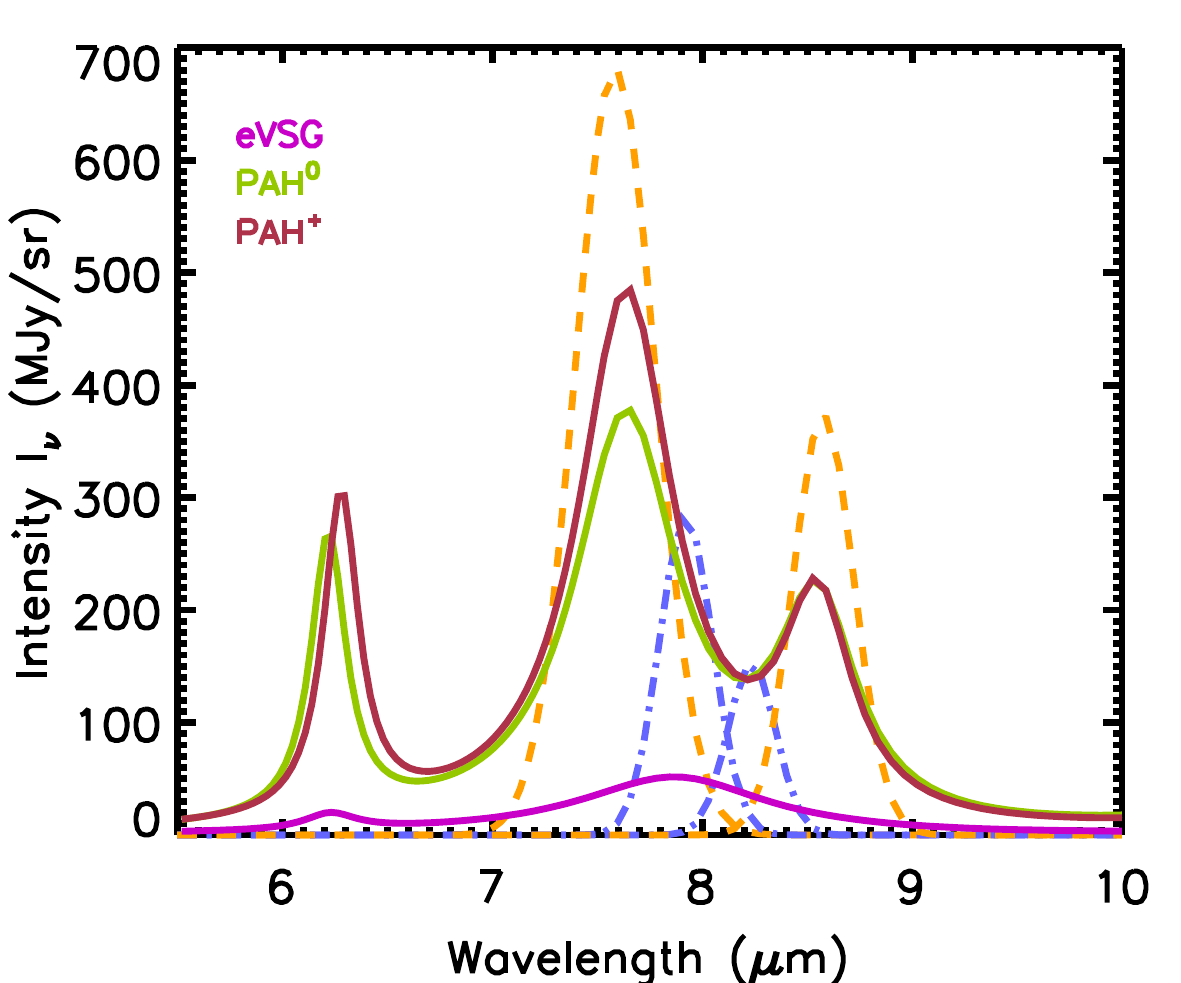}}
\caption{\label{fig_PAHTAT_decomp} A PAHTAT fit to a typical spectrum (left) and the comparison of its components to the four Gaussian components of the same spectrum (right). The orange dashed lines represent the G7.6 and G8.6 components and the blue dot-dashed lines the G7.8 and G8.2 components.} 
\end{figure}
%%%%%%%%%%%%%%%%%%%%%%%%%%%%%%%%%%%%%%%%%%%%%%%%%%

%%%%%%%%%%%%%%%%%%%%%%%%%%%%%%%%%%%%%%%%%%%%%%%%%%
\begin{figure*}[tb]
    \centering
\resizebox{13.6cm}{!}{%
  \includegraphics[angle=266.4]{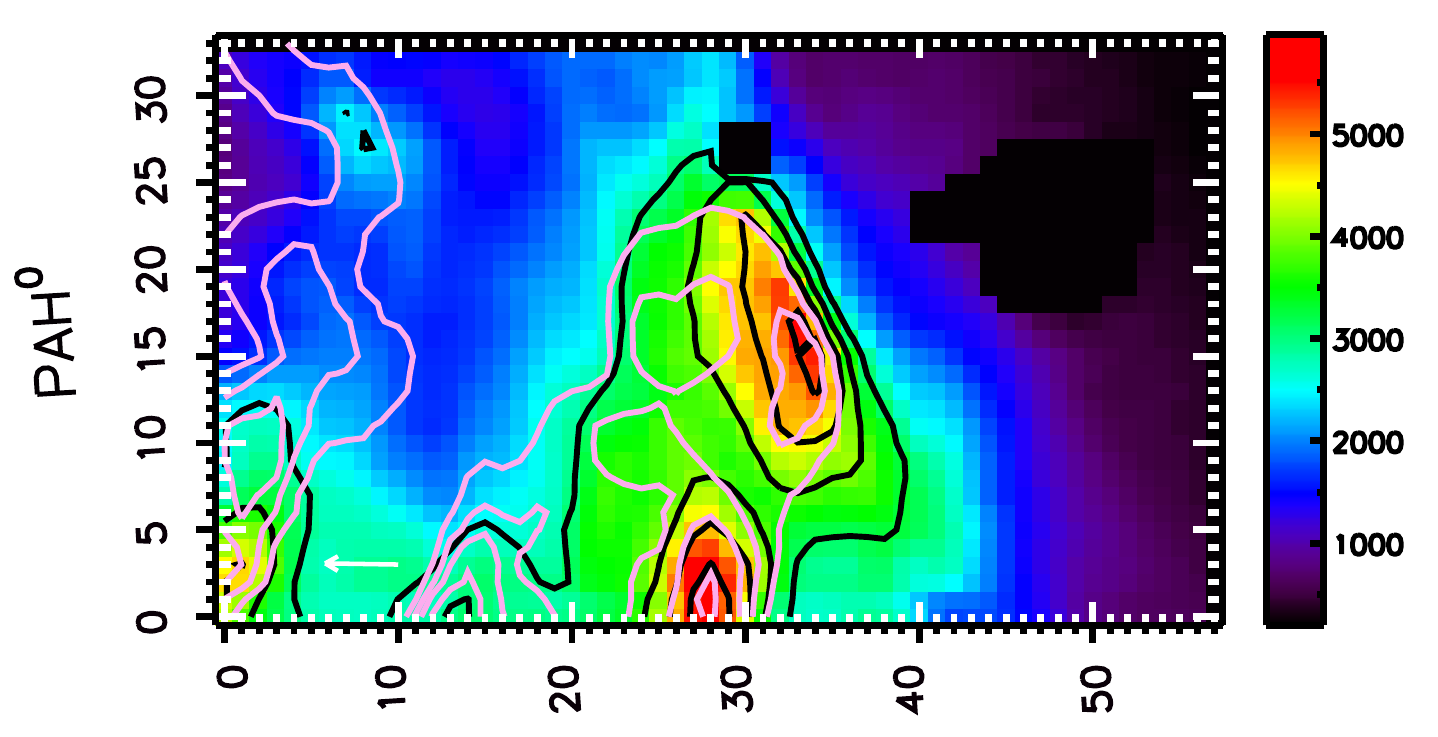}
  \includegraphics[angle=266.4]{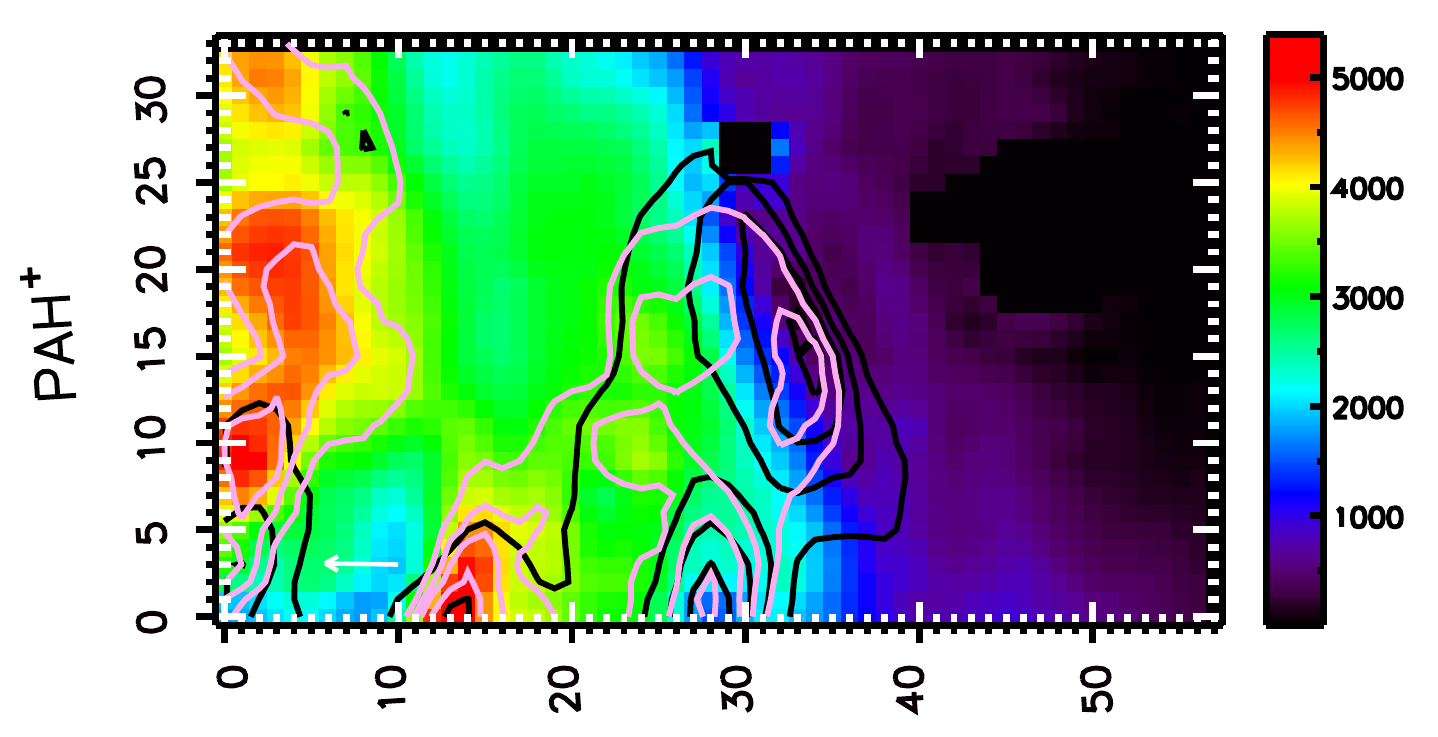}
    \includegraphics[angle=266.4]{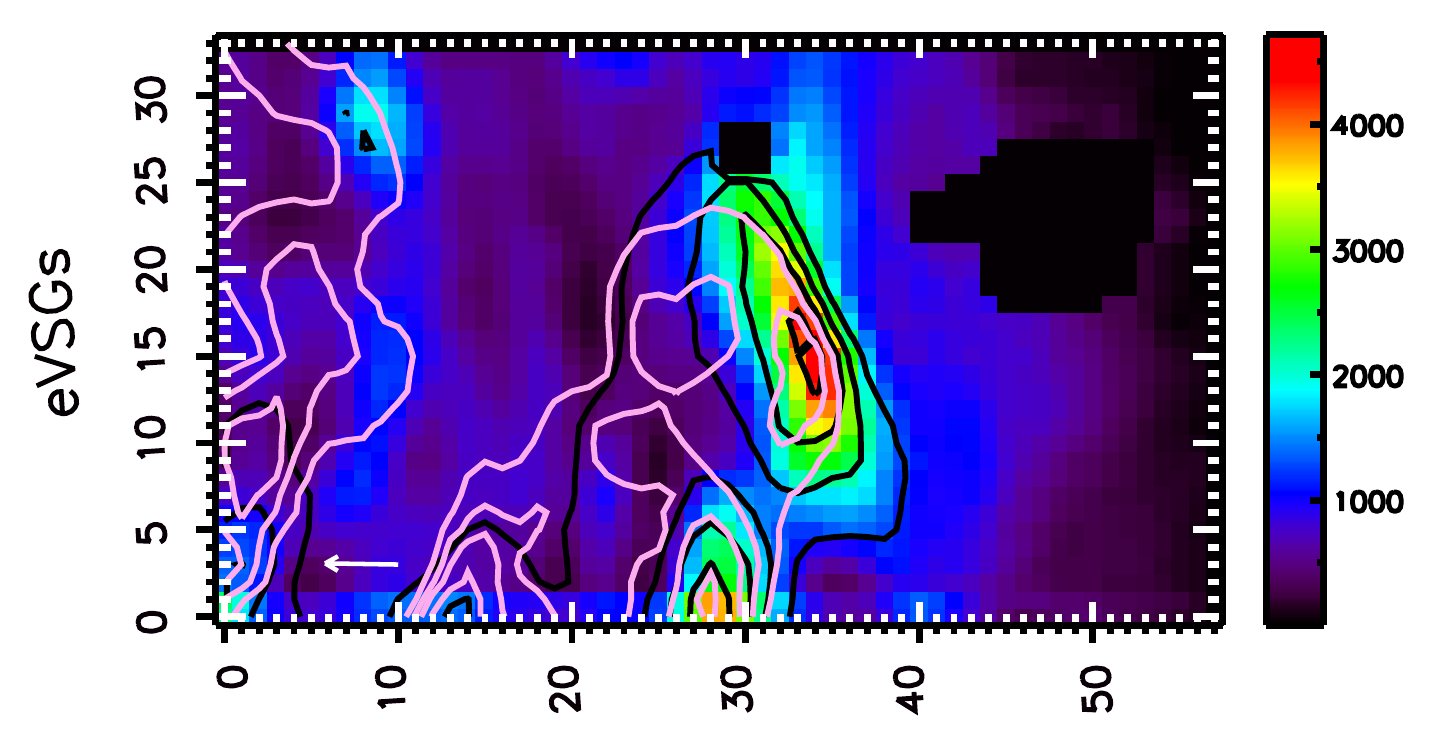}
    \includegraphics[angle=266.4]{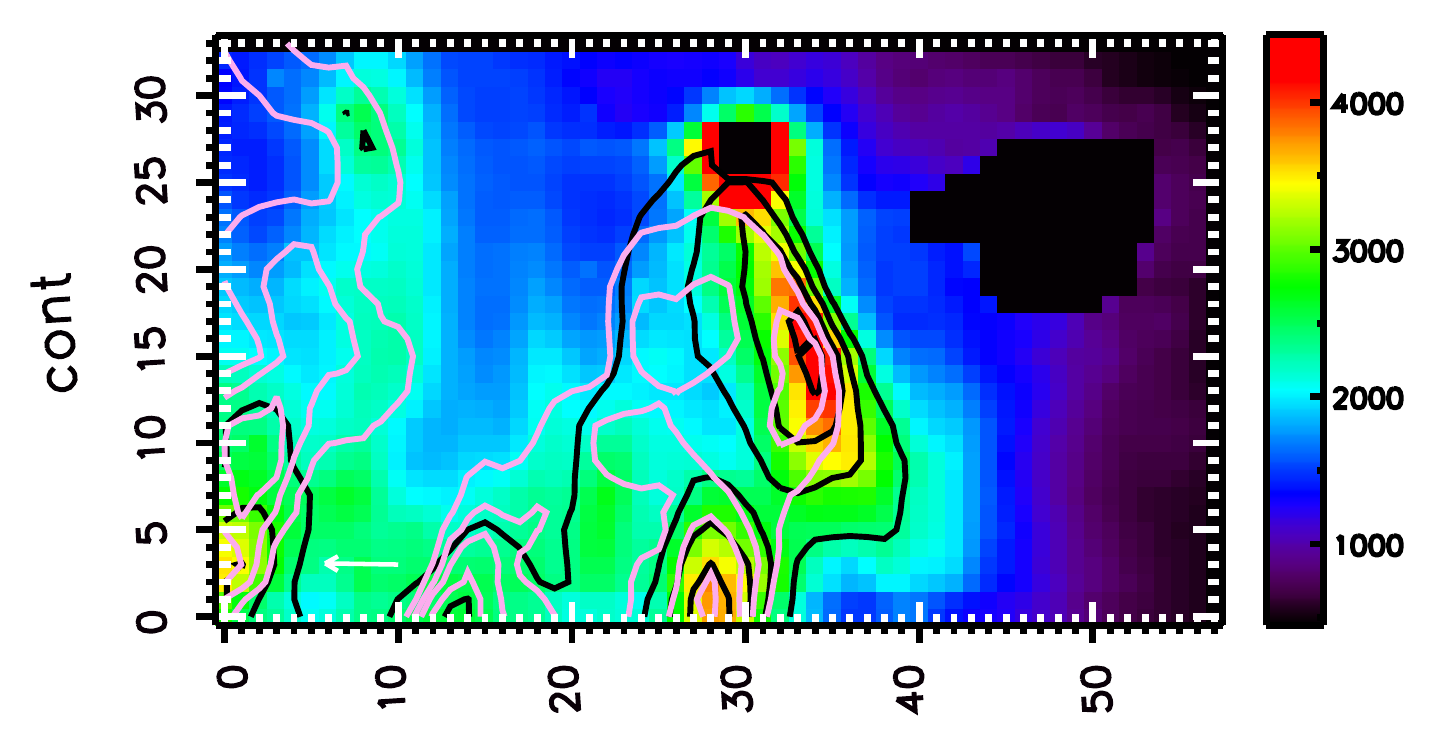}}
\resizebox{14.4cm}{!}{%
  \includegraphics[angle=274.1]{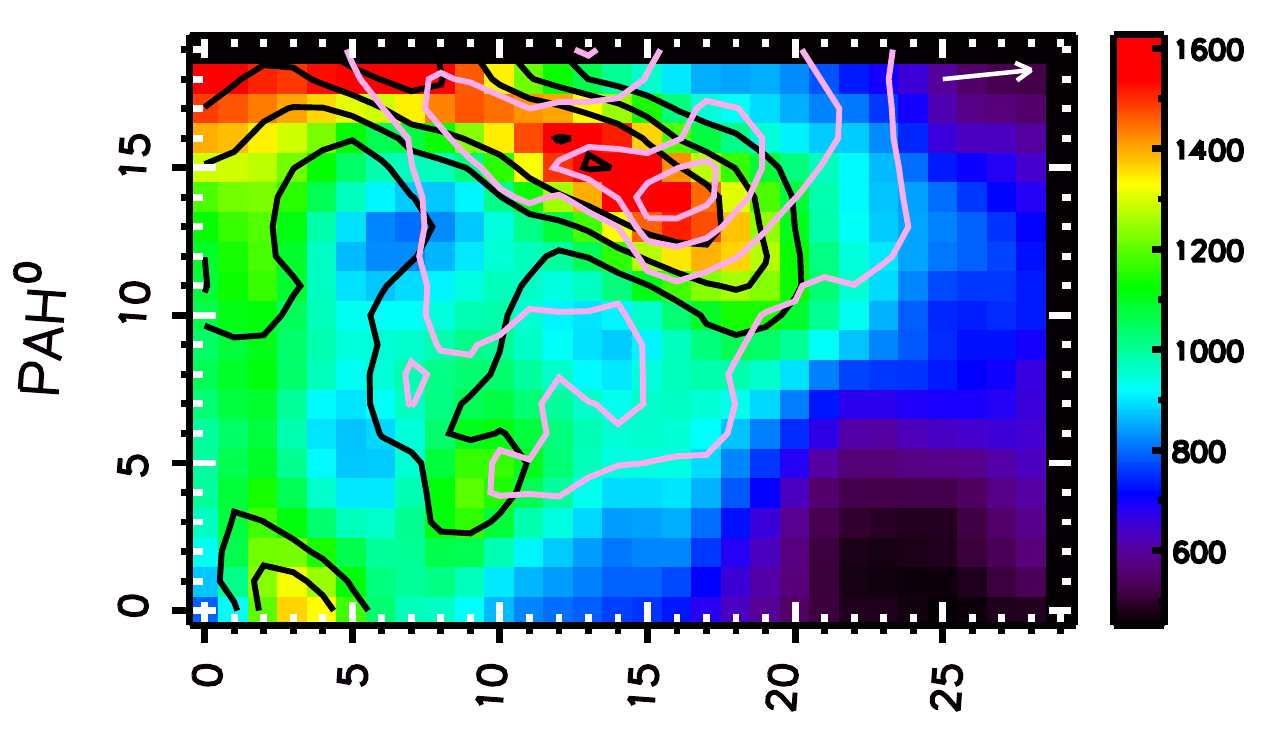}
  \includegraphics[angle=274.1]{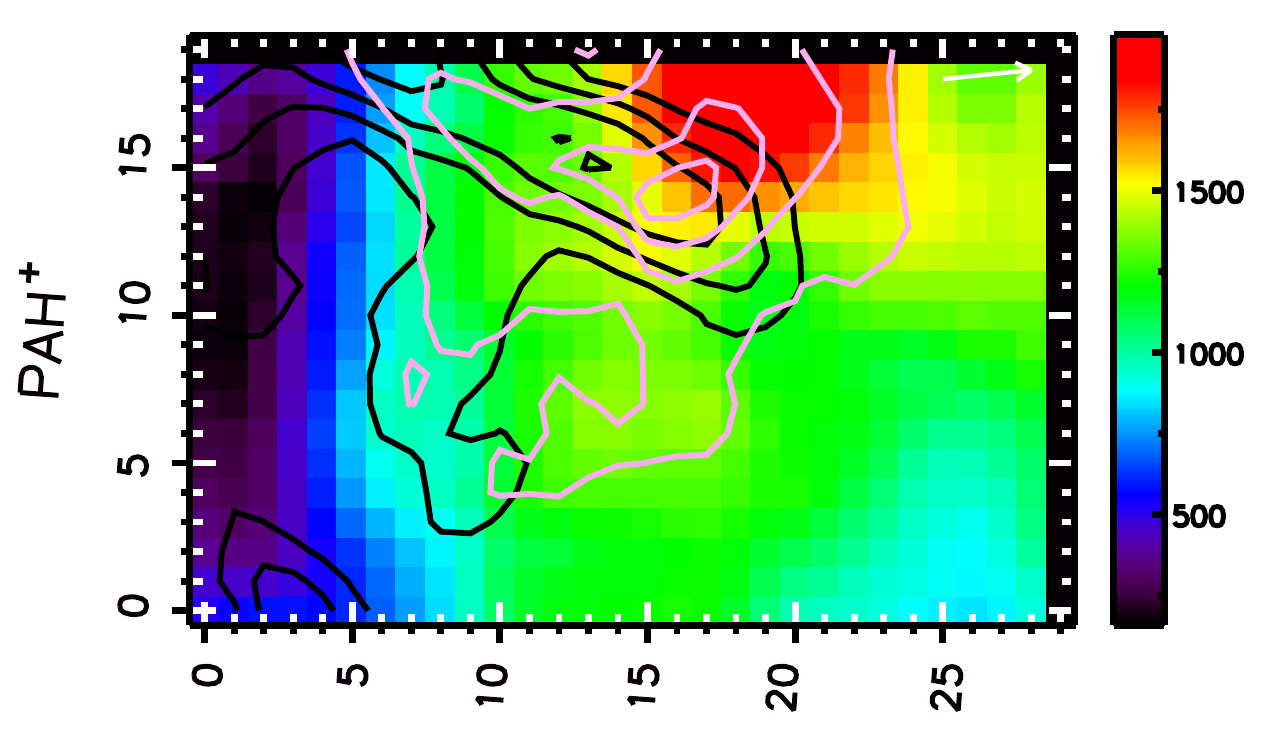}
     \includegraphics[angle=274.1]{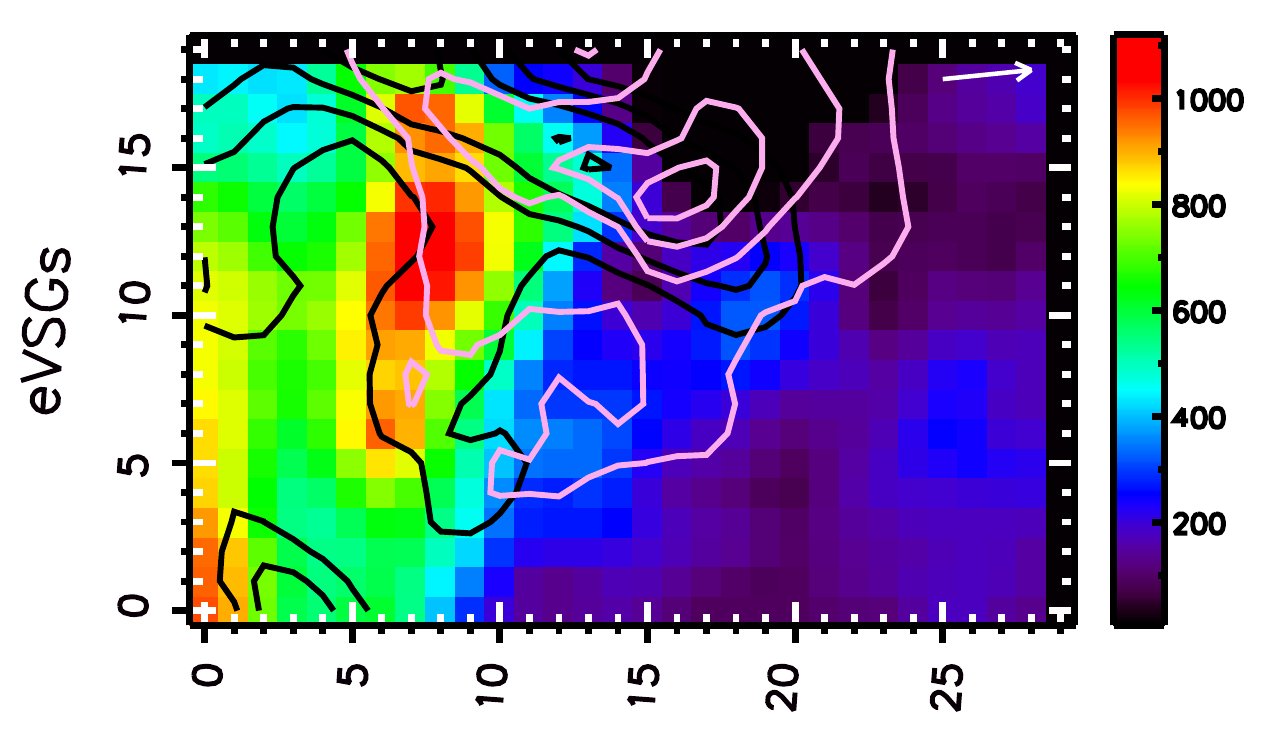}
      \includegraphics[angle=274.1]{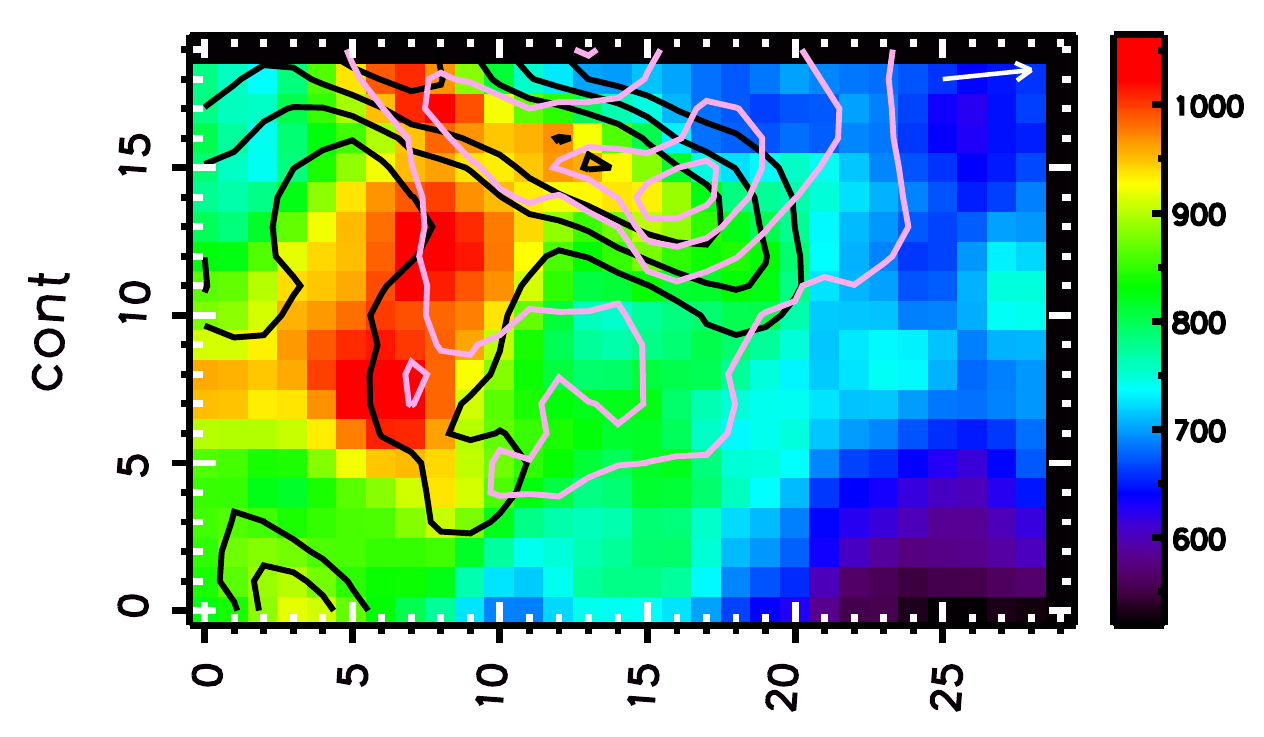}}
\caption{\label{fig_PAHTAT_maps} Spatial distribution of the integrated intensity from the PAHTAT components: PAH$^0$, PAH$^+$, 
eVSGs and the extinction corrected continuum for the south map (top panels) and north map (bottom panels). Map orientation, contours, units and symbols are the same as in Figs. \ref{fig_slmaps_s} and \ref{fig_slmaps_n}. The range in intensities of the color bar for the south continuum map is determined by excluding the immediate region of source C.}  
\end{figure*}
%%%%%%%%%%%%%%%%%%%%%%%%%%%%%%%%%%%%%%%%%%%%%%%%%%

%%%%%%%%%%%%%%%%%%%%%%%%%%%%%%%%%%%%%%%%%%%%%%%%%%
\begin{figure*}[tb]
    \centering
\resizebox{13.6cm}{!}{%
  \includegraphics{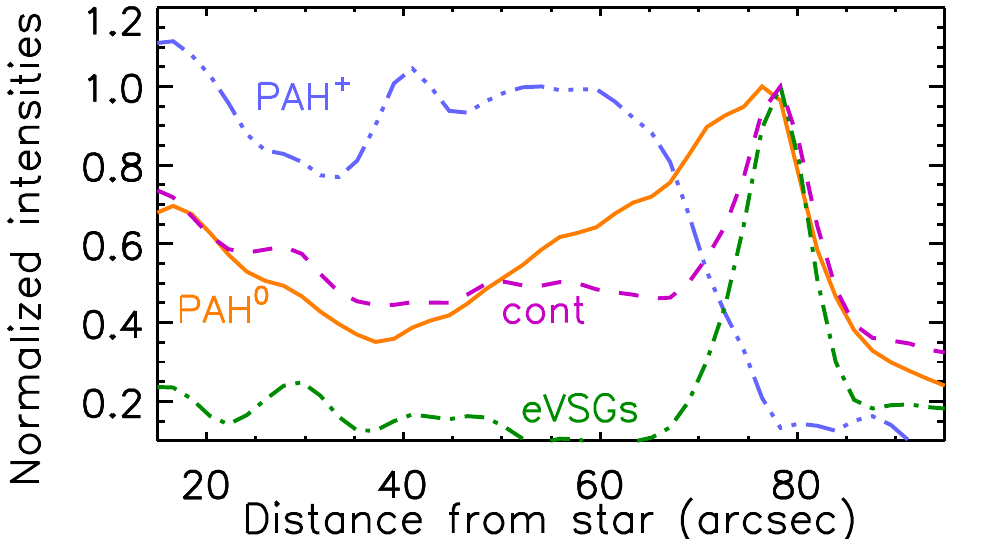}
   \includegraphics{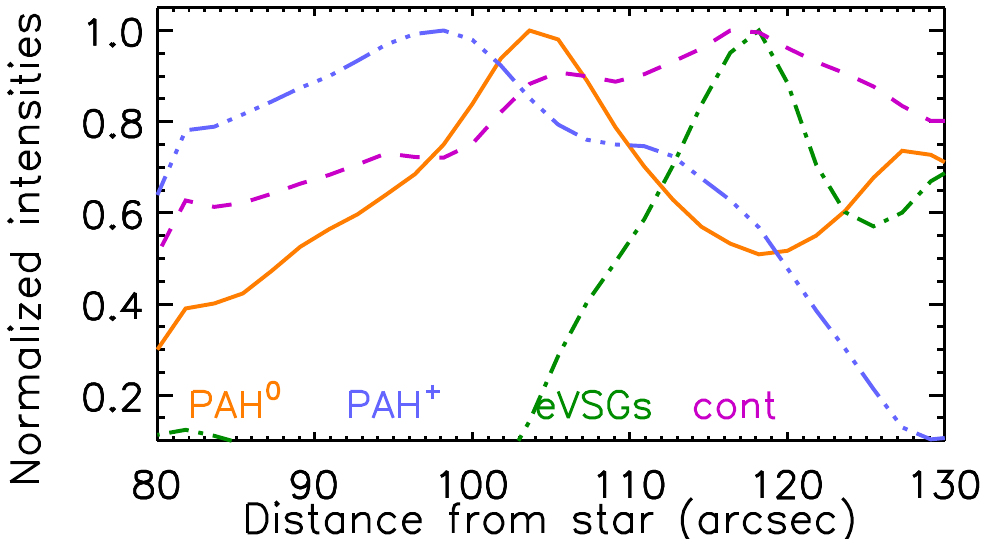}}
\caption{\label{linecuts_PAHTAT} Normalized feature/continuum intensity of the PAHTAT components along the same projected cut across the south (left) and north (right) FOV directed toward HD37093 as in Fig.~\ref{linecuts}. }
\end{figure*}
%%%%%%%%%%%%%%%%%%%%%%%%%%%%%%%%%%%%%%%%%%%%%%%%%%

\section{PAHTAT results}
\label{PAHTAT}

The BSS results \citep{Boissel:01, Rapacioli:05, Berne:07} have been employed to develop PAHTAT, a fitting program that allows the extraction of the distinct components found by the BSS analysis when no spectral map is available \citep{Pilleri:12}. Specifically, PAHTAT fits individual spectra assuming a combination of four templates (PAH$^0$, PAH$^+$, PAH$^x$ and eVSGs), three minor PAH bands, gas lines and a featureless dust continuum, while taking into account dust extinction.  As the PAH$^x$ template was only invoked to better fit highly excited regions \citep{Joblin:08}, we excluded this component from our fit. We obtain very good fits by using the PAHTAT tool (Fig.~\ref{fig_PAHTAT_decomp}). The largest discrepancy is found for the 6.2 \mum\, band and is not resolved when including the PAH$^x$ template. 

Fig.~\ref{fig_PAHTAT_decomp} shows a comparison of the PAHTAT and the four Gaussian components for a typical spectrum. Spectroscopically, the eVSG component is quite distinct from both the plateau emission and any of the four Gaussian components as defined in this paper (Figure~\ref{fig_PAHTAT_decomp}). In contrast, the G7.6 and the G8.6 component is more similar to respectively the 7.7 and 8.6 component in the PAH$^0$ and PAH$^+$ templates. 

By definition, the spatial morphology of the PAHTAT components (PAH$^0$, PAH$^+$, eVSGs) is unique (Figures \ref{fig_PAHTAT_maps} and \ref{linecuts_PAHTAT}) but they, of course, do resemble morphologies and line cuts presented in this paper. As expected, the morphology (and thus line cut) of the PAH$^0$ component mimics the 11.2 \mum\, PAH morphology, that of the PAH$^+$ resembles the spatial distribution of the 8.6 and 11.0 \mum\, PAH emission and that of the continuum component mimics the spatial distribution of the continuum emission. Interesting is the eVSG component. In the south map, its morphology is similar to that of the H$_2$ emission, very sharply peaked on the S ridge unlike the plateau emission. In the north map, it is most similar to that of the 8.1 \mum\, extreme. Compared to the plateau emission, it is more  narrowly peaked in the N ridge and has considerably less emission in the NW ridge.\\

\end{document}